\documentclass[11pt]{article}
\usepackage[english]{babel}
\usepackage{dsfont}
\usepackage{subcaption}
\usepackage{graphicx}
\usepackage{setspace}
\usepackage[margin=1in]{geometry}
\usepackage{amsmath}
\usepackage{amssymb}
\usepackage{amsthm}
\usepackage{natbib}
\usepackage{enumerate}
\usepackage{rotating}
\usepackage[title]{appendix}
\usepackage{makecell}
\usepackage{verbatim}

\usepackage{pgfplots}
\pgfplotsset{compat=1.18}
\definecolor{posblue}{HTML}{2E86AB}
\definecolor{negred}{HTML}{E63946}

\usepackage{etoolbox} %
\usepackage{color}
\usepackage[colorlinks,citecolor=blue]{hyperref}
\hypersetup{pdfauthor={Huang, Lamadon, Mogstad, Shaikh, and Wilhelm}}

\usepackage[nameinlink,noabbrev]{cleveref}
\usepackage{caption, floatrow}
\usepackage{hhline}
\usepackage{tabularx}
\usepackage[bottom]{footmisc}
\usepackage[final]{pdfpages}
\usepackage{booktabs}
\usepackage{float}
\usepackage{siunitx}
\usepackage{setspace}
\usepackage{multirow}
\usepackage{bm}
\usepackage{esvect}
\usepackage{tikz}
\usetikzlibrary{arrows.meta, positioning}

\newcommand{\LambdaWOneApprox}{0.70}
\newcommand{\LambdaWTwoApprox}{0.21}
\newcommand{\LambdaWThreeApprox}{0.09}
\newcommand{\EtaCOneApprox}{0.03}
\newcommand{\LambdaWTwoFoldedApprox}{0.18}
\newcommand{\LambdaWFourApprox}{0.03}

\makeatletter
\newcommand{\fn@body}[2]{%
  \begingroup
    \footnotesize
    \setlength{\parindent}{0pt}%
    \setlength{\parskip}{0pt}%
    \leftskip=0pt\relax
    \rightskip=0pt\relax
    \parfillskip=0pt plus 1fil\relax
    \noindent\textit{#1}\ #2\par
  \endgroup
}
\@ifundefined{floatfoot}
  {\newcommand{\figurenote}[2][Note:]{\par\vspace{0.6ex}\fn@body{#1}{#2}}}
  {\newcommand{\figurenote}[2][Note:]{\floatfoot{\fn@body{#1}{#2}}}}
\makeatother

\newcolumntype{Y}{>{\centering\arraybackslash}X}
\DeclareFloatVCode{myrowsep}{\vskip 4ex}

\usepackage{pgfplots}
\usepgfplotslibrary{groupplots}
\pgfplotsset{compat=1.18}

\def\qed{\rule{2mm}{2mm}}
\def\indep{\perp \!\!\!
\perp}

\usepackage[pagewise,mathlines]{lineno}
\mathchardef\dash="2D

\newtheorem{lemma}{Lemma}[section]
\newtheorem{definition}{Definition}[section]
\newtheorem{corollary}{Corollary}[section]
\newtheorem{proposition}{Proposition}[section]
\newtheorem{theorem}{Theorem}[section]
\theoremstyle{definition}
\newtheorem{example}{Example}[section]

\newtheorem{assumption}{Assumption}[section]

\newcommand{\ind}{\mathds{1}}

\renewcommand{\tilde }{\widetilde}

\makeatletter
\newcommand*\bigcdot{\,\mathpalette\bigcdot@{.5}}
\newcommand*\bigcdot@[2]{\mathbin{\vcenter{\hbox{\smathcal ebox{#2}{$\m@th#1\bullet$}}}}}
\makeatother

\AtEndEnvironment{remark}{~\qed}
\AtEndEnvironment{example}{~\qed}
\newenvironment{examplecontinued}[1]{%
  \par\addvspace{\baselineskip}%
  \noindent\textbf{Example~\ref{#1} (continued).}\quad\ignorespaces
}{%
  \ifhmode\unskip\nobreakspace\else\nobreak\noindent\fi\qed\par
}

\DeclareMathOperator{\var}{Var}
\DeclareMathOperator{\cov}{Cov}

\DeclareMathOperator{\tr}{Tr}
\DeclareMathOperator{\ATT}{ATT}
\DeclareMathOperator{\ATE}{ATE}
\DeclareMathOperator{\DID}{DID}

\renewcommand{\baselinestretch}{1.3}

\allowdisplaybreaks

\begin{document}
\begin{titlepage}
    \renewcommand{\baselinestretch}{1.1}

\author{
Xun Huang \\
University of Chicago \\
\and
Thibaut Lamadon\\
University of Chicago \\
\and
Magne Mogstad \\
University of Chicago \\
Statistics Norway \& NBER \\
\and
Azeem M.\ Shaikh\\
University of Chicago \\
\and
Daniel Wilhelm\\
LMU Munich \\
}

\bigskip

\title{\vspace{-.75in}Do Firm Effects Add Up?\thanks{Acknowledgments: Mogstad and Shaikh acknowledge financial support from NSF Grant SES-2419008; Wilhelm acknowledges financial support from the European Research Council (Consolidator Grant No. 101171720) and from an LMUexcellent grant.}}
\maketitle
\vspace{-0.3in}

\begin{spacing}{1}
\begin{abstract}
\noindent Empirical studies often use panel data to regress the wages paid to workers on fixed effects for the identities of the workers and the firms in which they work. Despite the ubiquity of these regressions, the appropriate interpretation of the resulting firm effect estimates remains elusive. We provide necessary and sufficient conditions for firm effect estimates to accurately describe the wage changes that accompany the mobility of workers between firms. We also show the additional assumptions needed for firm effects estimates to be interpreted as the causal wage effects of a randomly drawn worker moving between firms. These additional assumptions do not have testable implications, in contrast to the conditions needed for the estimates to be interpreted descriptively. The key to deriving these results is to show how firm effect estimates add up: the firm effect contrast for any pair of origin and destination firms first sums the wage changes of moving between firms within paths starting at the origin firms and ending in the destination firm, and then takes a positively weighted average of these sums across paths.
The paths will generally involve firms other than the origin and destination firms, and these intermediate firms may differ systematically and so can the workers moving between them. As a result, the conditions needed for the firm effects to be interpreted descriptively or causally are likely to fail. We therefore propose alternative estimators that can be given the same descriptive or causal interpretation under weaker conditions. We use Italian employer-employee data to show the empirical relevance of our theoretical results. This analysis examines several prominent conclusions about firm effects and their dispersion.
\end{abstract}
\end{spacing}



\end{titlepage}

\thispagestyle{empty}
\newpage

\section{Introduction}

Following \cite{Abowd:1999ii}, a large and influential literature in economics uses matched employer-employee data to regress wages paid to workers on fixed effects for the identities of the workers and the firms in which they work.\footnote{The work has produced several influential yet controversial conclusions, summarized in the review articles by \citet{Card:2018aa} and \cite{Kline:2024io}. These regressions have also been widely used in contexts other than firms and workers, including teachers and students \citep[e.g.,][]{Rockoff:2004aa}, hospitals and patients \citep[e.g.,][]{Finkelstein:2016aa}, and banks and firms \citep[e.g.,][]{Amiti:2018aa}.} However, despite the importance and widespread use of such regressions, the appropriate interpretation of the fixed effects corresponding to the firms in this regression remains unclear. We contribute to closing this gap in understanding by establishing the exact conditions under which these fixed effects can be interpreted descriptively or causally, by examining whether these conditions are testable, and by proposing alternative estimators that rely on weaker assumptions.

 To this end, we first derive a decomposition that reveals how firm effect estimates add up and the way in which they reflect the network structure of the set of firms that are connected through the mobility of workers.  We then use the understanding provided by this decomposition to answer three specific questions.  First, under what conditions do contrasts in firm effects accurately describe the wage changes accompanying moves of workers between firms, and are these conditions testable?  Second, under what additional conditions can they be interpreted as the causal wage effects of a randomly drawn worker moving between firms, and are these additional conditions testable?  Third, are there alternative estimators that can be given the same descriptive or causal interpretation under weaker conditions?

The decomposition that underlies our analysis shows that firm effect contrasts reflect wage changes accompanying moves between firms along chains of mobility that connect the origin firm to the destination firm.  These chains of mobility are formally described by signed paths, which consist of a sequence of firms together with the orientation of the mobility connection between each pair of firms in the sequence. 
We show that, after correctly adjusting for the wage growth of workers who choose not to move, firm effect contrasts can be written as a convex combination of the sums of difference-in-differences along suitable signed paths.  Our decomposition contrasts sharply with related non-convex decompositions in the literature; see, for example, \cite{Kline:2024io} and \cite{Goldsmith:2024io}.  It also corrects a claim in \cite{Kline:2024io} that convex path decompositions do not exist in general.  An important insight from our decomposition is that the chains of mobility will generally involve firms other than the origin and destination firms.

Using our decomposition, we provide the answer to our first question.  We derive the necessary and sufficient conditions for firm effect contrasts to accurately describe the wage changes accompanying moves of workers between firms.  In particular, we find that the firm effects contrasts can be given such a descriptive interpretation if and only if the same description can be given to the sum of difference-in-differences along any signed path between the pairs of firms. Put differently, the contrasts between firms $o$ and $d$ can be interpreted as describing the mean change in wages that accompanies the mobility of workers between these firms for all supported moves if and only if, for every such move, every signed path starting at $o$ and ending at $d$ can be given the same descriptive interpretation. This condition needs to hold no matter what firms other than $o$ or $d$ are involved in these paths. However, intermediate firms can differ systematically from firms $o$ and $d$ (e.g., by industry), and so can workers moving between them (e.g., by industry-specific experience).

Next, we examine commonly used heuristics based on event studies or binscatter plots for assessing whether the conditions for a descriptive interpretation hold.  We show that these diagnostics may lead one to conclude falsely that the conditions hold or conceal arbitrarily large violations of these conditions.  On the other hand, the conditions can be tested using methods developed in \cite{Anatolyev:2023aa}. The test allows us to examine whether firm effect contrasts can be interpreted as describing the wage changes accompanying moves of workers between firms.

With these results in mind, we revisit the Veneto Workers History data for 1999 and 2001 employed by \cite{Kline:2024io}.  We find that very few of the firm effect contrasts place any weight on the difference-in-differences corresponding to the origin and destination firm. In fact, even signed paths involving only comparable firms, i.e., firms in the same industry as the origin or destination firm, receive only 9.7\% weight on average.  The remaining 90.3\% of the weight is placed on signed paths that involve one or more non-comparable firms.  When restricted to only pairs of firms corresponding to supported moves, we find broadly similar patterns.
We further document that the corresponding signed path sums involving non-comparable firms differ meaningfully from the signed path sums involving only comparable firms.   
Applying the test developed in \cite{Anatolyev:2023aa}, we find these differences across signed path sums are statistically significant. We thereby conclude that the firm effect contrasts should not be interpreted descriptively as describing the wage changes accompanying moves of workers between firms.

Building upon this analysis, we provide the answer to our second question.  We then invoke an assumption of common trends in counterfactual wage growth among stayers and movers between any pair of firms. Under this assumption, we derive the additional conditions under which the firm effect contrasts can be interpreted as average causal wage effects of a randomly drawn worker moving between firms.  When restricted to only pairs of firms corresponding to supported moves, these additional conditions consist of mean independence of mobility and causal wage effects corresponding to supported moves.  When expanded to all pairs of firms, not just those corresponding to supported moves, we further require that average causal wage effects are additively decomposable in that the average causal wage effect between any pair of firms can be written as the sum of average causal wage effects between those two firms and any third firm. 

These conditions are sufficient, but meaningfully weaker than assumptions commonly invoked to justify the interpretation of firm effect contrasts as the average causal wage effects of a randomly drawn worker moving between firms. Moreover, we show that these conditions are necessary when we demand that the interpretation also holds for arbitrarily small changes to the distribution of mobility.  
In this sense, they are the necessary and sufficient conditions to ensure a causal interpretation of the firm effect contrasts that is robust to infinitesimal perturbations in the composition of workers in the sample of firms under study.
We emphasize that these conditions have no testable implications and are
required in addition to the conditions required for firm effect contrasts to accurately describe the wage changes accompanying moves of workers between firms. 

Using the same Veneto Workers History data for 1999 and 2001, we revisit two applications in which the firm effect contrasts have been given a causal interpretation: wage premiums for working in different industries \citep{Card:2024aa} and wage premiums for working in firms of different sizes \citep{Kline:2024io}. Industry and size premia are contrasts between two groups of firms rather than between two individual firms. Each premium is a weighted average of sums of difference-in-differences along signed paths that start at a firm in one group and end at a firm in the other. 
For every industry contrast, the large majority of the weight is placed on signed paths that pass through firms in other industries than those contrasted. 
A similar pattern is found for the firm-size contrasts. 
For example, the contrast between the two largest firm size groups mostly reflects signed paths that pass through relatively small firms. 
Because the cycle condition is rejected in our data, signed path sums between the same firms differ across paths, so the premiums depend on which paths receive weight. This suggests that the estimated industry and firm-size premia should not be interpreted either descriptively or causally.

Finally, we turn our attention to our third question.  Our decomposition suggests a natural, alternative estimator that averages the sums of difference-in-differences along a restricted subset of signed paths.  By selecting these signed paths appropriately, the resulting restricted path estimator may accurately describe the wage changes accompanying moves of workers between firms under conditions that are substantially weaker than required for firm effect contrasts to have the same interpretation.  We show that restrictions on heterogeneity across firms and workers determine which signed paths can be used to describe wage changes accompanying moves of workers between firms.  Interestingly, permissible signed paths may need to be restricted not only by the types of the involved firms or workers, but also the number and orientation of transitions between types of firms.  Applying our new estimator to the Veneto Workers History data, we see that modest amounts of heterogeneity across firms and workers leads to estimates of the wage changes accompanying moves of workers between firms that are meaningfully different than the firm effect contrast.

We also develop results that permit partial identification of one of the key objects computed using firm effects, the dispersion in average (counterfactual) pay across firms, under assumptions substantially weaker than required for the firm effect contrasts to have a causal interpretation.  The sharp bounds on this quantity can be computed using an optimization program with linear constraints that parameterize departures from common trends, mean independence of mobility and causal wage effects, and additive decomposability.  When applied to the Veneto Workers History data with no departures from these conditions, these bounds are empty, reflecting our earlier rejection of equality of signed path sums between any two pairs of firms.  By relaxing these conditions enough, the sharp bounds will include the estimate of dispersion in average pay across firms implied by the firm effects.  The minimal such relaxation provides a means of assigning an economic magnitude to the differences across signed path sums.  

The remainder of our paper is organized as follows.  In Section~\ref{sec: preliminaries}, we describe the regression first introduced by \cite{Abowd:1999ii} and the firm effects of interest.  Section \ref{sec: descriptive tool} develops our decomposition of the firm effect contrasts; required graph-theoretic language, including signed paths, are introduced as needed.  In Section~\ref{sec: did characterization section}, we characterize when the firm effect contrasts describe the wage changes associated with mobility.  We also discuss testing the conditions required for this interpretation of the firm effect contrasts, including commonly used diagnostic checks.  Finally, we introduce estimators based on a restricted subset of paths, and the way in which such estimators may describe the wage changes associated with mobility under substantially weaker conditions.  Section \ref{sec: causal interpretation} develops the additional conditions under which the firm effect contrasts can be interpreted causally.  In Sections \ref{sec: empirical results} and \ref{sec:causalempirical}, we apply these results to the Veneto Workers History data.  Section \ref{sec: firmeffects} develops our partial identification results and applies them to the same data.  

\section{Preliminaries}
\label{sec: preliminaries}
This section introduces the regression model used in \cite{Abowd:1999ii} and the estimand corresponding to the firm effects of interest.  Henceforth, we refer to this regression model as the AKM regression model.

\noindent {\bf AKM regression model.} Suppose there are $N$ workers employed at $K\geq 2$ firms in $T$ time periods. Denote by $k(i,t)$ the firm at which worker $i$ is employed in period $t$. \citet{Abowd:1999ii} considered the following regression model of (log) earnings $Y_{it}$ of worker $i$ in period $t$ on worker fixed effects $\mu_i$, firm fixed effects $\psi_k$, covariates $X_{it}$, and time-varying idiosyncratic errors $\epsilon_{it}$:
\begin{equation}\label{eq: AKM model}
	Y_{it} = X_{it}'\beta + \mu_i + \psi_{k(i,t)} + \epsilon_{it}.
\end{equation}

\noindent {\bf Two-period version in difference form.} To simplify the exposition, consider the special case in which there are $T=2$ time periods and no covariates other than a linear time trend, $X_{it}=t$. With two time periods, worker $i$ is employed at some firm $O_i$ (the ``origin'') in period $t=1$ and at a (possibly different) firm $D_i$ (the ``destination'') in period $t=2$, that is, $O_i=o$ if $k(i,1)=o$ and $D_i=d$ if $k(i,2)=d$.

For notational convenience, we suppress worker subscripts and let $\alpha_k := \psi_k - \psi_1$, $k=1,\ldots,K$, $\Delta := Y_{2}-Y_{1}$, and $\varepsilon := \epsilon_{2}-\epsilon_{1}$. Our choice of firm $1$ as the reference firm implies the normalization $\alpha_1=0$. The AKM regression model \eqref{eq: AKM model} can then be written in differenced form as
\begin{equation}\label{eq: AKM model differences simplified}
	\Delta = \beta + \sum_{j=2}^K \alpha_j (\ind\{D=j\} - \ind\{O=j\}) + \varepsilon.
\end{equation}
Let $I:=(\ind\{O=2 \},\ldots, \ind\{O=K\})'$ and $J:=(\ind\{D=2 \},\ldots, \ind\{D=K\})'$ denote the vectors of indicators for origin and destination firms.

\noindent {\bf The AKM estimand.} The AKM estimand is the ordinary least squares estimand $(\alpha_2,\ldots,\alpha_K)'$ associated with equation~\eqref{eq: AKM model differences simplified}. These coefficients are determined jointly with the time trend $\beta$ by the following equations:
\begin{equation}\label{eq: exogenous mobility}
E\left[
\left(
\Delta-\beta-\sum_{j=2}^{K}\alpha_j
\bigl(\ind\{D=j\}-\ind\{O=j\}\bigr)
\right)
\begin{pmatrix}
1\\ J-I
\end{pmatrix}
\right]=0.
\end{equation}
For the AKM estimand to be well defined, the regressors must not be multicollinear. Define
$$\mathbb{A}:=E\left[ \begin{pmatrix} 1\\ J-I \end{pmatrix}\begin{pmatrix} 1\\ J-I \end{pmatrix}' \right].$$
Whenever $\mathbb{A}$ is invertible, the solution of the normal equations in \eqref{eq: exogenous mobility} exists and is unique. 

Invertibility requires that firms are sufficiently connected by workers moving from one firm to another; in practice, it is ensured by restricting the set of firms to the maximally connected subset of firms \citep{Abowd:2002aa}. Lemma~\ref{lem: connectedness} in the Appendix \ref{app: auxiliary lemmas} formalizes the relationship between the invertibility condition and the connectedness of the firms.

\section{The Anatomy of the AKM Estimand}
\label{sec: descriptive tool}

\begin{figure}[t]
\centering
\begin{tikzpicture}[
  scale=0.95,
  nd/.style={circle,draw,thick,minimum size=8mm,inner sep=0pt,fill=white},
  arr/.style={thick,-{Stealth[length=2.4mm]}},
]
\node[nd] (n1) at (-3.2, 0.2) {$1$};
\node[nd] (n2) at (-1.0, 1.4) {$2$};
\node[nd] (n3) at ( 1.6, 1.4) {$3$};
\node[nd] (n4) at ( 4.2, 0.7) {$4$};
\node[nd] (n5) at ( 0.6,-1.6) {$5$};
\draw[arr] (n1) -- (n2);
\draw[arr] (n2) -- (n3);
\draw[arr] (n2) -- (n5);
\draw[arr] (n3) -- (n5);
\draw[arr] (n3) to[bend left=16] (n4);
\draw[arr] (n4) to[bend left=16] (n3);
\draw[arr] (n4) -- (n5);
\end{tikzpicture}
\caption{Directed graph of Example~\ref{ex: four-firm-illustration} with arrows indicating the supported moves in $\mathcal E$.}
\figurenote{Nodes represent the five firms in
  Example~\ref{ex: four-firm-illustration}. An arrow from firm $o$
  to firm $d$ indicates a supported move, with $o\neq d$ and
  $P_{od}>0$. Arrows indicate the direction of worker mobility.
  Stayers are omitted from the graph. The shares of movers and
  stayers are reported in Table~\ref{tab:pod-eight-firm}.}
\label{fig:network}
\end{figure}
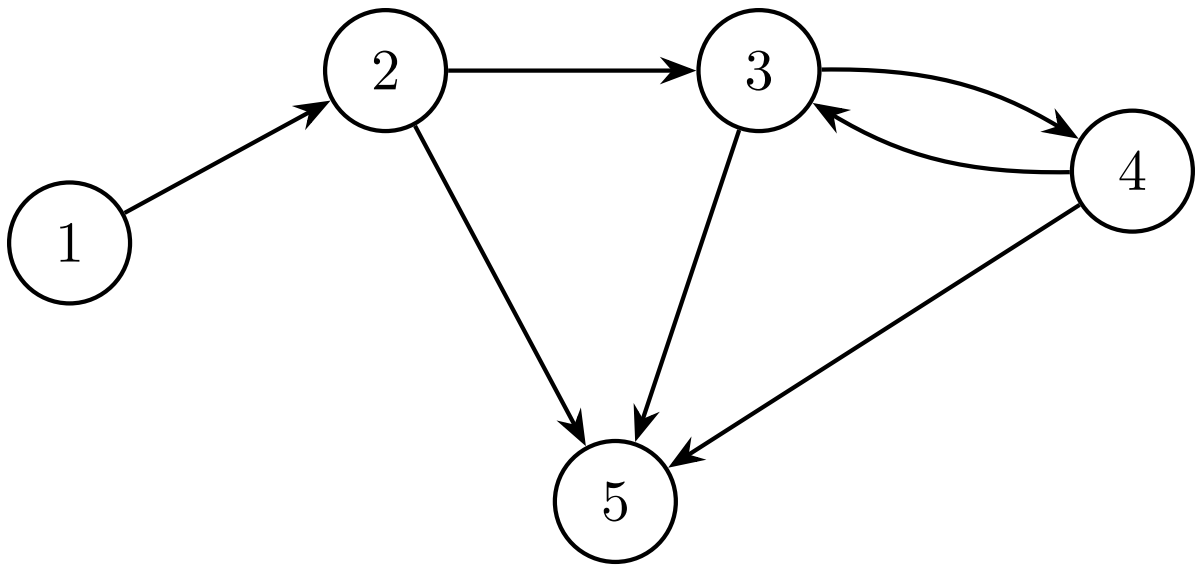

This section seeks to expose the anatomy of the AKM estimand. In what follows, we illustrate theoretical results with a running example consisting of five firms displayed in Figure~\ref{fig:network}, in which each arrow indicates a supported move, i.e., a pair of firms between which some workers move.

\subsection{Notation and regularity conditions}
\label{sec: notation and assumptions}

We denote the probability of a worker being employed by firm $o$ at $t=1$ and by firm $d$ at $t=2$ by $P_{od} := P\{O=o,D=d\}$ for $o,d\in\{1,\ldots,K\}$, and define the set of supported directed moves by
\[
\mathcal E:=\{(u,v)\in\{1,\ldots,K\}^2\colon u\neq v,\ P_{uv}>0\}.
\]
For any two firms $o$ and $d$ with $P_{od}>0$ and $P_{oo}>0$, the difference-in-differences contrast is defined as
\begin{equation}\label{eq: did definition}
\DID(o,d):=
E[\Delta\mid O=o,D=d]-E[\Delta\mid O=o,D=o].
\end{equation}
It compares the mean wage growth for those who move from firm $o$ to firm $d$ compared to those who stay at firm $o$.

\begin{table}[t!]
\centering
\begin{tabular}{@{}c|rrrrr|r@{}}
\toprule
$100\cdot P_{od}$ & $1$ & $2$ & $3$ & $4$ & $5$ & row sum \\
\midrule
$1$ & $2$ & $5$ & $0$ & $0$ & $0$ & $7$ \\
$2$ & $0$ & $18$ & $8$ & $0$ & $7$ & $33$ \\
$3$ & $0$ & $0$ & $14$ & $6$ & $5$ & $25$ \\
$4$ & $0$ & $0$ & $4$ & $8$ & $3$ & $15$ \\
$5$ & $0$ & $0$ & $0$ & $0$ & $20$ & $20$ \\
\midrule
column sum & $2$ & $23$ & $26$ & $14$ & $35$ & $100$ \\
\bottomrule
\end{tabular}
\caption{Joint distribution of $(O,D)$ in Example~\ref{ex: four-firm-illustration}.}
 \figurenote{This table reports the joint distribution of origin
  and destination firms in Example~\ref{ex: four-firm-illustration}.
  Rows index origin firms $o$ and columns index destination firms $d$.
  Entries are population shares $P_{od}=\Pr(O=o,D=d)$ expressed
  in percent. Diagonal entries correspond to stayers; off-diagonal
  entries correspond to movers. Row and column sums give the
  distributions of employment across firms in the first and
  second periods, respectively.}
\label{tab:pod-eight-firm}
\end{table}

We maintain the following assumption, which ensures that the AKM estimand and the difference-in-differences contrasts are well-defined.

\begin{assumption}\label{ass: invertibility of A}
	(i) The matrix $\mathbb{A}$ is invertible. (ii) For all $o,d\in\{1,\ldots,K\}$, $P_{od}>0$ implies $P_{oo}>0$. (iii) $\Delta$ has a finite first moment.
\end{assumption}

\noindent Part (i) ensures the existence and uniqueness of the AKM estimand. Part (ii) ensures that, for every move that occurs with positive probability, there is also a positive probability of staying in the origin firm. Together, parts (ii) and (iii) ensure that $\DID(o,d)$ is well-defined for all supported moves $(o,d)\in\mathcal E$.

\subsection{Signed paths: Definitions and examples}
\label{sec: paths definitions and examples}

The set of firms and supported moves defines a directed graph $G^{d}:=(\{1,\ldots,K\},\mathcal E)$. Our representation result will use information on how the firms are connected to each other, directly by supported moves and indirectly through a chain of mobility between firms.  Here, a chain of mobility refers to a sequence of supported moves, possibly involving one or more additional firms.  Signed paths and signed cycles describe this network structure:

\begin{definition}[signed paths and signed cycles]\label{def: signed path and signed cycle}
    A \emph{signed path} in $G^{d}$ from firm $j_0$ to firm $j_L$ is a tuple
    \begin{equation}
    W=(j_0,s_1,j_1,s_2,\ldots,s_L,j_L),
    \qquad L\geq 1,
    \label{eq: signed-path-definition}
    \end{equation}
    alternating firms $j_\ell\in\{1,\ldots,K\}$ and signs $s_\ell\in\{+1,-1\}$, such that the \emph{move used at step $\ell$},
    \[
    e_\ell:=
    \begin{cases}
    (j_{\ell-1},j_\ell) & \text{if }s_\ell=+1,\\
    (j_\ell,j_{\ell-1}) & \text{if }s_\ell=-1,
    \end{cases}
    \]
    is supported, $e_\ell\in\mathcal E$, for every $\ell=1,\ldots,L$. A signed path is called \emph{simple} if $j_0,j_1,\ldots,j_L$ are distinct. A \emph{signed cycle} is a signed path with $j_L=j_0$; it is called \emph{simple} if $j_0,j_1,\ldots,j_{L-1}$ are distinct.  A \emph{path} (respectively, \emph{cycle}) is a signed path (respectively, signed cycle) with $s_\ell=+1$ for every $\ell$.
\end{definition}

Next, we introduce signed path sums, which aggregate another quantity, such as the difference-in-differences contrasts, according to the chain of mobility described by a signed path.

\begin{definition}[signed path sum]
    For any vector $m\in\mathbb R^{|\mathcal E|}$ and signed path $W$, the signed path sum of $m$ along $W$ is given by:
\[
S(W,m):=\sum_{\ell=1}^L s_\ell\, m_{e_\ell} ,
\]
\end{definition}

\begin{figure}[t!]
\centering
\captionsetup[sub]{skip=15pt}
\tikzset{
  pnode/.style={circle,draw,thick,minimum size=7mm,
    inner sep=0pt,fill=white},
  pstep/.style={thick,-{Stealth[length=2.2mm]}},
  prev/.style={thick,dashed,-{Stealth[length=2.2mm]}},
  lab/.style={font=\scriptsize,inner sep=1pt,fill=white}
}

\begin{subfigure}{0.49\textwidth}
\centering
\begin{tikzpicture}[
  scale=0.7,
  ndfade/.style={circle,draw=gray!18,thick,minimum size=8mm,
    inner sep=0pt,fill=gray!6,text=gray!30},
  ndhi/.style={circle,draw,thick,minimum size=8mm,
    inner sep=0pt,fill=white},
  arrfade/.style={thick,gray!18,-{Stealth[length=2.4mm]}},
  arrhi/.style={thick,-{Stealth[length=2.4mm]}}
]
\node[ndfade] (n1) at (-3.2, 0.2) {$1$};
\node[ndhi]   (n2) at (-1.0, 1.4) {$2$};
\node[ndfade] (n3) at ( 1.6, 1.4) {$3$};
\node[ndfade] (n4) at ( 4.2, 0.7) {$4$};
\node[ndhi]   (n5) at ( 0.6,-1.6) {$5$};

\draw[arrfade] (n1) -- (n2);
\draw[arrfade] (n2) -- (n3);
\draw[arrfade] (n3) -- (n5);
\draw[arrfade] (n3) to[bend left=16] (n4);
\draw[arrfade] (n4) to[bend left=16] (n3);
\draw[arrfade] (n4) -- (n5);
\draw[arrhi] (n2) --
  node[lab,left=1pt] {$+(2,5)$} (n5);
\end{tikzpicture}
\caption{simple signed path $W_1^{25}$}
\end{subfigure}\hfill
\begin{subfigure}{0.49\textwidth}
\centering
\begin{tikzpicture}[
  scale=0.7,
  ndfade/.style={circle,draw=gray!18,thick,minimum size=8mm,
    inner sep=0pt,fill=gray!6,text=gray!30},
  ndhi/.style={circle,draw,thick,minimum size=8mm,
    inner sep=0pt,fill=white},
  arrfade/.style={thick,gray!18,-{Stealth[length=2.4mm]}},
  arrhi/.style={thick,-{Stealth[length=2.4mm]}}
]
\node[ndfade] (n1) at (-3.2, 0.2) {$1$};
\node[ndhi]   (n2) at (-1.0, 1.4) {$2$};
\node[ndhi]   (n3) at ( 1.6, 1.4) {$3$};
\node[ndfade] (n4) at ( 4.2, 0.7) {$4$};
\node[ndhi]   (n5) at ( 0.6,-1.6) {$5$};

\draw[arrfade] (n1) -- (n2);
\draw[arrfade] (n2) -- (n5);
\draw[arrfade] (n3) to[bend left=16] (n4);
\draw[arrfade] (n4) to[bend left=16] (n3);
\draw[arrfade] (n4) -- (n5);
\draw[arrhi] (n2) --
  node[lab,above] {$+(2,3)$} (n3);
\draw[arrhi] (n3) --
  node[lab,right=1pt] {$+(3,5)$} (n5);
\end{tikzpicture}
\caption{simple signed path $W_2^{25}$}
\end{subfigure}

\par\vspace{4em}

\begin{subfigure}{0.49\textwidth}
\centering
\begin{tikzpicture}[
  scale=0.7,
  ndfade/.style={circle,draw=gray!18,thick,minimum size=8mm,
    inner sep=0pt,fill=gray!6,text=gray!30},
  ndhi/.style={circle,draw,thick,minimum size=8mm,
    inner sep=0pt,fill=white},
  arrfade/.style={thick,gray!18,-{Stealth[length=2.4mm]}},
  arrhi/.style={thick,-{Stealth[length=2.4mm]}}
]
\node[ndfade] (n1) at (-3.2, 0.2) {$1$};
\node[ndhi]   (n2) at (-1.0, 1.4) {$2$};
\node[ndhi]   (n3) at ( 1.6, 1.4) {$3$};
\node[ndhi]   (n4) at ( 4.2, 0.7) {$4$};
\node[ndhi]   (n5) at ( 0.6,-1.6) {$5$};

\draw[arrfade] (n1) -- (n2);
\draw[arrfade] (n2) -- (n5);
\draw[arrfade] (n3) -- (n5);
\draw[arrfade] (n3) to[bend left=16] (n4);
\draw[arrhi] (n2) --
  node[lab,above] {$+(2,3)$} (n3);
\draw[arrhi] (n3) to[bend right=16]
  node[lab,below] {$-(4,3)$} (n4);
\draw[arrhi] (n4) --
  node[lab,above right=-1pt] {$+(4,5)$} (n5);
\end{tikzpicture}
\caption{signed path $W_3^{25}$}
\end{subfigure}\hfill
\begin{subfigure}{0.49\textwidth}
\centering
\begin{tikzpicture}[
  scale=0.7,
  ndfade/.style={circle,draw=gray!18,thick,minimum size=8mm,
    inner sep=0pt,fill=gray!6,text=gray!30},
  ndhi/.style={circle,draw,thick,minimum size=8mm,
    inner sep=0pt,fill=white},
  arrfade/.style={thick,gray!18,-{Stealth[length=2.4mm]}},
  arrhi/.style={thick,-{Stealth[length=2.4mm]}}
]
\node[ndfade] (n1) at (-3.2, 0.2) {$1$};
\node[ndfade] (n2) at (-1.0, 1.4) {$2$};
\node[ndhi]   (n3) at ( 1.6, 1.4) {$3$};
\node[ndhi]   (n4) at ( 4.2, 0.7) {$4$};
\node[ndfade] (n5) at ( 0.6,-1.6) {$5$};

\draw[arrfade] (n1) -- (n2);
\draw[arrfade] (n2) -- (n3);
\draw[arrfade] (n2) -- (n5);
\draw[arrfade] (n3) -- (n5);
\draw[arrfade] (n4) -- (n5);
\draw[arrhi] (n3) to[bend right=16]
  node[lab,below] {$-(4,3)$} (n4);
\draw[arrhi] (n4) to[bend right=16]
  node[lab,above] {$-(3,4)$} (n3);
\end{tikzpicture}
\caption{simple signed cycle $C_1$}
\end{subfigure}

\caption{The signed paths and signed cycle in the path-cycle
representation in Example~\ref{ex: four-firm-illustration}.}
\figurenote{Panels (a)--(c) show the three simple signed paths
from firm 2 to firm 5, and panel (d) shows a signed cycle.
Highlighted arrows indicate the direction of traversal. Each label
identifies the supported move and its sign: $+(o,d)$ contributes
$\DID(o,d)$ and $-(o,d)$ contributes $-\DID(o,d)$ to the signed sum.
Thus, traversing $(4,3)$ backwards differs from traversing $(3,4)$
forwards.}
\label{fig:path-cycle-subfigures}
\end{figure}
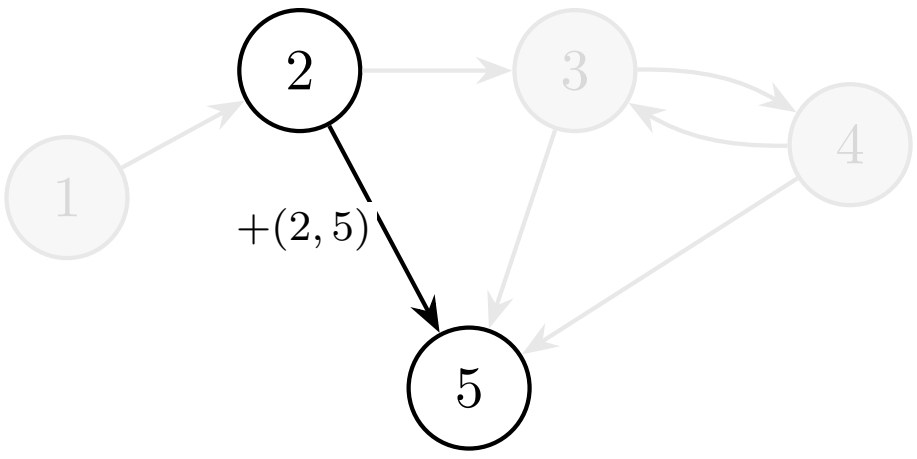
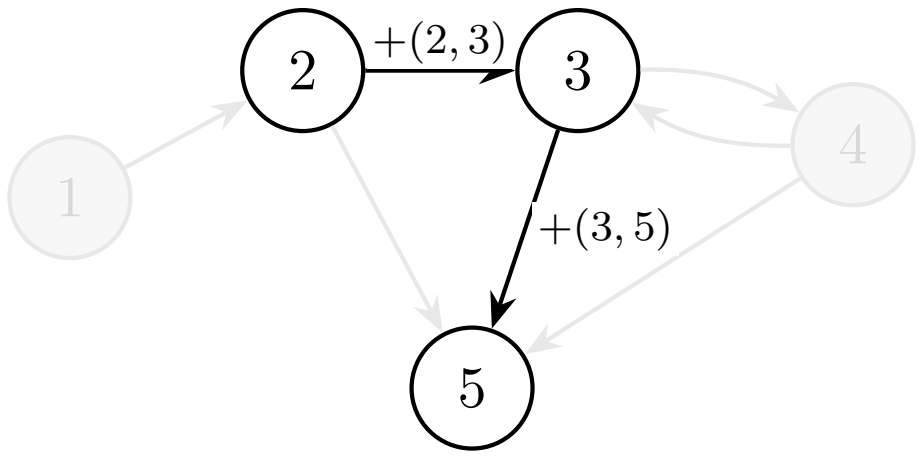
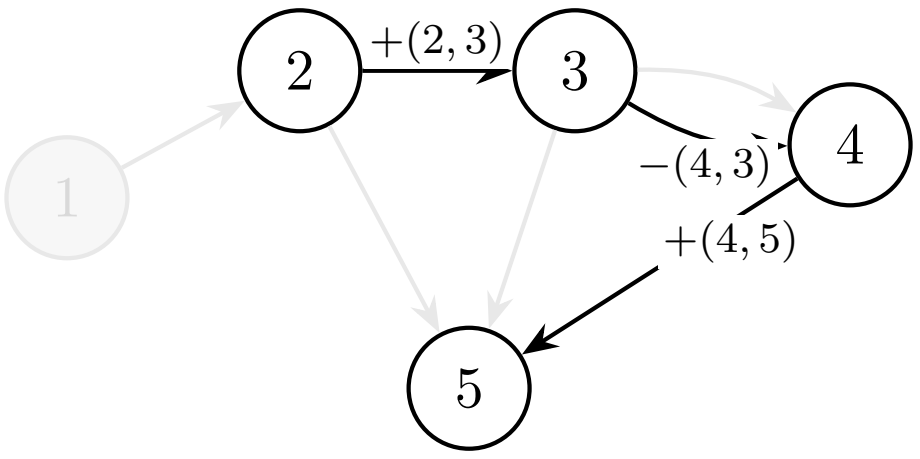
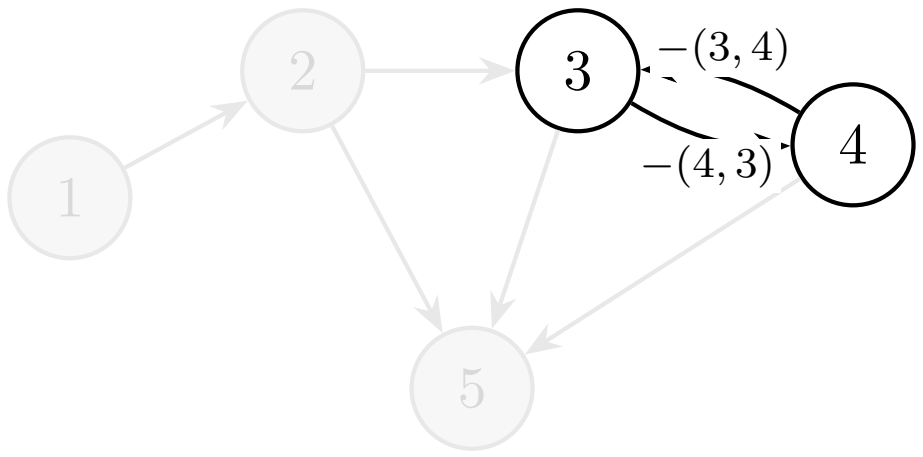

In what follows, when it causes no confusion, we may sometimes say signed path sum without specifying the quantity $m$ and/or the signed path $W$.  We may also sometimes say signed cycle sum instead of signed path sum when appropriate. Before continuing, we illustrate the preceding definitions using the following example:
\begin{example}
\label{ex: four-firm-illustration} In the example in Figure~\ref{fig:network}, the joint distribution of $(O,D)$ is given in Table~\ref{tab:pod-eight-firm} and the set of supported moves is
\[
\mathcal E=\{(1,2),(2,3),(2,5),(3,4),(3,5),(4,3),(4,5)\}.
\]
Throughout, we focus on the contrast between firms $2$ and $5$, which are connected both directly and through several indirect routes. Figure~\ref{fig:path-cycle-subfigures} displays three signed paths from firm $2$ to firm $5$ and one signed cycle:
\begin{align*}
W_1^{25}&:=(2,+1,5), & W_2^{25}&:=(2,+1,3,+1,5),\\
W_3^{25}&:=(2,+1,3,-1,4,+1,5), & C_1&:=(3,-1,4,-1,3).
\end{align*}
Each arrow points in the direction in which the signed path or signed cycle travels and is labeled by the supported move it uses together with the sign of the traversal. For instance, $W_3^{25}$ travels from firm $2$ to firm $3$ along the move $(2,3)$, then from firm $3$ to firm $4$ along the move $(4,3)$, and finally from firm $4$ to firm $5$ along the move $(4,5)$. The signed sum of the difference-in-differences contrasts along $W_3^{25}$ is given by $S(W_3^{25},\DID)=\DID(2,3)-\DID(4,3)+\DID(4,5)$.
\end{example}

\subsection{A representation result}
\label{sec: path decomposition}

To begin unpacking the anatomy of the AKM estimand, Lemma~\ref{lem: arbitrary no of firms} below expresses it as a linear combination of pairwise $\DID$s plus a time trend.\footnote{\citet[Section~3.1, equation~(3) and footnote~5]{Kline:2024io} provides a related sample counterpart to our population representation, expressing firm-effect estimates as linear combinations of covariate-adjusted mean wage changes associated with supported moves. Because his decomposition treats the covariate adjustment -- including the common time trend, which appears as an intercept in our differenced regression -- as given, its weights generally differ from ours, which incorporate the joint determination of the time trend and firm effects. Our representation also separates the difference-in-differences contrasts from the trend term $\tau_j$. His weights need not satisfy the zero-sum identity \eqref{eq: weights sum to zero}, although the corresponding contrast weights satisfy the unit-flow property defined below in \eqref{eq:kappa-unit-flow}.}
For two reasons, it is difficult to interpret this linear combination of $\DID$s, even descriptively. The first is that the weights on supported moves need not be nonnegative or sum to one, so the representation is not generally convex. The other reason is that not only $\DID(j,1)$ but also all other $\DID$s appear, which implies that $\alpha_j$ does not isolate the wage changes associated with moving between firms $1$ and $j$. 
Nevertheless, the lemma will prove to be a useful stepping stone to deriving a constructive representation result of the AKM estimand.  

\begin{lemma}[DID representation]\label{lem: arbitrary no of firms}
	If Assumption~\ref{ass: invertibility of A} holds, then the AKM estimand can be written as
	\begin{equation}\label{eq: alpha_j for arbitrary no of firms}
		\alpha_j = \tau_j + \sum_{(o,d)\in\mathcal E} \omega_{j,od} \DID(o,d), \qquad j=2,\ldots,K,
	\end{equation}
	where $e_j$ is the $K$-dimensional vector whose $j$-th element is equal to one and all other elements are equal to zero,
	\begin{align}
		\omega_{j,od} &:= e_j'\mathbb{A}^{-1}\nu_{od}\, P_{od}, \label{eq: omega definition}\\
		\nu_{od} &:= e_1 + \ind\{d>1\}e_d - \ind\{o>1\}e_o, \label{eq: nu definition}
	\end{align}
	the weights $\omega_{j,od}$ sum to zero, i.e.,
	\begin{equation}
		\sum_{(o,d):\,P_{od}>0} \omega_{j,od} = 0 , \label{eq: weights sum to zero}
	\end{equation}
	and the trend term $\tau_j$ is
	\begin{equation}
		\tau_j := \sum_{(o,d):\,P_{od}>0} \omega_{j,od}\,E[\Delta \mid O=o,D=o] . \label{eq: tau definition}
	\end{equation}
\end{lemma} 

Note that $\mathbb{A}$, and therefore also the weights $\omega_{j,od}$, depend only on the shares of movers and stayers $(P_{od}\colon o,d\in\{1,\ldots,K\})$. The term $\tau_j$ is a linear combination of origin-specific stayer trends. If the average time trend among the stayers at firm $o$, i.e., $E[\Delta | O=o,D=o]$, is the same for all firms $o$, then $\tau_j = 0$ by \eqref{eq: weights sum to zero}.

To derive a constructive representation result from Lemma~\ref{lem: arbitrary no of firms}, it is useful to consider the AKM estimand net of the trend term $\tau_j$: 
\begin{definition}[adjusted AKM estimand]\label{def: adjusted AKM estimand}
    The \emph{adjusted AKM estimand} is defined as the vector $(\widetilde \alpha_2,\ldots, \widetilde \alpha_K)'\in\mathbb{R}^{K-1}$ with $ \widetilde \alpha_j:= \alpha_j-\tau_j$, $j=2,\ldots,K$. 
\end{definition}
The adjusted AKM estimand contains the coefficients for the non-reference firms $2,\ldots,K$. For contrasts involving the reference firm, we extend the definitions by the conventions $\widetilde\alpha_1:=0$, $\tau_1:=0$, and $\omega_{1,od}:=0$ for all $o,d\in\{1,\ldots,K\}$.

In what follows, we will focus on pairwise differences in adjusted AKM estimands, i.e., $\tilde{\alpha}_k - \tilde{\alpha}_j$.  Below, we may also refer to these quantities more succinctly as adjusted AKM contrasts. From Lemma \ref{lem: arbitrary no of firms}, we immediately have that 
\begin{equation}
  \tilde{\alpha}_k - \tilde{\alpha}_j = \sum_{(o,d) \in \mathcal E} \kappa^{jk}_{od} \DID(o,d), ~\text{ where ~} \kappa^{jk}_{od} := \omega_{k,od} - \omega_{j,od}. \label{eq: adjusted AKM kappa contrast representation}
\end{equation}
As shown in Lemma~\ref{lem:left-out-invariance}, $\kappa^{jk}_{od}$, and therefore $\widetilde \alpha_k-\widetilde \alpha_j$, is invariant to the choice of the left-out firm; two regressions with different left-out firms yield the same pairwise differences.  For this reason, pairwise differences are the more natural object of analysis. 

Theorem \ref{thm: akm path decomposition} presents our representation result for these pairwise differences. The result reveals the way in which $\widetilde \alpha_k-\widetilde \alpha_j$ is intimately linked to the network structure of the connected set of firms: how firms are connected to each other, directly by supported moves and indirectly through chains of mobility that involve one or more additional firms. The theorem shows, in particular, that the adjusted AKM contrast $\tilde \alpha_k - \tilde \alpha_j$ represents the information in these moves by first summing the DIDs corresponding to moves in signed paths starting at firm $j$ and ending in firm $k$, and then takes a positively weighted average of these sums across the signed paths.

\begin{theorem}[signed path representation]\label{thm: akm path decomposition}
    Suppose Assumption~\ref{ass: invertibility of A} holds and fix two distinct firms $j,k\in\{1,\ldots,K\}$. Then there exist a finite nonempty collection $\mathcal W^{jk}$ of signed paths from $j$ to $k$ and positive constants $\{\lambda^{jk}(W)\}_{W\in\mathcal W^{jk}}$ such that
    \[
    \sum_{W\in\mathcal W^{jk}}\lambda^{jk}(W)=1
    \]
    and the pairwise differences in adjusted AKM estimands satisfy
    \begin{equation}\label{eq:akm-path-representation}
    \widetilde \alpha_k-\widetilde \alpha_j
    =
    \sum_{W\in\mathcal W^{jk}}
    \lambda^{jk}(W)S(W,\DID).
    \end{equation}
    The signed paths $\mathcal W^{jk}$ satisfy the following {\it no-cancellation condition}: for each $e \in \mathcal E$, the net number of traversals of $e$, i.e., forward traversals minus backward traversals, is either nonnegative for every $W\in\mathcal W^{jk}$ or nonpositive for every $W\in\mathcal W^{jk}$.
    The signed paths $\mathcal W^{jk}$ and constants $\{\lambda^{jk}(W)\}_{W\in\mathcal W^{jk}}$ can be constructed from knowledge of the mobility shares $(P_{od}\colon o,d\in\{1,\ldots,K\})$ alone.
\end{theorem}

The representation result in Theorem \ref{thm: akm path decomposition} contradicts the conjecture in 
\citet[Section~3.3.1, p.~143]{Kline:2024io} that the adjusted AKM contrasts ``cannot be written as a convex weighted average of direct and indirect contrasts.'' His conjecture is based on an example with four firms. His displayed weights are positive but do not sum to 1. However, the terms he labels indirect contrasts are not individually signed path sums between the two firms being compared. Our result shows that regrouping these terms into signed path sums of wage changes yields a convex representation of the same estimator. More generally, his contrast weights satisfy a corresponding version of the unit-flow property \eqref{eq:kappa-unit-flow}, so the existence of such a convex representation follows from the discussion following Lemma \ref{lem: cycle correction} below.

Three additional features of Theorem \ref{thm: akm path decomposition} are worth highlighting. First, the representation holds for every pair of distinct firms, whether or not $P_{jk}>0$ or $P_{kj} > 0$.  If $P_{jk} = P_{kj}=0$, then direct one-move paths are simply absent from $\mathcal W^{jk}$, and $\widetilde \alpha_k-\widetilde \alpha_j$ is built entirely from the $\DID$s in chains of mobility between $j$ and $k$ that involve one or more additional firms, including possibly firms that are not comparable (e.g., in terms of the industry to which they belong) with either firm $j$ or firm $k$. This case is particularly relevant empirically because most pairs of firms are not connected by a direct move.

Second, the no-cancellation condition can be viewed as a normalization of the path representation. The normalization rules out one source of redundancy: the same $\DID$ cannot enter different signed path sums with opposite net signs. Such offsetting uses could otherwise change the apparent importance of different chains of mobility while leaving the adjusted AKM contrast unchanged. Note that this normalization does not rule out cancellation within an individual path: a path may traverse the same supported move both forward and backward, producing offsetting terms in its signed path sum.

Even under this normalization, the representation need not be unique, but we emphasize that all representations are valid in that they reproduce the adjusted AKM contrasts and therefore accurately describe how the adjusted AKM contrasts depends on mobility shares and wage changes.  As a result, {\it any} of these representation could be used to assess the behavior of the adjusted AKM contrasts.  In our empirical analysis, however, we select among these representations one that minimizes the total weight on paths containing a firm whose industry differs from those of both endpoints. A large resulting weight therefore establishes that every such representation assigns substantial weight to these paths.

Third, the signed paths in $\mathcal W^{jk}$ need not be simple: a signed path can visit a firm more than once, so a path can, for instance, follow a simple signed path from $j$ to $k$ with a signed cycle inserted along the way. In Lemma~\ref{lem: cycle correction} below, proved in Appendix~\ref{app: proofs section descriptive}, we first write the adjusted AKM contrasts as a convex combination of simple signed path sums plus a cycle correction, by which we mean a nonnegative linear combination of simple signed cycle sums.  We state this result formally below, but, before doing so, note that empirically we find that the total weight in the cycle correction is close to zero, suggesting that nearly all signed paths in Theorem \ref{thm: akm path decomposition} are simple, at least in the Italian data we use.
\begin{lemma}[simple signed paths with cycle corrections]\label{lem: cycle correction}
Suppose Assumption~\ref{ass: invertibility of A} holds and fix two distinct firms $j,k\in\{1,\ldots,K\}$. Then there exists a finite nonempty collection $\mathcal W_s^{jk}$ of simple signed paths from $j$ to $k$; a finite, possibly empty collection $\mathcal C_s^{jk}$ of simple signed cycles; and positive constants $\{\lambda_s^{jk}(W)\}_{W\in\mathcal W_s^{jk}}$ and $\{\eta_s^{jk}(C)\}_{C\in\mathcal C_s^{jk}}$, such that
\[
\sum_{W\in\mathcal W_s^{jk}}\lambda_s^{jk}(W)=1
\]
and
\begin{equation}\label{eq: cycle correction representation}
\widetilde \alpha_k-\widetilde \alpha_j
=
\sum_{W\in\mathcal W_s^{jk}}\lambda_s^{jk}(W)S(W,\DID)
+
\sum_{C\in\mathcal C_s^{jk}}\eta_s^{jk}(C)S(C,\DID).
\end{equation}
The signed paths $\mathcal W_s^{jk}$, signed cycles $\mathcal C_s^{jk}$, and constants $\{\lambda_s^{jk}(W)\}_{W\in\mathcal W_s^{jk}}$ and $\{\eta_s^{jk}(C)\}_{C\in\mathcal C_s^{jk}}$ can be constructed from knowledge of the mobility shares $(P_{od}\colon o,d\in\{1,\ldots,K\})$ alone.
\end{lemma}

Once this representation is established, it is straightforward to ``fold'' the simple signed cycles into the simple signed paths and obtain the representation in Theorem \ref{thm: akm path decomposition} of the pairwise differences in AKM estimands in terms of signed path sums for possibly non-simple signed paths.  The key feature that permits the construction of the representation in \eqref{eq: cycle correction representation} is that the weights $\kappa_{od}^{jk}$ from \eqref{eq: adjusted AKM kappa contrast representation} satisfy the following property: 
\begin{equation}
\sum_{s:(r,s)\in\mathcal E}\kappa^{jk}_{rs}
-
\sum_{s:(s,r)\in\mathcal E}\kappa^{jk}_{sr}
=
\ind\{r=j\}-\ind\{r=k\}
\qquad\text{for every firm }r.
\label{eq:kappa-unit-flow}
\end{equation}
This property follows immediately from Lemma \ref{lem: properties of the weights} in Appendix \ref{app: auxiliary lemmas}.  The first sum in \eqref{eq:kappa-unit-flow} may be interpreted as ``flow'' leaving firm $r$, whereas the second sum may be interpreted as ``flow'' entering firm $r$; the difference is therefore the net outflow (if positive) or net inflow (if negative) at firm $r$.  From \eqref{eq:kappa-unit-flow}, we see that the net outflow at firm $j$ is one, the net inflow at firm $k$ is one, and the net flow at all other firms is zero.  For this reason, the property is sometimes referred to as the {\it unit-flow property}. The existence of the desired representation in \eqref{eq: cycle correction representation} then follows from a classical flow decomposition results in graph theory (e.g., \cite{Bertsekas:1998aa}).

We now illustrate Theorem \ref{thm: akm path decomposition} and the construction of the representation in Lemma \ref{lem: cycle correction} in our running example, emphasizing in particular how it hinges upon the unit-flow property.

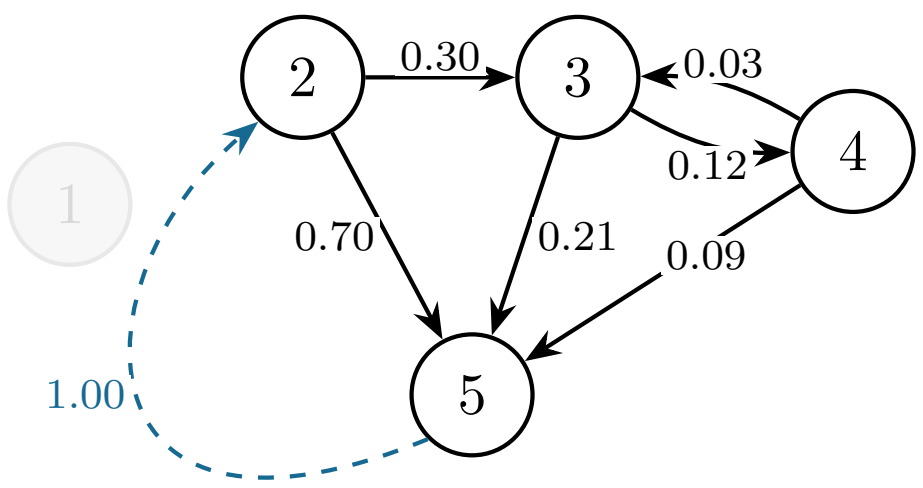
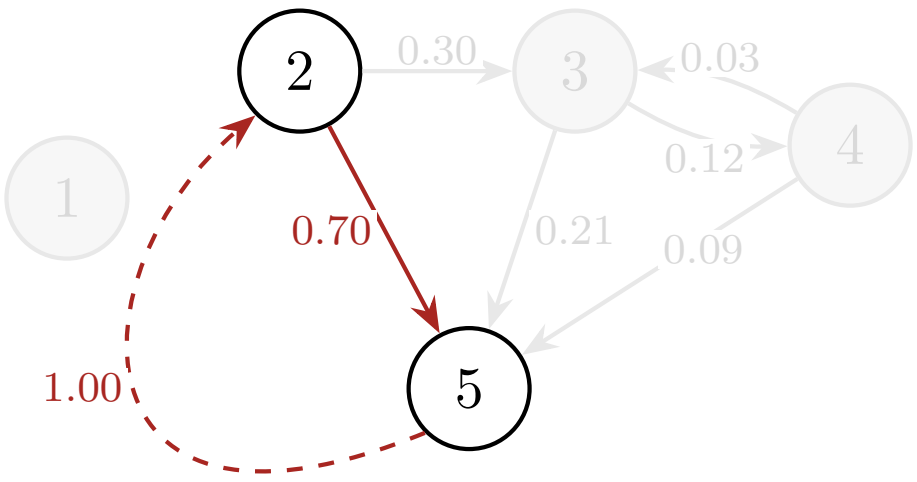
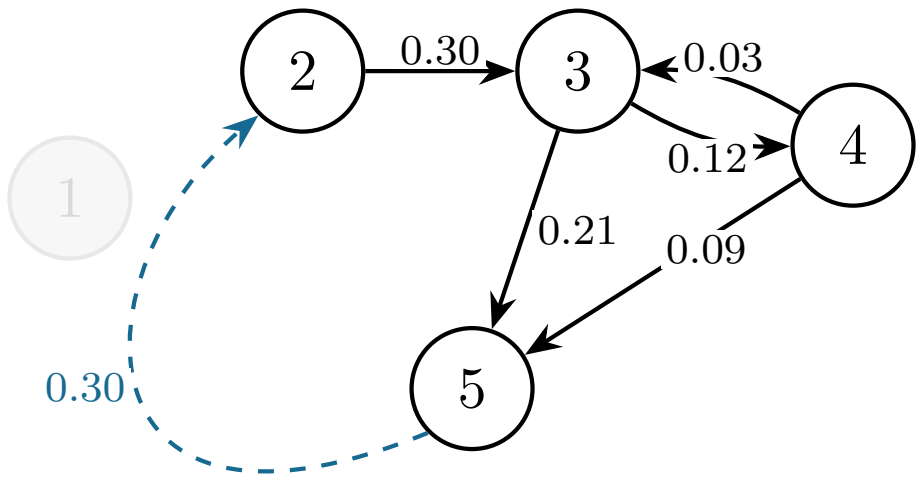
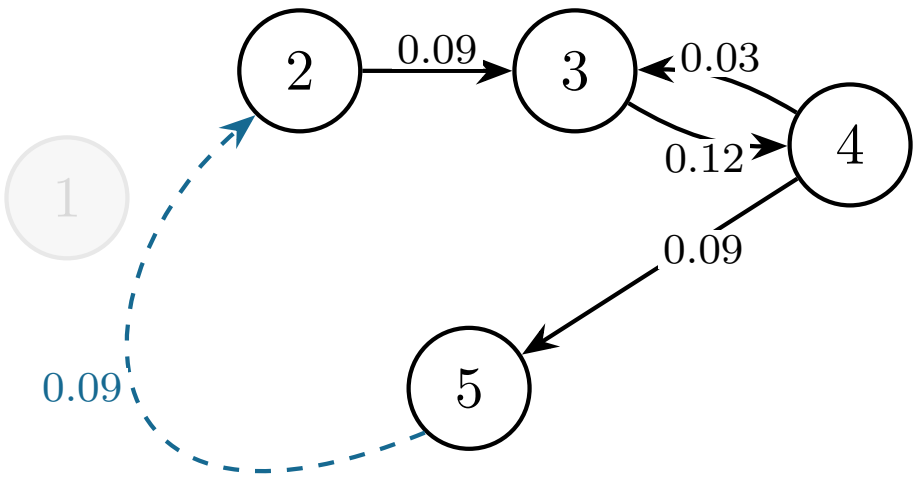
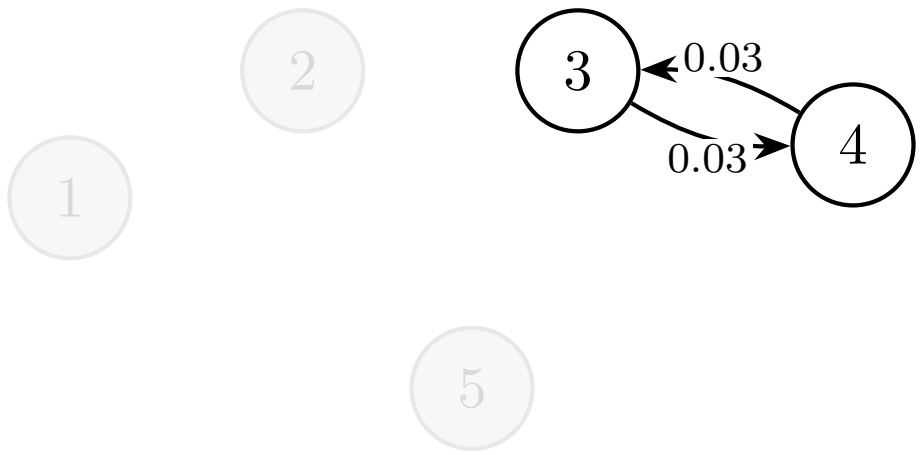
\begin{figure}[t!]
\centering
\definecolor{artificialBlue}{RGB}{22,105,145}
\definecolor{cycleRed}{RGB}{168,38,33}
\captionsetup[sub]{skip=15pt,justification=centering}
\tikzset{
  ndfade/.style={circle,draw=gray!18,thick,minimum size=8mm,
    inner sep=0pt,fill=gray!6,text=gray!30},
  ndhi/.style={circle,draw,thick,minimum size=8mm,
    inner sep=0pt,fill=white},
  arrfade/.style={thick,gray!18,-{Stealth[length=2.4mm]}},
  arrhi/.style={thick,-{Stealth[length=2.4mm]}},
  artificial/.style={arrhi,dashed,draw=artificialBlue},
  cyclehi/.style={arrhi,draw=cycleRed},
  artificialcycle/.style={artificial,draw=cycleRed},
  lab/.style={font=\scriptsize,inner sep=1pt,fill=white}
}

\begin{subfigure}{0.49\textwidth}
\centering
\begin{tikzpicture}[scale=0.7]
\path[use as bounding box] (-3.85,-2.8) rectangle (4.85,2.15);
\node[ndfade] (n1) at (-3.2, 0.2) {$1$};
\node[ndhi]   (n2) at (-1.0, 1.4) {$2$};
\node[ndhi]   (n3) at ( 1.6, 1.4) {$3$};
\node[ndhi]   (n4) at ( 4.2, 0.7) {$4$};
\node[ndhi]   (n5) at ( 0.6,-1.6) {$5$};
\draw[arrfade] (n1) -- node[lab,above,text=gray!30] {$0$} (n2);
\draw[arrhi] (n2) -- node[lab,above] {$0.30$} (n3);
\draw[arrhi] (n2) -- node[lab,left=1pt] {$0.70$} (n5);
\draw[arrhi] (n3) -- node[lab,right=1pt] {$0.21$} (n5);
\draw[arrhi] (n3) to[bend left=16]
  node[lab,above,text=cycleRed] {$-0.03$} (n4);
\draw[arrhi] (n4) to[bend left=16]
  node[lab,below,text=cycleRed] {$-0.12$} (n3);
\draw[arrhi] (n4) -- node[lab,above right=-1pt] {$0.09$} (n5);
\end{tikzpicture}
\caption{signed $\kappa^{25}$-weights\\
(unit flow from firm 2 to firm 5)}
\end{subfigure}\hfill
\begin{subfigure}{0.49\textwidth}
\centering
\begin{tikzpicture}[scale=0.7]
\path[use as bounding box] (-3.85,-2.8) rectangle (4.85,2.15);
\node[ndfade] (n1) at (-3.2, 0.2) {$1$};
\node[ndhi]   (n2) at (-1.0, 1.4) {$2$};
\node[ndhi]   (n3) at ( 1.6, 1.4) {$3$};
\node[ndhi]   (n4) at ( 4.2, 0.7) {$4$};
\node[ndhi]   (n5) at ( 0.6,-1.6) {$5$};
\draw[arrhi] (n2) -- node[lab,above] {$0.30$} (n3);
\draw[arrhi] (n2) -- node[lab,left=1pt] {$0.70$} (n5);
\draw[arrhi] (n3) -- node[lab,right=1pt] {$0.21$} (n5);
\draw[arrhi] (n3) to[bend right=16]
  node[lab,below] {$0.12$} (n4);
\draw[arrhi] (n4) to[bend right=16]
  node[lab,above] {$0.03$} (n3);
\draw[arrhi] (n4) -- node[lab,above right=-1pt] {$0.09$} (n5);
\draw[artificial] (n5.south west)
  .. controls (-3.2,-3.4) and (-3.2,-0.5) ..
  node[lab,pos=0.50,left=1pt,text=artificialBlue] {$1.00$} (n2.south west);
\end{tikzpicture}
\caption{reorient and add the artificial edge\\
(inflow equals outflow at every firm)}
\end{subfigure}

\par\vspace{2em}
\begin{subfigure}{0.49\textwidth}
\centering
\begin{tikzpicture}[scale=0.7]
\path[use as bounding box] (-3.85,-2.8) rectangle (4.85,2.15);
\node[ndfade] (n1) at (-3.2, 0.2) {$1$};
\node[ndhi]   (n2) at (-1.0, 1.4) {$2$};
\node[ndfade] (n3) at ( 1.6, 1.4) {$3$};
\node[ndfade] (n4) at ( 4.2, 0.7) {$4$};
\node[ndhi]   (n5) at ( 0.6,-1.6) {$5$};
\draw[arrfade] (n2) --
  node[lab,above,text=gray!30] {$0.30$} (n3);
\draw[arrfade] (n3) --
  node[lab,right=1pt,text=gray!30] {$0.21$} (n5);
\draw[arrfade] (n3) to[bend right=16]
  node[lab,below,text=gray!30] {$0.12$} (n4);
\draw[arrfade] (n4) to[bend right=16]
  node[lab,above,text=gray!30] {$0.03$} (n3);
\draw[arrfade] (n4) --
  node[lab,above right=-1pt,text=gray!30] {$0.09$} (n5);
\draw[cyclehi] (n2) -- node[lab,left=1pt,text=cycleRed] {$0.70$} (n5);
\draw[artificialcycle] (n5.south west)
  .. controls (-3.2,-3.4) and (-3.2,-0.5) ..
  node[lab,pos=0.50,left=1pt,text=cycleRed] {$1.00$} (n2.south west);
\end{tikzpicture}
\caption{extract the cycle $2\to5\to2$\\
(subtract $0.70$ from both edges)}
\end{subfigure}\hfill
\begin{subfigure}{0.49\textwidth}
\centering
\begin{tikzpicture}[scale=0.7]
\path[use as bounding box] (-3.85,-2.8) rectangle (4.85,2.15);
\node[ndfade] (n1) at (-3.2, 0.2) {$1$};
\node[ndhi]   (n2) at (-1.0, 1.4) {$2$};
\node[ndhi]   (n3) at ( 1.6, 1.4) {$3$};
\node[ndhi]   (n4) at ( 4.2, 0.7) {$4$};
\node[ndhi]   (n5) at ( 0.6,-1.6) {$5$};
\draw[arrhi] (n2) -- node[lab,above] {$0.30$} (n3);
\draw[arrhi] (n3) -- node[lab,right=1pt] {$0.21$} (n5);
\draw[arrhi] (n3) to[bend right=16]
  node[lab,below] {$0.12$} (n4);
\draw[arrhi] (n4) to[bend right=16]
  node[lab,above] {$0.03$} (n3);
\draw[arrhi] (n4) -- node[lab,above right=-1pt] {$0.09$} (n5);
\draw[artificial] (n5.south west)
  .. controls (-3.2,-3.4) and (-3.2,-0.5) ..
  node[lab,pos=0.50,left=1pt,text=artificialBlue] {$0.30$} (n2.south west);
\end{tikzpicture}
\caption{after the first extraction\\
(artificial edge: $0.30$)}
\end{subfigure}

\par\vspace{2em}
\begin{subfigure}{0.49\textwidth}
\centering
\begin{tikzpicture}[scale=0.7]
\path[use as bounding box] (-3.85,-2.8) rectangle (4.85,2.15);
\node[ndfade] (n1) at (-3.2, 0.2) {$1$};
\node[ndhi]   (n2) at (-1.0, 1.4) {$2$};
\node[ndhi]   (n3) at ( 1.6, 1.4) {$3$};
\node[ndhi]   (n4) at ( 4.2, 0.7) {$4$};
\node[ndhi]   (n5) at ( 0.6,-1.6) {$5$};
\draw[arrhi] (n2) -- node[lab,above] {$0.09$} (n3);
\draw[arrhi] (n3) to[bend right=16]
  node[lab,below] {$0.12$} (n4);
\draw[arrhi] (n4) to[bend right=16]
  node[lab,above] {$0.03$} (n3);
\draw[arrhi] (n4) -- node[lab,above right=-1pt] {$0.09$} (n5);
\draw[artificial] (n5.south west)
  .. controls (-3.2,-3.4) and (-3.2,-0.5) ..
  node[lab,pos=0.50,left=1pt,text=artificialBlue] {$0.09$} (n2.south west);
\end{tikzpicture}
\caption{after the second extraction\\
(artificial edge: $0.09$)}
\end{subfigure}\hfill
\begin{subfigure}{0.49\textwidth}
\centering
\begin{tikzpicture}[scale=0.7]
\path[use as bounding box] (-3.85,-2.8) rectangle (4.85,2.15);
\node[ndfade] (n1) at (-3.2, 0.2) {$1$};
\node[ndfade] (n2) at (-1.0, 1.4) {$2$};
\node[ndhi]   (n3) at ( 1.6, 1.4) {$3$};
\node[ndhi]   (n4) at ( 4.2, 0.7) {$4$};
\node[ndfade] (n5) at ( 0.6,-1.6) {$5$};
\draw[arrhi] (n3) to[bend right=16]
  node[lab,below] {$0.03$} (n4);
\draw[arrhi] (n4) to[bend right=16]
  node[lab,above] {$0.03$} (n3);
\end{tikzpicture}
\caption{after the third extraction\\
(remaining cycle: $0.03$)}
\end{subfigure}

\caption{Constructing the path--cycle decomposition in
Example~\ref{ex: four-firm-illustration}.}
\figurenote{Panel (a) shows the signed $\kappa^{25}_{jk}$-weights on the
supported moves, with negative weights in red. They have net outflow one at firm 2, net inflow one
at firm 5, and zero net flow elsewhere. Panel (b) reverses negative-weight
edges, takes their absolute weights, and adds the blue dashed artificial
edge $5\to2$ with flow one. Inflow then equals outflow at every firm.
Arrows in panels (b)--(f) indicate reoriented flow, not necessarily
observed mobility. Panel (c) highlights in red the cycle $2\to5\to2$; subtracting
$0.70$ from both edges gives panel (d). Panels (e) and (f) follow
extractions of $0.21$ around $2\to3\to5\to2$ and $0.09$ around
$2\to3\to4\to5\to2$, respectively. Each extraction subtracts the positive
minimum flow on a cycle from every edge. These three simple cycles each use the
artificial edge once. Deleting it gives the paths in
Figure~\ref{fig:path-cycle-subfigures}; their weights sum to one because
they exhaust its initial unit of flow. Panel (f) leaves a cycle with
positive coefficient $0.03$, which enters separately. All numbers are
rounded.}
\label{fig:flow-construction}
\end{figure}

\vspace{-.4cm}
\begin{examplecontinued}{ex: four-firm-illustration}
Recall from Lemma~\ref{lem: arbitrary no of firms} that
\begin{equation}
\label{eq:2to5akmcontrastkappadecomp}
\widetilde\alpha_5-\widetilde\alpha_2
=
\sum_{(o,d)\in\mathcal E}
\kappa^{25}_{od}\DID(o,d).
\end{equation}
Figure~\ref{fig:flow-construction} shows how to regroup this weighted sum to obtain the representation in Lemma~\ref{lem: cycle correction}. Panel~(a) displays the weight $\kappa^{25}_{od}$ assigned to each edge $(o,d)$. These weights satisfy the unit-flow property~\eqref{eq:kappa-unit-flow}: net outflow is one at firm~2, net inflow is one at firm~5, and net flow is zero elsewhere.\footnote{Because stayer-cell weights are omitted, the displayed weights need not sum to zero.}

Panel~(b) reverses negative-weight edges, assigns them their absolute weights, and adds an artificial edge $5\to2$ with weight one. Reversal preserves net flow, so \eqref{eq:kappa-unit-flow} implies that inflow now equals outflow at every firm. In the discussion below, we apply our earlier definitions of (signed) paths and (signed) cycles to the graph obtained after these actions.

Starting with any positive-flow edge in this augmented graph, we follow positive-flow edges until a firm first repeats. Because inflow equals outflow, a firm reached in this way always has a positive-flow edge on which to leave. Since there are finitely many firms, one must eventually repeat. The alternating sequence beginning with the first visit to the repeated firm and ending with the second is a simple cycle. Panel~(c) illustrates this for $2\to5\to2$. We assign to the cycle the smallest weight among its edges ($\approx 0.70$) and extract it by subtracting this amount from every edge of the cycle. The residual weights, shown in panel~(d), remain nonnegative. At each firm on the cycle, inflow and outflow decrease by the same amount; all other flows are unchanged. Inflow therefore still equals outflow at every firm. Each extraction reduces the number of positive-weight edges by at least one.  Repeating this process reduces all edge weights to zero after finitely many extractions. Panel~(e) follows extraction of $2\to3\to5\to2$ with weight $\approx 0.21$; panel~(f) follows extraction of $2\to3\to4\to5\to2$ with weight approximately $0.09$. Extracting the remaining cycle $3\to4\to3$, with weight $\approx 0.03$, completes the procedure.

Next, write each extracted cycle containing the artificial edge starting at firm~2, so that the artificial edge is last. Deleting that edge and the final occurrence of firm~2 leaves a simple path from firm~2 to firm~5. Accounting for the reversals between panels~(a) and~(b) gives simple signed paths in the original graph, with their corresponding extraction weights. These weights sum to one: each extracted simple cycle containing the artificial edge uses it exactly once, and together they reduce its initial weight of one to zero. Accounting for the reversals in cycles not containing the artificial edge likewise gives simple signed cycles in the original graph, with their corresponding weights.

To complete the construction, observe that each $\kappa^{25}_{od}$ equals the signed total of the weights of the simple signed paths and cycles using the supported move $(o,d)$. Multiplying these identities by $\DID(o,d)$ and summing gives, by \eqref{eq:2to5akmcontrastkappadecomp},
\begin{multline}
\widetilde\alpha_5-\widetilde\alpha_2
=
\underbrace{\lambda_s^{25}(W_1^{25})}
_{\approx\,\LambdaWOneApprox}
\cdot S(W_1^{25},\DID)
+
\underbrace{\lambda_s^{25}(W_2^{25})}
_{\approx\,0.21}
\cdot S(W_2^{25},\DID)
\\
+
\underbrace{\lambda_s^{25}(W_3^{25})}
_{\approx\,\LambdaWThreeApprox}
\cdot S(W_3^{25},\DID)
+
\underbrace{\eta_s^{25}(C_1)}
_{\approx\,\EtaCOneApprox}
\cdot S(C_1,\DID).
\label{eq:2to5pathcycle}
\end{multline}
This representation is of the desired form \eqref{eq: cycle correction representation} in Lemma~\ref{lem: cycle correction}.

Finally, fold the cycle $C_1$ into $W_2^{25}$ by defining the signed path
\[
W_4^{25} := (2,+1,3,-1,4,-1,3,+1,5).
\]
This path travels from firm $2$ to firm $3$, around $C_1$ back to firm $3$, and on to firm $5$. It is not simple because it visits firm $3$ twice. Since $S(W_4^{25},\DID)=S(W_2^{25},\DID)+S(C_1,\DID)$, substituting this identity into \eqref{eq:2to5pathcycle} and collecting terms yields the signed path representation in Theorem~\ref{thm: akm path decomposition}:
\begin{multline}
\widetilde \alpha_5-\widetilde \alpha_2 = \underbrace{\lambda^{25}(W_1^{25})}_{\LambdaWOneApprox} \cdot S(W_1^{25},\DID)  + \underbrace{\lambda^{25}(W_2^{25})}_{\LambdaWTwoFoldedApprox} \cdot S(W_2^{25},\DID) \\ + \underbrace{\lambda^{25}(W_3^{25})}_{\LambdaWThreeApprox} \cdot S(W_3^{25},\DID) + \underbrace{\lambda^{25}(W_4^{25})}_{\LambdaWFourApprox}\cdot S(W_4^{25},\DID), \label{eq:2to5pathdecomp}
\end{multline}
where $W_1^{25}$, $W_2^{25}$, and $W_3^{25}$ are simple signed paths.
\end{examplecontinued}

\section{AKM Estimand as a Descriptive Tool}
\label{sec: did characterization section}

The AKM estimand is sometimes viewed as a descriptive tool to summarize the wage changes that accompany the mobility of workers between firms \citep{Kline:2024io}. This section provides the necessary and sufficient conditions for the AKM estimand to be given such a descriptive interpretation.

\subsection{Necessary and sufficient conditions} \label{sec:neccsuff}
For any two firms $o$ and $d$ with $P_{od}>0$, the average wage change that accompanies the mobility of workers between these firms can be measured by $\DID(o,d)$ defined in equation~\eqref{eq: did definition}. Recall that $\DID(o,d)$ compares the mean wage change for those who move from firm $o$ to firm $d$ net of the wage change of those who stay at firm $o$. The motivation for netting out the wage change of these stayers is to account for changes in wages over time that would occur independently of workers moving from $o$ to $d$. This second difference could help isolate the wage change that accompanies the mobility of workers.

A key implication of Theorem~\ref{thm: akm path decomposition} is that further assumptions are necessary for $\widetilde \alpha_d-\widetilde \alpha_o$ to accurately measure $\DID(o,d)$. The problem is {\it not} how $\widetilde \alpha_d-\widetilde \alpha_o$ adjusts for the changes in wages amongst stayers over time; it is that $\widetilde \alpha_d-\widetilde \alpha_o$ is constructed from the $\DID$s in the chains of mobility between firms other than $o$ and $d$.\footnote{Exactly how $\tilde \alpha_d - \tilde \alpha_o$ adjust for changes in wages over time vary across studies. Many studies include a linear time trend in the AKM regression, as done in \eqref{eq: AKM model differences simplified}. However, it is also possible to exclude controls for calendar time, or to control for it more flexibly by including, e.g., interactions between the linear time trend and firm indicators in \eqref{eq: AKM model differences simplified}. Appendix~\ref{app: two-way} presents theoretical results for specifications of the AKM regression model without a calendar time effect and with flexible controls for time.  In each case, we find qualitatively similar conclusions to the ones we present in the main text for the specification in \eqref{eq: AKM model differences simplified}.} 

Theorem~\ref{thm: akm path decomposition} shows that $\tilde{\alpha}_d - \tilde{\alpha}_o$ first sums $\DID$s within the paths and then takes a positively weighted average of these sums across paths. The following theorem builds on this insight to show the necessary and sufficient conditions for $\tilde{\alpha}_d - \tilde{\alpha}_o$ to be given a descriptive interpretation as measuring $\DID(o,d)$.\footnote{
As explained in footnote~2, \cite{Kline:2024io} expresses firm-effect estimates
as linear combinations of covariate-adjusted mean wage changes associated
with supported moves. In Section~3.1.1 (p.~130), he shows that these
adjusted mean wage changes can be represented as differences in firm
effects if and only if their signed cycle sums vanish. This observation
is related to the equivalence between (i) and (iii) in Theorem \ref{thm: did characterization}, but the form of the adjustment matters when relating these cycle restrictions
to the regression estimand. For regression \eqref{eq: AKM model differences simplified}, the equivalence between
(i) and (iii) need not hold if $\widetilde{\alpha}$ is replaced by
$\alpha$ and $\operatorname{DID}(o,d)$ by
$E[\Delta \mid O=o,D=d]$. Our adjustments allow us to isolate the
restrictions across mobility paths required for the descriptive
interpretation, while allowing stayer wage trends to differ freely
across firms.
}

\begin{theorem}[descriptive interpretation conditions]
\label{thm: did characterization}
Suppose Assumption~\ref{ass: invertibility of A} holds. Then, the following are equivalent:
\begin{enumerate}[(i)]
    \item\label{item: did-recovery} The adjusted AKM estimand satisfies
    \begin{equation}
    \widetilde \alpha_d-\widetilde \alpha_o=\DID(o,d)
    \qquad
    \text{for all }(o,d)\in\mathcal E.
    \label{eq: did-recovery}
    \end{equation}

    \item\label{item: did-path-independence} For every supported move $(o,d)\in\mathcal E$ and every signed path $W$ from $o$ to $d$ in the graph $G^{d}$,
    \begin{equation}
        S(W,\DID)=\DID(o,d).
        \label{eq: did-path-independence}
    \end{equation}

    \item\label{item: did-cycles} For every simple signed cycle $C$ in the graph $G^{d}$,
    \begin{equation}
        S(C,\DID)=0.
        \label{eq: did-simple-cycles}
    \end{equation}
\end{enumerate}
\end{theorem}

Part (\ref{item: did-recovery}) states that the pairwise differences in adjusted AKM estimands recover the $\DID(o,d)$ for every supported move. This statement is equivalent to path independence in the sense described in part (\ref{item: did-path-independence}). Path independence means that any two signed paths with the same endpoints have the same signed sum of $\DID$s. Put differently, for all signed paths $W$ from $o$ to $d$, the signed sum $S(W,\DID)$ produces the same value, and this value is equal to $\DID(o,d)$.

The implication from (\ref{item: did-path-independence}) to (\ref{item: did-recovery}) follows directly from Theorem~\ref{thm: akm path decomposition}: the adjusted contrast $\widetilde \alpha_d-\widetilde \alpha_o$ is a positive weighted average of signed sums along signed paths from $o$ to $d$, each of which equals $\DID(o,d)$ by path independence, so the weighted average collapses to $\DID(o,d)$. The implication from (\ref{item: did-recovery}) to (\ref{item: did-path-independence}) follows upon observing that \eqref{eq: did-recovery} implies that the signed path sum $S(W,\DID)$ in \eqref{eq: did-path-independence} telescopes appropriately.  It is worth emphasizing, however, the strength of the implication from (\ref{item: did-recovery}) to (\ref{item: did-path-independence}).  It implies, in particular, that if, by coincidence, there is some pair of firms $(o,d) \in \mathcal E$ for which the weighted average of the signed path sums of the $\DID$s in \eqref{eq:akm-path-representation} equals $\DID(o,d)$ even though each individual signed path sums of the $\DID$s do not equal $\DID(o,d)$, then it must be that $\widetilde \alpha_k-\widetilde \alpha_j$ is not equal to $\DID(j,k)$ for at least one other pair of firms $(j,k) \in \mathcal E$. 

The key insight from the equivalence between (\ref{item: did-recovery}) and (\ref{item: did-path-independence})  is that $\widetilde \alpha_d-\widetilde \alpha_o$ can be interpreted as describing the mean change in wages that accompanies the mobility of workers between $o$ and $d$ for all supported moves if and only if, for every such move, each signed path starting at $o$ and ending at $d$ can be given the same descriptive interpretation. This conclusion holds no matter what firms other than $o$ or $d$ are involved in these paths. These intermediate firms can differ systematically from firms $o$ and $d$ (e.g., by industry) and so can the workers moving between them (e.g., by industry-specific experience).

Part (\ref{item: did-cycles}) restates path independence as a finite collection of restrictions that make no reference to the endpoints of any particular contrast: the signed sum of the $\DID$s in every simple cycle in the mobility network must vanish. Lemma~\ref{lem: potential and cycles} in Appendix \ref{app: proofs exact description} establishes the equivalence and shows the number of restrictions that are non-redundant. This formulation is particularly convenient for testing, as discussed in Section~\ref{sec: testable implications}. 

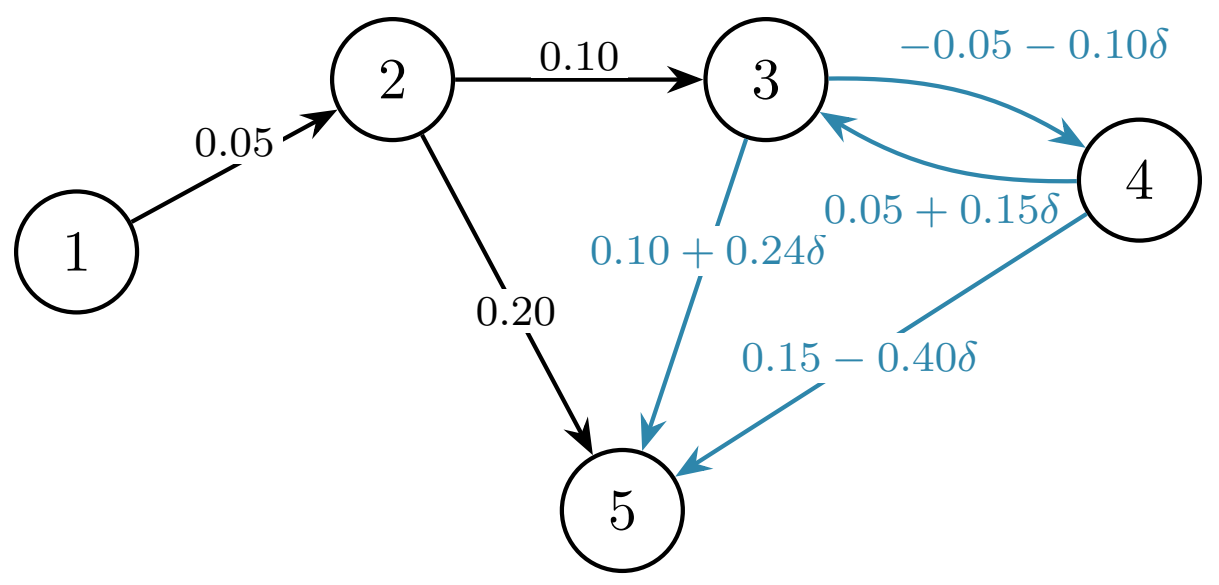
\begin{figure}[t]
\centering
\begin{tikzpicture}[
  scale=0.95,
  nd/.style={circle,draw,thick,minimum size=8mm,inner sep=0pt,fill=white},
  arr/.style={thick,-{Stealth[length=2.4mm]}},
  arrhi/.style={thick,posblue,-{Stealth[length=2.4mm]}},
  lab/.style={font=\scriptsize,inner sep=1.5pt,fill=white},
  labhi/.style={font=\scriptsize,inner sep=1.5pt,fill=white,text=posblue},
]
\node[nd] (n1) at (-3.2, 0.2) {$1$};
\node[nd] (n2) at (-1.0, 1.4) {$2$};
\node[nd] (n3) at ( 1.6, 1.4) {$3$};
\node[nd] (n4) at ( 4.2, 0.7) {$4$};
\node[nd] (n5) at ( 0.6,-1.6) {$5$};
\draw[arr]   (n1) -- node[lab,above] {$0.05$} (n2);
\draw[arr]   (n2) -- node[lab,above] {$0.10$} (n3);
\draw[arr]   (n2) -- node[lab,pos=0.55] {$0.20$} (n5);
\draw[arrhi] (n3) -- node[labhi,pos=0.36] {$0.10+0.24\delta$} (n5);
\draw[arrhi] (n3) to[bend left=16] node[labhi,above,pos=0.62,xshift=8pt,yshift=5.5pt] {$-0.05-0.10\delta$} (n4);
\draw[arrhi] (n4) to[bend left=16] node[labhi,below,yshift=-3pt] {$0.05+0.15\delta$} (n3);
\draw[arrhi] (n4) -- node[labhi,pos=0.55] {$0.15-0.40\delta$} (n5);
\end{tikzpicture}
\caption{The difference-in-differences contrasts of the running example}
\label{fig:did-values}
\figurenote{Arrows show supported moves in the network of Figure~\ref{fig:network}; labels are the corresponding values of $\DID(o,d)$. Blue arrows identify the four contrasts that vary with $\delta$. At $\delta=0$, each contrast equals $a_d-a_o$, where $a=(0,0.05,0.15,0.10,0.25)'$. The perturbation is constructed so that
the AKM estimands equal $a$ for every value of $\delta$.}
\end{figure}

\vspace{-.4cm}
\begin{examplecontinued}{ex: four-firm-illustration}
To illustrate Theorem~\ref{thm: did characterization}, we need to specify values for the $\DID$s. Figure \ref{fig:did-values} shows the chosen values of the $\DID$s for each pairwise difference with a supported move, while Table~\ref{tab:did-values} summarizes all the relevant information about the network of the five firms. The parameter $\delta$ changes the value of a subset of the $\DID$s. It will allow us to govern the violation, if any, of the necessary and sufficient conditions in Theorem~\ref{thm: did characterization}. 

Plugging the $\DID$ values into the running example, we see that 
\begin{align*}
 \widetilde \alpha_5-\widetilde \alpha_2 &= 
\LambdaWOneApprox\cdot \underbrace{S(W_1^{25},\DID)}_{0.20} +
\LambdaWTwoApprox\cdot \underbrace{S(W_2^{25},\DID)}_{0.20+0.24\delta} +
\LambdaWThreeApprox\cdot \underbrace{S(W_3^{25},\DID)}_{0.20-0.55\delta} +
\EtaCOneApprox\cdot \underbrace{S(C_1,\DID)}_{-0.05\delta} \\ &= 
\underbrace{\DID(2,5)}_{ 0.2 },
\end{align*}
from which it follows that the weighted average of the signed path sum equals $\DID(2,5)$, but the individual paths $W_2^{25}$ and $W_3^{25}$ do not have signed path sums equal to this value whenever $\delta$ is non-zero. Hence, part (ii)  of Theorem~\ref{thm: did characterization} is violated whenever $\delta$ is non-zero.  Part (iii) of the theorem is also violated whenever $\delta$ is non-zero since $C_1$ has a signed path sum that is not equal to zero whenever this is the case.  The same is true for other cycles.
For example, consider the cycle that goes through the direct edge $(2,5)$, and then through $(3,5)$ and $(2,3)$ in reverse:
\[
S( (2,+1,5,-1,3,-1,2), \DID ) = \underbrace{\DID(2,5)}_{0.20} - \underbrace{\DID(3,5)}_{0.10+0.24\delta} - \underbrace{\DID(2,3)}_{0.10} = -0.24\delta ,
\]

\begin{table}
\centering
\begin{tabular}{lrrrr}
\toprule
$(o,d)$ & $100\cdot P_{od}$ & $\widetilde \alpha_d-\widetilde \alpha_o$ & $\DID(o,d)$ & deviation \\
\midrule
$(1,2)$ & $5$ & $0.05$ & $0.05$ & $0$ \\
$(2,3)$ & $8$ & $0.10$ & $0.10$ & $0$ \\
$(2,5)$ & $7$ & $0.20$ & $0.20$ & $0$ \\
$(3,4)$ & $6$ & $-0.05$ & $-0.05-0.10\delta$ & $-0.10\delta$ \\
$(3,5)$ & $5$ & $0.10$ & $0.10+0.24\delta$ & $0.24\delta$ \\
$(4,3)$ & $4$ & $0.05$ & $0.05+0.15\delta$ & $0.15\delta$ \\
$(4,5)$ & $3$ & $0.15$ & $0.15-0.40\delta$ & $-0.40\delta$ \\
\bottomrule
\end{tabular}
\caption{Mobility shares, adjusted AKM contrasts, and difference-in-differences contrasts of the running example.}
\figurenote{Each row corresponds to a supported move in Figure~\ref{fig:network}. The table reports population mobility shares in percent, pairwise differences in adjusted AKM estimands, and difference-in-differences contrasts.
The final column reports $\DID(o,d)-(\widetilde\alpha_d-\widetilde\alpha_o)$. The parameter $\delta$ varies the difference-in-differences contrasts while leaving the pairwise differences in adjusted AKM estimands unchanged. At $\delta=0$, the two coincide for every supported move.}
\label{tab:did-values}
\end{table}

Part (\ref{item: did-recovery}) of the proposition tells us that when $\delta$ is non-zero, there has to be at least one pair $(o,d)$ with supported moves for which $\DID(o,d)$ is not accurately measured by $\widetilde \alpha_d-\widetilde \alpha_o$. In our example, the pairwise differences in adjusted AKM estimands fail to describe the mean wage change accompanying moves between 3 and 4, 3 and 5, 4 and 5, and 4 and 3. For example,
\begin{align*}
\widetilde \alpha_5-\widetilde \alpha_3 &=
0.44\cdot \underbrace{S(W^{35}_1,\DID)}_{0.10+0.24\delta} +
0.36\cdot \underbrace{S(W^{35}_2,\DID)}_{0.10} +
0.07\cdot \underbrace{S(W^{35}_3,\DID)}_{0.10-0.50\delta} +
0.12\cdot \underbrace{S(W^{35}_4,\DID)}_{0.10-0.55\delta} \\
& = 0.10 \neq \underbrace{\DID(3,5)}_{0.10+0.24\delta}
\end{align*}

The example underscores the strength of Theorem~\ref{thm: did characterization}: if the ``bias'' in the pairwise differences of AKM estimands coincidentally averages out to zero for one pair of firms, it will necessarily be present in the AKM estimates of one or more other pairs of firms unless the signed path sums of all simple cycles equals zero, which, in this case corresponds to $\delta = 0$.
\end{examplecontinued}

\subsection{A valid test and imperfect diagnostic checks}
\label{sec: testable implications}

\subsubsection{A joint test of all cycle restrictions}
\label{sec: joint cycle test}
The necessary and sufficient conditions in Theorem~\ref{thm: did characterization} have testable implications. The pairwise differences in adjusted AKM estimands can be given a descriptive interpretation as measuring the wage changes accompanying mobility if and only if any two signed paths with the same endpoints have the same signed path sum or, equivalently, if and only if the signed path sum of every simple signed cycle is zero:
\begin{equation}
\label{eq: path-independence-cycles}
S(C,\DID)=0 \text{ for all simple signed cycles } C \text{ in } G^d.
\end{equation}
For each simple signed cycle $C$, the left-hand side of this equality is simply a sum of $\DID$s, and it is therefore possible to test whether it equals zero. However, \eqref{eq: path-independence-cycles} stipulates that this restriction has to hold for all simple signed cycles $C$ in the graph $G^d$. 

Although \eqref{eq: path-independence-cycles} is stated as the signed path sum being zero for every simple signed cycle, Lemma~\ref{lem: potential and cycles} in Appendix \ref{app: proofs exact description} shows that it is, in fact, equivalent to the signed path sum being zero for only a particular subset of $|\mathcal E|-K+1$ such cycles. To construct this collection, choose $K-1$ edges in $\mathcal E$ connecting all $K$ firms when their directions are ignored. The selected edges form a so-called \emph{spanning tree}. For each supported move $(j,k)$ not selected for the spanning tree, traverse that move forward from $j$ to $k$. Both firms belong to the spanning tree, so there is a unique simple signed path in the spanning tree from $k$ back to $j$. Combining the move with this return path gives a simple signed cycle. This process yields one simple signed cycle for each of the $|\mathcal E|-K+1$ moves not selected for the spanning tree. The simple signed cycles obtained in this way are called the \emph{fundamental signed cycles} associated with the chosen spanning tree. Below we illustrate this construction in the context of our running example.

We can test the hypothesis that the signed path sums for all fundamental signed cycles are jointly equal to zero using the approach in \citet{Anatolyev:2023aa}, which was developed for testing many linear restrictions in a heteroskedastic linear regression model. To implement the required variance estimation, we restrict attention to moves observed for at least two workers. The test therefore concerns the cycle restrictions on the resulting mobility network. We apply this test in our empirical analysis in Section~\ref{sec: empirical results}.

\begin{table}
\centering
\begin{tabular}{llr}
\toprule
fundamental signed cycle & moves used & signed path sum \\
\midrule
$(4,+1,3,+1,4)$
& $(4,3),(3,4)$
& $0.05\delta$ \\
$(3,+1,5,-1,2,+1,3)$
& $(3,5),(2,5),(2,3)$
& $0.24\delta$ \\
$(4,+1,5,-1,2,+1,3,+1,4)$
& $(4,5),(2,5),(2,3),(3,4)$
& $-0.50\delta$ \\
\bottomrule
\end{tabular}
\caption{The three fundamental signed cycles and their signed path sums.}
\figurenote{In each row, the first move is not selected for the spanning tree and is traversed forward. The remaining moves provide the return path in the spanning tree. In the first column, $+1$ denotes traversal in the direction of a supported move and $-1$ denotes traversal against that direction. The second column identifies the supported moves in their order of traversal. The final column reports the signed sum of the difference-in-differences contrasts in Table~\ref{tab:did-values}.}
\label{tab:fundamental-cycles}
\end{table}

\vspace{-.4cm}
\begin{examplecontinued}{ex: four-firm-illustration}
As shown in Table~\ref{tab:did-values}, our example has seven supported mover edges and five firms, so the construction gives $7-5+1=3$ fundamental signed cycles. Choose the spanning tree consisting of $(1,2), (2,3), (2,5)$, and $(3,4)$. The supported moves not selected for this spanning tree are $(4,3)$, $(3,5)$, and $(4,5)$. The corresponding fundamental signed cycles and their signed path sums are reported in Table~\ref{tab:fundamental-cycles}. 

By Lemma~\ref{lem: potential and cycles} in Appendix \ref{app: proofs exact description}, \eqref{eq: path-independence-cycles} holds if and only if the three signed path sums in Table~\ref{tab:fundamental-cycles} are zero. The reported values therefore imply that the necessary and sufficient conditions in Theorem~\ref{thm: did characterization} hold in this example if and only if $\delta=0$.
\end{examplecontinued}

\subsubsection{Relationship between our test and event study diagnostic}  
A widely used diagnostic to assess the validity of the AKM estimand is to perform an event study \citep{Card:2013ee}. This diagnostic first assigns each firm to a group, defined by the mean wage of stayers (or some other time-invariant characteristic). Let $G_j$ be the group to which firm $j$ is assigned (e.g., quartiles in the mean wage distribution). Next, the diagnostic compares the mean wage change of moving between group $j$ and $k$ to the mean wage change of moving between group $k$ and $j$:
\begin{align}\label{eq:event-study}
    E[ \Delta | O \in G_j, D \in G_k] + E[ \Delta | O \in G_k, D \in G_j], 
\end{align}
It is argued that if \eqref{eq:event-study} is (not) sufficiently close to zero, then the AKM estimand can(not) be given a descriptive interpretation as measuring the wage changes accompanying mobility between group $j$ and $k$.

To compare the event study diagnostic to our test, it is useful to rewrite \eqref{eq:event-study}. To keep the expression simple, assume no time trend, i.e., $E[\Delta | O = o, D = o] = 0$ for all $o$, and that all $(o,d) \in \mathcal E$ moves have a reciprocal move $(d,o) \in \mathcal E$.  Under these assumptions, \eqref{eq:event-study} can be written as:
\begin{multline}\label{eq:event-study2}
  E\!\big[
    \underbrace{S((O,+1,D,+1,O),\DID)}
    _{= \DID(O,D)+\DID(D,O)}
    \,|\, O\in G_j,D\in G_k
  \big]
  \\
  +\underbrace{
    E[\DID(O,D)\mid O\in G_k,D\in G_j]
    -E[\DID(D,O)\mid O\in G_j,D\in G_k]
  }_{\text{compositional difference}}.
  \end{multline}
The first term averages cycle sums of length two, from $o$ to $d$ to $o$. When the 
conditions in Theorem~\ref{thm: did characterization} hold, then all signed cycle sums in \eqref{eq: did-simple-cycles} are zero, so the first term is zero. However, this first term being zero does not reflect all restrictions in \eqref{eq: did-simple-cycles}: while it reflects the implied symmetry in the $\DID$ of direct moves from $o$ to $d$ compared to $d$ to $o$, it does not capture restrictions stemming from the fact that the AKM estimand is almost entirely constructed from the $\DID$s in the chains of mobility between firms other than $o$ and $d$. Therefore, the event study cannot detect all violations of the conditions in Theorem~\ref{thm: did characterization}. Our test also examines cycles of length greater than two and thereby exploits all restrictions in \eqref{eq: did-simple-cycles}.

The compositional difference term reflects the mean wage change accompanying mobility between the groups $j$ and $k$ compared to the groups $k$ and $j$. This term may or may not be equal to zero when the conditions of Theorem~\ref{thm: did characterization} hold. 

In conclusion, \eqref{eq:event-study} being equal to zero does not imply that the AKM contrasts can be given a descriptive interpretation nor does a violation of that condition imply the reverse. Therefore, the event study is not a valid test of the conditions in Theorem~\ref{thm: did characterization} and, even when the compositional term is zero, the event study cannot detect all violations of the conditions. Appendix~\ref{app: event-study example} illustrates this point in our running example.

\subsubsection{Relationship between our test and a binscatter diagnostic}
\label{sec: binscatter diagnostic}

Another diagnostic compares binned averages of mover wage changes with the corresponding estimates from the AKM regression. \citet{Kline:2024io} reports such a plot for the Italian data and notes that ``the AKM model stipulates that, in the absence of noise, the dots should all lie on the dashed 45 degree line.'' Empirically, the fit is remarkably good, leading him to conclude that ``the AKM model provides a highly informative (albeit imperfect) summary of the expected wage effects of worker mobility.''  

We examine a DID-based version of this diagnostic, relating $\DID(o,d)$ to the adjusted AKM contrast $\widetilde\alpha_d-\widetilde\alpha_o$.\footnote{The edge effects plotted by \citet{Kline:2024io} subtract an estimated common year effect from mean wage changes, whereas our DIDs subtract the corresponding origin-stayer mean wage change. Under the dimension condition below, the same construction applies to the population counterpart of that diagnostic, using bins formed from unadjusted AKM contrasts: it preserves the firm coefficients, the common year coefficient, and every plotted bin average.} Specifically, group the supported mover edges into $B$ nonempty bins using fixed intervals of their adjusted AKM contrasts. Each plotted point reports the within-bin averages of $\widetilde\alpha_d-\widetilde\alpha_o$ and $\DID(o,d)$, weighted by the mobility shares $P_{od}$.

When the conditions in Theorem~\ref{thm: did characterization} hold, these bin averages lie on the 45-degree line. Therefore, departures from that line at the population level imply violations of those conditions. Agreement, however, need not establish that $\DID(o,d)=\widetilde\alpha_d-\widetilde\alpha_o$ on every supported move: positive and negative discrepancies can cancel within bins.

More generally, suppose Assumption~\ref{ass: invertibility of A} holds and let $r:=|\mathcal E|-K+1$ be the number of independent cycle restrictions. Proposition~\ref{prop:binscatter-cycle-blindness-appendix} in Appendix~\ref{app: proofs exact description} shows that, if $r>B$, suitably selected $r-B$ simple signed cycle sums can be assigned arbitrary values simultaneously while leaving the mobility shares, the conditional mean wage changes of stayers, the adjusted AKM estimands, and every plotted bin average unchanged. Agreement of the bin averages therefore neither establishes the cycle restrictions nor bounds the magnitude of their violations. The result concerns population quantities, so the failure does not depend on sampling noise. The dimension condition is sufficient for this failure; Appendix~\ref{app: binscatter example} illustrates the same failure in our running example, where $r=B$.

If each bin contained only one supported move, agreement with the 45-degree line would be equivalent to $\widetilde\alpha_d-\widetilde\alpha_o=\DID(o,d)$ for every $(o,d)\in\mathcal E$, and hence to all the cycle restrictions. Such a partition falls outside the dimension condition above. In practice, however, with few movers or origin-firm stayers, the individual $\DID$s can be imprecisely estimated, making an edge-by-edge plot difficult to assess. The joint test in Section~\ref{sec: joint cycle test} assesses the cycle restrictions while accounting for sampling uncertainty on the network for which the required variances can be estimated.

\subsection{Interpreting and relaxing the conditions}
\label{sec: interpreting conditions}

Theorem~\ref{thm: did characterization} shows that $\widetilde \alpha_d-\widetilde \alpha_o$ can be interpreted as describing the mean change in wages that accompanies the mobility of workers between $o$ and $d$ for every supported move $(o,d)\in\mathcal E$ if and only if, for every such move, each signed path starting at $o$ and ending at $d$ can be given the same descriptive interpretation. We now examine the restrictions on heterogeneity across firms and workers that are needed for this condition of path independence to hold. This examination motivates a natural alternative estimand that relies on weaker assumptions.

Let $c(j)$ describe time-invariant characteristics of firm $j$, while $g(o,d)$ describes characteristics of the set of workers moving from firm $o$ to firm $d$. For example, $c(j)$ could be the industry of firm $j$, and, in the case of gender, $g(o,d)$ could take three values: 0 if all movers are female; 1 if all movers are male; and 2 if the movers are mixed-gender.

Suppose that the conditional expectation of the observed wages as a function of the observed origin and destination firms for movers is given by:
\begin{align}
E[Y_{i1} \mid O_i = o,\, D_i = d] &= \mu_{od} + a_o, \label{eq:y1}\\
E[Y_{i2} \mid O_i = o,\, D_i = d] &= \mu_{od} + a_d
  + \tau + \Gamma\big(g(o,d),\, c(o),\, c(d)\big) , \label{eq:y2}
\end{align}
while the wage changes for stayers is simply assumed to be $\tau$. The difference between $\mu_{od}$ and $\mu_{od'}$ captures the mean gap in observed earnings in the first period between the workers who move from $o$ to $d$ compared to those who move from $o$ to $d'$.
The change in mean wages over time for the movers from $o$ to $d$ is given by $a_d + \Gamma\big(g(o,d),\, c(o),\, c(d)\big) + \tau - a_o$.

The AKM estimand recovers $\DID(o,d)$ for every supported move $(o,d)\in\mathcal E$ if and only if $S(W, \DID) = \DID(o, d)$ for every such move and every signed path $W$ from $o$ to $d$. To ensure that this condition holds, it is sufficient to assume $\Gamma(g,c(o),c(d))=0$ in (\ref{eq:y2}).
To help understand how this assumption restricts the permissible heterogeneity in wage changes across firms and workers, consider the hypothetical paths listed in Table~\ref{tab:paths}.

Each row of the table corresponds to a path $(o,+1,j,+1,d)$, starting at $o$, going through firm $j$ and ending at firm $d$, with $o$ and $d$ being in industry $A$ and $B$, respectively.
The first three columns show how the paths differ: according to the characteristic $c(j)$ of the intermediate firm $j$ (industry $B$ or $C$) and characteristics of the workers moving between the firms, summarized by $g(o,j)$ and $g(j,d)$ (1 if all movers between a given pair of firms are male and 0 if they are all female).
We assume that the direct move $(o,d)$ is made by male workers ($g(o,d) = 1$) and describe, in the fourth column, the discrepancy between $\DID(o,d)$ and the path sums on $(o,+1,j,+1,d)$:
\begin{equation}
\DID(o,j) + \DID(j,d)
= a_d - a_o + \Gamma\big(g(o,j), A, c(j)\big)
+ \Gamma\big(g(j,d), c(j), B\big) .
\label{eq:pathsum-third-firm}
\end{equation}

\noindent The last five columns show whether a given path sum is guaranteed to equal $\DID(o,d)$ under each specification of $\Gamma$.
The specifications combine three restrictions, shown in the bottom rows of the table.

Specification~(2) imposes $\Gamma(g,c,c)=0$ for all worker groups $g$
and industries $c$. Thus, for moves within an industry, mean wage
changes net of the common time trend depend only on the origin and
destination firm effects, with no additional variation across worker
groups. For moves across industries, however, the additional wage
change captured by $\Gamma$ may still vary across both industry pairs
and worker groups. The first path therefore gives
$\operatorname{DID}(o,d)$: its move from industry $A$ to industry $B$
involves workers in the same group as the direct movers, $g=1$, while
its remaining move is within industry $B$ and contributes no additional
$\Gamma$ term, regardless of the workers' group.

Specification~(3) additionally assumes that
$\Gamma(g,c,c')=\Gamma(c,c')$ for all worker groups $g$ and industries
$c,c'$. This rules out differences across worker groups in the
additional wage changes associated with moving between a given pair
of industries.
The second path therefore also gives
$\operatorname{DID}(o,d)$: the $\Gamma$ term for moving from industry
$A$ to industry $B$ is now the same for all groups, so workers making
this move need not belong to the same group as the direct movers.

Specification~(4) retains $\Gamma(g,c,c)=0$ for all worker groups $g$
and industries $c$ but allows the additional wage changes associated
with moving across industries to differ across worker groups.
Instead, it restricts how these changes vary across industry pairs
by requiring that there exist functions $\Gamma_1$ and $\Gamma_2$
such that
$\Gamma(g,c,c')=\Gamma_1(g,c)+\Gamma_2(g,c')$
for all worker groups $g$ and industries $c,c'$. Together, these
restrictions imply
\[
    \Gamma(g,A,C)+\Gamma(g,C,B)=\Gamma(g,A,B)
\]
for every worker group $g$ and all industries $A,B,C$. Thus, within
each group, the additional wage changes from moving from $A$ to $C$
and from $C$ to $B$ sum to the additional wage change from moving
directly from $A$ to $B$. The second path is no longer guaranteed
to give $\operatorname{DID}(o,d)$, because $\Gamma$ may differ across
groups, but the third path does. Both moves involve workers in the
same group as the direct movers, $g=1$, and the terms associated
with industry $C$ cancel because
$\Gamma_1(1,C)+\Gamma_2(1,C)=\Gamma(1,C,C)=0$.

\begin{table}[ht]
\centering
\small
\renewcommand{\arraystretch}{1.3}
\begin{tabular}{@{}cccl ccccc@{}}
\toprule
& & & & \multicolumn{5}{c}{Valid under specification} \\
\cmidrule(lr){5-9}
$g(o,j)$ & $c(j)$ & $g(j,d)$ & $\DID(o,j) + \DID(j,d) - \DID(o,d)$
& (1) & (2) & (3) & (4) & (5) \\
\midrule
1 & $B$ & any & $\Gamma(g(j,d),B,B)$ & $\times$ & \checkmark & \checkmark & \checkmark & \checkmark \\
0 & $B$ & any & $\Gamma(0,A,B) - \Gamma(1,A,B) + \Gamma(g(j,d),B,B)$ & $\times$ & $\times$ & \checkmark & $\times$ & \checkmark \\
\addlinespace
1 & $C$ & 1 & $\Gamma(1,A,C) + \Gamma(1,C,B) - \Gamma(1,A,B)$ & $\times$ & $\times$ & $\times$ & \checkmark & \checkmark \\
1 & $C$ & 0 & $\Gamma(1,A,C) + \Gamma(0,C,B) - \Gamma(1,A,B) $ & $\times$ & $\times$ & $\times$ & $\times$ & \checkmark \\
0 & $C$ & 1 & $\Gamma(0,A,C) + \Gamma(1,C,B) - \Gamma(1,A,B) $ & $\times$ & $\times$ & $\times$ & $\times$ & \checkmark \\
0 & $C$ & 0 & $\Gamma(0,A,C) + \Gamma(0,C,B) - \Gamma(1,A,B) $ & $\times$ & $\times$ & $\times$ & $\times$ & \checkmark \\
\midrule
\multicolumn{9}{@{}l}{\textit{Model specification: restrictions imposed on $\Gamma$}} \\
\multicolumn{4}{l}{$\Gamma(g,c,c\phantom{'})=0$} & --- & \checkmark & \checkmark & \checkmark & \checkmark \\
\multicolumn{4}{l}{$\Gamma(g,c,c')=\Gamma(c,c')$} & --- & --- & \checkmark & --- & \checkmark \\
\multicolumn{4}{l}{$\Gamma(g,c,c')=\Gamma_1(g,c)+\Gamma_2(g,c')$} & --- & --- & --- & \checkmark & \checkmark \\
\bottomrule
\end{tabular}
\caption{Validity of the path $o \to j \to d$ for $\DID(o,d)$, with
$c(o) = A$, $c(d) = B$ and $g(o,d) = 1$, under five specifications of
$\Gamma$. The bottom panel defines each specification by the restrictions it imposes on $\Gamma$: a checkmark indicates that a restriction is imposed and a dash that it is not. Specifications (2)--(5) impose $\Gamma(g,c,c)=0$, so the term $\Gamma(g(j,d),B,B)$ in the first two rows vanishes. Any means the row holds for both groups.}
\label{tab:paths}
\end{table}

Table~\ref{tab:paths} makes clear that choosing which paths to use to estimate the $\DID$ amounts to deciding which restrictions, if any, to impose on $\Gamma$. 
Specification (1) is one extreme: no restrictions are imposed, and none of the paths in the table give the $\DID(o,d)$. 
In practice, however, there are typically too few movers to precisely estimate the $\DID(o,d)$ solely by its sample counterpart. 
The other extreme is specification (5), which justifies the AKM estimand. 
Every path in the table gives the $\DID(o,d)$, but at the cost of strong restrictions on the permissible heterogeneity in wage changes across firms and workers.

Between these extremes lies a range of estimands that uses some, but not all, the paths. For distinct firms $o$ and $d$, fix a finite nonempty set $\mathcal R^{od}$ of distinct signed paths from $o$ to $d$ and define
\begin{equation}
\label{eq: restricted-path-estimand}
\theta_{od}^{\mathcal R}
 :=\frac{1}{|\mathcal R^{od}|}\sum_{W \in \mathcal R^{od}}S(W,\DID).
\end{equation}
For a supported move $(o,d)$, this estimand equals $\DID(o,d)$ whenever the maintained restrictions on $\Gamma$ make each selected path sum equal to the direct comparison, as illustrated in Table~\ref{tab:paths}.  In our empirical application, we restrict the set of paths to those compatible with specification (2), thereby allowing wage changes to differ by groups across industries, but not within industries.
The corresponding estimator replaces each $\DID$ along the selected paths by its sample counterpart and averages the resulting signed path sums. Theorem~\ref{thm: restricted-path-inference} in Appendix~\ref{app: restricted-path-inference} provides conditions for asymptotic normality as the number of workers increases, allowing the number of firms to grow. Importantly, these conditions do not require increasing worker counts at every firm or within every origin--destination cell.

\section{AKM as a Descriptive Tool: Empirical Results}
\label{sec: empirical results}
In this section, we will consider AKM as a descriptive tool in Italian matched employer--employee data.

\subsection{Italian data}
We use the same data and estimation sample as in \citet{Kline:2024io} and \citet{Kline:2020aa}. It is based on the Veneto Workers History data for the years 1999 and 2001. 

As in \citet{Kline:2024io} and \citet{Kline:2020aa}, we use data for the years 1999 and 2001. The estimation sample consists of the subsample of the leave-one-out connected set, which contains 1,320,034 person-year observations. The sample includes 42,492 firms and 660,017 workers. About 75 percent of these workers stay in the same firm in the two years, while around 25 percent switch employer. These movers are switching between 118,948 distinct pairs of origin and destination firms.

Following \citet{Kline:2024io} and \citet{Kline:2020aa}, we estimate the AKM regression model in \eqref{eq: AKM model differences simplified} on the leave-out-connected set. We use the change over time in the (log) daily wage of the worker as the dependent variable. The only time varying covariate is an indicator for the year being 2001. Tables \ref{tab:kss-counts} and \ref{tab:kss-variance} in Appendix \ref{app:data} show that both our estimation sample and AKM estimates are virtually identical to that of \citet{Kline:2024io} and KSS \citep{Kline:2020aa}.

\subsection{Path decomposition}

Table~\ref{tab:path-cross-industry} applies the path decomposition from Theorem~\ref{thm: akm path decomposition} to the Italian data. Panel A uses the leave-out connected set, while panel B considers the connected subset of the largest 5000 firms. Empirical studies using AKM often restrict attention to such large firms, hoping it will sidestep the problem of limited mobility bias \citep[see, e.g.,][]{Song:2019aa,Sorkin:2018aa}. The first three columns of the table sample pairs of firms with sampling weight exactly as in the analyzes of \citet{Kline:2024io} and \citet{Kline:2020aa}, i.e., proportional to the product of the firm sizes in the pair of firms. The last three columns assign zero sampling weight to pairs of firms for which there are no direct moves in the data. By doing so, it focuses attention on pairs of firms for which there exist $\DID$s.

One empirical insight from the first column of panel A is that the adjusted AKM contrasts assign little, if any, weight to direct paths to or from the origin and destination firms. This means that $\tilde \alpha_d - \tilde \alpha_o$ is rarely, if ever, informed by $\DID(o,d)$. Another empirical insight from this column is that the AKM estimand assigns virtually no weight to paths consisting only of firms comparable to the origin and destination firms.
Even if one coarsely groups firms into six different aggregate industries, we find that on average, merely 9.7 percent of the weight is on paths in which all firms are in the same industry as the origin and destination firms.  Table~\ref{tab:industry-counts} in Appendix~\ref{app:data} provides a description of these industries in terms of the number of firms and workers.

The remaining 90.3 percent of the weight is on paths in which one or more firms are in an industry different from the origin and destination firms. Hence, $\tilde \alpha_d - \tilde \alpha_o$ is almost entirely informed by paths that involve firms in industries other than those of firms $o$ and $d$. As shown in the second and third columns, there are many such paths, and they tend to be relatively long. Taken together, they contribute a lot to the AKM estimand. However, each of these individual paths tend to receive less weight than the paths in which the firms are in the same industry as the origin and destination firms.

The finding that the adjusted AKM contrasts assigns almost no weight to direct paths to or from the origin and destination firms has two possible explanations. One possibility is that, for most pairs of firms, there are no direct moves. Another possibility is that, even when there are direct moves, the AKM estimand assigns most weight to other paths. The results in the last three columns sort out these explanations by showing what happens if we assign zero sampling weight to pairs of firms for which there are no direct moves in the data. By construction, the AKM estimand then assigns more weight to direct paths between the origin and destination firms. However, it continues to assign the vast majority of the weight to paths in which one or more firms may be hard to compare to the origin and destination as they are in different industries.

In Panel B, we consider the connected subset of the 5,000 largest firms. Although this restriction may help reduce limited mobility bias, the average direct-path weight among sampled firm pairs with an observed direct-path does not materially change. 

\begin{table}[htbp]
\centering
\caption{Industry composition of the AKM path decomposition}
\label{tab:path-cross-industry}
\begingroup
\small
\setlength{\tabcolsep}{4pt}
\renewcommand{\arraystretch}{1.08}
\sisetup{group-separator={,},group-minimum-digits=4}

\begin{tabular*}{\textwidth}{@{\extracolsep{\fill}}lrrrrrr@{}}
\toprule
 & \multicolumn{3}{c}{All firm pairs}
 & \multicolumn{3}{c}{Pairs with a direct move} \\
\cmidrule(lr){2-4}\cmidrule(lr){5-7}
Path category
 & Weight & $\bar L$ & $\bar n$
 & Weight & $\bar L$ & $\bar n$ \\
\midrule

\multicolumn{7}{@{}l}{
  \emph{Panel A. Full sample}
} \\[3pt]

Direct path
 & 0.000 & 1.00 & 0.01
 & 0.048 & 1.00 & 1.0 \\

Reverse-direct path
 & 0.000 & 1.00 & 0.00
 & 0.020 & 1.00 & 0.35 \\

Indirect paths within endpoint industries
 & 0.097 & 10.91 & \num{3153}
 & 0.205 & 8.10 & \num{3151} \\

Indirect paths involving other industries
 & 0.903 & 13.64 & \num{73091}
 & 0.727 & 11.34 & \num{73060} \\

\cmidrule(lr){1-7}
All paths
 & 1.000 & 13.52 & \num{76244}
 & 1.000 & 11.20 & \num{76212} \\

\addlinespace[3pt]
Number of firm pairs
 & \multicolumn{3}{c}{100}
 & \multicolumn{3}{c}{100} \\

Mean cycle mass
 & \multicolumn{3}{c}{0.0003}
 & \multicolumn{3}{c}{0.0002} \\

\midrule

\multicolumn{7}{@{}l}{
  \emph{Panel B. Large-firm subsample}
} \\[3pt]

Direct path
 & 0.000 & 1.00 & 0.00
 & 0.080 & 1.00 & 1.0 \\

Reverse-direct path
 & 0.000 & 1.00 & 0.00
 & 0.019 & 1.00 & 0.25 \\

Indirect paths within endpoint industries
 & 0.149 & 8.68 & \num{1156}
 & 0.265 & 6.88 & \num{1134} \\

Indirect paths involving other industries
 & 0.851 & 10.35 & \num{16278}
 & 0.637 & 8.95 & \num{16285} \\

\cmidrule(lr){1-7}
All paths
 & 1.000 & 10.23 & \num{17434}
 & 1.000 & 8.82 & \num{17420} \\

\addlinespace[3pt]
Number of firm pairs
 & \multicolumn{3}{c}{100}
 & \multicolumn{3}{c}{100} \\

Mean cycle mass
 & \multicolumn{3}{c}{0.0001}
 & \multicolumn{3}{c}{0.0001} \\

\bottomrule

\end{tabular*}
\endgroup

\figurenote{Panel A uses the full leave-one-out (LOO) connected set in the 1999--2001 Italian data. Panel B uses the largest connected subset among the 5,000 largest firms in that set, ranked by person-year counts. Both column groups sample distinct firm pairs with weights proportional to $T_oT_d$, where $T_j$ is firm $j$'s person-year count. The second column group restricts eligibility to pairs with an observed move from $o$ to $d$. Direct and reverse-direct paths use $(o,d)$ forward and $(d,o)$ backward, respectively. Indirect paths are within endpoint industries if every intermediate firm belongs to either endpoint's industry; otherwise they involve other industries.
Weight and $\bar n$ report the mean category path weight and path count across sampled pairs. Within each path category, $\bar L$ averages pair-specific mean path lengths over firm pairs with at least one such path; the ``All paths'' row combines these category means using the corresponding mean path counts as weights.
For each pair, the table reports the path representation that places the least weight on paths involving other industries; the weights are those of this representation, while $\bar L$ and $\bar n$ describe one such representation.}
\end{table}

In Table~\ref{tab:firm-pair-premium}, we present and decompose the AKM estimates of a few randomly drawn pairs of large firms from the connected set of firms. The goal is to empirically illustrate how firm effects are measured in the data. We focus on pairs of firms for which there exist direct $\DID$s.

\begin{table}
\centering
\caption{Decomposition of AKM contrasts for selected firm pairs}
\label{tab:firm-pair-premium}
\begingroup
\footnotesize
\setlength{\tabcolsep}{3pt}
\renewcommand{\arraystretch}{1.12}

\begin{tabular*}{\textwidth}{@{\extracolsep{\fill}}rrrrrrrrrrrrrr@{}}
\toprule
 & \multicolumn{3}{c}{\shortstack{AKM contrast\\and components}}
 & \multicolumn{2}{c}{\shortstack{Direct\\path}}
 & \multicolumn{2}{c}{\shortstack{Reverse-direct\\path}}
 & \multicolumn{2}{c}{\shortstack{Indirect paths\\within endpoint\\industries}}
 & \multicolumn{2}{c}{\shortstack{Indirect paths\\involving other\\industries}}
 & \multicolumn{2}{c}{\shortstack{Restricted-\\path\\estimates}} \\
\cmidrule(lr){2-4}\cmidrule(lr){5-6}\cmidrule(lr){7-8}
\cmidrule(lr){9-10}\cmidrule(lr){11-12}\cmidrule(lr){13-14}
$\Delta\log w$
 & $\widehat\alpha_d-\widehat\alpha_o$
 & $\Delta\tau$
 & $\Delta\widetilde\alpha$
 & $\lambda$ & $\bar v$
 & $\lambda$ & $\bar v$
 & $\lambda$ & $\bar v$
 & $\lambda$ & $\bar v$
 & $\theta^{\mathcal R}_{od}$ & $\theta^{\mathcal R,g}_{od}$ \\
\midrule

$+0.875$ & $+0.509$ & $-0.305$ & $+0.814$
 & 0.041 & $+0.390$ & --- & ---
 & 0.168 & $+0.787$ & 0.791 & $+0.842$ & $+0.690$ & $+0.728$ \\

$+0.538$ & $+0.223$ & $-0.008$ & $+0.232$
 & 0.042 & $+0.353$ & --- & ---
 & 0.037 & $+0.317$ & 0.921 & $+0.223$ & $+0.226$ & $+0.165$ \\

$+0.084$ & $+0.115$ & $-0.018$ & $+0.133$
 & 0.048 & $+0.024$ & 0.024 & $+0.036$
 & 0.090 & $+0.137$ & 0.837 & $+0.142$ & $+0.091$ & $+0.024$ \\

$+0.221$ & $+0.091$ & $-0.035$ & $+0.126$
 & 0.093 & $+0.269$ & --- & ---
 & 0.163 & $+0.209$ & 0.743 & $+0.090$ & $+0.276$ & $+0.276$ \\

$+0.057$ & $+0.041$ & $-0.037$ & $+0.078$
 & 0.055 & $+0.459$ & --- & ---
 & 0.072 & $+0.020$ & 0.873 & $+0.059$ & $+0.076$ & $+0.076$ \\

$+0.088$ & $+0.036$ & $-0.011$ & $+0.047$
 & 0.254 & $+0.016$ & --- & ---
 & 0.265 & $+0.149$ & 0.481 & $+0.007$ & $+0.117$ & $+0.117$ \\

$-0.044$ & $+0.040$ & $+0.001$ & $+0.039$
 & 0.162 & $-0.008$ & --- & ---
 & 0.439 & $+0.003$ & 0.399 & $+0.098$ & $-0.078$ & $-0.078$ \\

$-0.489$ & $-0.151$ & $-0.037$ & $-0.114$
 & 0.040 & $-0.431$ & --- & ---
 & 0.190 & $-0.002$ & 0.770 & $-0.125$ & $-0.146$ & $-0.146$ \\

$+0.076$ & $-0.154$ & $+0.030$ & $-0.184$
 & 0.088 & $-0.361$ & --- & ---
 & 0.188 & $-0.280$ & 0.724 & $-0.138$ & $-0.001$ & $+0.088$ \\

$-0.116$ & $-0.302$ & $-0.010$ & $-0.292$
 & 0.086 & $-0.702$ & --- & ---
 & 0.460 & $-0.162$ & 0.453 & $-0.346$ & $-0.199$ & $-0.199$ \\

\bottomrule
\end{tabular*}
\endgroup

\figurenote{Each row reports a randomly sampled pair of large firms, $o$ and $d$, from the connected set of firms, with an observed move from $o$ to $d$ and stayers at both firms. $\Delta\log w$ is the destination-minus-origin difference in mean log wages. The estimated AKM contrast satisfies $\widehat\alpha_d-\widehat\alpha_o
=\Delta\tau+\Delta\widetilde\alpha$, where
$\Delta\tau=\widehat\tau_d-\widehat\tau_o$ is the estimated time-trend difference and $\Delta\widetilde\alpha$ is the estimated adjusted AKM contrast. Direct and reverse-direct paths use $(o,d)$ forward and $(d,o)$ backward, respectively. Indirect paths are within endpoint industries if every intermediate firm belongs to either endpoint's industry; otherwise they involve other industries. For each category, $\lambda$ is the total path weight and $\bar v$ is the path-weighted mean of the signed path sums of $\DID$s. The restricted-path estimates are the estimand in \eqref{eq: restricted-path-estimand}: the simple, unweighted average of the signed path sums of $\DID$s over a list of paths from the decomposition. For $\theta^{\mathcal R}_{od}$, the list contains the paths that stay within the endpoint industries and cross from the origin industry to the destination industry at most once, through a move traversed forward. For $\theta^{\mathcal R,g}_{od}$, that crossing must also be made by workers in the same gender group as the direct move from $o$ to $d$ (all female, all male, or mixed).
}
\end{table}

The first two columns report the observed difference in mean log wages between the destination and origin firms and the AKM estimate of $\alpha_d-\alpha_o$. The next two columns decompose the AKM estimate into estimates of its components: the time-trend difference $\tau_d-\tau_o$ and the adjusted AKM contrast $\widetilde\alpha_d-\widetilde\alpha_o$. In the following eight columns, we report $\lambda$, the total path weight, and $\bar v$, the path-weighted mean of the signed path sums of the $\DID$s for each of four path categories. The categories are direct paths, which traverse $(o,d)$ forward; reverse-direct paths, which traverse $(d,o)$ backward; indirect paths consisting only of firms in the industries of the origin or destination firms; and indirect paths involving firms in other industries.

The results show that, for most of these firm pairs, the majority of the path weight is on indirect paths involving firms in industries other than those of the origin and destination firms. The path-weighted means of the signed path sums of DIDs for these paths often differ substantially from those of the other path categories. In the absence of noise, such differences would violate the path-independence condition in Theorem~\ref{thm: did characterization} and imply that the adjusted AKM contrasts could not reproduce the corresponding DIDs simultaneously on all supported moves.

This finding raises concerns that previous studies giving AKM estimates such a descriptive interpretation may have drawn misleading conclusions. For example, several studies directly compare the beliefs of workers in a given firm about wage changes accompanying mobility to AKM estimates of the wage premium of the same firm \citep[see, e.g.,][]{Jaeger:2024aa}. Such comparisons only make sense if the AKM estimate actually describes the wage changes that accompany the mobility to or from that firm. Our findings suggest that it may not. Our results also cast doubt on whether it is sensible to add AKM estimates to the set of actual characteristics of the firms (such as industry or region), as done in German administrative data \citep{Bellmann:2020aa}. The problem is that an AKM estimate of a given firm may actually reflect the wage premiums of other firms.

Of course, individual AKM estimates estimates are noisy, and the results in Table~\ref{tab:firm-pair-premium} should be interpreted accordingly. It could be that different paths do not produce significantly different values.  To address this concern, we apply our statistical test of the cycle condition in \eqref{eq: path-independence-cycles} to the leave-one-out connected set used in Panel A of Table \ref{tab:path-cross-industry}.  We reject the null hypothesis that the signed sum in each of these cycles is zero at a significance level of 1 percent ($p$-value of 0.007). We therefore conclude that the AKM estimand cannot be given a descriptive interpretation as measuring the wage changes that accompany the mobility of workers between firms.

Given the rejection, it is natural to consider our weaker restricted path estimator. The last two columns of Table \ref{tab:firm-pair-premium} present the results from this estimator. The first of these columns invokes the assumption that $\Gamma(g,c(o),c(d))$ in \eqref{eq:y2} depends on the origin and destination industries jointly, but not on $g$, as in specification (3) of Table \ref{tab:paths}. We can then use any signed path that crosses industry exactly once, along a supported move traversed forward, and otherwise stays within an industry. In the last column, we let the gain from changing industry in addition differ by gender, as in specification (2) of Table \ref{tab:paths}. Then it is necessary to restrict the set of paths further, to those for which the single crossing is be made by workers in the same gender group as the direct move. The estimates are broadly similar across the two last columns, but often materially different from the adjusted AKM estimates. However, all these estimates are noisy and should be interpreted as an empirical illustration, as opposed to conclusive evidence about wages changes accompanying mobility between a specific pair of firms. 

\section{AKM for Causal Inference: Theory}
\label{sec: causal interpretation}
So far, we have focused on the anatomy of the AKM estimand and the conditions for it to be given a descriptive interpretation. We now shift attention to the additional assumptions needed for the AKM estimand to also be given a causal interpretation. We further show that these additional assumptions do not have testable implications, in contrast to the necessary and sufficient conditions for the AKM estimand to accurately describe the wage changes accompanying worker mobility.\footnote{Some studies \citep[e.g.][]{Sorkin:2026aa} seek to test these assumptions by using additional data, such as instruments or observed worker characteristics. Of course, a rejection could then reflect instrument invalidity or restrictions on how observed heterogeneity enters the potential outcome model.} 

\subsection{Potential outcomes and target parameters}
\label{sec: potential outcomes and reweightings}

\noindent{\bf Potential outcomes.} Define potential outcomes $Y_1(o)$, the log earnings of a worker in period $t=1$ employed at firm $o$, and $Y_2(o,d)$, the log earnings of a worker in period $t=2$, employed at firm $o$ in period $t=1$ and at firm $d$ in period $t=2$. The observed outcomes are the potential outcomes at the realized firms, $Y_1=Y_1(O)$ and $Y_2=Y_2(O,D)$. Differences of potential outcomes over time are denoted by $\Delta(o,d) := Y_2(o,d) - Y_1(o)$, so that $\Delta=\Delta(O,D)$. 

\noindent{\bf Target parameters.} Every target parameter we consider is a weighted average of the conditional average treatment effect of moving from $o$ to $d$ for individuals who moved from $j$ to $k$ or stayed at $j$: 
\begin{equation}
\ATE(o,d ; w):=\sum_{(j,k)\text{ s.t. }P_{jk} > 0 }w_{jk}\, E\bigl[Y_2(o,d)-Y_2(o,o)\bigm| O=j,D=k\bigr] ,
\label{eq: target definition}
\end{equation}
where the weights $w_{jk}$ are non-negative and sum to one over $(j,k)$ with $P_{jk}>0$.

Different weights correspond to different target parameters. One target parameter is the average effect of treatment on the treated (ATT) of moving from $o$ to $d$ for workers making this move, given by $\ATE(o,d; w^{\text{ATT}(o,d)})$ where $w^{\text{ATT}(o,d)}_{jk}:=\ind[j=o,k=d]$.  Another target parameter of interest is the average treatment effect (ATE) of moving from $o$ to $d$ for all movers and stayers, which is defined as $\ATE(o,d; w^\text{ATE})$ where $w_{jk}^\text{ATE} = P_{jk}$. We also consider the average effect of the treatment of moving from $o$ to $d$ for all movers (independently of the firms they moved between), given by $\ATE(o,d; w^\text{ATM})$ where $w_{jk}^\text{ATM}=P_{jk} \ind[ j \neq k] / \sum_{(u,v) \in \mathcal{E}} P_{uv}$.

It will also be useful to consider target parameters that assign more or less weight to certain $(o,d)$ pairs of firms based on some observable (time-invariant) firm characteristic (e.g., industry). Reweighting is governed by the choices of $q_{od}\geq 0$ for all $o,d$. The $q$-specific target parameters can then be defined as $\ATE(o,d; w^\text{ATE}(q))$ where $w_{jk}^\text{ATE}(q) = q_{jk} \cdot P_{jk} / \sum_{(u,v)} q_{uv} P_{uv}$, and  $\ATE(o,d; w^\text{ATM}(q))$ where $w_{jk}^\text{ATM}(q)=q_{jk} P_{jk} \ind[ j \neq k] / \sum_{(u,v) \in \mathcal{E}} q_{uv} P_{uv}$.

\noindent{\bf $q$-weighted AKM estimand.} The $q$-weighted AKM estimand is the ordinary least squares estimand $(\alpha_2(q),\ldots,\alpha_K(q))'$ associated with equation~\eqref{eq: AKM model differences simplified} under the reweighted distribution. Our choice of firm $1$ as the reference firm again implies the normalization $\alpha_1(q)=0$. These coefficients are determined jointly with the time trend $\beta(q)$ by the following equations:
\begin{equation}\label{eq: exogenous mobility q}
E_q\left[
\left(
\Delta-\beta(q)-\sum_{j=2}^{K}\alpha_j(q)
\bigl(\ind\{D=j\}-\ind\{O=j\}\bigr)
\right)
\begin{pmatrix}
1\\ J-I
\end{pmatrix}
\right]=0,
\end{equation}
where the expectation $E_q$ is taken with respect to $P_{od}$ reweighted by $q_{od}$.
For the $q$-weighted AKM estimand to be well defined, we assume the regressors are not perfectly collinear after reweighting, i.e., that the matrix
\[
\mathbb A_q := E_q \left[ \begin{pmatrix} 1\\ J-I \end{pmatrix}\begin{pmatrix} 1\\ J-I \end{pmatrix}' \right]
\]
is invertible.  Let $\tau_j(q)$ denote the trend term in \eqref{eq: tau definition} evaluated under the
reweighted distribution. Define
\[
    \widetilde{\alpha}_j(q)
    := \alpha_j(q) - \tau_j(q),
    \qquad j=2,\ldots,K,
\]
with $\widetilde{\alpha}_1(q)=\tau_1(q)=0$.

\subsection{Necessary and sufficient conditions}
\label{sec: ate characterization}

Throughout this section, we will impose the following common trend assumption:

\begin{assumption}[common trend]\label{ass: common trend}
	(i) $E|\Delta(o,d)|<\infty$ for all $o,d\in\{1,\ldots,K\}$. (ii) For all $(o,d)\in\mathcal E$, $$E[\Delta(o,o) | O=o,D=d] = E[\Delta(o,o)|O=o,D=o].$$
\end{assumption}

\noindent Together with Assumption \ref{ass: invertibility of A}(ii)-(iii),  Assumption \ref{ass: common trend} implies that $\DID(o,d)$ recovers the average effect of moving from $o$ to $d$ for the workers making this move:
\begin{equation}
\DID(o,d)=E\bigl[Y_2(o,d)-Y_2(o,o)\bigm| O=o,D=d\bigr]=:\ATT(o,d)
\qquad\text{for all }(o,d)\in\mathcal E .
\label{eq: did equals att}
\end{equation}
Under the common trend assumption, we can obtain representations of the adjusted AKM contrasts. Specifically, replacing $\DID(o,d)$ by $\ATT(o,d)$ in \eqref{eq: adjusted AKM kappa contrast representation} yields
\begin{equation}\label{eq: adjusted AKM kappa contrast representation ATT}
  \widetilde\alpha_d - \widetilde\alpha_o = \sum_{(j,k) \in \mathcal E} \kappa^{od}_{jk} \ATT(j,k),
\end{equation}
so that the adjusted AKM contrast can be interpreted as a linear combination of the $\ATT$s for the supported moves. Since, on the right-hand side, not only $\ATT(o,d)$ but also all other $\ATT(j,k)$, $(j,k)\in\mathcal E$, appear, this is similar to other ``contamination bias'' results in the literature (e.g., \cite{Hull:2018pd}, \cite{Chaisemartin:2023oo}, \cite{Goldsmith:2024io}).\footnote{\citet[Appendix~B, equation~(B.4)]{Goldsmith:2024io} provide a related decomposition for a finite population of workers with fixed firm assignments and stochastic potential outcomes. For the regression considered here, their assumptions on potential outcomes rule out dependence of $Y_2(o,d)$ on $o$ and impose common trends in potential outcomes at a common reference firm. Under these assumptions, their two-period decomposition expresses each firm coefficient as a linear combination of expected treatment effects for individual workers in both periods, measured relative to the reference firm. Interpreting $P_{jk}$ as the corresponding worker shares, our weights $\omega_{\ell,jk}$ aggregate the regression weights underlying their decomposition over workers with origin $j$ and destination $k$.}

Similarly, denoting by $\ATT\in\mathbb R^{|\mathcal E|}$ the vector with $e$-th element equal to $\ATT(o,d)$ for $e=(o,d)\in\mathcal E$, replacing $\DID(o,d)$ by $\ATT(o,d)$ in \eqref{eq:akm-path-representation} yields
\begin{equation}\label{eq:akm-path-representation ATT}
  \widetilde\alpha_d - \widetilde\alpha_o = \sum_{W\in\mathcal W^{od}} \lambda^{od}(W)S(W,\ATT),
\end{equation}
so that the adjusted AKM contrast can also be interpreted as a weighted average of sums of ATTs along signed paths.

Under the common trend assumption, the pairwise differences in adjusted AKM estimands identify the $\ATT$s if and only if the signed path sum of every signed cycle is zero, so the adjusted AKM estimand recovers the $\DID$s for every supported move; see Corollary~\ref{cor: appendix-att-recovery} in Appendix \ref{app: proofs section causal} for a formal statement. Although it is possible to empirically verify whether the adjusted AKM estimand recovers $\DID$s, the common trend assumption has no testable implications. Proposition~\ref{prop: appendix-observable-compatibility} in Appendix \ref{app: proofs section causal} shows, in particular, that every joint distribution of $(Y_1,Y_2,O,D)$ satisfying the mild regularity conditions in Assumption \ref{ass: invertibility of A}(ii)-(iii) is compatible with a joint distribution for $((Y_2(o,d),Y_1(o)) : 1 \leq o,d \leq K, O,D)$ that satisfies the common trend assumption.  Therefore, a causal interpretation of the adjusted AKM estimand as measuring $\ATT$s is necessarily based on conviction, not data.

Empirical studies that use AKM for causal inference rarely seek to recover ATTs, instead focusing on the so-called firm effects, which are typically defined as the effects of moving from firm $j$ to $k$ for a randomly drawn worker. In Theorem~\ref{prop:local-ate-recovery-supported-moves}, we invoke the common trend assumption and consider the additional conditions under which the adjusted AKM estimand can be given a causal interpretation as measuring such ATEs for all firm pairs for which there is a supported move.  In the theorem below, we use $\|q-\mathbf{1}\|_\infty$ to denote $\max_{o,d}|q_{od}-1|$, where $\mathbf{1}$ is the array with all
entries equal to one.

\begin{theorem}[ATE recovery for supported moves]
\label{prop:local-ate-recovery-supported-moves}

Suppose Assumptions~\ref{ass: invertibility of A} and~\ref{ass: common trend} hold. Fix $\rho \in (0,1)$. Then the following statements are equivalent:
\begin{enumerate}[(i)]
  \item\label{item:local-ate-edge-recovery} For every reweighting \(q\) satisfying $\lVert q-\mathbf 1\rVert_\infty<\rho$ and every supported move \((j,k)\in\mathcal E\),
  \[
      \widetilde\alpha_k(q)-\widetilde\alpha_j(q)
      =
      \ATE\bigl(j,k;w^{\mathrm{ATE}}(q)\bigr).
  \]

  \item\label{item:local-ate-edge-conditions} Both of the following conditions hold:
  \begin{enumerate}[(a)]
      \item For every supported move \((j,k)\in\mathcal E\) and every pair \((o,d)\) with \(P_{od}>0\),
      \[
      E\bigl[Y_2(j,k)-Y_2(j,j)\mid O=o,D=d\bigr]
      =
      E\bigl[Y_2(j,k)-Y_2(j,j)\bigr].
      \]

      \item Every signed cycle \(C\) in $G^d$ satisfies
      \[
          S(C,\DID)=0.
      \]
  \end{enumerate}
\end{enumerate}
\end{theorem}

It is useful to first discuss the sufficiency argument of the theorem. If conditions (a) and (b) in part (ii) of Theorem \ref{prop:local-ate-recovery-supported-moves} hold, then the theorem implies that it is possible to identify the ATEs for all pairs of firms for which there is a supported move in any population of stayers and movers reweighted by $q$. The possible reweightings include, of course, the unweighted population for which $w_{jk}$ equals $P_{jk}$. Conditions (a) and (b) in part (ii) are therefore sufficient for the adjusted AKM estimand $\widetilde\alpha_k-\widetilde\alpha_j$ to recover $E\bigl[Y_2(j,k)-Y_2(j,j)]$ for all firms $j,k$ with a supported move.

Condition (a) in part (ii) of Theorem \ref{prop:local-ate-recovery-supported-moves} is an exogeneity assumption. It states that the average causal effect of moving between any pair of firms $j$ to $k$ is the same for all groups of movers and stayers, regardless of which firms, if any, they moved between. Proposition~\ref{prop: appendix-observable-compatibility} in Appendix \ref{app: proofs section causal} shows that this additional condition also has no testable implications: every joint distribution of $(Y_1,Y_2,O,D)$ satisfying the mild regularity conditions in Assumption \ref{ass: invertibility of A}(ii)-(iii) is compatible with a joint distribution for $((Y_2(o,d),Y_1(o)) : 1 \leq o,d \leq K, O,D)$ that satisfies not only the common trend assumption, but also condition (a) in part (ii) of Theorem \ref{prop:local-ate-recovery-supported-moves}.  As previously noted, condition (b) in part (ii) of Theorem \ref{prop:local-ate-recovery-supported-moves} is, in contrast, testable.
Under condition (b), the adjusted AKM contrasts equal $\DID$s, which equal $\ATT$s under the common trend assumption; under condition (a), the $\ATT$s equal $\ATE$s.

The necessity argument of the theorem is more involved, and it is the reason why part (i) of Theorem \ref{prop:local-ate-recovery-supported-moves} is stated for every reweighting $q$ satisfying $\lVert q-\mathbf 1\rVert_\infty<\rho$. The issue is that part (i) does not necessarily imply part (ii) if one only considers a single population of movers and stayers reweighted by $q$. Consider, for example, the unweighted population for which $w_{jk}$ equals $P_{jk}$. For any value of $E\bigl[Y_2(j,k)-Y_2(j,j)]$, which is recovered by $\widetilde\alpha_k-\widetilde\alpha_j$ under part (i), there exist infinitely many combinations of $E\bigl[Y_2(j,k)-Y_2(j,j)\mid O=o,D=d\bigr]$, violating condition (a) in part (ii) of Theorem \ref{prop:local-ate-recovery-supported-moves}.  The theorem implies, however, that if one also considers populations of movers and stayers that are reweighted by $q$ satisfying $\lVert q-\mathbf 1\rVert_\infty<\rho$ for a fixed, but arbitrarily small, $\rho \in (0,1)$, then conditions (a) and (b) are not only sufficient, but also necessary.  In this sense, the theorem provides the necessary and sufficient conditions to ensure a causal interpretation of the adjusted AKM contrast in terms of ATEs that is robust to arbitrarily small changes in the composition of movers and stayers in the connected set.

\vspace{-.4cm}
\begin{examplecontinued}{ex: four-firm-illustration}
To illustrate Theorem~\ref{prop:local-ate-recovery-supported-moves}, we need to specify values for the conditional average treatment effects, $E\bigl[Y_2(j,k)-Y_2(j,j)\bigm| O=o,D=d\bigr]$. The first panel of Table~\ref{tab:ate-supported-values} displays these values for firm pairs for which there are supported moves. For these firms, $\DID$s are defined and equal to $\ATT$s under the assumption of common trend. They are displayed in bold on the diagonal of the table. The off-diagonal elements contain the conditional average treatment effects of moving from $j$ to $k$ for individuals who moved from $o$ to $d$. The second and third panels report the implied values for the unweighted and $q$-weighted ATEs and adjusted AKM contrasts.

There are two parameters in the table, $\delta$ and $\gamma$, which allow us to govern the violations, if any, of the necessary and sufficient conditions in Theorem~\ref{prop:local-ate-recovery-supported-moves}. The parameter $\delta$ changes the value of the conditional average treatment effects of moving from $j$ to $k$ in the same way, if at all, across all groups of movers $(o,d)$. If $\delta$ is non-zero, then, as illustrated in Section \ref{sec:neccsuff}, condition (b) in part (ii) fails. The parameter $\gamma$ creates a gap (of $\pm \gamma$) between the unconditional and the conditional average treatment effects of moving from $2$ to $3$, for individuals who moved from $1$ to $2$ and those who moved from $3$ to $5$. Therefore, condition (a) in part (ii) fails when $\gamma$ is non-zero.

\begin{table}
\centering
\caption{Conditional ATEs for the supported target moves, with the average effects implied by the original mobility distribution and by a nearby reweighting.}
\label{tab:ate-supported-values}
\begingroup
\footnotesize
\setlength{\tabcolsep}{1.4pt}
\medmuskip=1mu
\renewcommand{\arraystretch}{1.12}
\begin{tabular*}{\textwidth}{@{\extracolsep{\fill}}cccccccc@{}}
\toprule
& \multicolumn{7}{c}{$(j,k)$} \\
\cmidrule(lr){2-8}
$(o,d)$
& $(1,2)$ & $(2,3)$ & $(2,5)$ & $(3,4)$ & $(3,5)$ & $(4,3)$ & $(4,5)$ \\
\midrule
$(1,2)$ & $\bm{0.05}$ & $0.10+\gamma$ & $0.20$ & $-0.05-0.10\delta$ & $0.10+0.24\delta$ & $0.05+0.15\delta$ & $0.15-0.40\delta$ \\
$(2,3)$ & $0.05$ & $\bm{0.10}$ & $0.20$ & $-0.05-0.10\delta$ & $0.10+0.24\delta$ & $0.05+0.15\delta$ & $0.15-0.40\delta$ \\
$(2,5)$ & $0.05$ & $0.10$ & $\bm{0.20}$ & $-0.05-0.10\delta$ & $0.10+0.24\delta$ & $0.05+0.15\delta$ & $0.15-0.40\delta$ \\
$(3,4)$ & $0.05$ & $0.10$ & $0.20$ & $\bm{-0.05-0.10\delta}$ & $0.10+0.24\delta$ & $0.05+0.15\delta$ & $0.15-0.40\delta$ \\
$(3,5)$ & $0.05$ & $0.10-\gamma$ & $0.20$ & $-0.05-0.10\delta$ & $\bm{0.10+0.24\delta}$ & $0.05+0.15\delta$ & $0.15-0.40\delta$ \\
$(4,3)$ & $0.05$ & $0.10$ & $0.20$ & $-0.05-0.10\delta$ & $0.10+0.24\delta$ & $\bm{0.05+0.15\delta}$ & $0.15-0.40\delta$ \\
$(4,5)$ & $0.05$ & $0.10$ & $0.20$ & $-0.05-0.10\delta$ & $0.10+0.24\delta$ & $0.05+0.15\delta$ & $\bm{0.15-0.40\delta}$ \\
\midrule
\(\ATE\bigl(j,k;w^{\mathrm{ATE}}\bigr)\)
& $0.05$ & $0.10$ & $0.20$ & $-0.05-0.10\delta$ & $0.10+0.24\delta$ & $0.05+0.15\delta$ & $0.15-0.40\delta$ \\
\(\widetilde\alpha_k-\widetilde\alpha_j\)
& $0.05$ & $0.10$ & $0.20$ & $-0.05$ & $0.10$ & $0.05$ & $0.15$ \\
\midrule
\(\ATE\bigl(j,k;w^{\mathrm{ATE}}(q)\bigr)\)
& $0.05$ & $0.10-0.01\gamma$ & $0.20$ & $-0.05-0.10\delta$ & $0.10+0.24\delta$ & $0.05+0.15\delta$ & $0.15-0.40\delta$ \\
\(\widetilde\alpha_k(q)-\widetilde\alpha_j(q)\)
& $0.05-0.001\delta$ & $0.10-0.010\delta$ & $0.20+0.009\delta$ & $-0.05+0.005\delta$ & $0.10+0.019\delta$ & $0.05-0.005\delta$ & $0.15+0.015\delta$ \\
\bottomrule
\end{tabular*}
\endgroup

\figurenote{Columns index target moves $(j,k)$, and the first seven rows index workers' observed origin--destination cells $(o,d)$. Entries in the first panel are the conditional effects $E[Y_2(j,k)-Y_2(j,j)\mid O=o,D=d]$. Bold entries correspond to workers observed making the target move; under common trends, these equal the associated $\DID$ contrasts. Off-diagonal effects are specified for the illustration and are not identified by these comparisons. The ATE rows average over all twelve supported mover and stayer cells using weights $P_{od}$ in the second panel and $q_{od}P_{od}/\sum_{uv}q_{uv}P_{uv}$ in the third panel; omitted stayer cells have conditional effects equal to the target-move $\DID$. The adjacent rows report adjusted AKM contrasts under the same weights. The third panel sets $q_{35}=1.2$ and leaves all other values of $q_{od}$ equal to one. The parameters $\delta$ and $\gamma$ vary $\DID$ contrasts and off-diagonal conditional effects, respectively. Under the reweighting $q$, the coefficient on $\gamma$ is exactly $-1/101$; coefficients on $\delta$ are rounded to three decimal places.}
\end{table}

It is evident from the second panel of the table that $\delta=\gamma=0$ is sufficient for the adjusted AKM estimands to recover ATTs that are equal to the ATEs in the unweighted population for all pairs of firms $(j,k)$ with a supported move.  However, it is not necessary: when $\delta = 0$, the adjusted AKM estimand recovers the ATEs in the unweighted population even if $\gamma \neq 0$, which violates (a) in part (ii).
The reason is that the $\gamma$ deviations in the conditional average treatment effects cancel exactly. Although such cancellation could occur, it cannot happen for both the unweighted population (for which $w_{jk}$ equals $P_{jk}$) and the $q$-weighted population (for which $w_{35}$ is proportional to $1.2P_{35}$ and all other $w_{jk}$ are proportional to $P_{jk}$). This feature is evident from the third panel of the table. The $q$-weighted AKM contrast recovers the $q$-weighted ATEs only when $\delta=\gamma=0$. The $\gamma$ deviations in the conditional average treatment effects no longer cancel and, therefore, $\delta=0$ is insufficient. 
\end{examplecontinued}

Theorem~\ref{prop:local-ate-recovery-all-pairs} below extends the necessary and sufficient conditions for adjusted AKM contrasts to be given a causal interpretation to all firm pairs, including those for which there is no supported move, and thus the $\DID$s and $\ATT$s are not defined.\footnote{Existing work has discussed what economic models are compatible with a causal interpretation of the AKM estimand \citep[see e.g.][]{Eeckhout:2011aa,Lamadon:2022aa,Borovickova:2024aa}. Our necessary and sufficient conditions show the properties that economic models would need to satisfy for such compatibility.} Part (i) and condition (a) in part (ii) of the following theorem are the same as in Theorem~\ref{prop:local-ate-recovery-supported-moves} except that they now apply to every pair of firms. On the other hand, condition (b) in part (ii) of the following result is new to Theorem~\ref{prop:local-ate-recovery-all-pairs}. It is an additivity restriction on the ATEs: for a randomly drawn worker, the effect of moving from $j$ to $k$ is equal to the effect of moving from $j$ to an intermediate firm $\ell$ and from $\ell$ to $k$. The additivity condition is required to hold for all pairs $(j,k)$ and intermediate firm $\ell$. Under common trends and condition (a), the additivity restriction requires the ATE between any two distinct firms to equal the signed path sum of DIDs along every signed path connecting them. For pairs with a supported move, this requirement is equivalent to condition (b) in part (ii) of Theorem \ref{prop:local-ate-recovery-supported-moves}, requiring the signed path sum of DIDs for each signed cycle to be zero, which is testable in the data. For pairs without a supported move, equality to these path sums is a new restriction on potential outcomes.

\begin{theorem}[ATE recovery for all pairs]
  \label{prop:local-ate-recovery-all-pairs}

  Suppose Assumptions~\ref{ass: invertibility of A} and~\ref{ass: common trend} hold. Fix \(\rho>0\). Then the following statements are equivalent:
  \begin{enumerate}[(i)]
    \item\label{item:local-ate-all-pairs-recovery} For every reweighting $q$ satisfying $\|q-\mathbf{1}\|_\infty<\rho$ for which the target and adjusted AKM estimands are defined, and every pair of firms $j,k$,
    \[
        \widetilde\alpha_k(q)-\widetilde\alpha_j(q)
        =
        \ATE\bigl(j,k;w^{\mathrm{ATE}}(q)\bigr).
    \]
    
    \item\label{item:local-ate-all-pairs-conditions} Both of the following conditions hold:
    \begin{enumerate}[(a)]
        \item For all pairs of firms \(j,k\) and every pair \((o,d)\) with \(P_{od}>0\),
        \[
        E\bigl[Y_2(j,k)-Y_2(j,j)\mid O=o,D=d\bigr]
        =
        E\bigl[Y_2(j,k)-Y_2(j,j)\bigr].
        \]
    
        \item For all triples of firms \(j,k,\ell\):
        \[
            E\bigl[Y_2(j,k)-Y_2(j,j)\bigr] = E\bigl[Y_2(j,\ell)-Y_2(j,j)\bigr] + E\bigl[Y_2(\ell,k)-Y_2(\ell,\ell)\bigr].
        \]
    \end{enumerate}
  \end{enumerate}
  \end{theorem}

 The additional restrictions in Theorem \ref{prop:local-ate-recovery-all-pairs} again have no testable implications. Indeed, Proposition~\ref{prop: appendix-all-pairs-compatibility} in Appendix \ref{app: proofs section causal} shows that a joint distribution of $(Y_1,Y_2,O,D)$ satisfying the mild regularity conditions in Assumption \ref{ass: invertibility of A}(ii)-(iii) is compatible with a joint distribution for $((Y_2(o,d),Y_1(o)) : 1 \leq o,d \leq K, O,D)$ that satisfies the common trend assumption and conditions (a) and (b) in part (ii) of Theorem \ref{prop:local-ate-recovery-all-pairs} if and only if it satisfies the cycle restriction in condition (b) of part (ii) of Theorem \ref{prop:local-ate-recovery-supported-moves}.  In this sense, the cycle condition exhausts the testable implications of the common trends assumption and conditions (a) and (b) in part (ii) of Theorem \ref{prop:local-ate-recovery-all-pairs}.

We conclude this section with an illustration of Theorem \ref{prop:local-ate-recovery-all-pairs} using our running example:

\vspace{-.4cm}
\begin{examplecontinued}{ex: four-firm-illustration}
Consider the ATE between firms 2 and 4 and use condition (b) in part (ii) of Theorem \ref{prop:local-ate-recovery-all-pairs} to express it using intermediate firm $3$:
\begin{equation*}
    E\bigl[Y_2(2,4)-Y_2(2,2)\bigr] = E\bigl[Y_2(2,3)-Y_2(2,2)\bigr] + E\bigl[Y_2(3,4)-Y_2(3,3)\bigr].
\end{equation*}
The exogeneity condition tells us that the ATEs on supported moves are also equal to the ATT and to the DID. Finally, recall that condition (b) of part (ii) implies the cycle condition and hence the $\DID$ are equal to the adjusted contrast. We then get
\begin{align*}
E\bigl[Y_2(2,4)-Y_2(2,2)\bigr] 
    &= E\bigl[Y_2(2,3)-Y_2(2,2)\bigr] + E\bigl[Y_2(3,4)-Y_2(3,3)\bigr] \\
    &= \ATT(2,3) + \ATT(3,4)\\
    &= \DID(2,3) + \DID(3,4)\\
    &= \widetilde\alpha_3 - \widetilde\alpha_2 + \widetilde\alpha_4 - \widetilde\alpha_3\\
    &= \widetilde\alpha_4 - \widetilde\alpha_2.
\end{align*}
as required in part (i) of Theorem \ref{prop:local-ate-recovery-all-pairs}.
\end{examplecontinued}

A natural question is how the sufficient conditions in Theorem~\ref{prop:local-ate-recovery-all-pairs} compare to the assumptions usually invoked to justify a causal interpretation of the AKM estimand. As explained in \citet{Kline:2024io}, it is typically assumed that the model of potential earnings is given by:
\begin{align}
Y_{i1}(o)&=\mu_i+a_o+\epsilon_{i1}(o),
\label{eq: potential-wage-period-one}\\
Y_{i2}(o,d)&=\mu_i+a_d+\tau +\epsilon_{i2}(o,d),
\label{eq: potential-wage-period-two}
\end{align}

\noindent with the restrictions that the error terms do not depend on the firm assignments, $\epsilon_{i1}(o)=\epsilon_{i1}$ and $\epsilon_{i2}(o,d)=\epsilon_{i2}$, and the errors are mean independent of the firm assignments given the worker component, i.e., $E[\epsilon_{it}\mid\mu_i,O_i,D_i]=0$ for $t=1,2$. The first two restrictions impose constant firm effects: every worker gains $Y_{i2}(o,d)-Y_{i2}(o,o)=a_d-a_o$ from moving from $o$ to $d$. The last restriction is the exogenous mobility assumption: workers may sort into firms based on $\mu_i$, but not based on the errors. 

Although these restrictions are enough for AKM to receive a causal interpretation, they are meaningfully stronger than the sufficient conditions in Theorem~\ref{prop:local-ate-recovery-all-pairs}.
In particular, under constant treatment effects, exogenous mobility is clearly not needed. If every worker gains $a_k-a_j$ from moving from $j$ to $k$, then every conditional average treatment effect in \eqref{eq: target definition} equals $a_k-a_j$. Condition~(a) of Theorem~\ref{prop:local-ate-recovery-all-pairs} then holds, and so does condition~(b), because $a_k-a_j=(a_\ell-a_j)+(a_k-a_\ell)$. Under common trends, the adjusted AKM contrasts therefore recover the ATE for every pair of firms. With constant firm effects, common trends only require that, among workers starting at the same firm, the mean change in the errors does not depend on the destination. Exogenous mobility is stronger because it also rules out differences in the mean level of the errors across firm assignments, conditional on the worker component.

A researcher willing to assume constant firm effects can therefore replace exogenous mobility by common trends and still be able to recover the causal wage effects of moving between firms for a randomly drawn worker. 

\section{AKM for Causal Inference: Empirical results} \label{sec:causalempirical}
In this section, we revisit two applications in which AKM has been used to draw causal inferences.

\subsection{Industry premium}

We now use the Italian data to revisit the measurement and interpretation of wage premiums for working in different industries. Since \citet{Krueger:1988aa}, labor economists have been keenly interested in measuring the size of such premia.

\citet{Card:2024aa} argue that the wage premium of an industry should be measured as the weighted average of the AKM estimates for the firms in that industry, where the weight is proportional to the number of person-year observations in that firm. Let $\ell(j)$ denote the industry of firm $j$, and let the reference industry, construction, be denoted by $1$. Let $T_j$ denote the number of person-year observations for firm $j$ and define its within-industry weight as
\[
w_j:=\frac{T_j}{\sum_{k:\,\ell(k)=\ell(j)}T_k}.
\]
Letting $\bar\psi_\ell:=\sum_{j:\,\ell(j)=\ell}w_j\alpha_j$ for each industry $\ell$, the AKM-based industry wage premium for industry $\ell$ can then be expressed as
\begin{equation}\label{eq: industry-premium-weighting}
\bar\psi_\ell-\bar\psi_1
=\underbrace{\sum_{j:\,\ell(j)=\ell}w_j\tau_j
  -\sum_{j:\,\ell(j)=1}w_j\tau_j}_{=: \Delta\bar\tau}
  +\underbrace{\sum_{j:\,\ell(j)=\ell}\sum_{k:\,\ell(k)=1}
  w_jw_k\bigl(\widetilde\alpha_j-\widetilde\alpha_k\bigr)}_{=: \Delta\widetilde\alpha}.
\end{equation}
Since the weights sum to one within each industry, $\Delta\widetilde\alpha$ is a weighted average of $\widetilde\alpha_j-\widetilde\alpha_k$ with weight $w_jw_k$. \citet{Card:2024aa} argue that $\bar\psi_\ell-\bar\psi_1$ can be given the following causal interpretation: if workers were to move from a randomly selected job in the construction industry to a randomly selected job in industry $\ell$, then their wage would increase, on average, by $\bar\psi_\ell-\bar\psi_1$.

In Table~\ref{tab:indprem-paths-combined}, we present and decompose estimates of the industry wage premia proposed by \citet{Card:2024aa}. The estimation sample is the same as in \citet{Kline:2024io} and \citet{Kline:2020aa}. We focus on six broad industries, defined in the same way as in Section 5.  

The first three columns of the table report, for each industry, the mean (log) wage, the estimated wage premium, and the percent of the workforce. The results suggest that both mean wages and AKM estimates differ substantially by industry. For example, the AKM estimates for firms producing large or heavy products could be interpreted as suggesting an 11 percent wage gain of moving to a random job in this industry from a random job in the construction industry. 

The next two columns decompose the estimate of the industry premium into estimates of its components: the average time-trend and the average adjusted AKM contrast. In the last eight columns, we summarize four categories of paths: direct paths, reverse-direct paths, indirect paths confined to the endpoint industries, and indirect paths involving other industries. For each category, we report the total path weight $\lambda$ and the path-weighted mean $\bar v$ of the signed path sums of $\DID$s. The adjusted AKM contrast is the sum of $\lambda\bar v$ over the four categories.

The results show that a large share of the weight is on paths that involve firms in industries other than those of the origin and destination firms. Hence, the estimated industry firm premium is determined largely from the values of these paths. This feature would not be an issue if all paths between two firms had produced similar estimates of the effect of moving between these firms, but they do not: different paths between the same pair of firms give significantly different estimates. For this reason, the AKM-based industry wage premium should not be interpreted as capturing the mean wage impact of assigning workers to a randomly selected job in one industry versus another.

\begin{table}
\centering
\caption{Decomposition of AKM industry premia}
\label{tab:indprem-paths-combined}
\begingroup
\footnotesize
\setlength{\tabcolsep}{2pt}
\renewcommand{\arraystretch}{1.12}

\begin{tabular*}{\textwidth}{@{\extracolsep{\fill}}l*{13}{>{\raggedleft\arraybackslash}p{0.86cm}}@{}}
\toprule
 & \multicolumn{3}{c}{\shortstack{Industry\\statistics}}
 & \multicolumn{2}{c}{\shortstack{Premium\\components}}
 & \multicolumn{2}{c}{\shortstack{Direct\\paths}}
 & \multicolumn{2}{c}{\shortstack{Reverse-direct\\paths}}
 & \multicolumn{2}{c}{\shortstack{Indirect paths\\within endpoint\\industries}}
 & \multicolumn{2}{c}{\shortstack{Indirect paths\\involving other\\industries}} \\
\cmidrule(lr){2-4}\cmidrule(lr){5-6}\cmidrule(lr){7-8}\cmidrule(lr){9-10}
\cmidrule(lr){11-12}\cmidrule(lr){13-14}
Industry
 & $\log w$ & $\Delta\bar\psi$ & \% wf
 & $\Delta\bar\tau$ & $\Delta\widetilde\alpha$
 & $\lambda$ & $\bar v$
 & $\lambda$ & $\bar v$
 & $\lambda$ & $\bar v$
 & $\lambda$ & $\bar v$ \\
\midrule

      Heavy
        & 4.95 & 0.11 & 6.58 & 0.00 & 0.11
        & 0.05 & 0.04 & 0.05 & 0.07
        & 0.06 & 0.08 & 0.84 & 0.12 \\
      Service
        & 4.84 & 0.06 & 16.11 & $-$0.01 & 0.07
        & 0.04 & $-$0.03 & 0.03 & $-$0.03
        & 0.11 & 0.04 & 0.81 & 0.08 \\
      Trade
        & 4.78 & 0.05 & 17.20 & $-$0.01 & 0.06
        & 0.03 & 0.09 & 0.03 & 0.10
        & 0.10 & 0.08 & 0.84 & 0.06 \\
      Metal
        & 4.86 & 0.04 & 27.76 & $-$0.01 & 0.05
        & 0.08 & 0.08 & 0.07 & 0.07
        & 0.18 & 0.05 & 0.67 & 0.05 \\
      Light
        & 4.72 & 0.03 & 27.30 & 0.00 & 0.03
        & 0.04 & 0.07 & 0.04 & 0.06
        & 0.11 & 0.04 & 0.81 & 0.03 \\
      Construction
        & 4.79 & 0.00 & 5.05 & {---} & {---}
        & {---} & {---} & {---} & {---}
        & {---} & {---} & {---} & {---} \\

\bottomrule
\end{tabular*}

\endgroup

\figurenote{Construction is the reference industry. $\log w$ is mean log wages and \% wf is the person-year share. The premium $\Delta\bar\psi:=\bar\psi_\ell-\bar\psi_1$ is the sum of the average time-trend component, $\Delta\bar\tau$, and the average adjusted AKM contrast, $\Delta\widetilde\alpha$; all use firm person-year weights. The weights that $\Delta\widetilde\alpha$ places on the supported moves form a single flow from the firms of the reference industry to the firms of the row's industry, each firm weighted by its person-years. This flow is decomposed into signed paths, each starting at a firm of the reference industry, ending at a firm of the row's industry, and traversing every move in the direction of the move's weight. Among such decompositions, the table reports one that minimizes the total weight on paths visiting another industry (see Section~\ref{sec: path decomposition}). A direct (reverse-direct) path is one observed move between two such firms, traversed forward (backward). Indirect paths either remain within the two industries or visit another industry. For each category, $\lambda$ is the total weight of its paths and $\bar v$ the weight-weighted mean of their signed $\DID$ sums; the four $\lambda$ sum to one and $\sum\lambda\bar v=\Delta\widetilde\alpha$. Values of $\bar v$ are not comparable across categories, because the paths of different categories start and end at different firms. $\DID$s are computed using the origin-firm stayer mean wage change when available and otherwise using the panel-wide stayer mean wage change. Entries are rounded to two decimals. Dashes mark undefined reference-industry comparisons.}
\end{table}

\subsection{Firm size premium}
\citet{Kline:2024io} argues that examining how AKM estimates covary with observables can help to demystify the nature of these fundamentally unobservable objects. A prominent example of such an analysis is the estimation of so-called firm size premium, which compares AKM estimates in firms with more or less employees.\footnote{See, for example, \citet{Bloom:2018aa} who argue the firm size premium has been declining over time in the US.}

We now use the Italian data to revisit the measurement and interpretation of the wage premium for working in large firms. We begin by binning the firms into five groups by their number of distinct workers in 1999. Unlike in the industry-premium analysis, we then define the firm size wage premium as the simple average of the AKM estimates for firms in a given bin, weighting each firm equally. Let $\ell(j)$ denote the firm size bin of firm $j$, with bins ordered from smallest to largest, and let $n_\ell$ denote the number of firms in bin $\ell$. The within-bin weight of firm $j$ is
\[
w_j:=\frac{1}{n_{\ell(j)}}.
\]
Letting $\bar\psi_\ell:=\sum_{j:\,\ell(j)=\ell}w_j\alpha_j$ for each size bin $\ell$, the AKM-based wage premium for firm size bin $\ell>1$ relative to the next smaller bin can then be expressed as
\begin{equation}\label{eq: size-premium-weighting}
\bar\psi_\ell-\bar\psi_{\ell-1}
=\underbrace{\sum_{j:\,\ell(j)=\ell}w_j\tau_j
  -\sum_{j:\,\ell(j)=\ell-1}w_j\tau_j}_{=: \Delta\bar\tau}
  +\underbrace{\sum_{j:\,\ell(j)=\ell}\sum_{k:\,\ell(k)=\ell-1}
  w_jw_k\bigl(\widetilde\alpha_j-\widetilde\alpha_k\bigr)}_{=: \Delta\widetilde\alpha}.
\end{equation}
Since the weights sum to one within each firm size bin, $\Delta\widetilde\alpha$ is a weighted average of $\widetilde\alpha_j-\widetilde\alpha_k$ with weight $w_jw_k$. A possible causal interpretation of $\bar\psi_\ell-\bar\psi_{\ell-1}$ is that it measures the mean wage impact of moving workers from a randomly selected job in bin $\ell-1$ to a randomly selected job in bin $\ell$.

In Table~\ref{tab:path-size-values}, we present and decompose AKM-based firm size premia estimates. The estimation sample is the same as in \citet{Kline:2024io} and \citet{Kline:2020aa}. The first three columns of the table report, for each firm size bin, the mean (log) wage, the estimated firm size wage premium, and the percent of the workforce. The results suggest that both mean wages and AKM estimates increase monotonically in firm size. For example, the AKM-based firm size premia estimates for the largest firm size bin (1,001 or more workers) could be interpreted as suggesting an 8 percent wage gain from moving to a random job in this bin from a random job in the bin with a firm size range just below (201--1,000 workers).

The next two columns decompose the estimated firm-size premium into the average time-trend component and the average adjusted AKM contrast. In the last eight columns, we summarize four categories of paths: direct paths, reverse-direct paths, indirect paths confined to the endpoint size bins, and indirect paths involving other size bins. For each category, we report the total path weight $\lambda$ and the path-weighted mean $\bar v$ of the signed path sums of $\DID$s. The adjusted AKM contrast is the sum of $\lambda\bar v$ over the four categories.

The results show that a large share of the weight is on paths that involve firms in size bins other than those of the origin and destination firms, ranging from 37 percent for the two smallest size bins to 76 percent for the two largest. Hence, the firm size premium is determined to a large extent from the values of these paths. This feature is concerning since different paths between the same pair of firms produce significantly different estimates. For this reason, the AKM-based firm size wage premium should not be interpreted as capturing the mean wage impact of assigning workers to a randomly selected job in one firm size bin versus another.

\begin{table}
\centering
\caption{Decomposition of AKM firm-size premia}
\label{tab:path-size-values}
\begingroup
\footnotesize
\setlength{\tabcolsep}{2pt}
\renewcommand{\arraystretch}{1.12}

\begin{tabular*}{\textwidth}{@{\extracolsep{\fill}}l*{13}{>{\raggedleft\arraybackslash}p{0.86cm}}@{}}
\toprule
 & \multicolumn{3}{c}{\shortstack{Size-bin\\statistics}}
 & \multicolumn{2}{c}{\shortstack{Premium\\components}}
 & \multicolumn{2}{c}{\shortstack{Direct\\paths}}
 & \multicolumn{2}{c}{\shortstack{Reverse-direct\\paths}}
 & \multicolumn{2}{c}{\shortstack{Indirect paths\\within endpoint\\size bins}}
 & \multicolumn{2}{c}{\shortstack{Indirect paths\\involving other\\size bins}} \\
\cmidrule(lr){2-4}\cmidrule(lr){5-6}\cmidrule(lr){7-8}\cmidrule(lr){9-10}
\cmidrule(lr){11-12}\cmidrule(lr){13-14}
Size bin
 & $\log w$ & $\Delta\bar\psi$ & \% wf
 & $\Delta\bar\tau$ & $\Delta\widetilde\alpha$
 & $\lambda$ & $\bar v$
 & $\lambda$ & $\bar v$
 & $\lambda$ & $\bar v$
 & $\lambda$ & $\bar v$ \\
\midrule

      1001+
        & 5.08 & 0.08 & 8.70 & 0.00 & 0.08
        & 0.11 & 0.11 & 0.09 & $-$0.03
        & 0.05 & 0.06 & 0.76 & 0.09 \\
      201--1000
        & 4.92 & 0.04 & 15.16 & 0.01 & 0.03
        & 0.17 & 0.04 & 0.12 & 0.00
        & 0.08 & 0.04 & 0.64 & 0.03 \\
      51--200
        & 4.83 & 0.03 & 23.36 & 0.00 & 0.03
        & 0.23 & 0.02 & 0.16 & 0.02
        & 0.16 & 0.03 & 0.45 & 0.05 \\
      11--50
        & 4.76 & 0.05 & 35.36 & 0.00 & 0.05
        & 0.21 & 0.03 & 0.21 & 0.06
        & 0.21 & 0.05 & 0.37 & 0.06 \\
      1--10
        & 4.65 & {---} & 17.42 & {---} & {---}
        & {---} & {---} & {---} & {---}
        & {---} & {---} & {---} & {---} \\

\bottomrule
\end{tabular*}

\endgroup

\figurenote{Each size bin is compared with the next smaller bin; comparisons are undefined for the 1--10 bin. Firms are grouped by their number of distinct workers in 1999. $\log w$ is mean log wages and \% wf is the person-year share. The premium $\Delta\bar\psi:=\bar\psi_\ell-\bar\psi_{\ell-1}$ is the sum of the average time-trend component, $\Delta\bar\tau$, and the average adjusted AKM contrast, $\Delta\widetilde\alpha$; firm effects and time-trend components are averaged equally across firms within each bin. The weights that $\Delta\widetilde\alpha$ places on the supported moves form a single flow from the firms of the smaller bin to the firms of the row's bin, firms weighted equally. This flow is decomposed into signed paths, each starting at a firm of the smaller bin, ending at a firm of the row's bin, and traversing every move in the direction of the move's weight. Among such decompositions, the table reports one that minimizes the total weight on paths visiting another size bin (see Section~\ref{sec: path decomposition}). A direct (reverse-direct) path is one observed move between two such firms, traversed forward (backward). Indirect paths either remain within the two size bins or visit another size bin. For each category, $\lambda$ is the total weight of its paths and $\bar v$ the weight-weighted mean of their signed $\DID$ sums; the four $\lambda$ sum to one and $\sum\lambda\bar v=\Delta\widetilde\alpha$. Values of $\bar v$ are not comparable across categories, because the paths of different categories start and end at different firms. $\DID$s are computed using the origin-firm stayer mean wage change when available and otherwise using the panel-wide stayer mean wage change. Entries are rounded to two decimals. Dashes mark undefined comparisons.}
\end{table}

\section{Measuring dispersion in average pay across firms}
\label{sec: firmeffects}
So far, we have focused on the descriptive and causal interpretation of pairwise differences in adjusted AKM estimands, $\tilde{\alpha}_d - \tilde{\alpha}_o$. However, empirical studies often report estimates of the variance of AKM estimands as a measure of pay dispersion across firms. We now provide sufficient conditions under which the variance of adjusted AKM estimands measures dispersion across firms in the average wages they would pay the same population of workers. Next, we propose and apply a bounding approach that relies on weaker assumptions to measure this dispersion. The bounding approach also provides a way to quantify the extent to which the conditions in Theorem~\ref{prop:local-ate-recovery-all-pairs} are rejected. 

\subsection{Using the AKM estimand to measure dispersion in average pay across firms}
Theorem~\ref{prop:local-ate-recovery-all-pairs} provides sufficient conditions under which the variance of adjusted AKM estimands measures dispersion across firms in mean potential wages. Under these conditions, for any nonnegative firm weights $\omega_1,\ldots,\omega_K$ satisfying $\sum_{k=1}^K \omega_k=1$, we have
\begin{align}
\sum_{k=1}^K \omega_k \left( \widetilde\alpha_k  - \sum_{k'=1}^K \omega_{k'} \widetilde\alpha_{k'} \right)^2
&= \sum_{k=1}^K \omega_k \left( E[Y_2(1,k)] - \sum_{k'=1}^K \omega_{k'} E[Y_2(1,k')] \right)^2 \notag\\
&=: \var\nolimits_{\omega}\!\left(E[Y_2(1,k)]\right).
\end{align}
Here, the expectation averages over workers, holding the origin firm fixed at firm~1, and $\var_{\omega}$ denotes variance across destination firms with weights $\omega_k$. The equality follows from part (i) of Theorem~\ref{prop:local-ate-recovery-all-pairs} at $q=\mathbf 1$, which gives $\widetilde\alpha_k=E[Y_2(1,k)]-E[Y_2(1,1)]$; subtracting the common reference mean does not affect the variance.

A key implication is that it is not necessary to assume constant firm effects, i.e., $Y_2(j,k)-Y_2(j,j)=E[Y_2(j,k)-Y_2(j,j)]$ with probability one for every pair $j,k$, for the variance of adjusted AKM estimands to measure dispersion in mean potential wages across firms. The effects of moving between firms may vary across workers, provided the conditions in Theorem~\ref{prop:local-ate-recovery-all-pairs} hold. This insight could be useful for empirical work, given the substantial differences in estimated firm pay premia across groups of workers.\footnote{For example, see \citet{Card:2016aa} for differences by gender and \citet{Gerard:2021aa} for differences by race.}

\subsection{Bounding the dispersion in pay across firms}
\label{sec: firm-dispersion-bounds}
We now ask what can be learned about $\var_\omega(E[Y_2(1,k)])$ when the sufficient conditions for recovery of average treatment effects in Theorem~\ref{prop:local-ate-recovery-all-pairs} do not hold. We first characterize the population bounds under specified departures from these conditions and then estimate these bounds on the Italian data.

We maintain Assumption~\ref{ass: invertibility of A} and that potential outcomes have finite first moments. Let $\mathcal S:=\{(o,d):P_{od}>0\}$ include all supported mover and stayer origin-destination pairs, and write
\begin{equation}
\label{eq: dispersion-effect-definitions}
G_{jk| od}:=E[Y_2(j,k)-Y_2(j,j)\mid O=o,D=d],
\qquad
g_{jk}:=\sum_{(o,d)\in\mathcal S}P_{od}G_{jk| od}.
\end{equation}
Since $g_{1k}=E[Y_2(1,k)]-E[Y_2(1,1)]$, the target variance is
\[
V_\omega(G):=\sum_{k=1}^K \omega_k
\left(g_{1k}-\sum_{\ell=1}^K \omega_\ell g_{1\ell}\right)^2
=\var\nolimits_{\omega}\!\left(E[Y_2(1,k)]\right),
\]
where $G$ denotes the array $\{G_{jk|od}\colon j,k\in\{1,\ldots,K\},(o,d)\in\mathcal S\}$ and the weights $\omega_1,\ldots,\omega_K$ are the nonnegative firm weights introduced in the preceding subsection. The variance depends on $G$ through the terms $g_{1k}$, $k=1,\ldots,K$.

For a given vector $\varepsilon:=(\varepsilon^a_{jkod},\varepsilon^b_{jk\ell},\varepsilon^c_{o'd'}\colon j,k,\ell\in\{1,\ldots,K\}, (o,d)\in\mathcal S, (o',d')\in\mathcal E)$ of nonnegative constants and given firm weights $\omega_1,\ldots,\omega_K$, bounds on the variance can be constructed as the optimal values of the following minimization and maximization problems, provided the feasible set is nonempty:
\begin{align}
\underset{G}{\min\;/\;\max}\quad
& V_\omega(G)
\label{eq: dispersion-population-objective}\\
\text{s.t.}\quad
& G_{jj| od}=0
&& \qquad\forall j,\ (o,d)\in\mathcal S,
\label{eq: dispersion-diagonal}\\
& |G_{jk| od}-g_{jk}| \le \varepsilon^a_{jkod}
&& \qquad\forall j,k,\ (o,d)\in\mathcal S,
\label{eq: dispersion-mean-homogeneity}\\
& |g_{jk}-g_{j\ell}-g_{\ell k}| \le \varepsilon^b_{jk\ell}
&& \qquad\forall j,k,\ell,
\label{eq: dispersion-additivity}\\
& |G_{od| od}-\DID(o,d)|\le\varepsilon^c_{od}
&& \qquad\forall (o,d)\in\mathcal E.
\label{eq: dispersion-common-trends}
\end{align}
The restriction in \eqref{eq: dispersion-diagonal} is required by the definition of $G_{jk| od}$ in \eqref{eq: dispersion-effect-definitions}. The inequalities in \eqref{eq: dispersion-mean-homogeneity}--\eqref{eq: dispersion-common-trends} constrain deviations from the conditions in Theorem~\ref{prop:local-ate-recovery-all-pairs}: from exogeneity (condition (ii)(a)), from additivity (condition (ii)(b)), and from common trends (Assumption~\ref{ass: common trend}(ii)). The elements of $\varepsilon$ govern the magnitudes of allowed deviations.

For example, setting $\varepsilon^a_{jkod}=0.02$ for all $j,k,o,d$ allows the conditional average treatment effects of moving between $j$ and $k$ ($G_{jk|od}$) to differ from the average treatment effect of this move ($g_{jk}$) by at most 2 percentage points. The conditional average treatment effects of moving between $j$ and $k$ can therefore differ by at most 4 percentage points between groups of movers and stayers, regardless of which firms, if any, they moved between. The tolerances $\varepsilon^b_{jk\ell}$ govern the extent to which additivity is relaxed: they bound the difference between the average treatment effect of moving from $j$ to $k$ and the sum of the average treatment effects of moving from $j$ to $\ell$ and from $\ell$ to $k$. Setting all elements of $\varepsilon$ to zero recovers exogeneity, additivity, and common trends, without requiring treatment effects to be constant across workers.

The bounds on $\var_\omega(E[Y_2(1,k)])$ from the above program are sharp in the sense that every value in the resulting interval is attained by a joint distribution of potential outcomes and mobility that satisfies the maintained restrictions and is compatible with the observed joint distribution of $(Y_1,Y_2,O,D)$. Lemma~\ref{lem: dispersion-sharpness} in Appendix~\ref{app: dispersion-bounds} establishes sharpness by showing that every feasible array of conditional average treatment effects can be generated by such a distribution. Thus, for given values of $\varepsilon$ and firm weights $\omega_1,\ldots,\omega_K$, the observed data discipline the variance only through the $\DID$s and mobility probabilities $P_{od}$. Other features of the observed distribution cannot tighten the bounds without additional restrictions on potential outcomes.

The program is flexible enough to consider many possible relaxations depending on the choices of $\varepsilon$. We compare values of $\varepsilon$ componentwise: $\varepsilon'\leq\varepsilon$ if every element of $\varepsilon'$ is no larger than the corresponding element of $\varepsilon$. One possibility is to restrict attention to smallest (with respect to this partial order) values of $\varepsilon$ for which the minimum and maximum of the program are equal, i.e., the bounds on the variance collapse to a point. One could also search for smallest (again, with respect to this partial order) values of $\varepsilon$ for which the bounds on the variance cover the variance of the adjusted AKM estimands. The resulting values of $\varepsilon$ then offer a quantification of the extent to which the conditions in Theorem \ref{prop:local-ate-recovery-all-pairs} are violated. Such an analysis complements the test of the cycle condition (Section~\ref{sec: joint cycle test}) with a measure of the magnitude of the violations. 

It may be desirable to choose the same value of the tolerances within each of the constraints: for some values $\varepsilon^a$, $\varepsilon^b$, $\varepsilon^c$, set $\varepsilon^a_{jkod} = \varepsilon^a$, $\varepsilon^b_{jk\ell} = \varepsilon^b$, and  $\varepsilon^c_{od} = \varepsilon^c$ for all $j,k,o,d$. This choice is useful because it can be difficult to interpret and justify relaxations that vary across firm pairs for a given type of constraint. 
The resulting bounds on $\var_\omega(E[Y_2(1,k)])$ tell us what can be learned about the variance when each condition in Theorem~\ref{prop:local-ate-recovery-all-pairs} is relaxed in a uniform way across firm pairs. 

\vspace{-.4cm}
\begin{examplecontinued}{ex: four-firm-illustration}
To illustrate the bounding approach, we return to the running example with \(\delta=0.1\). As firm weights, we use the destination shares $\omega_k=\sum_{o=1}^K P_{ok}$, i.e., the column sums in Table~\ref{tab:pod-eight-firm}: $\omega=(0.02,0.23,0.26,0.14,0.35)$. Table~\ref{tab:bounds in example} presents the results from two relaxations of exogeneity (choosing $\varepsilon_{jkod}^a$), maintaining additivity and common trends in both cases (setting $\varepsilon_{jk\ell}^b=\varepsilon_{od}^c=0$ for all $j,k,\ell,o,d$).

The first two columns of Table~\ref{tab:bounds in example} report a minimal relaxation of exogeneity, i.e., values of $\varepsilon_{jkod}^a$ such that the upper and lower bounds on $\operatorname{Var}_{\omega}(E[Y_2(1,k)])$ are both equal to the variance of the adjusted AKM estimands; this happens to be possible in this example, but not in general. The resulting value of the variance is 0.0066. For workers making each supported move, the reported tolerance equals the absolute difference between the corresponding DID and adjusted AKM contrast in Table~\ref{tab:did-values}. These tolerances are zero for moves from firm 1 to firm 2, firm 2 to firm 3, and firm 2 to firm 5. For the remaining supported moves, they range from 0.0100 to 0.0400. For example, the DID for the move from firm 4 to firm 5 is 0.11, whereas the adjusted AKM contrast is 0.15. Under common trends, recovering this contrast as the average causal effect therefore requires the average effect for workers making this move to be 4 percentage points smaller than the average effect for the population. The minimal relaxation thus provides a means of assigning an economic magnitude to the violations of the conditions in Theorem~\ref{prop:local-ate-recovery-all-pairs}, complementing the rejection of the cycle condition.

\begin{table}
\centering
\begin{tabular}{lccc}
\toprule
& \multicolumn{2}{c}{Minimal relaxation}
& Uniform relaxation \\
\cmidrule(lr){2-3}
\cmidrule(lr){4-4}
$\operatorname{Var}_{\omega}(E[Y_2(1,k)])$
& \multicolumn{2}{c}{0.0066}
& $[0.0050,\,0.0080]$ \\

\addlinespace
& \multicolumn{3}{c}{$\varepsilon^a_{jkod}$} \\
\cmidrule(lr){2-4}
Move $(j,k)$ & Movers & Other groups & All groups \\
& $(o,d)=(j,k)$ & $(o,d)\neq(j,k)$ & \\
\midrule
$1\to2$ & 0      & 0      & 0.0263 \\
$2\to3$ & 0      & 0      & 0.0263 \\
$2\to5$ & 0      & 0      & 0.0263 \\
$3\to4$ & 0.0100 & 0.0006 & 0.0263 \\
$4\to3$ & 0.0150 & 0.0006 & 0.0263 \\
$3\to5$ & 0.0240 & 0.0013 & 0.0263 \\
$4\to5$ & 0.0400 & 0.0012 & 0.0263 \\
Pairs without movers
& \multicolumn{2}{c}{0} & 0.0263 \\
\bottomrule
\end{tabular}

\medskip
\begin{minipage}{\textwidth}
\footnotesize
\textit{Note:}
Example~\ref{ex: four-firm-illustration} with $\delta=0.1$ and exact DIDs. Additivity and common trends are maintained. The lower panel reports the exogeneity tolerances in \eqref{eq: dispersion-mean-homogeneity}. For each move $(j,k)$, ``other groups'' includes all movers and
stayers except workers making that move. The uniform relaxation sets every exogeneity tolerance to the smallest common value for which \eqref{eq: dispersion-population-objective}--\eqref{eq: dispersion-common-trends} is feasible, approximately 0.0263. Entries are rounded to four decimal places.
\end{minipage}

\caption{Relaxing exogeneity in Example~\ref{ex: four-firm-illustration}}
\label{tab:bounds in example}
\end{table}

The last column instead imposes the same exogeneity tolerance for every pair of firms and every group of movers and stayers. The smallest value for which the program is feasible is approximately \(\varepsilon^*=0.0263\). This relaxation allows the conditional average treatment effect of moving between any two firms to differ from the average treatment effect by at most approximately 2.63 percentage points, regardless of which firms, if any, the workers moved between. At smaller values, the program has no solution, so the observed DIDs are incompatible with the maintained restrictions.

At \(\varepsilon^*\), the bounds on \(\operatorname{Var}_{\omega}(E[Y_2(1,k)])\) are \([0.0050,0.0080]\). These bounds contain the variance of the adjusted AKM estimands, 0.0066, and remain relatively narrow: they range from approximately 76 to 122 percent of that value. Thus, substantial information about dispersion in average pay across firms remains available without requiring the average causal effect of a move to be the same for all groups of movers and stayers. In this example, allowing modest departures from exogeneity still rules out both no dispersion and dispersion substantially greater than that implied by the adjusted AKM estimands.

To understand why we obtain an interval rather than a point when the constraints become feasible (i.e., at $\varepsilon^*$), it is useful to observe that different path sums of DIDs become consistent with the admissible distributions of potential outcomes at different values of \(\varepsilon^a\). In our example, the most demanding cycle is the signed cycle $(4,+1,3,+1,5,-1,4)$, which moves from firm 4 to firm 3 and on to firm 5, and returns to firm 4 along the move $(4,5)$ traversed backward. This cycle determines when the complete program becomes feasible. Its signed path sum is
\[
S\bigl((4,+1,3,+1,5,-1,4),\DID\bigr)
=\DID(4,3)+\DID(3,5)-\DID(4,5)
=0.065+0.124-0.110
=0.079.
\]
Under additivity and common trends, this requires \(3\varepsilon^a\geq 0.079\), giving the threshold of approximately 0.0263. At this value, however, not all of the constraints bind. For example, the uniform relaxation also allows departures from exogeneity for the move from firm 1 to firm 2, which does not belong to any cycle. The average causal effects can therefore still vary across solutions to the program. As a result, the variance is bounded but not point identified.
\end{examplecontinued}

\subsection{Bounding in the data}
\label{sec: firm-dispersion-empirical}

We now apply the bounding approach to the Italian data. A natural estimator replaces the population DIDs in \eqref{eq: dispersion-common-trends} by their sample counterparts. With few movers between many firm pairs, however, the individual estimates can be too noisy for these constraints to yield useful bounds, and sampling discrepancies can make the program infeasible. Since the population DIDs are unknown, we instead minimize or maximize $V_\omega(G)$ jointly over $G$ and the DIDs, subject to the preceding restrictions and a bound on their discrepancy from the estimated DIDs.

Let $\mathcal F_\varepsilon$ be the set of pairs $(G,\DID)$ satisfying \eqref{eq: dispersion-diagonal}--\eqref{eq: dispersion-common-trends}. The estimation program optimizes over $(G,\DID)\in\mathcal F_\varepsilon$, subject to the additional restriction
\begin{equation}
\frac{
  1
}{
  \sum_{(o,d)\in \mathcal E}P_{od}
}\displaystyle\sum_{(o,d)\in \mathcal E} P_{od}
  \bigl(\DID(o,d)-\widehat{\DID}(o,d)\bigr)^2
\leq \varepsilon^d .
\label{eq:akm-empirical-discrepancy}
\end{equation}

The tolerances $\varepsilon$ govern the admissible departures from the substantive assumptions, whereas $\varepsilon^d$ governs the discrepancy attributed to sampling noise. For fixed $\varepsilon$, increasing $\varepsilon^d$ weakly widens the variance bounds whenever the program is feasible. 

To choose $\varepsilon^d$, we estimate the sampling variance of each DID. For a given move, this is the sample variance of the movers' wage changes divided by the number of movers, plus the same quantity for the origin-firm stayers. The sample variance for movers requires at least two movers; for moves with a single mover, we assume that single movers have, on average, the same individual wage-change variance as movers in moves with at least two movers. We then average these estimates over all supported moves using weights proportional to the sample mobility shares.

The sample is the largest connected set among the 5,000 largest firms, after dropping workers whose origin or destination firm has no stayers. It contains 4,549 firms and 356,619 workers, including 27,133 movers across 20,676 supported moves. We use sample destination shares as the firm weights, $\omega_k=\sum_o P_{ok}$, and take firm 1 to be the firm with the largest destination share. The preceding calculation gives $\varepsilon^d=0.0365$.

Table~\ref{tab:akm-empirical-relaxations} presents the results of relaxing exogeneity while maintaining additivity and common trends. The first two columns report a minimal relaxation that delivers a point estimate of
$\operatorname{Var}_{\omega}(E[Y_2(1,k)])$ equal to the variance of the adjusted AKM estimands, 0.0359. For 50.5 percent of supported moves, the tolerance for workers making that move is no larger than 0.01; for 78.3 percent, it is no larger than 0.02; and for 90.2 percent, it is no larger than 0.03. These values allow the average causal effect for workers making a particular move to differ from the average causal effect for the population by approximately one, two, and three percentage points, respectively. 

The average tolerance, weighted by the number of movers, is 0.0117, or approximately 1.17 percentage points, although some supported moves require substantially larger tolerances.  To put this into perspective, the corresponding median of the $\DID$s (which, under common trend, equals the ATTs) is about 0.01. In this way, the minimal relaxation assigns an economic magnitude to the violations of the conditions in Theorem \ref{prop:local-ate-recovery-all-pairs}. Alternatively, these results show that the usual interpretation of the variance of the adjusted AKM estimands can be justified under conditions weaker than those imposed in Theorem \ref{prop:local-ate-recovery-all-pairs}. In particular, exogeneity is not needed to learn about the dispersion in pay across firms. 

The last two columns instead relax exogeneity uniformly. The smallest common tolerance for which the program is feasible is approximately 0.0254. This value allows the conditional average treatment effect of moving between any two firms to differ from the average treatment effect by at most approximately 2.54 percentage points, regardless of which firms, if any, the workers moved between. At this tolerance, the reported lower bound on $\operatorname{Var}_{\omega}(E[Y_2(1,k)])$ is $0.0351$. In particular, the lower bound is nearly as large as the variance of the adjusted AKM estimands. Hence, under these conditions, our estimates suggest that $\operatorname{Var}_{\omega}(E[Y_2(1,k)])$ is at least 16 percent of the variance of log wages in 2001.\footnote{Even though we are considering the largest 5000 firms, one might be concerned about limited mobility bias. A considerable body of work is centered around such issues; see, for example, \citet{Andrews:2008aa}, \citet{Kline:2020aa}, and \citet{Bonhomme:2023aa}.}

The upper bound is $0.1135$, meaning that we cannot draw economically informative conclusions. One reason for this is that, unlike the lower bound, we cannot compute exactly the upper bound due to the size of the program. Although the upper bound is still valid, it could be that the sharp upper bound is smaller. 

\begin{table}[htbp]
\centering
\caption{Relaxing exogeneity in the Italian data}
\label{tab:akm-empirical-relaxations}
\begin{tabular}{lcccc}
\toprule
& \multicolumn{2}{c}{Minimal relaxation} & \multicolumn{2}{c}{Uniform relaxation} \\
\cmidrule(lr){2-3}\cmidrule(lr){4-5}
& own cell & other cells & own cell & other cells \\
& $\varepsilon^a_{odod}$ & $\varepsilon^a_{odo'd'}$ & $\varepsilon^a_{odod}$ & $\varepsilon^a_{odo'd'}$ \\
\midrule
$\operatorname{Var}_{\omega}(E[Y_2(1,k)])$
& \multicolumn{2}{c}{0.0359} & \multicolumn{2}{c}{$[0.0351,\,0.1135]$} \\
\addlinespace
& \multicolumn{4}{c}{Share of supported moves (\%)} \\
\cmidrule(lr){2-5}
$\varepsilon^a=0$        & 1.7  & 1.7   & 0.0   & 0.0   \\
$\varepsilon^a\leq0.001$ & 7.3  & 100.0 & 0.0   & 0.0   \\
$\varepsilon^a\leq0.005$ & 28.3 & 100.0 & 0.0   & 0.0   \\
$\varepsilon^a\leq0.01$  & 50.5 & 100.0 & 0.0   & 0.0   \\
$\varepsilon^a\leq0.02$  & 78.3 & 100.0 & 0.0   & 0.0   \\
$\varepsilon^a\leq0.03$  & 90.2 & 100.0 & 100.0 & 100.0 \\
$\varepsilon^a\leq0.05$  & 97.9 & 100.0 & 100.0 & 100.0 \\
$\varepsilon^a\leq0.1$   & 99.9 & 100.0 & 100.0 & 100.0 \\
\addlinespace
Largest tolerance        & 0.270  & $1.2\times10^{-6}$ & 0.0254 & 0.0254 \\
Mover-weighted mean      & 0.0117 & $6.4\times10^{-8}$ & 0.0254 & 0.0254 \\
\bottomrule
\end{tabular}

\medskip
\begin{minipage}{\textwidth}
\footnotesize
\textit{Note:}
Italian data, 1999--2001. Additivity and common trends are maintained, with $\varepsilon^d=0.0365$ in \eqref{eq:akm-empirical-discrepancy}. The lower panel summarizes $\varepsilon^a_{odod}$, the tolerance for workers making move $(o,d)$, and $\varepsilon^a_{odo'd'}$, the tolerance for the effect of move $(o,d)$ on workers in any other cell $(o',d')$. Under the minimal relaxation the other cells only offset the movers' deviation in the average effect, so $\varepsilon^a_{odo'd'}=P_{od}\,\varepsilon^a_{odod}/(1-P_{od})$ for all $(o',d')\neq(o,d)$; firm pairs without movers need no violation. Percentages weight supported moves equally; the mean weights them by their numbers of movers.
The uniform relaxation sets every $\varepsilon^a_{jkod}$ to approximately 0.0254. The reported variance interval contains the sharp bounds. The lower bound is attained to within $3\times10^{-6}$; the largest attained variance is 0.0369.
\end{minipage}
\end{table}

\section{Conclusion}
Much ink has been spilled about AKM regressions. 
A lot of focus has been on  uncertainty and estimation issues. Our focus is distinct (and primary to) the problem of statistical inference. We show what one can and cannot learn from AKM regressions if one had perfect knowledge of the population distribution of the data. This is the natural starting point for assessing the usefulness of these regressions, as the conclusions that one can draw under imperfect knowledge of the population distribution of the data are a subset of those that can be drawn under perfect knowledge.

Specifically, our paper provided necessary and sufficient conditions for firm effect estimates to accurately describe the wage changes that accompany the mobility of workers between firms. It also showed the additional assumptions needed for firm effects estimates to be interpreted as the causal wage effects of a randomly drawn worker moving between firms. These additional assumptions do not have testable implications, in contrast to the conditions needed for the estimates to accurately describe the wage changes accompanying worker mobility. 

The key to deriving these results is to show how firm effect estimates add up: after adjusting for the wage growth of workers who choose not to move, the firm effect contrast
for any pair of origin and destination firms can be written as a convex combination of the sums of
difference-in-differences along suitable signed paths starting at the
origin firm and ending at the destination firm.
The paths will generally involve firms other than the origin and destination firms, and these intermediate firms may differ systematically and so can the workers moving between them. As a result, the conditions are likely to fail for firm effect estimates to be interpreted descriptively, or causally, or both. We therefore proposed alternative estimators that can be given the same descriptive or causal interpretation under weaker conditions. 

We used Italian employer-employee data to show the empirical relevance of our theoretical results. This analysis re-examined several prominent studies about firm effects and their dispersion. Our findings suggest that the reported estimates are misinterpreted descriptively and causally. 

\newpage
\begin{spacing}{1.15}
\appendix{ \small 

\section{Proofs}

All theorems and lemmas have been formally checked by the Lean proof assistant, ensuring every logical step is correct. For completeness, we include proofs below. 

\subsection{Auxiliary results} 
\label{app: auxiliary lemmas}

\subsubsection{Properties of the AKM weights}

\begin{lemma}[Properties of the AKM weights]
    \label{lem: properties of the weights}
    Suppose Assumption~\ref{ass: invertibility of A} holds. Then, for any $j=2,\ldots,K$, the following two identities hold:
    \begin{enumerate}[(i)]
    \item $\sum_{o=1}^K\sum_{d=1}^K \omega_{j,od}=0$.

    \item For every \(k=1,\ldots,K\),
    \[
    \sum_{(u,v)\in\mathcal E}
    \omega_{j,uv}\bigl(\ind\{v=k\}-\ind\{u=k\}\bigr)
    =
    \ind\{j=k\}-\ind\{k=1\}.
    \]
    \end{enumerate}
\end{lemma}
\begin{proof}
    Let $X:=(1,J'-I')'\in\mathbb R^K$. Conditional on \(O=o,D=d\), the vector \(X\) is equal to $\nu_{od}=e_1+\ind\{d>1\}e_d-\ind\{o>1\}e_o$. Hence $\mathbb A=E[XX']$ and $\sum_{o=1}^K\sum_{d=1}^K P_{od}\nu_{od}=E[X]$.

    We first prove part (i). Since the first component of \(X\) is identically equal to one, $X'e_1=1$. Therefore
    \[
    \mathbb A e_1
    =
    E[XX']e_1
    =
    E[X(X'e_1)]
    =
    E[X].
    \]
    It follows that, for \(j=2,\ldots,K\),
    \[
    \sum_{o=1}^K\sum_{d=1}^K \omega_{j,od}
    =
    e_j'\mathbb A^{-1}
    \sum_{o=1}^K\sum_{d=1}^K P_{od}\nu_{od}
    =
    e_j'\mathbb A^{-1}E[X]
    =
    e_j'\mathbb A^{-1}\mathbb A e_1
    =
    e_j'e_1
    =
    0.
    \]

    Now, consider part (ii). First fix \(k=2,\ldots,K\). Since the \(k\)-th component of \(X\) is \(J_{k-1}-I_{k-1}\), we have $J_{k-1}-I_{k-1}=e_k'X$. Therefore,
    \begin{align*}
    \sum_{o=1}^K\sum_{d=1}^K
    \omega_{j,od}\bigl(\ind\{d=k\}-\ind\{o=k\}\bigr)
    &=
    e_j'\mathbb A^{-1}
    \sum_{o=1}^K\sum_{d=1}^K
    P_{od}\nu_{od}
    \bigl(\ind\{d=k\}-\ind\{o=k\}\bigr) \\
    &=
    e_j'\mathbb A^{-1}E[X(J_{k-1}-I_{k-1})] \\
    &=
    e_j'\mathbb A^{-1}E[XX']e_k \\
    &=
    e_j'\mathbb A^{-1}\mathbb A e_k \\
    &=
    e_j'e_k \\
    &=
    \ind\{j=k\}.
    \end{align*}

    Because \(P_{od}=0\) implies \(\omega_{j,od}=0\), and because the term \(\ind\{d=k\}-\ind\{o=k\}\) vanishes on diagonal cells \(o=d\), the same identity may be written as
    \begin{equation}\label{eq: weights sum to indicator}
    \sum_{(o,d)\in\mathcal E}
    \omega_{j,od}\bigl(\ind\{d=k\}-\ind\{o=k\}\bigr)
    =
    \ind\{j=k\},
    \qquad k=2,\ldots,K.
    \end{equation}

    It remains only to handle \(k=1\). For each supported move \((o,d)\in\mathcal E\),
    \[
    \sum_{k=1}^K\bigl(\ind\{d=k\}-\ind\{o=k\}\bigr)=0.
    \]
    Hence
    \[
    \sum_{k=1}^K
    \sum_{(o,d)\in\mathcal E}
    \omega_{j,od}\bigl(\ind\{d=k\}-\ind\{o=k\}\bigr)
    =0.
    \]
    Using the already established identities for \(k=2,\ldots,K\), i.e. \eqref{eq: weights sum to indicator}, we obtain
    \[
    \sum_{(o,d)\in\mathcal E}
    \omega_{j,od}\bigl(\ind\{d=1\}-\ind\{o=1\}\bigr)
    =
    -\sum_{k=2}^K \ind\{j=k\}
    =
    -1.
    \]
    Since \(j\neq 1\), this is exactly
    \[
    \ind\{j=1\}-\ind\{1=1\}.
    \]
    Thus part (ii) holds for all \(k=1,\ldots,K\).
\end{proof}

In the following lemma we show the relationship of invertibility of \(\mathbb A\) (Assumption~\ref{ass: invertibility of A}(i)) and connectedness of the undirected graph defined by the supported moves. The lemma adapts the standard rank characterization of a graph's incidence matrix \citep[Exercise~5.10, p.~209]{Bertsekas:1998aa} to our design, which also includes an intercept. Define the undirected graph \(G^{u}:=(\{1,\ldots,K\},\mathcal U)\) with \(\mathcal U:=\bigl\{\{u,v\}: u\ne v,\; P_{uv}+P_{vu}>0\bigr\}\). The graph is \emph{connected} if for every \(o,d\in \{1,\ldots,K\}\), $o\neq d$, there exists an undirected path in \(G^{u}\) from \(o\) to \(d\). For some $C\subset \{1,\ldots,K\}$ and some $\mathcal V\subset \mathcal U$, the subgraph $G:=(C,\mathcal V)$ is called a \emph{component} of $G^u$ if it is a maximal connected subgraph. This means that there is no other subgraph $(C',\mathcal V')$ with $C\subset C'$ that is connected.

\begin{lemma}[connectedness]\label{lem: connectedness}
    \begin{enumerate}[(i)]
        \item If Assumption~\ref{ass: invertibility of A}(i) holds, then \(G^{u}\) is connected.
        \item If Assumption~\ref{ass: invertibility of A}(ii) holds and \(G^{u}\) is connected, then Assumption~\ref{ass: invertibility of A}(i) holds.
    \end{enumerate}
\end{lemma}

\begin{proof}
    First, we prove part (i). With $\nu_{od}=e_1+\ind\{d>1\}e_d-\ind\{o>1\}e_o$, we have $X:=(1,J'-I')'=\nu_{OD}$ and $\mathbb A=E[XX']=\sum_{o=1}^K\sum_{d=1}^K P_{od}\nu_{od}\nu_{od}'$.

    For any vector $b:=(b_1,\ldots,b_K)'\in\mathbb R^K$ with \(b_1:=0\), and for any $c\in \mathbb R$, define \(\theta\in\mathbb R^K\) as $\theta:=(c,b_2,\ldots,b_K)'$. Then, for every \(o,d\in \{1,\ldots,K\}\),
    \[
    \theta'\nu_{od}
    =
    c+b_d-b_o .
    \]
    Suppose, towards a contradiction, that \(G^{u}\) is not connected. Let $G:=(C,\mathcal V)$ be a component of \(G^{u}\) that does not contain firm \(1\). Such a component exist because, by Exercise 11, p. 30, in \cite{diestel2025graph}, firm 1 must belong to at least one component of $G^u$, and since $G^u$ is not connected, there must exist another nonempty component not containing firm 1. Now, consider particular values for \(c\) and \(b_1,\ldots,b_K\) as follows: let \(c:=0\) and \(b_k:=\ind\{k\in C\}\) for \(k=1,\ldots,K\). Because \(C\) is nonempty and does not contain firm \(1\), the vector \(\theta=(0,b_2,\ldots,b_K)'\) is nonzero. Then, for every \(o,d\in \{1,\ldots,K\}\), \(o\neq d\),
    \[
    \theta'\nu_{od} = b_d-b_o.
    \]
    Consider any \(o,d\in \{1,\ldots,K\}\) with \(P_{od}>0\). If \(o=d\), then trivially \(b_d-b_o=0\). If \(o\ne d\), then \(\{o,d\}\in\mathcal U\), so \(o\) and \(d\) belong to the same connected component of \(G^{u}\), i.e., either both are in \(C\) or both are in \(\{1,\ldots,K\}\setminus C\). Hence \(b_d=b_o\). Therefore, for every \((o,d)\) with \(P_{od}>0\),
    \[
    \theta'\nu_{od}
    =
    c+b_d-b_o
    =
    0 .
    \]
    It follows that
    \begin{equation}\label{eq: theta'Atheta=0}
    \theta'\mathbb{A}\theta = \sum_{o=1}^K\sum_{d=1}^K P_{od}(\theta'\nu_{od})^2 = 0 .
    \end{equation}
    But \(\mathbb{A}\) is symmetric and positive semidefinite by definition. If \(\mathbb{A}\) were nonsingular, then it would be positive definite, so \(\theta'\mathbb{A}\theta>0\) for every nonzero \(\theta\), a contradiction. Thus \(G^{u}\) must be connected.

    We now show part (ii). Suppose \(G^u\) is connected and \(\theta=(c,b_2,\ldots,b_K)'\) satisfies $\theta'\mathbb{A}\theta=0$. Then, we want to show that $\theta=0$. \eqref{eq: theta'Atheta=0} implies that \(\theta'\nu_{od}=0\) for all \(o,d\) with \(P_{od}>0\), i.e. \(\theta=(c,b_2,\ldots,b_K)'\) satisfies
    \[
    c+b_d-b_o=0
    \quad\text{for all }(o,d)\text{ with }P_{od}>0.
    \]
    Since \(G^u\) is connected, there exists at least one off-diagonal supported transition: for some \(u\ne v\), $P_{uv}+P_{vu}>0$. If \(P_{uv}>0\), Assumption~\ref{ass: invertibility of A}(ii) implies \(P_{uu}>0\); if \(P_{vu}>0\), it implies \(P_{vv}>0\). Hence, there exists some firm \(r\) with \(P_{rr}>0\). Applying the displayed equation to the supported diagonal cell \((r,r)\) gives
    \[
    c+b_r-b_r=c=0.
    \]
    Now take any edge \(\{u,v\}\in \mathcal U\). If \(P_{uv}>0\), then
    \[
    0=c+b_v-b_u=b_v-b_u.
    \]
    If \(P_{vu}>0\), then
    \[
    0=c+b_u-b_v=b_u-b_v.
    \]
    Thus \(b_u=b_v\) along every edge of \(G^u\). Since \(G^u\) is connected, all \(b_k\)'s are equal. Because \(b_1=0\), it follows that
    \[
    b_1=b_2=\cdots=b_K=0.
    \]
    Together with \(c=0\), this implies \(\theta=0\). Therefore \(\mathbb A\) is positive definite and hence invertible.
\end{proof}

\begin{corollary}\label{cor: signed-path-between-firms}
    Define the directed graph $G^{d}:=(\{1,\ldots,K\},\mathcal E)$. If Assumption~\ref{ass: invertibility of A}(i) holds, then for any two distinct firms \(u,v\in \{1,\ldots,K\}\), there exists a signed path in $G^{d}$ (as defined in Section~\ref{sec: descriptive tool}) from \(u\) to \(v\).
\end{corollary}

\begin{proof}
    By Lemma~\ref{lem: connectedness}, the undirected graph \(G^{u}\) is connected. Therefore, for any two distinct firms \(u,v\in \{1,\ldots,K\}\), there exists a path in $G^{u}$ from \(u\) to \(v\)
    \[
    u=j_0,j_1,\ldots,j_L=v
    \]
    with \(L\geq 1\) such that, for each \(\ell=1,\ldots,L\),
    \[
    P_{j_{\ell-1},j_\ell}+P_{j_\ell,j_{\ell-1}}>0 .
    \]
    For each \(\ell\), choose a supported directed move \(e_\ell=(u_\ell,v_\ell)\in\mathcal E\) whose endpoints satisfy \( \{u_\ell,v_\ell\}=\{j_{\ell-1},j_\ell\}\). Then
    \[
    W=(j_0,s_1,j_1,\ldots,s_L,j_L),
    \qquad s_\ell=\begin{cases}+1,&e_\ell=(j_{\ell-1},j_\ell),\\-1,&e_\ell=(j_\ell,j_{\ell-1}),\end{cases}
    \]
    is a signed path in $G^{d}$ from \(u\) to \(v\).
\end{proof}

\subsubsection{A classical flow decomposition}
\label{app: classical-flow-decomposition}

In this section, we state a classical decomposition result that is used in the proof of Lemma~\ref{lem: cycle correction}. The result is stated for a finite directed multigraph. Formally, a finite directed multigraph is a tuple
\[
G=(V,E,t,h),
\]
where $V$ is a finite set of vertices, $E$ is a finite set, possibly empty, of directed edges, and $t,h:E\to V$ assign to each edge $e\in E$ its tail $t(e)\in V$ and head $h(e)\in V$. We assume throughout that there are no loops, i.e., $t(e)\neq h(e)$ for every $e\in E$. Distinct edges may have the same tail and head. An ordinary directed graph is the special case in which each edge can be identified with its ordered pair of endpoints $e=(u,v)$, with $t(e)=u$ and $h(e)=v$.

A path in $G$ is a sequence
\[
W=(v_0,e_1,v_1,\ldots,e_L,v_L)
\]
such that $t(e_\ell)=v_{\ell-1}$ and $h(e_\ell)=v_\ell$ for each $\ell=1,\ldots,L$. A path is called simple if $v_0,\ldots,v_L$ are distinct. A cycle is a path with $v_L=v_0$; it is called simple if $v_0,\ldots,v_{L-1}$ are distinct. For a path $W=(v_0,e_1,v_1,\ldots,e_L,v_L)$, define its \emph{unsigned incidence vector} $\pi^+(W)\in\mathbb R^E$ as the vector with $e$-th element equal to
\begin{equation}\label{eq: unsigned incidence vector definition}
\pi^+_e(W):=\sum_{\ell=1}^L\ind\{e_\ell=e\}.
\end{equation}
This vector counts how often each directed edge $e\in E$ occurs in the path. For a simple path or a simple cycle, no edge occurs more than once, and hence $\pi^+(W)$ has entries in $\{0,1\}$.

For a vector $x\in\mathbb R^E$, which assigns a real-valued weight to each edge in $E$, define the net inflow induced by $x$ at vertex $k\in V$ by
\[
b_k(x):=\sum_{e\in E:\,h(e)=k}x_e-\sum_{e\in E:\,t(e)=k}x_e.
\]
Thus a positive value of $b_k(x)$ means that, under the edge weights $x$, more mass enters $k$ than leaves $k$.

The following theorem is a well-known result in graph theory, sometimes referred to as the ``flow decomposition theorem,'' that decomposes a nonnegative weight vector, or flow, $x\in\mathbb R^E$ into a sum of weights along simple paths and simple cycles. The version below is a specialization of Proposition~1.1 in \citet{Bertsekas:1998aa} to a nonnegative flow with one ``source'' and one ``sink'' and a generalization to directed multigraphs following his discussion on p.~3. For completeness, we also give a proof.

\begin{theorem}[classical flow decomposition]\label{thm:flow-decomposition}
Let $G=(V,E,t,h)$ be a finite directed multigraph and let $x\in\mathbb R^E$ be nonnegative. Then there exist a finite collection, possibly empty, of simple paths $\mathcal W$ in $G$, a finite collection, possibly empty, of simple cycles $\mathcal C$ in $G$, and positive constants $\{\lambda(W)\}_{W\in\mathcal W}$ and $\{\eta(C)\}_{C\in\mathcal C}$ such that
\[
x=\sum_{W\in\mathcal W}\lambda(W)\pi^+(W)+\sum_{C\in\mathcal C}\eta(C)\pi^+(C).
\]
In this decomposition, a sum over an empty collection is defined to be zero.
\end{theorem}

\begin{proof}
Let $y^0=x$. If $y^0=0$, there is nothing to prove. Otherwise, suppose that $y^t\neq 0$ has been constructed. Let
\[
\operatorname{supp}(y^t):=\{e\in E:y^t_e>0\}.
\]
Choose an edge $e_1\in\operatorname{supp}(y^t)$ and set
\[
v_0:=t(e_1),\qquad v_1:=h(e_1).
\]
Starting from the walk
\[
(v_0,e_1,v_1),
\]
extend it as follows. Suppose we have constructed
\[
(v_0,e_1,v_1,\ldots,e_k,v_k)
\]
with $v_0,\ldots,v_k$ distinct. If there is no edge $e\in\operatorname{supp}(y^t)$ with $t(e)=v_k$, then stop and call
\[
W=(v_0,e_1,v_1,\ldots,e_k,v_k)
\]
the extracted object. This is a simple path.

If there exists an edge $e_{k+1}\in\operatorname{supp}(y^t)$ with $t(e_{k+1})=v_k$, choose one such edge and set $w:=h(e_{k+1})$. If $w\notin\{v_0,\ldots,v_k\}$, append $e_{k+1}$ and $w$ to the walk and continue. If $w=v_r$ for some $r\in\{0,\ldots,k\}$, then stop and call \(C=(v_r,e_{r+1},v_{r+1},\ldots,e_k,v_k,e_{k+1},v_r)\) the extracted object. Since $v_0,\ldots,v_k$ were distinct, $C$ is a simple cycle. Because $V$ is finite, this extension procedure terminates after finitely many steps. Thus at iteration $t$, we obtain either a simple path or a simple cycle $Q_t$, all of whose edges lie in $\operatorname{supp}(y^t)$.

Define
\[
\alpha_t:=\min\{y^t_e:e\text{ is an edge of }Q_t\}.
\]
Since every edge of $Q_t$ lies in $\operatorname{supp}(y^t)$, we have $\alpha_t>0$. Set
\[
y^{t+1}:=y^t-\alpha_t\pi^+(Q_t).
\]
Because $Q_t$ is simple, its unsigned incidence vector has entries only in $\{0,1\}$. Hence $y^{t+1}\geq 0$. Moreover, at least one edge of $Q_t$ attains the minimum defining $\alpha_t$, so at least one positive residual edge becomes zero. Therefore
\[
|\operatorname{supp}(y^{t+1})|<|\operatorname{supp}(y^t)|.
\]
Since $E$ is finite, the strict support decrease can occur only finitely many times, say $T$ times. The algorithm therefore reaches a residual $y^T$ with $\operatorname{supp}(y^T)=\emptyset$, and hence $y^T=0$.

For each $t=0,\ldots,T-1$, the update rule gives
\[
y^{t+1}=y^t-\alpha_t\pi^+(Q_t).
\]
Summing these identities telescopically yields
\[
x=y^0=\sum_{t=0}^{T-1}\alpha_t\pi^+(Q_t).
\]
Finally, collect all extracted simple paths as $\mathcal W$ and all extracted simple cycles as $\mathcal C$, with the corresponding coefficients denoted by $\lambda(W)>0$, $W\in\mathcal W$, and $\eta(C)>0$, $C\in\mathcal C$. This gives the desired decomposition.
\end{proof}

\begin{corollary}[cycle decomposition of a circulation]\label{cor:circulation-decomposition}
Let $G=(V,E,t,h)$ be a finite directed multigraph and let $x\in\mathbb R^E$ be nonnegative and satisfy
\[
b_k(x)=0\qquad\text{for every }k\in V.
\]
Then there exist a finite collection, possibly empty, of simple cycles $\mathcal C$ in $G$ and positive constants $\{\eta(C)\}_{C\in\mathcal C}$ such that
\[
x=\sum_{C\in\mathcal C}\eta(C)\pi^+(C).
\]
\end{corollary}

\begin{proof}
Use the constructive algorithm from the proof of Theorem~\ref{thm:flow-decomposition}, starting from $y^0=x$. We prove by induction on $t$ that, whenever $y^t$ is constructed,
\[
b_k(y^t)=0\qquad\text{for every }k\in V.
\]
This is true at $t=0$ by assumption. Suppose it is true at step $t$ and that $y^t\neq 0$. In the extension procedure from the proof of Theorem~\ref{thm:flow-decomposition}, consider a constructed walk \((v_0,e_1,v_1,\ldots,e_L,v_L)\), where every edge $e_1,\ldots,e_L$ belongs to $\operatorname{supp}(y^t)$. If the path-stopping case were to occur at $v_L$, then there would be no edge $e\in\operatorname{supp}(y^t)$ with $t(e)=v_L$. Hence
\[
\sum_{e\in E:\,t(e)=v_L}y^t_e=0.
\]
But the last edge $e_L$ satisfies $h(e_L)=v_L$ and $y^t_{e_L}>0$, so
\[
\sum_{e\in E:\,h(e)=v_L}y^t_e\geq y^t_{e_L}>0.
\]
Therefore
\[
b_{v_L}(y^t)=\sum_{e\in E:\,h(e)=v_L}y^t_e-\sum_{e\in E:\,t(e)=v_L}y^t_e>0,
\]
contradicting the induction hypothesis. Thus the path-stopping case cannot occur. The object extracted at step $t$ must be a simple cycle; denote it by $C_t$.

For every directed cycle $C$ and every vertex $k\in V$,
\[
b_k(\pi^+(C))=0,
\]
because, along a cycle, each visit to $k$ contributes exactly one entering edge and one leaving edge. Hence the update
\[
y^{t+1}=y^t-\alpha_t\pi^+(C_t)
\]
satisfies $b_k(y^{t+1})=0$ for every $k\in V$. The support-decrease argument in the proof of Theorem~\ref{thm:flow-decomposition} shows that the algorithm terminates after finitely many steps with some $y^T=0$. Since every extracted object is a simple cycle, the telescoping identity from the proof of Theorem~\ref{thm:flow-decomposition} gives
\[
x=\sum_{C\in\mathcal C}\eta(C)\pi^+(C)
\]
for a finite collection $\mathcal C$ of simple cycles and positive coefficients $\{\eta(C)\}_{C\in\mathcal C}$.
\end{proof}

\begin{corollary}[decomposition of a one-unit directed flow]\label{cor:unit-flow-decomposition}
Let $G=(V,E,t,h)$ be a finite directed multigraph and let $x\in\mathbb R^E$ be nonnegative. Fix distinct vertices $o,d\in V$. Suppose that
\[
b_k(x)=\ind\{d=k\}-\ind\{o=k\}\qquad\text{for every }k\in V.
\]
Then there exist a finite nonempty collection $\mathcal W$ of simple directed paths from $o$ to $d$, a finite collection, possibly empty, of simple directed cycles $\mathcal C$, and positive constants $\{\lambda(W)\}_{W\in\mathcal W}$ and $\{\eta(C)\}_{C\in\mathcal C}$ such that
\[
\sum_{W\in\mathcal W}\lambda(W)=1
\]
and
\[
x=\sum_{W\in\mathcal W}\lambda(W)\pi^+(W)+\sum_{C\in\mathcal C}\eta(C)\pi^+(C).
\]
\end{corollary}

\begin{proof}
Add one artificial directed edge $e_0\notin E$ from $d$ to $o$. Form the augmented directed multigraph
\(\overline G=(V,\overline E,\overline t,\overline h)\), where \(\overline E:=E\cup\{e_0\}\), \(\overline t\) and \(\overline h\) agree with \(t\) and \(h\) on \(E\), and \(\overline t(e_0)=d\) and \(\overline h(e_0)=o\).
Define \(\overline x\in\mathbb R^{\overline E}\) by \(\overline x_e:=x_e\), \(e\in E\), and \(\overline x_{e_0}:=1\).
Let $\overline b_k$ denote the map defined as $b_k$ but on the augmented graph $\overline G$. For every $k\in V$,
\[
\overline b_k(\overline x)=b_k(x)+\ind\{o=k\}-\ind\{d=k\}=0.
\]
By Corollary~\ref{cor:circulation-decomposition}, there is a finite collection $\mathcal R$ of simple cycles in $\overline G$ and positive constants $\{\gamma(R)\}_{R\in\mathcal R}$ such that
\[
\overline x=\sum_{R\in\mathcal R}\gamma(R)\pi^+(R).
\]
Separate these cycles according to whether they contain the artificial edge $e_0$. If $R$ contains $e_0$, then, because $R$ is simple, it contains $e_0$ exactly once. Rotate the cycle so that it is written as \(R=(d,e_0,o,f_1,w_1,\ldots,f_M,w_M)\), where \(w_M=d\). Deleting the artificial edge $e_0$ leaves a simple directed path \(W_R=(o,f_1,w_1,\ldots,f_M,w_M)\) from $o$ to $d$ in the original graph $G$. If $R$ does not contain $e_0$, then $R$ is already a simple directed cycle in the original graph $G$.

Restricting the decomposition of $\overline x$ to the original edge set $E$ gives
\[
x=\sum_{R\in\mathcal R:\,e_0\in R}\gamma(R)\pi^+(W_R)+\sum_{R\in\mathcal R:\,e_0\notin R}\gamma(R)\pi^+(R).
\]
Looking instead at the coordinate $e_0$ gives
\[
1=\overline x_{e_0}=\sum_{R\in\mathcal R:\,e_0\in R}\gamma(R),
\]
because every simple cycle containing $e_0$ uses it exactly once. Hence the paths obtained by deleting $e_0$ have positive coefficients that sum to one. In particular, there is at least one such path. Relabeling these paths as $\mathcal W$ and the remaining cycles as $\mathcal C$ yields the result.
\end{proof}

\subsection{Proofs for Section~\ref{sec: descriptive tool}}
\label{app: proofs section descriptive}

\begin{proof}[Proof of Lemma~\ref{lem: arbitrary no of firms}]
Since $\mathbb{A}$ is invertible, the solution $(\beta,\alpha_2,\ldots,\alpha_K)'$ to \eqref{eq: AKM model differences simplified} and \eqref{eq: exogenous mobility} can be written as
\[
\begin{pmatrix}\beta\\\alpha_2\\\vdots\\\alpha_K \end{pmatrix} = \mathbb{A}^{-1} E\left[ \begin{pmatrix}
1\\ J-I \end{pmatrix} E(\Delta \mid O,D ) \right]
\]
and thus
\begin{equation}
\alpha_j = \sum_{d=1}^K e_j'\mathbb{A}^{-1} e_d \left(E\left[ \begin{pmatrix}
1\\ J-I \end{pmatrix} E(\Delta \mid O,D ) \right]\right)_d.
\label{eq: alpha_j expression 1}
\end{equation}
We have
\[
\left(E\left[ \begin{pmatrix}
1\\ J-I \end{pmatrix} E(\Delta \mid O,D ) \right]\right)_1 = \sum_{1\leq o,d\leq K\colon P_{od}>0} P_{od} E[\Delta \mid O=o,D=d]
\]
and, for $d=2,\ldots,K$,
\begin{align*}
\left(E\left[ \begin{pmatrix}
1\\ J-I \end{pmatrix} E(\Delta \mid O,D ) \right]\right)_d
&= \sum_{o=1\colon P_{od}>0}^K P_{od} E[\Delta \mid O=o,D=d] \\
&\quad - \sum_{o=1\colon P_{do}>0}^K P_{do} E[\Delta \mid O=d,D=o].
\end{align*}
Substituting these two expressions into \eqref{eq: alpha_j expression 1} yields
\begin{align*}
\alpha_j &= e_j'\mathbb{A}^{-1} e_1 \sum_{1\leq o,d\leq K\colon P_{od}>0} P_{od} E[\Delta \mid O=o,D=d]\\
&\quad + \sum_{d=2}^K e_j'\mathbb{A}^{-1} e_d \sum_{o=1\colon P_{od}>0}^K P_{od} E[\Delta \mid O=o,D=d]\\
&\quad - \sum_{d=2}^K e_j'\mathbb{A}^{-1} e_d \sum_{o=1\colon P_{do}>0}^K P_{do} E[\Delta \mid O=d,D=o].
\end{align*}
Collecting terms in front of $E[\Delta \mid O=o,D=d]$ yields
\begin{equation}
\alpha_j = \sum_{1\leq o,d\leq K\colon P_{od}>0} \omega_{j,od} E[\Delta \mid O=o,D=d].
\label{eq: alpha_j decomposition step before DID}
\end{equation}
The desired expression in \eqref{eq: alpha_j for arbitrary no of firms} then follows after adding and subtracting
\[
\sum_{1\leq o,d\leq K\colon P_{od}>0} \omega_{j,od}E[\Delta \mid O=o,D=o].
\]
Finally, $\sum_{1\leq o,d\leq K}\omega_{j,od} =0$ is shown in Lemma~\ref{lem: properties of the weights}(i).
\end{proof}

The following proofs use the signed incidence vector associated with a signed path. It is the signed analogue of the unsigned incidence vector $\pi^+$ in \eqref{eq: unsigned incidence vector definition}: each step contributes with the sign of its traversal rather than with a nonnegative count. Specifically, for a signed path $W$ as in \eqref{eq: signed-path-definition}, the \emph{signed incidence vector} $\pi(W)\in\mathbb R^{|\mathcal E|}$ is the vector of dimension $|\mathcal E|$ with elements $\pi_e(W)$ indexed by supported moves $e\in\mathcal E$ and given by
\begin{equation}\label{eq: signed incidence vector definition}
\pi_e(W):=\sum_{\ell\colon\,e_\ell=e}s_\ell ,
\end{equation}
where $e_\ell$ is the move used at step $\ell$ and $s_\ell$ its sign.

Thus $\pi_e(W)$ is the net number of times the signed path traverses the supported move $e$: each traversal along the move contributes $+1$ and each traversal against it contributes $-1$. If the signed path uses each supported move at most once, then $\pi_e(W)\in\{-1,0,1\}$. The same definition applies to signed cycles. When every step of $W$ is traversed along its own direction, $\pi(W)$ coincides with the unsigned incidence vector $\pi^+(W)$ of Section~\ref{app: classical-flow-decomposition} computed on $G^d$. Collecting the steps that use the same supported move gives
\[
S(W,m)=\sum_{e\in\mathcal E}\pi_e(W)\, m_e
\qquad\text{for every }m\in\mathbb R^{|\mathcal E|},
\]
so the signed sum of Section~\ref{sec: descriptive tool} is the inner product of $m$ with the signed incidence vector.

\begin{proof}[Proof of Lemma~\ref{lem: cycle correction}]
Let $V_K:=\{1,\ldots,K\}$ and fix distinct firms $j,k\in V_K$. To simplify notation, write $\kappa^{jk}_e:=\kappa^{jk}_{uv}$ for $e=(u,v)\in\mathcal E$.
Let $G^d:=(V_K,\mathcal E)$ be the directed graph defined by the supported moves in $\mathcal E$, where each $e=(u,v)\in\mathcal E$ has tail $t(e)=u$ and head $h(e)=v$.

The proof proceeds in four steps. First, we construct an auxiliary directed multigraph $G^+$ in which an edge is weighted by $|\kappa^{jk}_e|$. Second, we show that $(|\kappa^{jk}_e|\colon e\in\mathcal E)$ represents a one-unit directed flow from $j$ to $k$ in $G^+$. Third, we apply Corollary~\ref{cor:unit-flow-decomposition} to $G^+$ to obtain a decomposition of $(|\kappa^{jk}_e|\colon e\in\mathcal E)$ in terms of paths and cycles in $G^+$. Fourth, we translate the path and cycle decomposition back into paths and cycles in the original graph $G^d$.

\textbf{Step 1:} We construct an auxiliary finite directed multigraph \(G^+=(V_K,\mathcal A,t^+,h^+)\) whose directed edges are labelled by supported moves in the original graph. For every $e=(u,v)\in\mathcal E$ with $\kappa^{jk}_e\neq 0$, introduce one auxiliary directed edge $a_e$, labelled by $e$, with
\begin{equation*}
    t^+(a_e) = \begin{cases}
        u & \text{if } \kappa^{jk}_e>0, \\
        v & \text{if } \kappa^{jk}_e<0.
    \end{cases}\qquad\text{and}\qquad
    h^+(a_e) = \begin{cases}
        v & \text{if } \kappa^{jk}_e>0, \\
        u & \text{if } \kappa^{jk}_e<0.
    \end{cases}
\end{equation*}
Then $\mathcal{A}$ collects all such auxiliary directed edges:
$\mathcal{A}:=\{a_e\colon e\in\mathcal E,\ \kappa^{jk}_e\neq 0\}$. The auxiliary graph is a directed multigraph because distinct moves $e_1,e_2\in\mathcal E$ may induce distinct directed edges $a_{e_1},a_{e_2}\in\mathcal A$ in the graph $G^+$ with the same tail and head, i.e. $t^+(a_{e_1})=t^+(a_{e_2})$ and $h^+(a_{e_1})=h^+(a_{e_2})$.

\textbf{Step 2:} Define $x\in\mathbb R^{\mathcal A}$ by $x_{a_e}:=|\kappa^{jk}_e|$ for $a_e\in\mathcal A$. Denote its net inflow at vertex $r$ in the auxiliary graph by
\[
b_r^+(x):=\sum_{a\in\mathcal A:\,h^+(a)=r}x_a-\sum_{a\in\mathcal A:\,t^+(a)=r}x_a.
\]
We now show that $x$ is a one-unit directed flow from $j$ to $k$, i.e. for all $r\in\{1,\ldots,K\}$,
\begin{equation}\label{eq:one-unit-flow}
b_r^+(x)=\ind\{k=r\}-\ind\{j=r\}.
\end{equation}
To this end, first notice that, for each $e\in\mathcal E$ with $\kappa^{jk}_e\neq 0$,
\[
\ind\{h^+(a_e)=r\}|\kappa^{jk}_e|-\ind\{t^+(a_e)=r\}|\kappa^{jk}_e|
=
\kappa^{jk}_e\bigl(\ind\{h(e)=r\}-\ind\{t(e)=r\}\bigr).
\]
If $\kappa^{jk}_e>0$, then $a_e$ has the same direction as $e$; if $\kappa^{jk}_e<0$, then $a_e$ has the reverse direction and $|\kappa^{jk}_e|=-\kappa^{jk}_e$. Therefore, summing over $e$ gives, for every $r\in V_K$,
\begin{align*}
b_r^+(x)
&=\sum_{e\in\mathcal E:\,\kappa^{jk}_e\neq 0}
\left[
\ind\{h^+(a_e)=r\}|\kappa^{jk}_e|-\ind\{t^+(a_e)=r\}|\kappa^{jk}_e|
\right] \\
&=\sum_{e\in\mathcal E}\kappa^{jk}_e\bigl(\ind\{h(e)=r\}-\ind\{t(e)=r\}\bigr) \\
&=\sum_{(u,v)\in\mathcal E}\kappa_{uv}^{jk}\bigl(\ind\{v=r\}-\ind\{u=r\}\bigr).
\end{align*}
We next evaluate the final sum. For $r\in\{2,\ldots,K\}$, unsupported moves have zero weights and diagonal terms satisfy $\ind\{v=r\}-\ind\{u=r\}=0$. Therefore, using Lemma~\ref{lem: properties of the weights}(ii) and recalling that $\omega_{1,uv}=0$, we obtain, for all $r\in\{2,\ldots,K\}$,
\begin{align*}
\sum_{(u,v)\in\mathcal E}\kappa_{uv}^{jk}\bigl(\ind\{v=r\}-\ind\{u=r\}\bigr)
&=\sum_{u=1}^K\sum_{v=1}^K(\omega_{k,uv}-\omega_{j,uv})\bigl(\ind\{v=r\}-\ind\{u=r\}\bigr) \\
&=\ind\{k=r\}-\ind\{j=r\}.
\end{align*}
Thus, it remains to show that \eqref{eq:one-unit-flow} holds also for $r=1$. However, this follows immediately from the following two facts. First,
\[
\sum_{r=1}^K b_r^+(x)=0
\]
because each auxiliary directed edge is counted once with sign $+1$ and once with sign $-1$. Second,
\[
\sum_{r=1}^K\bigl(\ind\{k=r\}-\ind\{j=r\}\bigr)=0.
\]
Therefore, we have shown that \eqref{eq:one-unit-flow} holds for all $r\in\{1,\ldots,K\}$.

\textbf{Step 3:} By Corollary~\ref{cor:unit-flow-decomposition} applied to the auxiliary directed multigraph $G^+$, there exist a finite nonempty collection $\mathcal W^+$ of simple directed paths from $j$ to $k$, a finite collection, possibly empty, $\mathcal C^+$ of simple directed cycles, and positive coefficients $\{\lambda^+(\widetilde W)\}_{\widetilde W\in\mathcal W^+}$ and $\{\eta^+(\widetilde C)\}_{\widetilde C\in\mathcal C^+}$ such that
\[
\sum_{\widetilde W\in\mathcal W^+}\lambda^+(\widetilde W)=1
\]
and
\begin{equation}\label{eq:auxiliary-unit-flow-decomposition}
x=\sum_{\widetilde W\in\mathcal W^+}\lambda^+(\widetilde W)\pi^+(\widetilde W)+\sum_{\widetilde C\in\mathcal C^+}\eta^+(\widetilde C)\pi^+(\widetilde C),
\end{equation}
where $\pi^+(\widetilde W)\in\mathbb R^{\mathcal A}$ is the unsigned incidence vector \eqref{eq: unsigned incidence vector definition} of the directed path $\widetilde W$ in the auxiliary multigraph $G^+$, with coordinates indexed by the auxiliary edges $a\in\mathcal A$.

\textbf{Step 4:} We now translate paths and cycles in the auxiliary graph $G^+$ back into signed paths and signed cycles in the original graph $G^d$. Consider a path
\(\widetilde W=(j_0,a_{e_1},j_1,\ldots,a_{e_L},j_L)\), where \(j_0=j\) and \(j_L=k\).
from $j$ to $k$ in the auxiliary graph $G^+$. Because $a_{e_\ell}$ is labelled by the move $e_\ell\in\mathcal E$, $\widetilde W$ induces the signed path
\(\Phi(\widetilde W):=(j_0,s_1,j_1,\ldots,s_L,j_L)\), where \(s_\ell:=\operatorname{sign}(\kappa^{jk}_{e_\ell})\).
from $j$ to $k$ in the notation of Definition~\ref{def: signed path and signed cycle}; its move at step $\ell$ is $e_\ell$. Since $\widetilde W$ is simple, $\Phi(\widetilde W)$ is a simple signed path. Similarly, each simple directed cycle $\widetilde C\in\mathcal C^+$ induces a simple signed cycle, denoted $\Phi(\widetilde C)$, in the original graph $G^d$. Let $\mathcal W_s^{jk}$ be the finite collection of distinct signed paths in $G^d$ induced by paths in $\mathcal W^+$, and let $\mathcal C_s^{jk}$ be the finite collection of distinct signed cycles in $G^d$ induced by cycles in $\mathcal C^+$. Define constants $\lambda_s^{jk}(W)$ and $\eta_s^{jk}(C)$ for $G^d$ corresponding to the coefficients $\lambda^+(\widetilde W)$ and $\eta^+(\widetilde C)$ in the auxiliary graph $G^+$ by summing over all paths or cycles in the auxiliary graph $G^+$ that induce the same signed object in $G^d$:
\[
\lambda_s^{jk}(W):=\sum_{\widetilde W\in\mathcal W^+:\,\Phi(\widetilde W)=W}\lambda^+(\widetilde W),
\qquad W\in\mathcal W_s^{jk},
\]
and
\[
\eta_s^{jk}(C):=\sum_{\widetilde C\in\mathcal C^+:\,\Phi(\widetilde C)=C}\eta^+(\widetilde C),
\qquad C\in\mathcal C_s^{jk}.
\]
The constants $\{\lambda_s^{jk}(W)\}_{W\in\mathcal W_s^{jk}}$ and $\{\eta_s^{jk}(C)\}_{C\in\mathcal C_s^{jk}}$ inherit the properties that they are positive and that $\{\lambda_s^{jk}(W)\}_{W\in\mathcal W_s^{jk}}$ sum to one from the corresponding constants in the auxiliary graph.

It remains to verify the representation of the AKM contrasts in \eqref{eq: cycle correction representation}. Fix $e\in\mathcal E$. If $\kappa^{jk}_e=0$, no auxiliary edge labelled by $e$ was introduced, and no induced signed path or signed cycle uses $e$; hence the $e$-th coordinate of the claimed decomposition is zero on both sides. If $\kappa^{jk}_e\neq 0$, let \(s_e:=\operatorname{sign}(\kappa^{jk}_e)\in\{-1,1\}\). For every auxiliary path or cycle $R$, the signed incidence vector of the induced signed path or cycle $\Phi(R)$ in $G^d$ and the unsigned incidence vector of $R$ in $G^+$ are related by
\[
\pi_e(\Phi(R))=s_e\cdot\pi^+_{a_e}(R),
\]
because $a_e$ is directed in the same direction as $e$ when $s_e=1$ and in the reverse direction when $s_e=-1$. The $a_e$-coordinate of the unsigned decomposition in \eqref{eq:auxiliary-unit-flow-decomposition} gives
\[
|\kappa^{jk}_e|=\sum_{\widetilde W\in\mathcal W^+}\lambda^+(\widetilde W)\pi^+_{a_e}(\widetilde W)+\sum_{\widetilde C\in\mathcal C^+}\eta^+(\widetilde C)\pi^+_{a_e}(\widetilde C).
\]
Multiplying by $s_e$ yields the corresponding coordinate of the signed decomposition,
\[
\kappa^{jk}_e=\sum_{W\in\mathcal W_s^{jk}}\lambda_s^{jk}(W)\pi_e(W)+\sum_{C\in\mathcal C_s^{jk}}\eta_s^{jk}(C)\pi_e(C).
\]
Since this holds for every $e\in\mathcal E$, we have the vector identity
\begin{equation}\label{eq:kappa-simple-decomposition}
\kappa^{jk}=\sum_{W\in\mathcal W_s^{jk}}\lambda_s^{jk}(W)\pi(W)+\sum_{C\in\mathcal C_s^{jk}}\eta_s^{jk}(C)\pi(C).
\end{equation}
Here $\kappa^{jk}\in\mathbb R^{|\mathcal E|}$ is the vector with $e$-th element equal to $\kappa^{jk}_e$. The construction also shows that, for every $e$ with $\kappa^{jk}_e\neq 0$, all nonzero coordinates $\pi_e(W)$ and $\pi_e(C)$ have the sign of $\kappa^{jk}_e$; if $\kappa^{jk}_e=0$, all these coordinates are zero. Taking the inner product of both sides with the vector $\DID\in\mathbb R^{|\mathcal E|}$ gives
\[
\sum_{(u,v)\in\mathcal E}\kappa_{uv}^{jk}\DID(u,v)
=
\sum_{W\in\mathcal W_s^{jk}}\lambda_s^{jk}(W)S(W,\DID)+\sum_{C\in\mathcal C_s^{jk}}\eta_s^{jk}(C)S(C,\DID).
\]
Combining this identity with the edge-level representation \eqref{eq: adjusted AKM kappa contrast representation} proves the desired representation of $\widetilde\alpha_k-\widetilde\alpha_j$.

The construction of $G^+$, the auxiliary weights $x$, and the resulting decomposition uses only the signs and magnitudes of the coordinates of $\kappa^{jk}$. With any fixed deterministic rule for choosing the auxiliary paths, cycles, and coefficients, the signed paths, signed cycles, and coefficients are therefore chosen as functions only of $\{\kappa^{jk}_{uv}\}_{(u,v)\in\mathcal E}$, and hence only of the mobility shares.
\end{proof}

The proofs of Theorems~\ref{thm: akm path decomposition} and~\ref{thm: did characterization} use the following operations on signed paths in the graph $G^d$ (Definition~\ref{def: signed path and signed cycle}). For a signed path $W=(j_0,s_1,j_1,\ldots,s_L,j_L)$, define its reversal by \(\overline W=(j_L,-s_L,j_{L-1},\ldots,-s_1,j_0)\).
It uses the same moves in reverse order and direction, so $\pi(\overline W)=-\pi(W)$. For signed paths $W=(j_0,s_1,j_1,\ldots,s_L,j_L)$ and $W'=(j_L,s'_1,j'_1,\ldots,s'_{L'},j'_{L'})$, where $W'$ starts at the firm at which $W$ ends, define their concatenation by
\[
W\oplus W':=(j_0,s_1,j_1,\ldots,s_L,j_L,s'_1,j'_1,\ldots,s'_{L'},j'_{L'}).
\]
Then, $\pi(W\oplus W')=\pi(W)+\pi(W')$. For a signed cycle $C$, denote by $C^n$ the signed cycle that traverses $C$ $n$ times. Then, such a repeated signed cycle satisfies $\pi(C^n)=n\pi(C)$. Rotating the starting firm of a signed cycle leaves its incidence vector unchanged.

\begin{proof}[Proof of Theorem~\ref{thm: akm path decomposition}]
Fix distinct firms $j,k$. We start with the decomposition constructed in Lemma~\ref{lem: cycle correction} and then fold the cycles into the paths one at a time.

To this end, initially, set $\mathcal W:=\mathcal W_s^{jk}$, $\mathcal C:=\mathcal C_s^{jk}$, $\lambda:=\lambda_s^{jk}$, and $\eta:=\eta_s^{jk}$, where $\mathcal W_s^{jk}$, $\mathcal C_s^{jk}$, $\lambda_s^{jk}$, and $\eta_s^{jk}$ are as in Lemma~\ref{lem: cycle correction}. If $\mathcal C$ is empty, no folding is needed. Otherwise, fix a cycle $C\in\mathcal C$ and a path $W^*\in\mathcal W$, and write $W^*=(j_0,s_1,j_1,\ldots,s_L,j_L)$ with $j_0=j$ and $j_L=k$.

Next, let $n$ be the smallest integer $n\geq 1$ such that $\eta(C)/n\leq\lambda(W^*)$. We construct a signed path $W^\circ$ from $j$ to $k$ satisfying
\[
\pi(W^\circ)=\pi(W^*)+n\pi(C)
\]
by distinguishing two cases. First, suppose $C$ passes through one of the firms $j_0,\ldots,j_L$ on $W^*$. Let $m$ be the smallest index such that $j_m$ lies on $C$ and replace $C$ by its rotation based at $j_m$. Define $W^\circ$ as the tuple obtained from $W^*$ by inserting the steps of $C^{n}$ after position $m$: the path $W^\circ$ follows $W^*$ up to $j_m$, traverses the rotated cycle $n$ times, returning each time to $j_m$, and then continues along the remaining steps of $W^*$. Since consecutive firms agree at each insertion point, $W^\circ$ is a signed path from $j$ to $k$, and since every step of $W^*$ and of $C^{n}$ appears exactly once in $W^\circ$, its incidence vector is $\pi(W^\circ)=\pi(W^*)+n\pi(C)$.

Second, suppose $C$ does not pass through any firm on $W^*$. Let $c$ denote the firm at which $C$ is based. The firms $j_0$ and $c$ are then distinct, and Corollary~\ref{cor: signed-path-between-firms} yields a signed path $R$ from $j_0$ to $c$. Define the signed path \(W^\circ:=R\oplus C^{n}\oplus\overline R\oplus W^*\) from $j$ to $k$. By the properties of reversal and concatenation stated above, \(\pi(W^\circ)=\pi(R)+n\pi(C)-\pi(R)+\pi(W^*)=\pi(W^*)+n\pi(C)\).

Now define a new collection $\mathcal W''$ of signed paths from $j$ to $k$ with weights $\lambda''$ as follows: assign to $W^\circ$ the weight $\eta(C)/n$, decrease the weight of $W^*$ to $\lambda(W^*)-\eta(C)/n$ and remove $W^*$ from the collection if this quantity is zero, and leave all other paths and weights unchanged. (If $W^\circ$ coincides with a path already in the collection, add the weights.) All weights in the new collection are positive and
\[
\sum_{W\in\mathcal W''}\lambda''(W)=\sum_{W\in\mathcal W}\lambda(W)=1.
\]
Moreover,
\begin{align*}
\sum_{W\in\mathcal W''}\lambda''(W)\pi(W)
&=\sum_{W\in\mathcal W}\lambda(W)\pi(W)-\frac{\eta(C)}{n}\pi(W^*)+\frac{\eta(C)}{n}\bigl(\pi(W^*)+n\pi(C)\bigr)\\
&=\sum_{W\in\mathcal W}\lambda(W)\pi(W)+\eta(C)\pi(C).
\end{align*}
Replacing $\mathcal W,\lambda$ by $\mathcal W'',\lambda''$ and deleting $C$ therefore preserves \eqref{eq:kappa-simple-decomposition}, with the remaining cycles in place of $\mathcal C_s^{jk}$. Repeating this construction removes every cycle after finitely many steps. Denote the resulting collection of signed paths by $\mathcal W^{jk}$ and its positive weights by $\lambda^{jk}$. The path weights still sum to one and
\[
\kappa^{jk}=\sum_{W\in\mathcal W^{jk}}\lambda^{jk}(W)\pi(W).
\]

The no-cancellation condition is preserved at each step. Indeed, the paths and cycles initially have the coordinate signs established in the proof of Lemma~\ref{lem: cycle correction}. Since $\pi(W^\circ)=\pi(W^*)+n\pi(C)$, every nonzero coordinate of the new path has the same sign as $\kappa^{jk}_e$, and coordinates with $\kappa^{jk}_e=0$ remain zero. Connector traversals cancel within $W^\circ$. Thus, for each move $e$, all final paths have nonnegative net traversals if $\kappa^{jk}_e>0$, nonpositive net traversals if $\kappa^{jk}_e<0$, and zero net traversals if $\kappa^{jk}_e=0$.

Taking the inner product with $\DID$ and using \eqref{eq: adjusted AKM kappa contrast representation} proves \eqref{eq:akm-path-representation}. If the initial cycle collection is empty, the decomposition from Lemma~\ref{lem: cycle correction} already has all these properties. Every choice made in the construction follows a fixed deterministic rule. Since the initial decomposition depends only on the mobility shares and connectors use only the supported graph, the resulting paths and weights also depend only on the mobility shares.
\end{proof}

\subsection{Invariance to the left-out firm}
\label{app: invariance-to-left-out-firm}

In this subsection, we consider AKM regressions as in \eqref{eq: AKM model differences simplified}, except that an arbitrary firm $\ell$ is left out (rather than firm 1):
\begin{equation}
    \Delta
    =
    \beta^{(\ell)}+
    \sum_{m\neq \ell}\alpha_m^{(\ell)}
    \bigl(\ind\{D=m\}-\ind\{O=m\}\bigr)+\varepsilon^{(\ell)},
    \qquad
        E\left[\varepsilon^{(\ell)} \begin{pmatrix}
           1\\ J^{(\ell)}-I^{(\ell)}
           \end{pmatrix}   \right]=0,
    \label{eq:left-out-regression}
\end{equation}
where $I^{(\ell)}:=(\ind\{O=o \}\colon o\neq \ell)'$ and $J^{(\ell)}:=(\ind\{D=d \}\colon d\neq \ell)'$. Define
$$\mathbb{A}^{(\ell)}:=E\left[ \begin{pmatrix} 1\\ J^{(\ell)}-I^{(\ell)} \end{pmatrix}\begin{pmatrix} 1\\ J^{(\ell)}-I^{(\ell)} \end{pmatrix}' \right].$$
The AKM estimand is now the vector containing the coefficients $\alpha_j^{(\ell)}$, $j\neq \ell$ that satisfy \eqref{eq:left-out-regression}, with the normalization $\alpha_{\ell}^{(\ell)}=0$.

Let $\omega_{m,od}^{(\ell)}$, $\tau_m^{(\ell)}$, $\widetilde\alpha_m^{(\ell)}:=\alpha_m^{(\ell)}-\tau_m^{(\ell)}$, and $\kappa_{od}^{jk,(\ell)}:=\omega_{k,od}^{(\ell)}-\omega_{j,od}^{(\ell)}$ denote the analogues of the objects in \eqref{eq: omega definition}, \eqref{eq: tau definition}, Definition~\ref{def: adjusted AKM estimand}, and \eqref{eq: adjusted AKM kappa contrast representation} for the regression that leaves out firm $\ell$. For this choice of reference firm, the same conventions give $\omega_{\ell,od}^{(\ell)}=\tau_\ell^{(\ell)}=\widetilde\alpha_\ell^{(\ell)}=0$.

\begin{lemma}[Invariance to the left-out firm]
\label{lem:left-out-invariance}
Fix two left-out firms $\ell_1,\ell_2\in\{1,\ldots,K\}$ and suppose that $\mathbb A^{(\ell_1)}$ and $\mathbb A^{(\ell_2)}$ are invertible. Then, for every pair of distinct firms $j,k\in\{1,\ldots,K\}$,
\begin{equation}
\widetilde\alpha_k^{(\ell_1)}-\widetilde\alpha_j^{(\ell_1)}
=
\widetilde\alpha_k^{(\ell_2)}-\widetilde\alpha_j^{(\ell_2)}.
\label{eq:adjusted-contrast-invariance}
\end{equation}
Moreover, for every supported move $(o,d)\in\mathcal E$,
\begin{equation}
\kappa_{od}^{jk,(\ell_1)}
=
\kappa_{od}^{jk,(\ell_2)}.
\label{eq:kappa-invariance}
\end{equation}
\end{lemma}

\begin{proof}
We first show that, for all $j,k\in\{1,\ldots,K\}$,
\begin{equation}
    \alpha_k^{(\ell_1)}-\alpha_j^{(\ell_1)}
    =
    \alpha_k^{(\ell_2)}-\alpha_j^{(\ell_2)}.
    \label{eq:unadjusted-contrast-invariance}
\end{equation}
Abbreviate $Z_k:=\ind\{D=k\}-\ind\{O=k\}$ for $k=1,\ldots,K$. Because $\sum_{k=1}^K Z_k=0$, the linear span generated by $(1, J^{(\ell)}-I^{(\ell)})$ is the same for every $\ell$. More specifically, the uncorrelatedness of the residual and regressors in \eqref{eq:left-out-regression} implies that
$$E\left[\varepsilon^{(\ell)} Z_{\ell}  \right]=-E\left[\varepsilon^{(\ell)} \sum_{k\neq \ell}Z_k  \right]=0.$$
Therefore, the residual is in fact uncorrelated with all variables $Z_k$:
\begin{equation}
    E\left[\varepsilon^{(\ell)} Z_k  \right]=0 \quad \forall k\in\{1,\ldots,K\}.
    \label{eq:uncorrelatedness-residual-regressors}
\end{equation}

From the regression leaving out $\ell_1$, we obtain coefficients $\alpha_k^{(\ell_1)}$, $k\neq \ell_1$, $\alpha_{\ell_1}^{(\ell_1)}=0$, and the intercept $\beta^{(\ell_1)}$. Similarly, from the regression leaving out $\ell_2$, we obtain coefficients $\alpha_k^{(\ell_2)}$, $k\neq \ell_2$, $\alpha_{\ell_2}^{(\ell_2)}=0$, and the intercept $\beta^{(\ell_2)}$. Now, re-parametrize the coefficients from the first regression by defining
$$\bar{\alpha}_k:=\alpha_k^{(\ell_1)}-\alpha_{\ell_2}^{(\ell_1)} \quad \forall k\neq\ell_2\qquad\text{and}\qquad \bar{\beta}:=\beta^{(\ell_1)}.$$
Then, the fitted value under these re-parametrized coefficients is
\begin{align*}
\bar\beta+\sum_{k\neq \ell_2}\bar\alpha_k Z_k
&=\beta^{(\ell_1)}+\sum_{k=1}^K(\alpha_k^{(\ell_1)}-\alpha_{\ell_2}^{(\ell_1)})Z_k \\
&=\beta^{(\ell_1)}+\sum_{k=1}^K \alpha_k^{(\ell_1)}Z_k-\alpha_{\ell_2}^{(\ell_1)}\sum_{k=1}^K Z_k \\
&=\beta^{(\ell_1)}+\sum_{k=1}^K \alpha_k^{(\ell_1)}Z_k.
\end{align*}
Thus the re-parametrized coefficients lead to the same fitted value and residual $\varepsilon^{(\ell_1)}$ as the regression that leaves out $\ell_1$. Since, by \eqref{eq:uncorrelatedness-residual-regressors}, that residual $\varepsilon^{(\ell_1)}$ is orthogonal to the constant and to every $Z_k$, it is in particular orthogonal to the regressors in the regression leaving out firm $\ell_2$. Therefore $(\bar\beta,(\bar\alpha_k)_{k\neq\ell_2})$ solves the normal equations for the regression that leaves out $\ell_2$. By invertibility of $\mathbb A^{(\ell_2)}$, the solution is unique, so
\begin{equation*}
\beta^{(\ell_2)}=\beta^{(\ell_1)},
\qquad
\alpha_k^{(\ell_2)}=\alpha_k^{(\ell_1)}-\alpha_{\ell_2}^{(\ell_1)},
\quad k=1,\ldots,K,
\end{equation*}
where the equality for the omitted firm uses the convention $\alpha_{\ell_2}^{(\ell_2)}=0$. Equation \eqref{eq:unadjusted-contrast-invariance} immediately follows.

The above argument also applies if, in the regression, $\Delta$ is replaced by any other random variable with finite mean. In particular, replace $\Delta$ by $\ind\{O=o,D=d\}$ for some $(o,d)\in\mathcal E$. Then, applying the raw normal-equation representation \eqref{eq: alpha_j decomposition step before DID} to this new regression, we obtain AKM estimands $\alpha_k^{(\ell)}$, $k\neq \ell$, and weights $\omega_{k,uv}^{(\ell)}$, $k\neq \ell$, $u,v\in\{1,\ldots,K\}$, that satisfy
\begin{equation}
    \alpha_k^{(\ell)}
    =
    \sum_{(u,v)\in\mathcal E}\omega_{k,uv}^{(\ell)}E[\ind\{O=o,D=d\}\mid O=u,D=v],
    \qquad k=1,\ldots,K,
    \label{eq:cell-mean-representation}
\end{equation}
recalling that $\omega_{\ell,rs}^{(\ell)}=0$. Then the only nonzero conditional mean contributing to the sum in \eqref{eq:cell-mean-representation} is the one for $(u,v)=(o,d)$, so
\[
\alpha_k^{(\ell)}-\alpha_j^{(\ell)}
=
\omega_{k,od}^{(\ell)}-\omega_{j,od}^{(\ell)}
=
\kappa_{od}^{jk,(\ell)}.
\]
However, \eqref{eq:unadjusted-contrast-invariance} applies not only to the regression with outcome $\Delta$: an analogous equation holds for the new outcome $\ind\{O=o,D=d\}$. Hence,
\(\kappa_{od}^{jk,(\ell_1)}=\kappa_{od}^{jk,(\ell_2)}\),
which proves \eqref{eq:kappa-invariance}.

Finally, the analogue of \eqref{eq: adjusted AKM kappa contrast representation} under left-out firm $\ell$ gives, for the original outcome $\Delta$,
\begin{equation}\label{eq:adjusted-contrast-invariance-left-out}
    \widetilde\alpha_k^{(\ell)}-\widetilde\alpha_j^{(\ell)}
=
\sum_{(o,d)\in\mathcal E}\kappa_{od}^{jk,(\ell)}\DID(o,d).
\end{equation}
The difference-in-difference contrasts $\DID(o,d)$ do not depend on the choice of left-out firm. Since the weight $\kappa_{od}^{jk,(\ell_1)}$ is equal to $\kappa_{od}^{jk,(\ell_2)}$, $(o,d)\in\mathcal E$, the right-hand side of \eqref{eq:adjusted-contrast-invariance-left-out} with $\ell=\ell_1$ is equal to the right-hand side of \eqref{eq:adjusted-contrast-invariance-left-out} with $\ell=\ell_2$. This proves \eqref{eq:adjusted-contrast-invariance}.
\end{proof}

\subsection{Proofs for Section~\ref{sec: did characterization section}}
\label{app: proofs exact description}

\begin{lemma}[Recovery of firm levels]\label{lem: potential to levels}
Suppose Assumption~\ref{ass: invertibility of A} holds and there is $a\in\mathbb R^K$ such that $\DID(u,v)=a_v-a_u$ for all $(u,v)\in\mathcal E$. Then $\widetilde\alpha_j=a_j-a_1$ for every firm $j\in\{1,\ldots,K\}$.
\end{lemma}
\begin{proof}
Let $j\ne1$. By Lemma~\ref{lem: arbitrary no of firms} and Definition~\ref{def: adjusted AKM estimand}, $\widetilde\alpha_j=\sum_{(u,v)\in\mathcal E}\omega_{j,uv}\DID(u,v)$. Substituting $\DID(u,v)=a_v-a_u$ and applying Lemma~\ref{lem: properties of the weights}(ii) gives
\[
\widetilde\alpha_j
=\sum_{(u,v)\in\mathcal E}\omega_{j,uv}(a_v-a_u)
=\sum_{k=1}^K a_k
  \sum_{(u,v)\in\mathcal E}\omega_{j,uv}
  (\ind\{v=k\}-\ind\{u=k\})
=a_j-a_1.
\]
For $j=1$ both sides are zero.
\end{proof}

Recall that the undirected graph $G^u$ is defined before Lemma~\ref{lem: connectedness} and the directed graph $G^d$ in Section~\ref{sec: paths definitions and examples}.

\begin{lemma}[level differences and cycles]\label{lem: potential and cycles}
Suppose the graph $G^u$ is connected. 
\begin{enumerate}
  \item Let $m\in\mathbb R^{|\mathcal E|}$, and write $m_{uv}:=m_{(u,v)}$ for its entry at the move $(u,v)\in\mathcal E$. The following are equivalent:
    \begin{enumerate}[(i)]
    \item For some $a\in\mathbb R^K$, $m_{uv}=a_v-a_u$ holds for all $(u,v)\in\mathcal E$;
    \item $S(C,m)=0$ for every signed cycle in $G^d$;
    \item $S(C,m)=0$ for every simple signed cycle in $G^d$.
    \end{enumerate}

  \item The vectors $m\in\mathbb R^{|\mathcal E|}$ satisfying 1.(i)-1.(iii) form a linear subspace of $\mathbb R^{|\mathcal E|}$ of dimension $K-1$. 
  
  \item There exist simple signed cycles $C_1,\ldots,C_{|\mathcal E|-K+1}$ such that:
    \begin{enumerate}[(a)]
    \item $S(C_j,m)=0$ for all $j=1,\ldots,|\mathcal E|-K+1$ implies 1.(i);
    \item the linear forms $m\mapsto S(C_j,m)$, $j=1,\ldots,|\mathcal E|-K+1$, are linearly independent.
    \end{enumerate}
\end{enumerate}
\end{lemma}
\begin{proof} The proof proceeds in four steps.

  {\bf Step 1: 1.(i) $\Rightarrow$ 1.(ii).} Suppose 1.(i) holds. Let $W=(j_0,s_1,j_1,\ldots,s_L,j_L)$ be a signed path, and let $e_\ell$ be the move used at step $\ell$ (as in Definition~\ref{def: signed path and signed cycle}). For each step,
  \[
  s_\ell m_{e_\ell}
  =
  \begin{cases}
  m_{j_{\ell-1}j_\ell}=a_{j_\ell}-a_{j_{\ell-1}},&s_\ell=+1,\\
  -m_{j_\ell j_{\ell-1}}=-(a_{j_{\ell-1}}-a_{j_\ell}),&s_\ell=-1.
  \end{cases}
  \]
  Consequently,
  \begin{align*}
  S(W,m)
  =
  \sum_{\ell=1}^{L}(a_{j_\ell}-a_{j_{\ell-1}})
  =
  a_{j_L}-a_{j_0}.
  \end{align*}
  For a signed cycle, $j_L=j_0$, so 1.(i) implies 1.(ii). 
  
  {\bf Step 2: 1.(ii) $\Rightarrow$ 1.(iii).} This is immediate because every simple signed cycle is a signed cycle.
  
  {\bf Step 3: 1.(iii) $\Rightarrow$ 1.(i).} Suppose 1.(iii) holds. A finite connected undirected graph contains a spanning tree\footnote{A tree is a connected subgraph without cycles. A spanning tree is a tree that contains all vertices.}; the spanning tree has a unique simple path between any two vertices and has $K-1$ edges \cite[Theorem~1.5.1 and Corollary~1.5.2, p.~14]{diestel2025graph}. Choose such a tree in $G^u$ and denote by $\mathcal T\subseteq\mathcal E$ the set of moves along the edges in the tree; if for a given edge in the tree moves in both directions are in $\mathcal{E}$, then pick one of them to be included in $\mathcal T$. Then, $\mathcal T$ contains $|\mathcal T|=K-1$ moves. For $j\neq k$, let $W_{\mathcal T}^{jk}$ denote the unique simple signed path from $j$ to $k$ traversing only moves in $\mathcal T$. It exists and is unique because the tree contains a unique simple path between $j$ and $k$, and each edge of the tree corresponds to exactly one move in $\mathcal T$, which determines the sign of the corresponding step. Define
  \[
  a_1:=0,
  \qquad
  a_j:=S(W_{\mathcal T}^{1j},m),\quad j=2,\ldots,K.
  \]
  For the tree-path comparisons in this proof, let $W_{\mathcal T}^{11}$ denote the zero-length path at firm~1, with $S(W_{\mathcal T}^{11},m):=0$. This local convention covers a tree edge incident to the root and does not change the nonempty-path convention in Definition~\ref{def: signed path and signed cycle}.

  Fix $(u,v)\in\mathcal T$. Removing its undirected edge separates the tree into the component containing $u$ and the component containing $v$. If firm $1$ is in the component containing $u$, then the path $W_{\mathcal T}^{1v}$ is equal to the path $W_{\mathcal T}^{1u}$ followed by $(u,v)$ forward, and hence
  \[
  a_v=a_u+m_{uv}.
  \]
  If firm $1$ is in the component containing $v$, then the path $W_{\mathcal T}^{1u}$ is equal to $W_{\mathcal T}^{1v}$ followed by the move $(u,v)$ backward, and hence
  \[
  a_u=a_v-m_{uv}.
  \]
  Thus
  \begin{equation}\label{eq: potential identity on tree}
    m_{uv}=a_v-a_u\qquad\text{for every }(u,v)\in\mathcal T.
  \end{equation}
  For each $e=(u,v)\in\mathcal E\setminus\mathcal T$, let $C_e$ traverse $e$ forward and then along the signed path $W_{\mathcal T}^{vu}$ from $v$ to $u$. The path through the tree has no repeated vertices and does not use the move $e$, so $C_e$ is a simple signed cycle. Every step of $W_{\mathcal T}^{vu}$ uses a move in $\mathcal T$, on which \eqref{eq: potential identity on tree} holds. The telescoping calculation in Step~1 uses the identity $m_{uv}=a_v-a_u$ only on the moves used by the signed path, so it applies to $W_{\mathcal T}^{vu}$ and gives $S(W_{\mathcal T}^{vu},m)=a_u-a_v$. Hence, by 1.(iii),
  \[
  0=S(C_e,m)=m_{uv} + S(W_{\mathcal T}^{vu},m)=m_{uv}+a_u-a_v.
  \]
  Therefore $m_{uv}=a_v-a_u$ holds not only for $(u,v)\in\mathcal T$, but for all $(u,v)\in\mathcal E$, which establishes 1.(i). The construction also shows that 1.(iii) holding for all simple signed cycles of the form $C_e$, rather than all signed cycles, is sufficient for 1.(i).

  {\bf Step 4: Dimensions and independent fundamental-cycle restrictions.} In this step, we prove parts 2. and 3. of the lemma. Define the \emph{signed incidence matrix} $\mathbb D\in\mathbb R^{|\mathcal E|\times K}$ by
  \[
  \mathbb D_{(u,v),r}:=\ind\{v=r\}-\ind\{u=r\},
  \qquad (u,v)\in\mathcal E,\quad r=1,\ldots,K.
  \]
  Then $\mathbb D a=(a_v-a_u\colon (u,v)\in\mathcal E)$, so the set of vectors $m\in\mathbb R^{|\mathcal E|}$ satisfying 1.(i) equals the image of $\mathbb D$, denoted by $\operatorname{im}\mathbb D$. Notice that $\mathbb D a=0$ if and only if $a_u=a_v$ for every $(u,v)\in\mathcal E$. Since $G^u$ is connected, any two firms are joined by a path in $G^u$, along which consecutive entries of $a$ are then equal. Hence $\mathbb D a=0$ if and only if all elements of $a$ are equal. Thus, the kernel of $\mathbb D$ is the set of constant vectors: \( \ker\mathbb D=\{a\in\mathbb R^K:a_1=\cdots=a_K\}\).
  In consequence, the kernel and image of $\mathbb D$ have dimensions \(\dim(\ker\mathbb D)=1\) and \(\dim(\operatorname{im}\mathbb D)=\operatorname{rank}\mathbb D=K-1\). This establishes part 2. of the lemma.
  
  Define the circulation space by
  \[
  Z:=\left\{z\in\mathbb R^{|\mathcal E|}:
  \sum_{(u,v)\in\mathcal E}z_{uv}
  \bigl(\ind\{v=r\}-\ind\{u=r\}\bigr)=0
  \text{ for every }r=1,\ldots,K\right\}
  =\ker\mathbb D'.
  \]
  For any signed cycle $C=(j_0,s_1,j_1,\ldots,s_L,j_L)$, write its step-$\ell$ move as $e_\ell=(u_\ell,v_\ell)$. The signed-step definition gives
  \[
  s_\ell\bigl(\ind\{v_\ell=r\}-\ind\{u_\ell=r\}\bigr)
  =\ind\{j_\ell=r\}-\ind\{j_{\ell-1}=r\}.
  \]
  Using the signed incidence vector defined in \eqref{eq: signed incidence vector definition},
  \begin{align*}
  (\mathbb D'\pi(C))_r
  &=\sum_{(u,v)\in\mathcal E}\pi_{uv}(C)
  \bigl(\ind\{v=r\}-\ind\{u=r\}\bigr)\\
  &=\sum_{\ell=1}^{L}s_\ell
  \bigl(\ind\{v_\ell=r\}-\ind\{u_\ell=r\}\bigr)\\
  &=\sum_{\ell=1}^{L}
  \bigl(\ind\{j_\ell=r\}-\ind\{j_{\ell-1}=r\}\bigr)\\
  &=\ind\{j_L=r\}-\ind\{j_0=r\}=0.
  \end{align*}
  Consider the set of simple signed cycles $C_e$, $e\in\mathcal E\setminus\mathcal T$, that was constructed in Step 3. The above equality establishes, in particular, that $\pi(C_e)$ belongs to $Z$ for every $e\in\mathcal E\setminus\mathcal T$. Moreover, each $C_e$ uses its own non-tree move forward and no other non-tree move, so
  \[
  \pi_f(C_e)=\ind\{f=e\},
  \qquad \forall e,f\in\mathcal E\setminus\mathcal T.
  \]
  If for some vector $\gamma\in\mathbb R^{\mathcal E\setminus\mathcal T}$, we have $\sum_{e\in\mathcal E\setminus\mathcal T}\gamma_e\pi(C_e)=0$, then its coordinate at a non-tree move $f \in\mathcal E\setminus\mathcal T$ gives
  \[
  0=\sum_{e\in\mathcal E\setminus\mathcal T}\gamma_e\pi_f(C_e)=\gamma_f.
  \]
  Hence the vectors $\pi(C_e)$, $e\in\mathcal E\setminus\mathcal T$, are linearly independent. Since
  \[
  \dim Z
  =|\mathcal E|-\operatorname{rank}\mathbb D'
  =|\mathcal E|-K+1
  =|\mathcal E\setminus\mathcal T|,
  \]
  they form a basis of $Z$. Finally, $S(C_e,m)=\pi(C_e)'m$, so
  \[
  S(C_e,m)=\pi(C_e)'m=0,\qquad e\in\mathcal E\setminus\mathcal T,
  \]
  forms a set of $|\mathcal E|-K+1$ independent linear restrictions on $m$. Step 3 already showed that these are sufficient for 1.(i). Enumerating the cycles $C_e$, $e\in\mathcal E\setminus\mathcal T$, as $C_1,\ldots,C_{|\mathcal E|-K+1}$ therefore yields cycles with properties 3.(a) and 3.(b). This establishes part 3. of the lemma.
\end{proof}

\begin{proof}[Proof of Theorem~\ref{thm: did characterization}] We prove the equivalence result in three steps.

  {\bf Step 1: (i) $\Rightarrow$ (ii) and (iii).} Suppose (i) holds and recall the convention $\widetilde\alpha_1=0$. Then $\DID(u,v)=\widetilde\alpha_v-\widetilde\alpha_u$ for all $(u,v)\in\mathcal E$. The telescoping calculation in Step~1 of the proof of Lemma~\ref{lem: potential and cycles}, with $m=\DID$ and $a=\widetilde\alpha$, gives
  \[
  S(W,\DID)=\widetilde\alpha_{j_L}-\widetilde\alpha_{j_0}
  \]
  for every signed path $W=(j_0,s_1,j_1,\ldots,s_L,j_L)$. If $j_0=o$, $j_L=d$, and $(o,d)\in\mathcal E$, the last expression equals $\DID(o,d)$ by (i), which proves (ii). If $W$ is a signed cycle, then $j_L=j_0$ and its sum is zero. Thus, (iii) holds.

  {\bf Step 2: (iii) $\Rightarrow$ (i).} Suppose (iii) holds. By Assumption~\ref{ass: invertibility of A}(i) and Lemma~\ref{lem: connectedness}(i), $G^u$ is connected, so Lemma~\ref{lem: potential and cycles} gives $a\in\mathbb R^K$ satisfying
  \[
  \DID(u,v)=a_v-a_u\qquad \forall (u,v)\in\mathcal E.
  \]
  Assumption~\ref{ass: invertibility of A} and Lemma~\ref{lem: potential to levels} then imply $\widetilde\alpha_j=a_j-a_1$ for every firm $j$. Thus, for every $(o,d)\in\mathcal E$,
  \[
  \widetilde\alpha_d-\widetilde\alpha_o
  =(a_d-a_1)-(a_o-a_1)
  =a_d-a_o=\DID(o,d),
  \]
  which proves (i).

  {\bf Step 3: (ii) $\Rightarrow$ (iii).} Suppose (ii) holds, take any signed cycle $C=(j_0,s_1,j_1,\ldots,s_L,j_L)$ in $G^d$, and let $u:=j_0$ be the firm at which $C$ starts. The move used at its first step has $u$ as an endpoint: $e_1=(u,v)$ if $s_1=+1$ and $e_1=(v,u)$ if $s_1=-1$, where $v:=j_1\neq u$. We use the concatenation $\oplus$ defined in Appendix~\ref{app: proofs section descriptive} before the proof of Theorem~\ref{thm: akm path decomposition}. Since $\pi(W\oplus W')=\pi(W)+\pi(W')$ and $S(W,m)=\pi(W)'m$, signed sums are additive under concatenation: $S(W\oplus W',m)=S(W,m)+S(W',m)$. If $e_1=(u,v)$, which belongs to $\mathcal E$, then $C\oplus(u,+1,v)$ is a signed path from $u$ to $v$. Condition (ii) and additivity of signed sums imply
  \[
  \DID(u,v)
  =
  S(C\oplus(u,+1,v),\DID)
  =
  S(C,\DID)+\DID(u,v),
  \]
  so $S(C,\DID)=0$. If $e_1=(v,u)$, which belongs to $\mathcal E$, then $(v,+1,u)\oplus C$ is a signed path from $v$ to $u$, and
  \[
  \DID(v,u)
  =
  S((v,+1,u)\oplus C,\DID)
  =
  \DID(v,u)+S(C,\DID),
  \]
  again giving $S(C,\DID)=0$. Both applications of (ii) have supported endpoints; the concatenated paths need not be simple, as permitted by (ii). Thus every signed cycle has zero sum, which proves (iii).
\end{proof}

\noindent{\bf Notation for the binscatter and alternative distributions.} Proposition~\ref{prop:binscatter-cycle-blindness-appendix} below formalises the claim in Section~\ref{sec: binscatter diagnostic} that agreement of the binned averages with the 45-degree line does not imply the cycle restrictions. The bins defined in this paragraph are the population counterpart of those described there. Let $P$ denote the joint law of $(Y_1,Y_2,O,D)$ on $\mathbb R^2\times\{1,\ldots,K\}^2$, and let $E$ denote expectation under $P$. Fix an integer $B\geq1$ and a labelled partition of the supported moves satisfying
\[
\mathcal E=\bigcup_{b=1}^{B}\mathcal E_b,\qquad \mathcal E_b\neq\emptyset\quad(b=1,\ldots,B),\qquad \mathcal E_b\cap\mathcal E_{b'}=\emptyset\quad(b\neq b').
\]
The population binscatter uses mobility-share weights within each of these bins, with horizontal and vertical coordinates in bin $b$ given by, respectively,
\[
\frac{\sum_{(u,v)\in\mathcal E_b}P_{uv}(\widetilde\alpha_v-\widetilde\alpha_u)}{\sum_{(u,v)\in\mathcal E_b}P_{uv}},
\qquad 
\frac{\sum_{(u,v)\in\mathcal E_b}P_{uv}\DID(u,v)}{\sum_{(u,v)\in\mathcal E_b}P_{uv}}.
\]
Each denominator is positive because $\mathcal E_b$ is a nonempty subset of $\mathcal E$. The partition is held fixed when comparing distributions. If the bins are instead produced by a fixed deterministic rule depending only on the mobility shares and adjusted AKM contrasts, preserving those inputs yields the same labelled partition. 

For any alternative joint law $P^{\star}$ on the same coordinate space satisfying Assumption~\ref{ass: invertibility of A}, let $E^{\star}$ denote expectation under $P^{\star}$, and analogously add a superscript $\star$ for population quantities evaluated under the alternative law. In particular,
\[
P_{uv}^{\star}:=P^{\star}\{O=u,D=v\},\qquad u,v\in\{1,\ldots,K\},
\]
and, whenever $u\neq v$, $P_{uv}^{\star}>0$, and $P_{uu}^{\star}>0$,
\[
\DID^{\star}(u,v):=E^{\star}[\Delta\mid O=u,D=v]-E^{\star}[\Delta\mid O=u,D=u],\qquad \Delta=Y_2-Y_1.
\]
The coefficients $\beta^{\star},\alpha_2^{\star},\ldots,\alpha_K^{\star}$ solve the normal equations in \eqref{eq: exogenous mobility} under $P^{\star}$, and $\tau_k^{\star}$ is the trend adjustment in \eqref{eq: tau definition} evaluated under $P^{\star}$. Thus
\[
\widetilde\alpha_k^{\star}:=\alpha_k^{\star}-\tau_k^{\star}\quad(k=2,\ldots,K),\qquad \alpha_1^{\star}=\tau_1^{\star}=\widetilde\alpha_1^{\star}:=0.
\]
The quantities without a superscript continue to refer to the original law $P$.

\begin{proposition}[Binscatter blindness to cycle violations]
\label{prop:binscatter-cycle-blindness-appendix}
Suppose Assumption~\ref{ass: invertibility of A} holds under $P$ and $r:=|\mathcal E|-K+1>B$. There exist $r-B$ distinct simple signed cycles $C_1,\ldots,C_{r-B}$ in $G^d$ such that, for every $(c_1,\ldots,c_{r-B})\in\mathbb R^{r-B}$, there exists a joint law $P^{\star}$ satisfying Assumption~\ref{ass: invertibility of A} and all of the following equalities simultaneously:
\begin{enumerate}[(i)]
\item The selected cycle sums attain the prescribed values:
\[
S(C_\ell,\DID^{\star})=c_\ell,\qquad \ell=1,\ldots,r-B.
\]
\item The mobility shares and supported stayer means are preserved:
\begin{align*}
P_{uv}^{\star}&=P_{uv},\qquad u,v\in\{1,\ldots,K\},\\
E^{\star}[\Delta\mid O=u,D=u]&=E[\Delta\mid O=u,D=u],\qquad u\in\{1,\ldots,K\}:P_{uu}>0.
\end{align*}
\item The common time coefficient, unadjusted firm coefficients, and adjusted firm coefficients are preserved:
\[
\beta^{\star}=\beta,\qquad \alpha_k^{\star}=\alpha_k,\qquad \widetilde\alpha_k^{\star}=\widetilde\alpha_k,\qquad k=1,\ldots,K.
\]
\item Both coordinates of every bin average are preserved: for every $b=1,\ldots,B$,
\begin{align*}
\frac{\sum_{(u,v)\in\mathcal E_b}P_{uv}^{\star}(\widetilde\alpha_v^{\star}-\widetilde\alpha_u^{\star})}{\sum_{(u,v)\in\mathcal E_b}P_{uv}^{\star}}&=\frac{\sum_{(u,v)\in\mathcal E_b}P_{uv}(\widetilde\alpha_v-\widetilde\alpha_u)}{\sum_{(u,v)\in\mathcal E_b}P_{uv}},\\
\frac{\sum_{(u,v)\in\mathcal E_b}P_{uv}^{\star}\DID^{\star}(u,v)}{\sum_{(u,v)\in\mathcal E_b}P_{uv}^{\star}}&=\frac{\sum_{(u,v)\in\mathcal E_b}P_{uv}\DID(u,v)}{\sum_{(u,v)\in\mathcal E_b}P_{uv}}.
\end{align*}
\end{enumerate}
\end{proposition}

\begin{proof}[Proof of Proposition~\ref{prop:binscatter-cycle-blindness-appendix}]
The proof first constructs perturbation directions satisfying the regression and bin restrictions. It then selects $r-B$ simple signed cycles whose responses to those directions form an invertible matrix. For arbitrary prescribed cycle sums, this matrix determines an alternative law $P^{\star}$. The remaining steps verify its regularity and each equality in (i)--(iv).

\noindent{\bf Step 1: Constructing directions that satisfy the regression and bin restrictions.} By Assumption~\ref{ass: invertibility of A}(i) and Lemma~\ref{lem: connectedness}(i), $G^u$ is connected. Retain the signed incidence matrix and circulation space from the proof of Lemma~\ref{lem: potential and cycles}:
\[
\mathbb D_{(u,v),k}:=\ind\{v=k\}-\ind\{u=k\},\qquad \mathbb D\in\mathbb R^{|\mathcal E|\times K},\qquad Z:=\ker\mathbb D'.
\]
That proof establishes $\dim Z=|\mathcal E|-K+1=r$ and constructs distinct simple signed cycles $F_1,\ldots,F_r$ in $G^d$ such that $\pi(F_1),\ldots,\pi(F_r)$ form a basis of $Z$. Define
\[
Z_B:=\left\{z\in Z:\sum_{(u,v)\in\mathcal E_b}z_{uv}=0\text{ for every }b=1,\ldots,B\right\},
\]
and consider the linear map $T: Z \to \mathbb R^B$ given by
\[
Tz=\begin{bmatrix}
\sum_{(u,v)\in\mathcal E_1}z_{uv} & \cdots & \sum_{(u,v)\in\mathcal E_B}z_{uv}
\end{bmatrix}'
\] 
so that $\ker T=Z_B$ and ${\rm rank} T \le B$. Hence rank--nullity applied to $T$ gives
\[
\dim Z_B\geq\dim Z-B=r-B>0.
\]
Choose linearly independent vectors $z^{(1)},\ldots,z^{(r-B)}\in Z_B$. Since the bins partition $\mathcal E$, for every $h=1,\ldots,r-B$,
\[
\sum_{(u,v)\in\mathcal E}z^{(h)}_{uv}=\sum_{b=1}^{B}\sum_{(u,v)\in\mathcal E_b}z^{(h)}_{uv}=0.
\]
Combining this identity with $\mathbb D'z^{(h)}=0$ and \eqref{eq: nu definition} yields
\begin{align*}
\sum_{(u,v)\in\mathcal E}z^{(h)}_{uv}\nu_{uv}
&=e_1\sum_{(u,v)\in\mathcal E}z^{(h)}_{uv}+\sum_{k=2}^{K}e_k\sum_{(u,v)\in\mathcal E}z^{(h)}_{uv}\bigl(\ind\{v=k\}-\ind\{u=k\}\bigr)\\
&=e_1\cdot0+\sum_{k=2}^{K}e_k(\mathbb D'z^{(h)})_k=0.
\end{align*}
Thus the bin equations eliminate the intercept moment, and the circulation equations eliminate the firm-indicator moments. A change of $z^{(h)}_{uv}/P_{uv}$ in the mean wage change on move $(u,v)$ has probability-weighted change $z^{(h)}_{uv}$, so these identities will preserve the regression and the vertical bin averages.

\noindent{\bf Step 2: Selecting the cycles before specifying their target values.} Define $\mathbb M\in\mathbb R^{r\times(r-B)}$ by
\[
\mathbb M_{gh}:=\sum_{(u,v)\in\mathcal E}\pi_{uv}(F_g)\frac{z^{(h)}_{uv}}{P_{uv}},\qquad g=1,\ldots,r,\quad h=1,\ldots,r-B.
\]
Every denominator is positive by the definition of $\mathcal E$. The entry $\mathbb M_{gh}$ is the change in the sum on $F_g$ produced by the $h$th perturbation direction. To choose $r-B$ cycle sums that can be varied independently, it suffices to prove that $\mathbb M$ has column rank $r-B$.

Suppose $\mathbb Mx=0$ for some $x=(x_1,\ldots,x_{r-B})'\in\mathbb R^{r-B}$. Since $\sum_{h=1}^{r-B}x_hz^{(h)}\in Z_B\subseteq Z$, there exist $\gamma_1,\ldots,\gamma_r\in\mathbb R$ such that
\[
\sum_{h=1}^{r-B}x_hz^{(h)}=\sum_{g=1}^{r}\gamma_g\pi(F_g).
\]
Substitution of this representation into the following nonnegative quadratic form gives
\begin{align*}
\sum_{(u,v)\in\mathcal E}\frac{\left(\sum_{h=1}^{r-B}x_hz^{(h)}_{uv}\right)^2}{P_{uv}}
&=\sum_{(u,v)\in\mathcal E}\left(\sum_{g=1}^{r}\gamma_g\pi_{uv}(F_g)\right)\frac{\sum_{h=1}^{r-B}x_hz^{(h)}_{uv}}{P_{uv}}\\
&=\sum_{g=1}^{r}\gamma_g\sum_{h=1}^{r-B}x_h\sum_{(u,v)\in\mathcal E}\pi_{uv}(F_g)\frac{z^{(h)}_{uv}}{P_{uv}}\\
&=\sum_{g=1}^{r}\gamma_g\sum_{h=1}^{r-B}x_h\mathbb M_{gh}\\
&=\sum_{g=1}^{r}\gamma_g(\mathbb Mx)_g=0.
\end{align*}
Each summand on the left is nonnegative and has a strictly positive denominator. Hence
\[
\sum_{h=1}^{r-B}x_hz^{(h)}_{uv}=0\quad\text{for every }(u,v)\in\mathcal E,\qquad\text{so}\qquad \sum_{h=1}^{r-B}x_hz^{(h)}=0.
\]
Linear independence of the selected directions implies $x=0$. Therefore $\ker\mathbb M=\{0\}$ and $\operatorname{rank}\mathbb M=r-B$.

Equality of row and column rank gives $r-B$ linearly independent rows. Choose their distinct indices $g_1,\ldots,g_{r-B}$ and define
\[
C_\ell:=F_{g_\ell}\quad(\ell=1,\ldots,r-B),\qquad \mathbb M_*:=\bigl(\mathbb M_{g_\ell h}\bigr)_{\ell,h=1}^{r-B}.
\]
The matrix $\mathbb M_*\in\mathbb R^{(r-B)\times(r-B)}$ has linearly independent rows and is therefore invertible. The selected $C_1,\ldots,C_{r-B}$ are distinct simple signed cycles in $G^d$. All these choices depend only on the original law and its bins.

\noindent{\bf Step 3: Constructing the alternative law and verifying regularity and (ii).} Fix any $(c_1,\ldots,c_{r-B})\in\mathbb R^{r-B}$ and set
\[
x:=\mathbb M_*^{-1}\begin{pmatrix}c_1-S(C_1,\DID)\\ \vdots\\ c_{r-B}-S(C_{r-B},\DID)\end{pmatrix}.
\]
The vector $x$ is finite because $\mathbb M_*$ is invertible and the original DIDs are finite by Assumption~\ref{ass: invertibility of A}(ii)--(iii). On the probability space carrying the original variables, define
\[
Y_2^{\star}:=
\begin{cases}
Y_2+\displaystyle\sum_{h=1}^{r-B}x_h\frac{z^{(h)}_{OD}}{P_{OD}}, &(O,D)\in\mathcal E,\\
Y_2, &(O,D)\notin\mathcal E.
\end{cases}
\]
Let $P^{\star}$ be the joint law of $(Y_1,Y_2^{\star},O,D)$ under $P$. Thus $\Delta$ remains the coordinate function $Y_2-Y_1$ under $P^{\star}$, and its distribution under $P^{\star}$ is the distribution of $Y_2^{\star}-Y_1$ under $P$. The piecewise definition uses division only on supported mover cells and leaves every stayer unchanged.

The distribution of $(O,D)$ is unchanged, so
\[
P_{uv}^{\star}=P_{uv},\qquad u,v\in\{1,\ldots,K\}.
\]
Consequently,
\[
\mathbb A^{\star}=\sum_{u,v=1}^{K}P_{uv}^{\star}\nu_{uv}\nu_{uv}'=\sum_{u,v=1}^{K}P_{uv}\nu_{uv}\nu_{uv}'=\mathbb A,
\]
which is invertible. Moreover,
\[
P_{uv}^{\star}>0\ \Longrightarrow\ P_{uv}>0\ \Longrightarrow\ P_{uu}>0\ \Longrightarrow\ P_{uu}^{\star}>0.
\]
These identities verify parts (i) and (ii) of Assumption~\ref{ass: invertibility of A} under $P^{\star}$. For part (iii), the triangle inequality gives
\begin{align*}
E^{\star}|\Delta|
&=E|Y_2^{\star}-Y_1|\\
&\leq E|\Delta|+\sum_{(u,v)\in\mathcal E}P_{uv}\left|\sum_{h=1}^{r-B}x_h\frac{z^{(h)}_{uv}}{P_{uv}}\right|\\
&\leq E|\Delta|+\sum_{(u,v)\in\mathcal E}P_{uv}\sum_{h=1}^{r-B}|x_h|\frac{|z^{(h)}_{uv}|}{P_{uv}}\\
&=E|\Delta|+\sum_{h=1}^{r-B}|x_h|\sum_{(u,v)\in\mathcal E}|z^{(h)}_{uv}|<\infty.
\end{align*}
The last inequality follows from Assumption~\ref{ass: invertibility of A}(iii), finiteness of $\mathcal E$, and finiteness of the selected vectors and $x$. Hence $P^{\star}$ satisfies the entire assumption.

For every $(u,v)\in\mathcal E$, the definition of $P^{\star}$ and the constant shift on that cell imply
\begin{align*}
E^{\star}[\Delta\mid O=u,D=v]
&=E[Y_2^{\star}-Y_1\mid O=u,D=v]\\
&=E\left[\Delta+\sum_{h=1}^{r-B}x_h\frac{z^{(h)}_{uv}}{P_{uv}}\ \middle|\ O=u,D=v\right]\\
&=E[\Delta\mid O=u,D=v]+\sum_{h=1}^{r-B}x_h\frac{z^{(h)}_{uv}}{P_{uv}}.
\end{align*}
For every $u$ with $P_{uu}>0$, the shift vanishes on the stayer cell, giving
\[
E^{\star}[\Delta\mid O=u,D=u]=E[Y_2^{\star}-Y_1\mid O=u,D=u]=E[\Delta\mid O=u,D=u].
\]
Together with preservation of the shares, this establishes (ii). Subtracting the unchanged origin-stayer mean from the mover-cell mean yields
\begin{align*}
\DID^{\star}(u,v)
&=E^{\star}[\Delta\mid O=u,D=v]-E^{\star}[\Delta\mid O=u,D=u]\\
&=E[\Delta\mid O=u,D=v]+\sum_{h=1}^{r-B}x_h\frac{z^{(h)}_{uv}}{P_{uv}}-E[\Delta\mid O=u,D=u]\\
&=\DID(u,v)+\sum_{h=1}^{r-B}x_h\frac{z^{(h)}_{uv}}{P_{uv}},\qquad (u,v)\in\mathcal E.
\end{align*}
This identity determines the effect of the constructed law on both the bin averages and the selected cycle sums.

\noindent{\bf Step 4: Preserving the regression coefficients in (iii).} Because $\mathbb A^{\star}=\mathbb A$, subtraction of the normal equations in \eqref{eq: exogenous mobility} under the two laws gives
\begin{align*}
\mathbb A\begin{pmatrix}\beta^{\star}-\beta\\ \alpha_2^{\star}-\alpha_2\\ \vdots\\ \alpha_K^{\star}-\alpha_K\end{pmatrix}
&=\sum_{u,v:P_{uv}>0}P_{uv}\nu_{uv}\bigl(E^{\star}[\Delta\mid O=u,D=v]-E[\Delta\mid O=u,D=v]\bigr)\\
&=\sum_{(u,v)\in\mathcal E}P_{uv}\nu_{uv}\sum_{h=1}^{r-B}x_h\frac{z^{(h)}_{uv}}{P_{uv}}\\
&=\sum_{h=1}^{r-B}x_h\sum_{(u,v)\in\mathcal E}z^{(h)}_{uv}\nu_{uv}=0.
\end{align*}
The second equality uses preservation of the stayer means, and the last equality is the moment identity from Step~1. Invertibility of $\mathbb A$ therefore gives
\[
\beta^{\star}=\beta,\qquad \alpha_k^{\star}=\alpha_k\quad(k=2,\ldots,K).
\]
The reference-firm equality $\alpha_1^{\star}=\alpha_1=0$ holds by normalization.

For $k=2,\ldots,K$, \eqref{eq: omega definition} and preservation of the shares imply
\[
\omega_{k,uv}^{\star}=e_k'(\mathbb A^{\star})^{-1}\nu_{uv}P_{uv}^{\star}=e_k'\mathbb A^{-1}\nu_{uv}P_{uv}=\omega_{k,uv}.
\]
The trend definition in \eqref{eq: tau definition} and the unchanged supported stayer means consequently give
\begin{align*}
\tau_k^{\star}
&=\sum_{u,v:P_{uv}^{\star}>0}\omega_{k,uv}^{\star}E^{\star}[\Delta\mid O=u,D=u]\\
&=\sum_{u,v:P_{uv}>0}\omega_{k,uv}E[\Delta\mid O=u,D=u]=\tau_k.
\end{align*}
Thus
\[
\widetilde\alpha_k^{\star}=\alpha_k^{\star}-\tau_k^{\star}=\alpha_k-\tau_k=\widetilde\alpha_k,\qquad k=2,\ldots,K,
\]
and $\widetilde\alpha_1^{\star}=\widetilde\alpha_1=0$ by normalization. This proves (iii).

\noindent{\bf Step 5: Preserving the plotted coordinates in (iv).} The partition is fixed by the setup. If it is generated by the specified deterministic rule, Steps~3 and~4 preserve all its inputs and therefore yield the same bin membership. For every $b=1,\ldots,B$, the bin remains nonempty and its total share is
\[
\sum_{(u,v)\in\mathcal E_b}P_{uv}^{\star}=\sum_{(u,v)\in\mathcal E_b}P_{uv}>0.
\]
Preservation of the shares and adjusted coefficients gives the horizontal-coordinate equality
\[
\frac{\sum_{(u,v)\in\mathcal E_b}P_{uv}^{\star}(\widetilde\alpha_v^{\star}-\widetilde\alpha_u^{\star})}{\sum_{(u,v)\in\mathcal E_b}P_{uv}^{\star}}
=\frac{\sum_{(u,v)\in\mathcal E_b}P_{uv}(\widetilde\alpha_v-\widetilde\alpha_u)}{\sum_{(u,v)\in\mathcal E_b}P_{uv}}.
\]
For the vertical coordinate, the DID identity from Step~3 and the defining equations of $Z_B$ yield
\begin{align*}
\frac{\sum_{(u,v)\in\mathcal E_b}P_{uv}^{\star}\DID^{\star}(u,v)}{\sum_{(u,v)\in\mathcal E_b}P_{uv}^{\star}}
&=\frac{\sum_{(u,v)\in\mathcal E_b}P_{uv}\left(\DID(u,v)+\sum_{h=1}^{r-B}x_hz^{(h)}_{uv}/P_{uv}\right)}{\sum_{(u,v)\in\mathcal E_b}P_{uv}}\\
&=\frac{\sum_{(u,v)\in\mathcal E_b}P_{uv}\DID(u,v)}{\sum_{(u,v)\in\mathcal E_b}P_{uv}}+\frac{\sum_{h=1}^{r-B}x_h\sum_{(u,v)\in\mathcal E_b}z^{(h)}_{uv}}{\sum_{(u,v)\in\mathcal E_b}P_{uv}}\\
&=\frac{\sum_{(u,v)\in\mathcal E_b}P_{uv}\DID(u,v)}{\sum_{(u,v)\in\mathcal E_b}P_{uv}}.
\end{align*}
This proves (iv).

\noindent{\bf Step 6: Attaining all the prescribed cycle sums in (i).} The signed-incidence identity following \eqref{eq: signed incidence vector definition}, the DID identity from Step~3, and the definition of $\mathbb M_*$ give, for $\ell=1,\ldots,r-B$,
\begin{align*}
S(C_\ell,\DID^{\star})
&=\sum_{(u,v)\in\mathcal E}\pi_{uv}(C_\ell)\DID^{\star}(u,v)\\
&=\sum_{(u,v)\in\mathcal E}\pi_{uv}(C_\ell)\left(\DID(u,v)+\sum_{h=1}^{r-B}x_h\frac{z^{(h)}_{uv}}{P_{uv}}\right)\\
&=S(C_\ell,\DID)+\sum_{h=1}^{r-B}x_h\sum_{(u,v)\in\mathcal E}\pi_{uv}(C_\ell)\frac{z^{(h)}_{uv}}{P_{uv}}\\
&=S(C_\ell,\DID)+\sum_{h=1}^{r-B}(\mathbb M_*)_{\ell h}x_h\\
&=S(C_\ell,\DID)+c_\ell-S(C_\ell,\DID)=c_\ell.
\end{align*}
The last equality follows from the choice of $x$ in Step~3. The cycles were selected in Step~2 before the target vector was specified, and the same law $P^{\star}$ satisfies all the equalities in (i)--(iv). This proves the proposition.
\end{proof}

\subsubsection{Inference for restricted-path averages}
\label{app: restricted-path-inference}

This subsection gives conditions for inference on the restricted-path estimand in \eqref{eq: restricted-path-estimand}. The sampling framework allows both the worker distribution and the number of firms to vary with the number of workers.

\noindent {\bf Sampling and population quantities.} Let $N$ denote the number of workers. We will develop an asymptotic theory under which $N$ tends to infinity. We consider data-generating processes that drift with $N$ and therefore index population quantities by $N$. For instance, let $K_N$ denote the number of firms. Denote by $U_N$ an auxiliary randomization. Let $P_N$ denote the joint law of $((\Delta_i,O_i,D_i)_{i=1}^N,U_N)$, and write $E_N$, $\var_N$, and $\cov_N$ for expectation, variance, and covariance under $P_N$.

\begin{assumption}[Sampling]\label{ass: restricted-path-sampling}
For every $N$, the following conditions hold:
\begin{enumerate}[(i)]
\item $K_N$ is a deterministic integer with $2\leq K_N<\infty$, and $(\Delta_i,O_i,D_i)$, $i=1,\ldots,N$, are independent and identically distributed with values in $\mathbb R\times\{1,\ldots,K_N\}^2$.
\item $E_N[|\Delta_i|]<\infty$ for every $i=1,\ldots,N$.
\item $U_N\indep(\Delta_i,O_i,D_i)_{i=1}^N$ under $P_N$.
\end{enumerate}
\end{assumption}

Under Assumption~\ref{ass: restricted-path-sampling}, the counterparts of the mobility shares, supported moves, and directed mobility graph in Sections~\ref{sec: notation and assumptions} and~\ref{sec: paths definitions and examples} are
\[
P_{N,uv}:=P_N\{O_i=u,D_i=v\},\qquad \mathcal E_N:=\{(u,v)\in\{1,\ldots,K_N\}^2\colon u\neq v,\ P_{N,uv}>0\},
\]
and $G_N^d:=(\{1,\ldots,K_N\},\mathcal E_N)$. Signed path sums use the convention in Section~\ref{sec: paths definitions and examples}. For $(u,v)\in\mathcal E_N$ with $P_{N,uu}>0$, the counterpart of \eqref{eq: did definition} is
\begin{equation}\label{eq: restricted-path-population-did}
\DID_N(u,v):=E_N[\Delta_i\mid O_i=u,D_i=v]-E_N[\Delta_i\mid O_i=u,D_i=u].
\end{equation}

\noindent {\bf Selected paths and sample contrasts.} For each $N$, let $o,d$ denote the endpoint firm labels, and define
\begin{equation}\label{eq: restricted-path-design}
\mathcal G_N:=\sigma((O_i,D_i)_{i=1}^N,U_N).
\end{equation}

Let $\mathcal R_N^{od}$ be a finite, $\mathcal G_N$-measurable random set of distinct signed paths from $o$ to $d$ in $G_N^d$. For $W\in\mathcal R_N^{od}$, let $L(W)$ denote its length and $e_\ell=(u_\ell,v_\ell)$ its move at step $\ell$, with sign $s_\ell$, as in Definition~\ref{def: signed path and signed cycle}. For every required mover or origin-stayer cell
\[
(u,v)\in\bigcup_{\ell=1}^{L(W)}\{(u_\ell,v_\ell),(u_\ell,u_\ell)\},
\]
let $\mathcal I_{N,W}(u,v)$ denote the worker subset assigned to path $W$ and cell $(u,v)$. Note that repeated uses of a cell within $W$ use the same subset, and subsets assigned to different paths may be disjoint or overlap.

\begin{assumption}[Path and worker selection]\label{ass: restricted-path-selection}
For every $N$, the following conditions hold:
\begin{enumerate}[(i)]
\item The labels $o,d$ are deterministic and satisfy $o,d\in\{1,\ldots,K_N\}$ and $o\neq d$. The set $\mathcal R_N^{od}$ is $\mathcal G_N$-measurable and, $P_N$-almost surely, is a finite collection of signed paths from $o$ to $d$ in $G_N^d$.
\item $P_N$-almost surely, $P_{N,u_\ell u_\ell}>0$ for every path $W\in\mathcal R_N^{od}$ and every $\ell=1,\ldots,L(W)$.
\item The worker subsets assigned to the required cells of the selected paths are $\mathcal G_N$-measurable. $P_N$-almost surely, for every entry $W\in\mathcal R_N^{od}$ and every cell $(u,v)\in\bigcup_{\ell=1}^{L(W)}\{(u_\ell,v_\ell),(u_\ell,u_\ell)\}$,
\[
\emptyset\neq\mathcal I_{N,W}(u,v)\subseteq\{i\in\{1,\ldots,N\}:O_i=u,D_i=v\}.
\]
\end{enumerate}
\end{assumption}

For each move $(u,v)$ used by a path $W\in\mathcal R_N^{od}$, define
\begin{equation}\label{eq: restricted-path-sample-did}
\widehat{\DID}_{N,W}(u,v):=\frac{\sum_{i\in\mathcal I_{N,W}(u,v)}\Delta_i}{|\mathcal I_{N,W}(u,v)|}-\frac{\sum_{i\in\mathcal I_{N,W}(u,u)}\Delta_i}{|\mathcal I_{N,W}(u,u)|},
\end{equation}
and let
\begin{equation}\label{eq: restricted-path-used-workers}
\mathcal J_N:=\bigcup_{W\in\mathcal R_N^{od}}\ \bigcup_{\ell=1}^{L(W)}\left(\mathcal I_{N,W}(u_\ell,v_\ell)\cup\mathcal I_{N,W}(u_\ell,u_\ell)\right)
\end{equation}
be the set of workers used by the selected paths. The choice $\mathcal I_{N,W}(u,v)=\{i\in\{1,\ldots,N\}:O_i=u,D_i=v\}$ uses all workers in each required cell and gives the sample counterpart described in Section~\ref{sec: interpreting conditions}. Assumption~\ref{ass: restricted-path-selection} permits selection based on the observed mobility network, deterministic firm or edge characteristics, and the independent randomization $U_N$. 

\noindent {\bf Restricted-path estimand and estimator.} On $\{|\mathcal R_N^{od}|\geq2\}$, define
\begin{align}
\theta_{od,N}^{\mathcal R}&:=\frac{1}{|\mathcal R_N^{od}|}\sum_{W\in\mathcal R_N^{od}}S(W,\DID_N),\label{eq: restricted-path-inference-target}\\
\widehat\theta_{od,N}^{\mathcal R}&:=\frac{1}{|\mathcal R_N^{od}|}\sum_{W\in\mathcal R_N^{od}}S(W,\widehat{\DID}_{N,W}).\label{eq: restricted-path-inference-estimator}
\end{align}
The first quantity is the estimand in \eqref{eq: restricted-path-estimand} for the selected set $\mathcal R_N^{od}$; it is $\mathcal G_N$-measurable and may be random because the set may depend on the observed mobility network. The estimator replaces each population contrast by its sample counterpart. Its variance estimator is the dispersion of the estimated path sums divided by the number of paths:
\begin{equation}\label{eq: restricted-path-inference-variance}
\widehat V_N:=\frac{\sum_{W\in\mathcal R_N^{od}}\{S(W,\widehat{\DID}_{N,W})-\widehat\theta_{od,N}^{\mathcal R}\}^2}{|\mathcal R_N^{od}|(|\mathcal R_N^{od}|-1)}.
\end{equation}
For $i=1,\ldots,N$, define the coefficients associated with the path sums and their average by
\begin{align}
b_{N,Wi}&:=\sum_{\ell=1}^{L(W)}s_\ell\left\{\frac{\ind\{i\in\mathcal I_{N,W}(u_\ell,v_\ell)\}}{|\mathcal I_{N,W}(u_\ell,v_\ell)|}-\frac{\ind\{i\in\mathcal I_{N,W}(u_\ell,u_\ell)\}}{|\mathcal I_{N,W}(u_\ell,u_\ell)|}\right\},\label{eq: restricted-path-worker-coefficients}\\
b_{N,i}&:=\frac{1}{|\mathcal R_N^{od}|}\sum_{W\in\mathcal R_N^{od}}b_{N,Wi}.\notag
\end{align}
These coefficients are $\mathcal G_N$-measurable and vanish for $i\notin\mathcal J_N$.

\noindent {\bf Conditional variance and path heterogeneity.} On the same event, define the covariance entries for $W,W'\in\mathcal R_N^{od}$ and the quantities $V_N$ and $H_N$ by
\begin{align}
(\Sigma_N)_{W,W'}&:=\cov_N\left(S(W,\widehat{\DID}_{N,W}),S(W',\widehat{\DID}_{N,W'})\,\middle|\,\mathcal G_N\right),\label{eq: restricted-path-covariance}\\
V_N&:=\var_N(\widehat\theta_{od,N}^{\mathcal R}\mid\mathcal G_N),\label{eq: restricted-path-conditional-variance}\\
H_N&:=\frac{\sum_{W\in\mathcal R_N^{od}}\{S(W,\DID_N)-\theta_{od,N}^{\mathcal R}\}^2}{|\mathcal R_N^{od}|(|\mathcal R_N^{od}|-1)}.\label{eq: restricted-path-heterogeneity}
\end{align}
Thus $\Sigma_N$ is the $|\mathcal R_N^{od}|\times|\mathcal R_N^{od}|$ conditional covariance matrix of the sample path sums, $V_N$ is the conditional variance of their average, and $H_N$ measures dispersion in the corresponding population path sums. Write $\tr(\Sigma_N)$ and $\lambda_{\max}(\Sigma_N)$ for the trace and largest eigenvalue of $\Sigma_N$.

On $\{|\mathcal R_N^{od}|<2\}$, set $\theta_{od,N}^{\mathcal R},\widehat\theta_{od,N}^{\mathcal R},\widehat V_N,V_N,H_N,b_{N,Wi},b_{N,i}$ to zero and let $\Sigma_N$ be the zero matrix indexed by the selected set. For an empty set, set $\tr(\Sigma_N)=\lambda_{\max}(\Sigma_N)=0$. Maxima over empty collections of nonnegative terms and ratios with zero denominators are also set to zero. Let $\Phi$ denote the standard normal distribution function and $z_{1-\alpha/2}:=\Phi^{-1}(1-\alpha/2)$ for $\alpha\in(0,1)$. All convergence statements below are with respect to the sequence of sampling laws $P_N$ as $N\to\infty$.

\begin{theorem}[Inference for restricted-path averages]\label{thm: restricted-path-inference}
Suppose that Assumptions~\ref{ass: restricted-path-sampling} and~\ref{ass: restricted-path-selection} hold and that:
\begin{enumerate}[(i)]
\item There is a constant $1\leq C<\infty$, independent of $N$, such that, almost surely, every $i\in\mathcal J_N$ satisfies
\begin{equation}\label{eq: restricted-path-fourth-moment}
E_N[(\Delta_i-E_N[\Delta_i\mid\mathcal G_N])^4\mid\mathcal G_N]\leq C\{\var_N(\Delta_i\mid\mathcal G_N)\}^2<\infty.
\end{equation}
\item The number of selected paths and the conditional worker contributions satisfy
\[
|\mathcal R_N^{od}|\xrightarrow{p}\infty,\qquad V_N\xrightarrow{p}0,\qquad P_N\{V_N>0\}\longrightarrow1,
\]
and
\begin{equation}\label{eq: restricted-path-worker-share}
\frac{\max_{i\in\mathcal J_N}\{b_{N,i}^2\var_N(\Delta_i\mid\mathcal G_N)\}}{V_N}\xrightarrow{p}0.
\end{equation}
\item The conditional covariance matrix satisfies
\begin{equation}\label{eq: restricted-path-dependence}
\frac{|\mathcal R_N^{od}|^2V_N}{\tr(\Sigma_N)}\xrightarrow{p}1,\qquad \frac{\lambda_{\max}(\Sigma_N)}{\tr(\Sigma_N)}\xrightarrow{p}0.
\end{equation}
\end{enumerate}
Then
\begin{align}
\widehat\theta_{od,N}^{\mathcal R}-\theta_{od,N}^{\mathcal R}&\xrightarrow{p}0,\label{eq: restricted-path-consistency}\\
\frac{\widehat\theta_{od,N}^{\mathcal R}-\theta_{od,N}^{\mathcal R}}{\sqrt{V_N}}&\xrightarrow{d}\mathcal N(0,1),\qquad \frac{\widehat V_N}{V_N+H_N}\xrightarrow{p}1.\label{eq: restricted-path-normal-limit}
\end{align}
Moreover, $P_N\{\widehat V_N>0\}\to1$ and, for every fixed $\alpha\in(0,1)$,
\begin{equation}\label{eq: restricted-path-coverage}
\liminf_{N\to\infty}P_N\left\{\left|\widehat\theta_{od,N}^{\mathcal R}-\theta_{od,N}^{\mathcal R}\right|\leq z_{1-\alpha/2}\sqrt{\widehat V_N}\right\}\geq1-\alpha.
\end{equation}
If $H_N/V_N\xrightarrow{p}0$, then
\begin{equation}\label{eq: restricted-path-exactness}
\frac{\widehat V_N}{V_N}\xrightarrow{p}1,\qquad \frac{\widehat\theta_{od,N}^{\mathcal R}-\theta_{od,N}^{\mathcal R}}{\sqrt{\widehat V_N}}\xrightarrow{d}\mathcal N(0,1),
\end{equation}
and the probability in \eqref{eq: restricted-path-coverage} converges to $1-\alpha$. A sufficient condition for $H_N/V_N\xrightarrow{p}0$ is
\[
P_N\{S(W,\DID_N)=S(W',\DID_N)\text{ for every }W,W'\in\mathcal R_N^{od}\}\longrightarrow1.
\]
\end{theorem}

The theorem allows $K_N$ to increase with $N$ without a separate rate restriction. Its proof uses the following two lemmas: the first gives the conditional representation and normal approximation for the estimator, and the second controls the conditional mean and variance of $\widehat V_N$.

\begin{lemma}[Conditional representation and normal approximation]
\label{lem: restricted-path-worker-representation}
Suppose that Assumptions~\ref{ass: restricted-path-sampling} and~\ref{ass: restricted-path-selection} hold and that \eqref{eq: restricted-path-fourth-moment} holds. Almost surely on $\{|\mathcal R_N^{od}|\geq2\}$,
\begin{align}
E_N[S(W,\widehat{\DID}_{N,W})\mid\mathcal G_N]&=S(W,\DID_N),\qquad E_N[\widehat\theta_{od,N}^{\mathcal R}\mid\mathcal G_N]=\theta_{od,N}^{\mathcal R},\label{eq: restricted-path-conditional-unbiasedness}\\
S(W,\widehat{\DID}_{N,W})-S(W,\DID_N)&=\sum_{i\in\mathcal J_N}b_{N,Wi}\{\Delta_i-E_N[\Delta_i\mid\mathcal G_N]\},\label{eq: restricted-path-worker-representation}\\
\widehat\theta_{od,N}^{\mathcal R}-\theta_{od,N}^{\mathcal R}&=\sum_{i\in\mathcal J_N}b_{N,i}\{\Delta_i-E_N[\Delta_i\mid\mathcal G_N]\},\label{eq: restricted-path-estimator-representation}\\
(\Sigma_N)_{W,W'}&=\sum_{i\in\mathcal J_N}b_{N,Wi}b_{N,W'i}\var_N(\Delta_i\mid\mathcal G_N),\label{eq: restricted-path-worker-covariance}\\
V_N&=\sum_{i\in\mathcal J_N}b_{N,i}^2\var_N(\Delta_i\mid\mathcal G_N)=\frac{\sum_{W,W'\in\mathcal R_N^{od}}(\Sigma_N)_{W,W'}}{|\mathcal R_N^{od}|^2},\label{eq: restricted-path-worker-variance}
\end{align}
for all paths $W,W'\in\mathcal R_N^{od}$. Moreover,
\begin{equation}
\label{eq: restricted-path-projection-share}
\max_{1\leq i\leq N}\var_N\bigl(E_N[\widehat\theta_{od,N}^{\mathcal R}\mid\Delta_i,\mathcal G_N]\mid\mathcal G_N\bigr)=\max_{i\in\mathcal J_N}\{b_{N,i}^2\var_N(\Delta_i\mid\mathcal G_N)\}.
\end{equation}
On $\{|\mathcal R_N^{od}|\geq2,\ V_N>0\}$,
\begin{equation}
\label{eq: restricted-path-normal-approximation}
\begin{aligned}
&\sup_{x\in\mathbb R}\left|P_N\left\{\frac{\widehat\theta_{od,N}^{\mathcal R}-\theta_{od,N}^{\mathcal R}}{\sqrt{V_N}}\leq x\,\middle|\,\mathcal G_N\right\}-\Phi(x)\right|\\
&\qquad\leq\min\left\{1,58\sqrt{C}\left(\frac{\max_{i\in\mathcal J_N}\{b_{N,i}^2\var_N(\Delta_i\mid\mathcal G_N)\}}{V_N}\right)^{1/2}\right\}.
\end{aligned}
\end{equation}
\end{lemma}

\begin{proof}[Proof of Lemma~\ref{lem: restricted-path-worker-representation}]
Conditioning on $\mathcal G_N$ fixes the selected paths and worker subsets. By Assumption~\ref{ass: restricted-path-sampling}(i) and (iii), the random variables $\Delta_1,\ldots,\Delta_N$ are conditionally independent given $\mathcal G_N$, and moreover,
\[
E_N[\Delta_i\mid\mathcal G_N]=E_N[\Delta_i\mid O_i,D_i],\qquad i=1,\ldots,N,
\]
almost surely. Since each $\mathcal I_{N,W}(u,v)$ is $\mathcal G_N$-measurable, the conditional mean of the sample average of $\Delta_i$ over $\mathcal I_{N,W}(u,v)$ satisfies
\[
E_N\left[\frac{\sum_{i\in\mathcal I_{N,W}(u,v)}\Delta_i}{|\mathcal I_{N,W}(u,v)|}\,\middle|\,\mathcal G_N\right]=\frac{\sum_{i\in\mathcal I_{N,W}(u,v)}E_N[\Delta_i\mid O_i=u,D_i=v]}{|\mathcal I_{N,W}(u,v)|}=E_N[\Delta_i\mid O_i=u,D_i=v].
\]
Subtraction and summation over signed steps give \eqref{eq: restricted-path-conditional-unbiasedness}. Expanding the assigned cell averages gives
\begin{align*}
S(W,\widehat{\DID}_{N,W})
&=
\sum_{i\in\mathcal J_N}\sum_{\ell=1}^{L(W)}s_{\ell}\left\{\frac{\ind\{i\in\mathcal I_{N,W}(u_{\ell},v_{\ell})\}}{|\mathcal I_{N,W}(u_{\ell},v_{\ell})|}-\frac{\ind\{i\in\mathcal I_{N,W}(u_{\ell},u_{\ell})\}}{|\mathcal I_{N,W}(u_{\ell},u_{\ell})|}\right\}\Delta_i
=
\sum_{i\in\mathcal J_N}b_{N,Wi}\Delta_i,
\\[1em]
\widehat\theta_{od,N}^{\mathcal R}
&=
\frac{1}{|\mathcal R_N^{od}|}\sum_{W\in\mathcal R_N^{od}}
\left(\sum_{i\in\mathcal J_N}b_{N,Wi}\Delta_i\right)
=
\sum_{i\in\mathcal J_N}\left(
  \frac{1}{|\mathcal R_N^{od}|}\sum_{W\in\mathcal R_N^{od}}b_{N,Wi}
\right)\Delta_i
=
\sum_{i\in\mathcal J_N}b_{N,i}\Delta_i.
\end{align*}
Subtracting the conditional means proves \eqref{eq: restricted-path-worker-representation}--\eqref{eq: restricted-path-estimator-representation}. Conditional independence and the finite second moments implied by \eqref{eq: restricted-path-fourth-moment} yield
\begin{align*}
(\Sigma_N)_{W,W'}&=\sum_{i,j\in\mathcal J_N}b_{N,Wi}b_{N,W'j}\cov_N(\Delta_i,\Delta_j\mid\mathcal G_N)=\sum_{i\in\mathcal J_N}b_{N,Wi}b_{N,W'i}\var_N(\Delta_i\mid\mathcal G_N),\\
V_N&=\sum_{i\in\mathcal J_N}b_{N,i}^2\var_N(\Delta_i\mid\mathcal G_N)=\frac{\sum_{W,W'\in\mathcal R_N^{od}}\sum_{i\in\mathcal J_N}b_{N,Wi}b_{N,W'i}\var_N(\Delta_i\mid\mathcal G_N)}{|\mathcal R_N^{od}|^2}.
\end{align*}
These identities prove \eqref{eq: restricted-path-worker-covariance}--\eqref{eq: restricted-path-worker-variance}. For $i\in\mathcal J_N$, conditioning the estimator representation on $(\Delta_i,\mathcal G_N)$ gives
\begin{align*}
E_N[\widehat\theta_{od,N}^{\mathcal R}\mid\Delta_i,\mathcal G_N]
&=\theta_{od,N}^{\mathcal R}+b_{N,i}\{\Delta_i-E_N[\Delta_i\mid\mathcal G_N]\}+\sum_{\substack{j\in\mathcal J_N\\j\neq i}}b_{N,j}\{E_N[\Delta_j\mid\Delta_i,\mathcal G_N]-E_N[\Delta_j\mid\mathcal G_N]\}\\
&=\theta_{od,N}^{\mathcal R}+b_{N,i}\{\Delta_i-E_N[\Delta_i\mid\mathcal G_N]\}.
\end{align*}
For $i\notin\mathcal J_N$, the same conditional expectation equals $\theta_{od,N}^{\mathcal R}$. Taking conditional variances proves \eqref{eq: restricted-path-projection-share}.

It remains to establish the normal approximation. Conditional Cauchy--Schwarz and \eqref{eq: restricted-path-fourth-moment} give, for each $i\in\mathcal J_N$,
\begin{align*}
E_N[|b_{N,i}\{\Delta_i-E_N[\Delta_i\mid\mathcal G_N]\}|^3\mid\mathcal G_N]
&\leq |b_{N,i}|^3\{\var_N(\Delta_i\mid\mathcal G_N)\}^{1/2}\{E_N[(\Delta_i-E_N[\Delta_i\mid\mathcal G_N])^4\mid\mathcal G_N]\}^{1/2}\\
&\leq\sqrt{C}\{b_{N,i}^2\var_N(\Delta_i\mid\mathcal G_N)\}^{3/2}.
\end{align*}
On $\{V_N>0\}$, the summands in \eqref{eq: restricted-path-estimator-representation}, divided by $\sqrt{V_N}$, are conditionally independent and centered, have finite third absolute moments, and have total conditional variance one. Theorem~1.1 of \citet{raic2019explicit}, applied in dimension one to the half-lines $(-\infty,x]$, therefore gives
\begingroup
\allowdisplaybreaks
\begin{align*}
&\sup_{x\in\mathbb R}\left|P_N\left\{\frac{\widehat\theta_{od,N}^{\mathcal R}-\theta_{od,N}^{\mathcal R}}{\sqrt{V_N}}\leq x\,\middle|\,\mathcal G_N\right\}-\Phi(x)\right|
\\
\leq&
\frac{58\sqrt{C}}{V_N^{3/2}}\sum_{i\in\mathcal J_N}\{b_{N,i}^2\var_N(\Delta_i\mid\mathcal G_N)\}^{3/2}
\\
\leq&
\frac{58\sqrt{C}}{V_N^{3/2}}\left(\max_{i\in\mathcal J_N}\{b_{N,i}^2\var_N(\Delta_i\mid\mathcal G_N)\}\right)^{1/2}\sum_{i\in\mathcal J_N}b_{N,i}^2\var_N(\Delta_i\mid\mathcal G_N)
\\
=&
58\sqrt{C}\left(\frac{\max_{i\in\mathcal J_N}\{b_{N,i}^2\var_N(\Delta_i\mid\mathcal G_N)\}}{V_N}\right)^{1/2},
\end{align*}
\endgroup
in which the constant is obtained by $42\cdot1^{1/4}+16=58$. A difference of distribution functions has absolute value at most one, which supplies the truncation in \eqref{eq: restricted-path-normal-approximation}.
\end{proof}

\begin{lemma}[Conditional moments of the path dispersion]
\label{lem: restricted-path-dispersion-moments}
Suppose that Assumptions~\ref{ass: restricted-path-sampling} and~\ref{ass: restricted-path-selection} hold and that \eqref{eq: restricted-path-fourth-moment} holds. On $\{|\mathcal R_N^{od}|\geq2\}$,
\begin{equation}
\label{eq: restricted-path-exact-variance-mean}
E_N[\widehat V_N\mid\mathcal G_N]=H_N+\frac{\tr(\Sigma_N)-|\mathcal R_N^{od}|V_N}{|\mathcal R_N^{od}|(|\mathcal R_N^{od}|-1)}
\end{equation}
and
\begin{equation}
\label{eq: restricted-path-variance-bound}
\var_N(\widehat V_N\mid\mathcal G_N)\leq\frac{2(C+2)\tr(\Sigma_N^2)}{|\mathcal R_N^{od}|^2(|\mathcal R_N^{od}|-1)^2}+\frac{8\lambda_{\max}(\Sigma_N)H_N}{|\mathcal R_N^{od}|(|\mathcal R_N^{od}|-1)}.
\end{equation}
\end{lemma}

\begin{proof}[Proof of Lemma~\ref{lem: restricted-path-dispersion-moments}]
All calculations condition on $\mathcal G_N$ and are on $\{|\mathcal R_N^{od}|\geq2\}$. The proof first centers the worker representation across paths, then bounds the quadratic and linear terms in the resulting dispersion identity.

\noindent{\bf Step 1: Centering and the conditional mean.} Define the centering matrix, indexed by the paths in $\mathcal R_N^{od}$, by
\[
\mathbb C_N:=I_{|\mathcal R_N^{od}|}-\frac{1}{|\mathcal R_N^{od}|}\mathbf1\mathbf1'
\qquad\text{with}\qquad
(\mathbb C_N)_{W,W'}:=\ind\{W=W'\}-\frac{1}{|\mathcal R_N^{od}|},
\]
where $\mathbb I_{|\mathcal R_N^{od}|}$ denotes the $|\mathcal R_N^{od}| \times |\mathcal R_N^{od}|$ identity matrix, and $\mathbf1$ denotes the vector of $|\mathcal R_N^{od}|$ ones. Observe that being a projection matrix onto the orthogonal complement of $\mathbf1$, $\mathbb C_N$ satisfies $\mathbb C_N=\mathbb C_N'$, $\mathbb C_N^2=\mathbb C_N$, and $\mathbb C_N \mathbf1=0$. 

Next observe that by Lemma~\ref{lem: restricted-path-worker-representation} \eqref{eq: restricted-path-worker-representation} and \eqref{eq: restricted-path-estimator-representation},
\[
S(W,\widehat{\DID}_{N,W})-\widehat\theta_{od,N}^{\mathcal R}=S(W,\DID_N)-\theta_{od,N}^{\mathcal R}+\sum_{i\in\mathcal J_N}(b_{N,Wi}-b_{N,i})\{\Delta_i-E_N[\Delta_i\mid\mathcal G_N]\}.
\]
Since $\sum_{W\in\mathcal R_N^{od}}\{S(W,\DID_N)-\theta_{od,N}^{\mathcal R}\}=0$, squaring and summing yields
\begingroup 
\allowdisplaybreaks
\begin{align}
|\mathcal R_N^{od}|(|\mathcal R_N^{od}|-1)\widehat V_N
&=
|\mathcal R_N^{od}|(|\mathcal R_N^{od}|-1)H_N
+
\sum_{W\in\mathcal R_N^{od}}\left(\sum_{i\in\mathcal J_N}(b_{N,Wi}-b_{N,i})\{\Delta_i-E_N[\Delta_i\mid\mathcal G_N]\}\right)^2
\notag\\
&\qquad+
2\sum_{W\in\mathcal R_N^{od}}\{S(W,\DID_N)-\theta_{od,N}^{\mathcal R}\}
\sum_{i\in\mathcal J_N}b_{N,Wi}\{\Delta_i-E_N[\Delta_i\mid\mathcal G_N]\},
\label{eq: restricted-path-dispersion-identity}
\end{align}
\endgroup
in which the last term has conditional mean zero. 

For the quadratic term above, note that the covariance matrix of the centered path errors is $\mathbb C_N\Sigma_N\mathbb C_N$, with entries
\begingroup 
\allowdisplaybreaks
\begin{align*}
(\mathbb C_N\Sigma_N\mathbb C_N)_{W,W'}
&=
\sum_{U,V\in\mathcal R_N^{od}}
(\mathbb C_N)_{W,U}(\Sigma_N)_{U,V}(\mathbb C_N)_{V,W'}
\\
&=
\sum_{U,V\in\mathcal R_N^{od}}(\mathbb C_N)_{W,U}(\mathbb C_N)_{W',V}
\sum_{i\in\mathcal J_N}b_{N,Ui}b_{N,Vi}\var_N(\Delta_i\mid\mathcal G_N)
\\
&=
\sum_{i\in\mathcal J_N}(b_{N,Wi}-b_{N,i})(b_{N,W'i}-b_{N,i})\var_N(\Delta_i\mid\mathcal G_N),
\end{align*}
\endgroup
where the second equality follows from Lemma~\ref{lem: restricted-path-worker-representation} \eqref{eq: restricted-path-worker-covariance}. Consequently, the quadratic term has conditional expectation
\begingroup 
\allowdisplaybreaks
\begin{align*}
&E_N\left[
  \sum_{W\in\mathcal R_N^{od}}\left(\sum_{i\in\mathcal J_N}(b_{N,Wi}-b_{N,i})\{\Delta_i-E_N[\Delta_i\mid\mathcal G_N]\}\right)^2
  \ \middle|\ \mathcal G_N
\right]
\\
=&
\sum_{W\in\mathcal R_N^{od}}\left(
  \sum_{i\in\mathcal J_N}(b_{N,Wi}-b_{N,i})^2\var_N(\Delta_i\mid\mathcal G_N)
  +
  \sum_{i \neq j}(b_{N,Wi}-b_{N,i})(b_{N,Wj}-b_{N,j})\cov_N(\Delta_i,\Delta_j\mid\mathcal G_N)
\right)
\\
=&
\sum_{W\in\mathcal R_N^{od}}
\sum_{i\in\mathcal J_N}(b_{N,Wi}-b_{N,i})^2\var_N(\Delta_i\mid\mathcal G_N)
\\
=&
\sum_{W\in\mathcal R_N^{od}} (\mathbb C_N\Sigma_N\mathbb C_N)_{W,W}
\\
=&
\tr(\mathbb C_N\Sigma_N\mathbb C_N)
\\
=&
\tr(\mathbb C_N\Sigma_N)
\\
=&
\tr(\Sigma_N)-\frac{1}{|\mathcal R_N^{od}|}\sum_{W,W'\in\mathcal R_N^{od}}(\Sigma_N)_{W,W'}
\\
=&
\tr(\Sigma_N)-|\mathcal R_N^{od}|V_N,
\end{align*}
\endgroup
where the last equality follows from Lemma~\ref{lem: restricted-path-worker-representation} \eqref{eq: restricted-path-worker-variance}. Since $H_N$ is $\mathcal G_N$-measurable, combining the above and taking conditional expectation of \eqref{eq: restricted-path-dispersion-identity} proves \eqref{eq: restricted-path-exact-variance-mean}.

\noindent{\bf Step 2: Bounding the quadratic term.} Conditional independence and centering imply, for $i_1,i_2,i_3,i_4\in\mathcal J_N$,
\[
\begin{aligned}
E_N\left[\prod_{a=1}^{4}\{\Delta_{i_a}-E_N[\Delta_{i_a}\mid\mathcal G_N]\}\,\middle|\,\mathcal G_N\right]
=\begin{cases}
E_N[(\Delta_i-E_N[\Delta_i\mid\mathcal G_N])^4\mid\mathcal G_N],&i_1=i_2=i_3=i_4=i,\\
\var_N(\Delta_i\mid\mathcal G_N)\var_N(\Delta_j\mid\mathcal G_N),&\text{permuting }(i,i,j,j),\ i\neq j,\\
0,&\text{otherwise}.
\end{cases}
\end{aligned}
\]
In the last case at least one index occurs once, and its centered factor has conditional mean zero. Expanding the square of the quadratic term and grouping these cases gives
\begingroup 
\allowdisplaybreaks
{\small
\begin{align*}
&\var_N\left(
  \sum_{W\in\mathcal R_N^{od}}\left[
    \sum_{i\in\mathcal J_N}(b_{N,Wi}-b_{N,i})\{\Delta_i-E_N[\Delta_i\mid\mathcal G_N]\}
  \right]^2\,\middle|\,\mathcal G_N
\right)
\\
=&
\var_N\left(
  \sum_{i\in\mathcal J_N}\left(
    \sum_{W\in\mathcal R_N^{od}}(b_{N,Wi}-b_{N,i})^2
  \right)
  \{\Delta_i-E_N[\Delta_i\mid\mathcal G_N]\}^2
  \right.
  \\
  &\hspace{1em}+\left.
  2\sum_{i,j\in\mathcal J_N, i<j}\left(
    \sum_{W\in\mathcal R_N^{od}}(b_{N,Wi}-b_{N,i})(b_{N,Wj}-b_{N,j})
  \right)
  \{\Delta_i-E_N[\Delta_i\mid\mathcal G_N]\}\{\Delta_j-E_N[\Delta_j\mid\mathcal G_N]\}
  \ \middle|\ \mathcal G_N
\right)
\\
=&
\sum_{i\in\mathcal J_N}\left(
  \sum_{W\in\mathcal R_N^{od}}(b_{N,Wi}-b_{N,i})^2
\right)^2
\var_N(\{\Delta_i-E_N[\Delta_i\mid\mathcal G_N]\}^2\mid\mathcal G_N)
\\
&\hspace{1em}+
4\sum_{\substack{i,j\in\mathcal J_N\\ i<j}}\left(
  \sum_{W\in\mathcal R_N^{od}}(b_{N,Wi}-b_{N,i})(b_{N,Wj}-b_{N,j})
\right)^2
\var_N(\{\Delta_i-E_N[\Delta_i\mid\mathcal G_N]\}\{\Delta_j-E_N[\Delta_j\mid\mathcal G_N]\}\mid\mathcal G_N)
\\
=&
\sum_{i\in\mathcal J_N}\left(
  \sum_{W\in\mathcal R_N^{od}}(b_{N,Wi}-b_{N,i})^2
\right)^2
\left\{E_N[(\Delta_i-E_N[\Delta_i\mid\mathcal G_N])^4\mid\mathcal G_N]-\var_N(\Delta_i\mid\mathcal G_N)^2\right\}
\\
&\hspace{1em}+
4\sum_{\substack{i,j\in\mathcal J_N\\i<j}}\left(
  \sum_{W\in\mathcal R_N^{od}}(b_{N,Wi}-b_{N,i})(b_{N,Wj}-b_{N,j})
\right)^2
\var_N(\Delta_i\mid\mathcal G_N)\var_N(\Delta_j\mid\mathcal G_N)
\\
\leq& 
C\sum_{i\in\mathcal J_N}\left(
  \sum_{W\in\mathcal R_N^{od}}(b_{N,Wi}-b_{N,i})^2
\right)^2
\var_N(\Delta_i\mid\mathcal G_N)^2
\\
&\hspace{1em}+
2\sum_{\substack{i,j\in\mathcal J_N\\i\neq j}}\left(
  \sum_{W\in\mathcal R_N^{od}}(b_{N,Wi}-b_{N,i})(b_{N,Wj}-b_{N,j})
\right)^2
\var_N(\Delta_i\mid\mathcal G_N)\var_N(\Delta_j\mid\mathcal G_N)
\\
\leq&
(C+2)\sum_{i,j\in\mathcal J_N}\left(
  \sum_{W\in\mathcal R_N^{od}}(b_{N,Wi}-b_{N,i})(b_{N,Wj}-b_{N,j})
\right)^2
\var_N(\Delta_i\mid\mathcal G_N)\var_N(\Delta_j\mid\mathcal G_N),
\end{align*}
}
\endgroup
in which the first inequality follows from \eqref{eq: restricted-path-fourth-moment}. Reordering finite sums in the squared covariance entries gives
\begingroup 
\allowdisplaybreaks
\begin{align*}
\tr\{(\mathbb C_N\Sigma_N\mathbb C_N)^2\}
&=
\sum_{W,W'\in\mathcal R_N^{od}}\left(\sum_{i\in\mathcal J_N}(b_{N,Wi}-b_{N,i})(b_{N,W'i}-b_{N,i})\var_N(\Delta_i\mid\mathcal G_N)\right)^2
\\
&=
\sum_{i,j\in\mathcal J_N}\left(\sum_{W\in\mathcal R_N^{od}}(b_{N,Wi}-b_{N,i})(b_{N,Wj}-b_{N,j})\right)^2
\var_N(\Delta_i\mid\mathcal G_N)\var_N(\Delta_j\mid\mathcal G_N).
\end{align*}
\endgroup
The fourth-moment bound therefore implies that the quadratic-term variance is at most $(C+2)\tr\{(\mathbb C_N\Sigma_N\mathbb C_N)^2\}$. Next observe that
\begin{align*}
\tr(\Sigma_N^2)-\tr\{(\mathbb C_N\Sigma_N\mathbb C_N)^2\}
=
\tr\{\Sigma_N(\mathbb I_{|\mathcal R_N^{od}|}-\mathbb C_N)\Sigma_N\}
+
\tr\{\mathbb C_N\Sigma_N(\mathbb I_{|\mathcal R_N^{od}|}-\mathbb C_N)\Sigma_N\mathbb C_N\}\ge0,
\end{align*}
where the last inequality holds since both matrices inside the traces are positive semidefinite, and thus
\begingroup 
\allowdisplaybreaks
\begin{align}
\label{eq: restricted-path-var-bound1}
\var_N\left(
  \sum_{W\in\mathcal R_N^{od}}\left[
    \sum_{i\in\mathcal J_N}(b_{N,Wi}-b_{N,i})\{\Delta_i-E_N[\Delta_i\mid\mathcal G_N]\}
  \right]^2\,\middle|\,\mathcal G_N
\right)
\le 
(C+2)\tr(\Sigma_N^2).
\end{align}
\endgroup

\noindent{\bf Step 3: Bounding the linear term and combining the bounds.} The covariance formula in Lemma~\ref{lem: restricted-path-worker-representation} \eqref{eq: restricted-path-worker-covariance} gives
\begin{align}
&\var_N\left(
  \sum_{W\in\mathcal R_N^{od}}\{S(W,\DID_N)-\theta_{od,N}^{\mathcal R}\}\sum_{i\in\mathcal J_N}b_{N,Wi}\{\Delta_i-E_N[\Delta_i\mid\mathcal G_N]\}
  \,\middle|\,\mathcal G_N
\right)
\notag\\
=&
\sum_{W,W'\in\mathcal R_N^{od}}
\{S(W,\DID_N)-\theta_{od,N}^{\mathcal R}\}\{S(W',\DID_N)-\theta_{od,N}^{\mathcal R}\}
\notag\\
&\hspace{4em}\times
\cov_N\left(
  \sum_{i\in\mathcal J_N}b_{N,Wi}\{\Delta_i-E_N[\Delta_i\mid\mathcal G_N]\},
  \sum_{j\in\mathcal J_N}b_{N,W'j}\{\Delta_j-E_N[\Delta_j\mid\mathcal G_N]\}
  \,\middle|\,\mathcal G_N
\right)
\notag\\
=&
\sum_{W,W'\in\mathcal R_N^{od}}
\{S(W,\DID_N)-\theta_{od,N}^{\mathcal R}\}\{S(W',\DID_N)-\theta_{od,N}^{\mathcal R}\}
\sum_{i\in\mathcal J_N}b_{N,Wi}b_{N,W'i}\var_N(\Delta_i\mid\mathcal G_N)
\notag\\
=&
\sum_{W,W'\in\mathcal R_N^{od}}\{S(W,\DID_N)-\theta_{od,N}^{\mathcal R}\}(\Sigma_N)_{W,W'}\{S(W',\DID_N)-\theta_{od,N}^{\mathcal R}\}
\notag\\
\leq&
\lambda_{\max}(\Sigma_N)\sum_{W\in\mathcal R_N^{od}}\{S(W,\DID_N)-\theta_{od,N}^{\mathcal R}\}^2
\notag\\
=&
\lambda_{\max}(\Sigma_N)|\mathcal R_N^{od}|(|\mathcal R_N^{od}|-1)H_N,
\label{eq: restricted-path-var-bound2}
\end{align}
in which the inequality follows from the spectral bound for the positive semidefinite matrix $\Sigma_N$. Taking conditional variances on both sides of \eqref{eq: restricted-path-dispersion-identity} and applying the variance bounds \eqref{eq: restricted-path-var-bound1} and \eqref{eq: restricted-path-var-bound2} to the conditionally centered quadratic and linear terms, respectively, yield
\[
|\mathcal R_N^{od}|^2(|\mathcal R_N^{od}|-1)^2\var_N(\widehat V_N\mid\mathcal G_N)\leq2(C+2)\tr(\Sigma_N^2)+8\lambda_{\max}(\Sigma_N)|\mathcal R_N^{od}|(|\mathcal R_N^{od}|-1)H_N,
\]
hence proving \eqref{eq: restricted-path-variance-bound}.
\end{proof}

\begin{proof}[Proof of Theorem~\ref{thm: restricted-path-inference}]
The two lemmas give a conditional normal approximation and conditional moment bounds for $\widehat V_N$. The proof first converts the normal approximation to an unconditional limit, then establishes the variance ratio and the coverage conclusions.

Let $\mathcal H_N:=\{|\mathcal R_N^{od}|\geq2,\ V_N>0\}$. Condition (ii) of Theorem~\ref{thm: restricted-path-inference} gives
\[
P_N(\mathcal H_N^c)\leq P_N\{|\mathcal R_N^{od}|<2\}+P_N\{V_N=0\}\longrightarrow0.
\]
All conditional calculations below are on $\mathcal H_N$. On this event $\tr(\Sigma_N)>0$, since $\Sigma_N$ is positive semidefinite and $\tr(\Sigma_N)=0$ would imply $\Sigma_N=0$ and hence $V_N=0$. The following elementary implication will convert conditional bounds to unconditional limits: if $R_N\geq0$ and $R_N\xrightarrow{p}0$, then, for every $\eta\in(0,1)$,
\[
E_N[\min\{1,R_N\}]\leq\eta+P_N\{R_N>\eta\},\qquad\text{so}\qquad E_N[\min\{1,R_N\}]\longrightarrow0,
\]
in which the last conclusion follows by taking the limit superior and then letting $\eta\downarrow0$.

\noindent{\bf Step 1: Consistency and asymptotic normality.} Conditional unbiasedness in Lemma~\ref{lem: restricted-path-worker-representation}, conditional Chebyshev, and iterated expectations give, for every $\delta>0$,
\[
P_N\{|\widehat\theta_{od,N}^{\mathcal R}-\theta_{od,N}^{\mathcal R}|>\delta\}\leq P_N(\mathcal H_N^c)+E_N\left[\min\left\{1,\frac{V_N}{\delta^2}\right\}\right]\longrightarrow0,
\]
in which the last convergence uses $V_N\xrightarrow{p}0$ from condition (ii) of Theorem~\ref{thm: restricted-path-inference}. The normal-approximation bound in the same lemma similarly implies
\begin{align*}
&\sup_{x\in\mathbb R}\left|P_N\left\{\frac{\widehat\theta_{od,N}^{\mathcal R}-\theta_{od,N}^{\mathcal R}}{\sqrt{V_N}}\leq x\right\}-\Phi(x)\right|\\
&\qquad\leq P_N(\mathcal H_N^c)+E_N\left[\min\left\{1,58\sqrt{C}\left(\frac{\max_{i\in\mathcal J_N}\{b_{N,i}^2\var_N(\Delta_i\mid\mathcal G_N)\}}{V_N}\right)^{1/2}\right\}\right]\longrightarrow0,
\end{align*}
in which the convergence follows from \eqref{eq: restricted-path-worker-share} and the preceding bounded-expectation implication. This proves consistency \eqref{eq: restricted-path-consistency} and the asymptotic normality part of \eqref{eq: restricted-path-normal-limit}.

\noindent{\bf Step 2: The variance ratio.} The objective is to show that $\widehat V_N/(V_N+H_N)\xrightarrow{p}1$. The argument first establishes that the conditional expectation of $\widehat V_N$ is asymptotically equivalent to $V_N+H_N$, and then shows that its conditional fluctuations are negligible relative to the same quantity.

\noindent{\it Conditional bias.} On $\mathcal H_N$, the first restriction in \eqref{eq: restricted-path-dependence} and $|\mathcal R_N^{od}|\xrightarrow{p}\infty$ imply
\[
\frac{\tr(\Sigma_N)}{|\mathcal R_N^{od}|^2V_N}\xrightarrow{p}1,\qquad \frac{\tr(\Sigma_N)}{|\mathcal R_N^{od}|(|\mathcal R_N^{od}|-1)V_N}=\frac{|\mathcal R_N^{od}|}{|\mathcal R_N^{od}|-1}\frac{\tr(\Sigma_N)}{|\mathcal R_N^{od}|^2V_N}\xrightarrow{p}1.
\]
By Lemma~\ref{lem: restricted-path-dispersion-moments} \eqref{eq: restricted-path-exact-variance-mean}, the difference between the conditional expectation and $V_N+H_N$ is
\[
E_N[\widehat V_N\mid\mathcal G_N]-(V_N+H_N)=\frac{\tr(\Sigma_N)-|\mathcal R_N^{od}|^2V_N}{|\mathcal R_N^{od}|(|\mathcal R_N^{od}|-1)}.
\]
Since $H_N\geq0$ and $V_N>0$ on $\mathcal H_N$,
\begin{align*}
\frac{|E_N[\widehat V_N\mid\mathcal G_N]-(V_N+H_N)|}{V_N+H_N}
&=\frac{|\tr(\Sigma_N)-|\mathcal R_N^{od}|^2V_N|}{|\mathcal R_N^{od}|(|\mathcal R_N^{od}|-1)(V_N+H_N)}\\
&\leq\frac{|\mathcal R_N^{od}|}{|\mathcal R_N^{od}|-1}\left|\frac{\tr(\Sigma_N)}{|\mathcal R_N^{od}|^2V_N}-1\right|\xrightarrow{p}0.
\end{align*}
Thus the conditional bias is asymptotically negligible relative to $V_N+H_N$.

\noindent{\it Conditional fluctuation.} It remains to control $\widehat V_N-E_N[\widehat V_N\mid\mathcal G_N]$. Because $\Sigma_N$ is positive semidefinite, its eigenvalues are nonnegative and satisfy
\[
\tr(\Sigma_N^2)=\sum_{a=1}^{|\mathcal R_N^{od}|}\lambda_a(\Sigma_N)^2\leq\lambda_{\max}(\Sigma_N)\sum_{a=1}^{|\mathcal R_N^{od}|}\lambda_a(\Sigma_N)=\lambda_{\max}(\Sigma_N)\tr(\Sigma_N).
\]
Moreover, $V_N>0$ and $H_N\geq0$ imply
\[
(V_N+H_N)^2\geq V_N^2,\qquad (V_N+H_N)^2\geq4V_NH_N,
\]
where the second inequality follows from $(V_N-H_N)^2\geq0$. Applying these inequalities to the variance bound in Lemma~\ref{lem: restricted-path-dispersion-moments} \eqref{eq: restricted-path-variance-bound} gives
\begin{align*}
\frac{\var_N(\widehat V_N\mid\mathcal G_N)}{(V_N+H_N)^2}
&\leq
\frac{2(C+2)\tr(\Sigma_N^2)}{|\mathcal R_N^{od}|^2(|\mathcal R_N^{od}|-1)^2V_N^2}+\frac{2\lambda_{\max}(\Sigma_N)}{|\mathcal R_N^{od}|(|\mathcal R_N^{od}|-1)V_N}
\\
&\leq
2(C+2)\frac{\lambda_{\max}(\Sigma_N)}{\tr(\Sigma_N)}\left\{\frac{\tr(\Sigma_N)}{|\mathcal R_N^{od}|(|\mathcal R_N^{od}|-1)V_N}\right\}^2
+
2\frac{\lambda_{\max}(\Sigma_N)}{\tr(\Sigma_N)}\frac{\tr(\Sigma_N)}{|\mathcal R_N^{od}|(|\mathcal R_N^{od}|-1)V_N}
\\
&\xrightarrow{p}0.
\end{align*}
The limit follows because the ratios involving $\tr(\Sigma_N)/\{|\mathcal R_N^{od}|(|\mathcal R_N^{od}|-1)V_N\}$ converge in probability to one, whereas $\lambda_{\max}(\Sigma_N)/\tr(\Sigma_N)\xrightarrow{p}0$ by \eqref{eq: restricted-path-dependence}.

For every $\delta>0$, conditional Chebyshev's inequality and the law of iterated expectations therefore give
\begin{align*}
P_N\left\{\mathcal H_N,\ \frac{|\widehat V_N-E_N[\widehat V_N\mid\mathcal G_N]|}{V_N+H_N}>\delta\right\}
\leq 
E_N\left[\ind\{\mathcal H_N\}\min\left\{1,\frac{\var_N(\widehat V_N\mid\mathcal G_N)}{\delta^2(V_N+H_N)^2}\right\}\right]
\longrightarrow0,
\end{align*}
in which the last convergence follows from the preceding conditional variance bound and the bounded-expectation implication established before Step~1.

\noindent{\it Combining the bounds.} On $\mathcal H_N$, the triangle inequality gives
\[
\left|\frac{\widehat V_N}{V_N+H_N}-1\right|\leq\frac{|\widehat V_N-E_N[\widehat V_N\mid\mathcal G_N]|}{V_N+H_N}+\frac{|E_N[\widehat V_N\mid\mathcal G_N]-(V_N+H_N)|}{V_N+H_N}.
\]
The first term converges to zero in probability by the conditional fluctuation argument, and the second by the conditional bias argument. Since $P_N(\mathcal H_N^c)\to0$, it follows that
\[
\frac{\widehat V_N}{V_N+H_N}\xrightarrow{p}1.
\]
This proves the variance-ratio conclusion in \eqref{eq: restricted-path-normal-limit}.

\noindent{\bf Step 3: Positivity, coverage, and studentization.} Step~1 establishes asymptotic normality when the estimator is normalized by its conditional standard deviation $\sqrt{V_N}$, while Step~2 establishes consistency of $\widehat V_N$ for $V_N+H_N$. The remaining task is to combine these conclusions to obtain conservative inference, and then establish conditions under which the same inference is asymptotically exact.

\noindent{\it Positivity and a lower bound for the variance estimator.} Since $H_N\geq0$, the variance-ratio conclusion of Step~2 implies that, for every fixed $\eta\in(0,1)$,
\begin{align*}
P_N\{\widehat V_N<(1-\eta)V_N\}
&\leq P_N(\mathcal H_N^c)+P_N\left\{\mathcal H_N,\ \frac{\widehat V_N}{V_N+H_N}<1-\eta\right\}\\
&\leq P_N(\mathcal H_N^c)+P_N\left\{\left|\frac{\widehat V_N}{V_N+H_N}-1\right|>\eta\right\}\longrightarrow0.
\end{align*}
On $\mathcal H_N$, $V_N>0$, so $\widehat V_N=0$ implies $\widehat V_N<(1-\eta)V_N$. Consequently,
\[
P_N\{\widehat V_N=0\}\leq P_N(\mathcal H_N^c)+P_N\{\widehat V_N<(1-\eta)V_N\}\longrightarrow0,
\]
and since $\widehat V_N\geq0$ by definition, this establishes $P_N\{\widehat V_N>0\}\to1$. 

\noindent{\it Conservative coverage.} Fix $\alpha\in(0,1)$ and $\eta\in(0,1)$. On $\mathcal H_N\cap\{\widehat V_N\geq(1-\eta)V_N\}$,
\[
\frac{|\widehat\theta_{od,N}^{\mathcal R}-\theta_{od,N}^{\mathcal R}|}{\sqrt{V_N}}\leq z_{1-\alpha/2}\sqrt{1-\eta}\quad\Longrightarrow\quad|\widehat\theta_{od,N}^{\mathcal R}-\theta_{od,N}^{\mathcal R}|\leq z_{1-\alpha/2}\sqrt{\widehat V_N}.
\]
Therefore,
\begin{align*}
&P_N\left\{|\widehat\theta_{od,N}^{\mathcal R}-\theta_{od,N}^{\mathcal R}|\leq z_{1-\alpha/2}\sqrt{\widehat V_N}\right\}\\
&\qquad\geq P_N\left\{\frac{|\widehat\theta_{od,N}^{\mathcal R}-\theta_{od,N}^{\mathcal R}|}{\sqrt{V_N}}\leq z_{1-\alpha/2}\sqrt{1-\eta}\right\}-P_N(\mathcal H_N^c)-P_N\{\widehat V_N<(1-\eta)V_N\}.
\end{align*}
By the asymptotic normality established in Step~1 and continuity of the standard normal distribution function,
\[
\liminf_{N\to\infty}P_N\left\{|\widehat\theta_{od,N}^{\mathcal R}-\theta_{od,N}^{\mathcal R}|\leq z_{1-\alpha/2}\sqrt{\widehat V_N}\right\}\geq2\Phi(z_{1-\alpha/2}\sqrt{1-\eta})-1.
\]
Since this inequality holds for every $\eta\in(0,1)$, letting $\eta\downarrow0$ gives
\[
\liminf_{N\to\infty}P_N\left\{|\widehat\theta_{od,N}^{\mathcal R}-\theta_{od,N}^{\mathcal R}|\leq z_{1-\alpha/2}\sqrt{\widehat V_N}\right\}\geq1-\alpha,
\]
which proves \eqref{eq: restricted-path-coverage}.

\noindent{\it Asymptotically exact inference.} Suppose additionally that $H_N/V_N\xrightarrow{p}0$, then on $\mathcal H_N$, 
\[
\frac{\widehat V_N}{V_N}=\frac{\widehat V_N}{V_N+H_N}\left(1+\frac{H_N}{V_N}\right),
\]
in which the first factor converges in probability to one by Step~2, and the second by the additional assumption. Since $P_N(\mathcal H_N^c)\to0$, it follows that
\[
\frac{\widehat V_N}{V_N}\xrightarrow{p}1.
\]
On $\mathcal H_N\cap\{\widehat V_N>0\}$, whose probability tends to one, the studentized statistic satisfies
\[
\frac{\widehat\theta_{od,N}^{\mathcal R}-\theta_{od,N}^{\mathcal R}}{\sqrt{\widehat V_N}}=\frac{\widehat\theta_{od,N}^{\mathcal R}-\theta_{od,N}^{\mathcal R}}{\sqrt{V_N}}\left(\frac{\widehat V_N}{V_N}\right)^{-1/2}\xrightarrow{d}\mathcal N(0,1),
\]
where the convergence follows from Step~1, which then implies
\[
P_N\left\{|\widehat\theta_{od,N}^{\mathcal R}-\theta_{od,N}^{\mathcal R}|\leq z_{1-\alpha/2}\sqrt{\widehat V_N}\right\}\longrightarrow1-\alpha.
\]
This proves \eqref{eq: restricted-path-exactness} and the exact asymptotic coverage conclusion.

\noindent{\it Sufficient condition for exactness.} Finally, suppose that the population path sums coincide across all selected paths with probability tending to one. On the event
\[
\{S(W,\DID_N)=S(W',\DID_N)\text{ for every }W,W'\in\mathcal R_N^{od}\},
\]
the definition of $\theta_{od,N}^{\mathcal R}$ implies $S(W,\DID_N)=\theta_{od,N}^{\mathcal R}$ for every selected path $W$, and hence $H_N=0$ whenever $|\mathcal R_N^{od}|\geq2$. Thus, for every $\delta>0$,
\[
P_N\{H_N/V_N>\delta\}\leq P_N(\mathcal H_N^c)+P_N\left\{\exists\,W,W'\in\mathcal R_N^{od}:S(W,\DID_N)\neq S(W',\DID_N)\right\}\longrightarrow0.
\]
Therefore $H_N/V_N\xrightarrow{p}0$, establishing the sufficient condition stated in the theorem.
\end{proof}

\subsection{Proofs for Section~\ref{sec: causal interpretation}}
\label{app: proofs section causal}

For nonnegative multipliers $q=(q_{od})_{o,d=1}^K$ satisfying
$\sum_{u,v}q_{uv}P_{uv}>0$, define the reweighted cell probabilities
and expectation operator by
\begin{equation}
\label{eq: reweighted-cell-law}
\begin{aligned}
    P_{od}(q)
    &:= \frac{q_{od}P_{od}}{\sum_{u,v}q_{uv}P_{uv}}, \\
    E_q[Z]
    &:= \frac{E[q_{OD}Z]}{\sum_{u,v}q_{uv}P_{uv}},
\end{aligned}
\end{equation}
where $Z$ is any integrable random variable.
For the proofs below, we call such a reweighting \emph{admissible}
if $\mathbb A_q$ is invertible and
\begin{equation}
\label{eq: reweighted-stayer-support}
    P_{od}(q)>0 \implies P_{oo}(q)>0
    \qquad \text{for all } o,d\in\{1,\ldots,K\}.
\end{equation}
Under the maintained moment conditions, admissibility specifies
the reweightings for which the target and adjusted AKM estimands
in Theorem \ref{prop:local-ate-recovery-all-pairs} are defined.
Under Assumption \ref{ass: invertibility of A}, every reweighting satisfying
$\|q-\mathbf{1}\|_\infty<1$ is admissible.

Write $\mathcal S=\{(u,v):P_{uv}>0\}$, $\mathcal D=\{(u,u):P_{uu}>0\}$, and, for a reweighting $q$, $\mathcal S(q):=\{(u,v):P_{uv}(q)>0\}$ for the set of cells supported under $q$, including stayer cells; the set $\mathcal E(q)$ of moves supported under $q$ consists of its off-diagonal elements. Define
\[
G_{jk| uv}:=E[Y_2(j,k)-Y_2(j,j)\mid O=u,D=v],\qquad
g_{jk}:=\sum_{(u,v)\in\mathcal S}P_{uv}G_{jk| uv}.
\]
Under Assumption~\ref{ass: common trend}(i), these means are finite, and by the law of total expectation, $g_{jk}=E[Y_2(j,k)-Y_2(j,j)]$. Reweighting leaves conditional means unchanged on the cells supported under $q$: for every reweighting $q$, every random variable $X$ with $E|X|<\infty$, including functions of the potential outcomes, and every $(u,v)\in\mathcal S(q)$,
\begin{equation}\label{eq: reweighted-cell-means}
E_q[X\mid O=u,D=v]:=\frac{E_q[X\ind\{O=u,D=v\}]}{P_{uv}(q)}=\frac{q_{uv}E[X\ind\{O=u,D=v\}]}{q_{uv}P_{uv}}=E[X\mid O=u,D=v].
\end{equation}
Moreover, $P_{uv}(\mathbf1)=P_{uv}$ and
\[
\ATE(j,k;w^{\mathrm{ATE}}(q))
=\sum_{(u,v)\in\mathcal S}P_{uv}(q)G_{jk| uv}.
\]
For the regression calculations, write $m_{uv}:=E[\Delta\mid O=u,D=v]$ for $(u,v)\in\mathcal S$. We first prove two auxiliary facts for a single fixed pair $j,k$; neither requires recovery for any other pair.

\begin{lemma}[Continuation of recovery from local weights]\label{lem: local continuation}
Suppose Assumptions~\ref{ass: invertibility of A} and~\ref{ass: common trend}(i) hold and that, for a fixed pair $j,k$ and some $\rho>0$,
\[
\widetilde\alpha_k(q)-\widetilde\alpha_j(q)
=\ATE(j,k;w^{\mathrm{ATE}}(q))
\]
for every admissible reweighting $q$ with $\lVert q-\mathbf1\rVert_\infty<\rho$. Then the equality holds for every reweighting $q$ such that the undirected graph on $\{1,\ldots,K\}$ with an edge $\{u,v\}$ whenever $u\neq v$ and $P_{uv}(q)+P_{vu}(q)>0$ is connected and $q_{uu}>0$ for every $(u,u)\in\mathcal D$.
\end{lemma}
\begin{proof}
Fix a reweighting $q$ satisfying the conditions in the conclusion. The proof connects $q$ to the original population and clears the regression denominator to obtain a polynomial that vanishes by local recovery.

{\bf Step 1: A normalized segment of admissible reweightings.} Define the normalized version of $q$ by
\[
\widehat q_{uv}:=\frac{q_{uv}}{\sum_{x,y}q_{xy}P_{xy}}.
\]
Then
\[
\sum_{u,v}\widehat q_{uv}P_{uv}=1,
\qquad
P_{uv}(\widehat q)=\widehat q_{uv}P_{uv}=P_{uv}(q),
\]
and, by \eqref{eq: reweighted-cell-law}, $E_{\widehat q}=E_q$. Thus $q$ and $\widehat q$ define the same reweighted law, so the regression, the stayer adjustment, and the target coincide under $q$ and $\widehat q$. Set
\[
q(t):=(1-t)\mathbf1+t\widehat q,
\qquad 0\leq t\leq1.
\]
For every $t\in[0,1]$, $q(t)$ is a reweighting, and
\begin{align*}
\sum_{u,v}q_{uv}(t)P_{uv}
&=
(1-t)\sum_{u,v}P_{uv}+t\sum_{u,v}\widehat q_{uv}P_{uv}=1,\\
P_{uv}(q(t))
&=
q_{uv}(t)P_{uv}
=
(1-t)P_{uv}+tP_{uv}(q).
\end{align*}
We show that $q(t)$ is admissible for every $t\in[0,1]$. For $t<1$, $q_{uv}(t)\geq1-t>0$, so the cells supported under $q(t)$ are exactly those in $\mathcal S$, and the undirected graph of supported moves under $q(t)$ is $G^u$, which is connected by Assumption~\ref{ass: invertibility of A}(i) and Lemma~\ref{lem: connectedness}(i). At $t=1$, this graph is connected by the choice of $q$. For every $t\in[0,1]$,
\[
P_{uv}(q(t))>0\ \Longrightarrow\ P_{uv}>0
\ \Longrightarrow\ P_{uu}>0,
\]
where the last implication is Assumption~\ref{ass: invertibility of A}(ii). If $t<1$, then $q_{uu}(t)>0$; if $t=1$, then $q_{uu}(1)=\widehat q_{uu}>0$ because $(u,u)\in\mathcal D$. Hence $P_{uu}(q(t))>0$ whenever $P_{uv}(q(t))>0$, which is \eqref{eq: reweighted-stayer-support}. Lemma~\ref{lem: connectedness}(ii), applied to the cell probabilities $P_{uv}(q(t))$, therefore gives
\[
\det\mathbb A_{q(t)}\neq0\qquad(0\leq t\leq1).
\]
Moreover, $E_{q(t)}|\Delta|=E[q_{OD}(t)|\Delta|]\leq\max_{u,v}q_{uv}(t)\,E|\Delta|<\infty$ by Assumption~\ref{ass: invertibility of A}(iii), and the target is finite because all $G_{jk| uv}$ are finite by Assumption~\ref{ass: common trend}(i). Thus, for every $t\in[0,1]$, $q(t)$ is admissible, and the reweighted law satisfies Assumption~\ref{ass: invertibility of A}.

{\bf Step 2: The adjusted coefficients as rational functions.} By \eqref{eq: reweighted-cell-means}, for every $t\in[0,1]$ and every $(u,v)$ with $P_{uv}(q(t))>0$,
\[
E_{q(t)}[\Delta\mid O=u,D=v]=m_{uv},\qquad
E_{q(t)}[Y_2(j,k)-Y_2(j,j)\mid O=u,D=v]=G_{jk| uv}.
\]
The reweighted design and adjusted moment vector are
\[
\mathbb A_{q(t)}
=
\sum_{(u,v)\in\mathcal S}P_{uv}(q(t))\nu_{uv}\nu_{uv}',
\qquad
b^{\mathrm{DID}}(q(t))
:=
\sum_{(u,v)\in\mathcal S}P_{uv}(q(t))\nu_{uv}(m_{uv}-m_{uu}),
\]
of dimensions $K\times K$ and $K$, respectively. For $\ell=2,\ldots,K$, the representation \eqref{eq: alpha_j decomposition step before DID} in the proof of Lemma~\ref{lem: arbitrary no of firms} and the definition of $\tau_\ell(q)$ in Section~\ref{sec: potential outcomes and reweightings}, both evaluated under the reweighted law, yield
\begin{align*}
\widetilde\alpha_\ell(q(t))
&=
\alpha_\ell(q(t))-\tau_\ell(q(t))
\\
&=
\sum_{(u,v)\in\mathcal S}\omega_{\ell,uv}(q(t))m_{uv}
-
\sum_{(u,v)\in\mathcal S}\omega_{\ell,uv}(q(t))m_{uu}
\\
&=
\sum_{(u,v)\in\mathcal S}e_\ell'\mathbb A_{q(t)}^{-1}\nu_{uv}P_{uv}(q(t))(m_{uv}-m_{uu})
\\
&=
e_\ell'\mathbb A_{q(t)}^{-1}b^{\mathrm{DID}}(q(t)).
\end{align*}
The sums run over the original support, on which all conditional means are defined; a cell deleted by $q(t)$ has coefficient $\omega_{\ell,uv}(q(t))=0$. The first coordinate of $\mathbb A_{q(t)}^{-1}b^{\mathrm{DID}}(q(t))$ is an intercept, whereas $\widetilde\alpha_1(q(t))=0$ by normalization.

Substituting $P_{uv}(q(t))=(1-t)P_{uv}+tP_{uv}(q)$,
\begin{align*}
\mathbb A_{q(t)}
&=
\sum_{(u,v)\in\mathcal S}
\bigl((1-t)P_{uv}+tP_{uv}(q)\bigr)\nu_{uv}\nu_{uv}',
\\
b^{\mathrm{DID}}(q(t))
&=
\sum_{(u,v)\in\mathcal S}
\bigl((1-t)P_{uv}+tP_{uv}(q)\bigr)\nu_{uv}(m_{uv}-m_{uu}),
\\
\ATE(j,k;w^{\mathrm{ATE}}(q(t)))
&=
\sum_{(u,v)\in\mathcal S}
\bigl((1-t)P_{uv}+tP_{uv}(q)\bigr)G_{jk| uv}.
\end{align*}
Each displayed expression is affine in $t$. The affine formulas for the matrix and moment vector also define algebraic extensions to real $t$ outside $[0,1]$; no reweighted population is asserted there. Let $\operatorname{adj}$ denote the adjugate matrix, and define the polynomial
\begin{align*}
p(t)&:=
\sum_{\ell=2}^{K}
\bigl(\ind\{k=\ell\}-\ind\{j=\ell\}\bigr)
e_\ell'\operatorname{adj}(\mathbb A_{q(t)})\,b^{\mathrm{DID}}(q(t))
\\
&\hspace{4em}-
\det\mathbb A_{q(t)} \sum_{(u,v)\in\mathcal S}
\bigl((1-t)P_{uv}+tP_{uv}(q)\bigr)G_{jk| uv}.
\end{align*}
Each entry of the adjugate is a determinant of a $(K-1)\times(K-1)$ matrix with affine entries, and therefore has degree at most $K-1$. The first term has degree at most $K$, and the second at most $K+1$, so $p$ has degree at most $K+1$ unless it is the zero polynomial. On $[0,1]$, the adjugate form of Cramer's rule gives
\[
\operatorname{adj}(\mathbb A_{q(t)})
=
\det\mathbb A_{q(t)}\,\mathbb A_{q(t)}^{-1}.
\]
The coefficient identity above and the normalization at firm $1$ therefore imply, for $t\in[0,1]$,
\[
p(t)=\det\mathbb A_{q(t)}\Bigl\{
\widetilde\alpha_k(q(t))-\widetilde\alpha_j(q(t))
-
\ATE(j,k;w^{\mathrm{ATE}}(q(t)))\Bigr\}.
\]

{\bf Step 3: Local recovery implies recovery at the endpoint.} Since
\[
\lVert q(t)-\mathbf1\rVert_\infty=t\lVert\widehat q-\mathbf1\rVert_\infty,
\]
the positive number
\[
t_0:=\min\left\{1,
\frac{\rho}{1+\lVert\widehat q-\mathbf1\rVert_\infty}\right\}
\]
satisfies, for $0\leq t\leq t_0$,
\[
\lVert q(t)-\mathbf1\rVert_\infty
\le
\frac{\rho\lVert\widehat q-\mathbf1\rVert_\infty}
{1+\lVert\widehat q-\mathbf1\rVert_\infty}
<
\rho.
\]
All these reweightings are admissible by Step~1, so local recovery gives $p(t)=0$ throughout $[0,t_0]$. A nonzero polynomial of degree at most $K+1$ cannot have more than $K+1$ distinct roots: by the factor theorem, the product of the corresponding distinct linear factors would divide it. Thus $p$ is identically zero. Evaluating at $t=1$ gives
\[
0=\det\mathbb A_{\widehat q}
\bigl\{\widetilde\alpha_k(\widehat q)-\widetilde\alpha_j(\widehat q)
-
\ATE(j,k;w^{\mathrm{ATE}}(\widehat q))\bigr\}.
\]
The determinant is nonzero, so recovery holds under $\widehat q$. Since $q$ and $\widehat q$ define the same reweighted law, recovery also holds under $q$.
\end{proof}

\begin{lemma}[Cell constancy from recovery]\label{lem: cell constancy}
Under the hypotheses of Lemma~\ref{lem: local continuation}, $G_{jk| uv}=g_{jk}$ for every $(u,v)\in\mathcal S$.
\end{lemma}
\begin{proof}
The proof first establishes constancy on each support consisting of a spanning tree and all original stayer cells, and then uses a common stayer cell to identify the same constant across these supports.

{\bf Step 1: Tree-supported populations.} The graph $G^u$ is connected by Assumption~\ref{ass: invertibility of A}(i) and Lemma~\ref{lem: connectedness}(i). Since $\sum_{u,v}P_{uv}=1$, some $P_{uv}$ is positive; Assumption~\ref{ass: invertibility of A}(ii) then gives $P_{uu}>0$. Hence $\mathcal D\neq\emptyset$. Fix $r$ with $(r,r)\in\mathcal D$; then $(r,r)\in\mathcal S(q)$ for every reweighting $q$ constructed below.

Choose $\mathcal T\subseteq\mathcal E$ by choosing one supported directed move for each edge of an undirected spanning tree of $G^u$. Every specified supported move can belong to such a set. Indeed, start with that move and its two endpoints. If an unreached firm remains, a path in $G^u$ from the reached set to that firm contains an edge with one reached and one unreached endpoint. Add such an edge, choosing a supported orientation, and its unreached endpoint. Each addition preserves connectedness and cannot create a cycle because it adds a new firm. After $K-2$ additions, the resulting tree spans all firms and still contains the specified move with its original orientation.

For the fixed set $\mathcal T$, the construction in Step~3 of the proof of Lemma~\ref{lem: potential and cycles}, applied to $m=\DID$, gives $a^{\mathcal T}\in\mathbb R^K$ with $a_1^{\mathcal T}=0$ and, by \eqref{eq: potential identity on tree},
\[
\DID(u,v)=a_v^{\mathcal T}-a_u^{\mathcal T}
\qquad\text{for every }(u,v)\in\mathcal T.
\]
That construction applies to any spanning tree of $G^u$ and uses the values of $m$ only on the moves in $\mathcal T$. Set $S_{\mathcal T}:=\mathcal T\cup\mathcal D$ and take any reweighting $q$ that is positive on $S_{\mathcal T}$ and zero on $\mathcal S\setminus S_{\mathcal T}$. Values of $q$ outside $\mathcal S$ do not affect the reweighted law. The total weighted mass is positive, the moves supported under $q$ are exactly those in $\mathcal T$, so $\mathcal E(q)=\mathcal T$, and $\mathcal D\subseteq\mathcal S(q)$. Moreover,
\[
P_{uv}(q)>0\ \Longrightarrow\ P_{uv}>0
\ \Longrightarrow\ (u,u)\in\mathcal D\subseteq S_{\mathcal T}
\ \Longrightarrow\ P_{uu}(q)>0,
\]
which is \eqref{eq: reweighted-stayer-support}. The undirected graph of moves supported under $q$ is the spanning tree, so it is connected, and Lemma~\ref{lem: connectedness}(ii), applied to the cell probabilities $P_{uv}(q)$, shows that $\mathbb A_q$ is invertible. Thus $q$ is admissible. Moreover, $E_q|\Delta|\leq\max_{u,v}q_{uv}E|\Delta|/\sum_{x,y}q_{xy}P_{xy}<\infty$ by Assumption~\ref{ass: invertibility of A}(iii), so the reweighted law satisfies Assumption~\ref{ass: invertibility of A}. Since $q$ satisfies the conditions in the conclusion of Lemma~\ref{lem: local continuation}, that lemma gives recovery under $q$.

By \eqref{eq: reweighted-cell-means}, the conditional means of $\Delta$ under $q$ coincide with $m_{uv}$ on $\mathcal S(q)$. Hence the DID contrasts under the reweighted law equal $\DID(u,v)$ for every $(u,v)\in\mathcal E(q)=\mathcal T$, and they satisfy the displayed identity on every move supported under $q$. Lemma~\ref{lem: potential to levels}, applied to the reweighted law, gives
\[
\widetilde\alpha_\ell(q)=a_\ell^{\mathcal T}-a_1^{\mathcal T}=a_\ell^{\mathcal T}
\qquad(\ell=1,\ldots,K),
\]
and hence
\[
\widetilde\alpha_k(q)-\widetilde\alpha_j(q)
=
a_k^{\mathcal T}-a_j^{\mathcal T}.
\]
The right-hand side does not depend on the positive multipliers on $S_{\mathcal T}$. Combining this identity with recovery yields
\[
\sum_{(u,v)\in S_{\mathcal T}}P_{uv}(q)G_{jk| uv}
=a_k^{\mathcal T}-a_j^{\mathcal T}
\]
for every such $q$. This step requires the identity $\DID(u,v)=a_v^{\mathcal T}-a_u^{\mathcal T}$ only on $\mathcal T$, not on the original move set $\mathcal E$.

{\bf Step 2: Varying two cell probabilities.} Fix distinct cells $(u_1,v_1),(u_2,v_2)\in S_{\mathcal T}$. Both original cell probabilities are positive. As in Step~1 of the proof of Lemma~\ref{lem: local continuation}, rescaling $q$ by a positive constant does not change the reweighted law, so we may choose the multipliers above normalized so that
\[
\sum_{(u,v)\in S_{\mathcal T}}q_{uv}P_{uv}=1.
\]
For $0<h<q_{u_1v_1}P_{u_1v_1}$, define
\[
q'_{u_1v_1}:=q_{u_1v_1}-\frac{h}{P_{u_1v_1}},
\qquad
q'_{u_2v_2}:=q_{u_2v_2}+\frac{h}{P_{u_2v_2}},
\]
and leave all other multipliers unchanged. Then
\[
q'_{u_1v_1}
=\frac{q_{u_1v_1}P_{u_1v_1}-h}{P_{u_1v_1}}>0,
\qquad q'_{u_2v_2}>0,
\]
so $\mathcal S(q')=\mathcal S(q)$, and
\[
\sum_{(u,v)\in S_{\mathcal T}}q'_{uv}P_{uv}
=
\sum_{(u,v)\in S_{\mathcal T}}q_{uv}P_{uv}-h+h=1.
\]
Consequently,
\[
P_{u_1v_1}(q')=P_{u_1v_1}(q)-h,
\qquad
P_{u_2v_2}(q')=P_{u_2v_2}(q)+h,
\]
with every other cell probability unchanged. By Step~1, both $q$ and $q'$ satisfy the recovery identity with the same $a^{\mathcal T}$. Subtracting gives
\begin{align*}
0&=\sum_{(u,v)\in S_{\mathcal T}}
\bigl(P_{uv}(q')-P_{uv}(q)\bigr)G_{jk| uv}
\\
&=
-h G_{jk| u_1v_1}+h G_{jk| u_2v_2}
\\
&=
h\bigl(G_{jk| u_2v_2}-G_{jk| u_1v_1}\bigr).
\end{align*}
Since $h>0$, the two cell means are equal. The two cells were arbitrary, and $(r,r)\in S_{\mathcal T}$, so
\[
G_{jk| uv}=G_{jk| rr}\qquad\text{for every }(u,v)\in S_{\mathcal T}.
\]

{\bf Step 3: Extending constancy to the original support.} By Step~1, every supported mover cell belongs to a set $\mathcal T$ of the type constructed there, and every supported stayer cell belongs to every $S_{\mathcal T}$. The same cell $(r,r)$ belongs to all these sets. Therefore $G_{jk| uv}=G_{jk| rr}$ for every $(u,v)\in\mathcal S$. Finally,
\[
g_{jk}
=
\sum_{(u,v)\in\mathcal S}P_{uv}G_{jk| uv}
=
G_{jk| rr}\sum_{(u,v)\in\mathcal S}P_{uv}
=
G_{jk| rr}.
\]
Thus $G_{jk| uv}=g_{jk}$ for every $(u,v)\in\mathcal S$.
\end{proof}

\begin{proof}[Proof of Theorem~\ref{prop:local-ate-recovery-supported-moves}]
Fix $\rho\in(0,1)$ as in the statement. Since $\rho<1$, Assumption \ref{ass: invertibility of A} implies that every reweighting $q$
with $\|q-\mathbf{1}\|_\infty<\rho$ is admissible. Recall the definitions of $\mathcal S$, $G_{jk| uv}$, and $g_{jk}$ at the beginning of this subsection; in particular, $g_{jk}=E[Y_2(j,k)-Y_2(j,j)]$. The proof first obtains cell mean constancy from local recovery and uses recovery at the original shares to establish zero sums on all signed cycles. Conversely, the zero-cycle condition supplies firm levels whose differences equal the DIDs on supported moves; mean constancy fixes every reweighted target at the corresponding level difference, and Lemma~\ref{lem: potential to levels}, applied to the reweighted law, gives the same value for the adjusted AKM contrast.

{\bf Step 1: (i) $\Rightarrow$ (ii)(a).} Suppose (i) holds and fix $(j,k)\in\mathcal E$. Since every reweighting $q$ with $\lVert q-\mathbf1\rVert_\infty<\rho$ is admissible, condition (i) supplies the hypothesis of Lemma~\ref{lem: local continuation} for this pair and the same $\rho$, and Assumptions~\ref{ass: invertibility of A} and~\ref{ass: common trend}(i) hold. Lemma~\ref{lem: cell constancy} therefore gives
\[
G_{jk| uv}=g_{jk}\qquad\text{for every }(u,v)\in\mathcal S.
\]
Since $(j,k)\in\mathcal E$ was arbitrary, this proves (ii)(a). Moreover, \eqref{eq: did equals att} and this constancy give
\[
\DID(j,k)=E[Y_2(j,k)-Y_2(j,j)\mid O=j,D=k]=G_{jk| jk}=g_{jk}.
\]
Here the final equality uses (ii)(a), not common trends alone.

{\bf Step 2: (i) $\Rightarrow$ (ii)(b).} The original reweighting $q=\mathbf1$ is admissible by Assumption~\ref{ass: invertibility of A}, and $\lVert\mathbf1-\mathbf1\rVert_\infty=0<\rho$. Hence, by (i) and Step~1, for every $(j,k)\in\mathcal E$,
\[
\widetilde\alpha_k-\widetilde\alpha_j=\ATE(j,k;w^{\mathrm{ATE}}(\mathbf1))
=\sum_{(u,v)\in\mathcal S}P_{uv}G_{jk| uv}
=g_{jk}=\DID(j,k).
\]
Thus condition 1.(i) of Lemma~\ref{lem: potential and cycles} holds with $m=\DID$ and $a=\widetilde\alpha$. The graph $G^u$ is connected by Assumption~\ref{ass: invertibility of A}(i) and Lemma~\ref{lem: connectedness}(i), so the implication 1.(i) $\Rightarrow$ 1.(ii) of Lemma~\ref{lem: potential and cycles} gives $S(C,\DID)=0$ for every signed cycle $C$ in $G^d$. This includes cycles with repeated firms or repeated moves, which proves (ii)(b).

{\bf Step 3: (ii) $\Rightarrow$ (i).} Conversely, suppose (ii)(a)--(b) hold. The graph $G^u$ is connected by Assumption~\ref{ass: invertibility of A}(i) and Lemma~\ref{lem: connectedness}(i). By (ii)(b) and the implication 1.(ii) $\Rightarrow$ 1.(i) of Lemma~\ref{lem: potential and cycles}, applied with $m=\DID$, there exists $a\in\mathbb R^K$ such that
\[
\DID(u,v)=a_v-a_u\qquad\text{for every }(u,v)\in\mathcal E.
\]
For each $(j,k)\in\mathcal E$, (ii)(a) evaluated at the cell $(j,k)$ and \eqref{eq: did equals att} give
\[
g_{jk}=G_{jk| jk}=\DID(j,k)=a_k-a_j.
\]

Now fix a reweighting $q$ with $\lVert q-\mathbf1\rVert_\infty<\rho$. Since $\rho<1$, every entry of $q$ is positive, so $\mathcal S(q)=\mathcal S$ and $\mathcal E(q)=\mathcal E$. We verify that the reweighted law satisfies Assumption~\ref{ass: invertibility of A}. Since $q$ is admissible, $\mathbb A_q$ is invertible and \eqref{eq: reweighted-stayer-support} holds. Moreover, $E_q|\Delta|\leq\max_{u,v}q_{uv}E|\Delta|/\sum_{x,y}q_{xy}P_{xy}<\infty$ by Assumption~\ref{ass: invertibility of A}(iii). By \eqref{eq: reweighted-cell-means}, the conditional means of $\Delta$ under $q$ equal $m_{uv}$ for every $(u,v)\in\mathcal S(q)=\mathcal S$. Hence the DID contrasts under the reweighted law equal $\DID(u,v)=a_v-a_u$ for every $(u,v)\in\mathcal E(q)=\mathcal E$. Lemma~\ref{lem: potential to levels}, applied to the reweighted law, gives
\[
\widetilde\alpha_r(q)=a_r-a_1\qquad(r=1,\ldots,K).
\]
For every $(j,k)\in\mathcal E$, condition (ii)(a) and the preceding identification of $g_{jk}$ give
\begin{align*}
\ATE(j,k;w^{\mathrm{ATE}}(q))&=\sum_{(u,v)\in\mathcal S}P_{uv}(q)G_{jk| uv}
=\sum_{(u,v)\in\mathcal S}P_{uv}(q)g_{jk}
=(a_k-a_j)\sum_{(u,v)\in\mathcal S}P_{uv}(q)\\
&=a_k-a_j
=(a_k-a_1)-(a_j-a_1)
=\widetilde\alpha_k(q)-\widetilde\alpha_j(q),
\end{align*}
where $\sum_{(u,v)\in\mathcal S}P_{uv}(q)=1$ because $P_{uv}(q)=0$ for $(u,v)\notin\mathcal S$. This proves (i).
\end{proof}

\begin{proof}[Proof of Theorem~\ref{prop:local-ate-recovery-all-pairs}]
The proof proceeds in three steps.

{\bf Step 1: (i) $\Rightarrow$ (ii)(a).} Suppose (i) holds and fix any pair of firms $j,k$. The recovery identity in (i), restricted to this fixed pair, gives the hypothesis of Lemma~\ref{lem: local continuation}. Since Assumptions~\ref{ass: invertibility of A} and~\ref{ass: common trend}(i) hold, Lemma~\ref{lem: cell constancy} therefore implies, for every $(o,d)\in\mathcal S$,
\begin{align*}
E[Y_2(j,k)-Y_2(j,j)\mid O=o,D=d]&=\sum_{(u,v)\in\mathcal S}P_{uv}E[Y_2(j,k)-Y_2(j,j)\mid O=u,D=v]\\
&=E[Y_2(j,k)-Y_2(j,j)],
\end{align*}
where the first equality is Lemma~\ref{lem: cell constancy} and the second is the law of total expectation.

{\bf Step 2: (i) $\Rightarrow$ (ii)(b).} The reweighting $q=\mathbf1$ is admissible by Assumption~\ref{ass: invertibility of A}, and $\lVert\mathbf1-\mathbf1\rVert_\infty=0<\rho$, so condition (i) holds at $q=\mathbf1$. Hence for every pair $j,k$,
\begin{align*}
E[Y_2(j,k)-Y_2(j,j)]&=\sum_{(u,v)\in\mathcal S}P_{uv}E[Y_2(j,k)-Y_2(j,j)\mid O=u,D=v]\\
&=
\ATE(j,k;w^{\mathrm{ATE}}(\mathbf1))
\\
&=\widetilde\alpha_k-\widetilde\alpha_j,
\end{align*}
in which the first line is the law of total expectation, with finite terms by Assumption~\ref{ass: common trend}(i), and the second line follows from $w_{uv}^{\mathrm{ATE}}(\mathbf1)=P_{uv}$ and the definition of the target. Consequently, for every triple of firms $j,k,\ell$,
\begin{align*}
E[Y_2(j,k)-Y_2(j,j)]&=\widetilde\alpha_k-\widetilde\alpha_j\\
&=(\widetilde\alpha_\ell-\widetilde\alpha_j)+(\widetilde\alpha_k-\widetilde\alpha_\ell)\\
&=E[Y_2(j,\ell)-Y_2(j,j)]+E[Y_2(\ell,k)-Y_2(\ell,\ell)].
\end{align*}

{\bf Step 3: (ii) $\Rightarrow$ (i).} Conversely, suppose (ii)(a)--(b) hold. To apply Lemma~\ref{lem: potential to levels}, construct $a\in\mathbb R^K$ by
\[
a_r:=E[Y_2(1,r)-Y_2(1,1)],\qquad r=1,\ldots,K.
\]
Condition (ii)(b), applied to the initial firm $1$, terminal firm $k$, and intermediate firm $j$, yields
\[
E[Y_2(1,k)-Y_2(1,1)]=E[Y_2(1,j)-Y_2(1,1)]+E[Y_2(j,k)-Y_2(j,j)],
\]
and rearranging then gives, for every pair $j,k$,
\begin{equation}
\label{eq: all-pairs-mean-potential}
E[Y_2(j,k)-Y_2(j,j)]=E[Y_2(1,k)-Y_2(1,1)]-E[Y_2(1,j)-Y_2(1,1)]=a_k-a_j.
\end{equation}
Under Assumptions~\ref{ass: invertibility of A} and~\ref{ass: common trend}, equation~\eqref{eq: did equals att} holds, and hence for every supported move $(u,v)\in\mathcal E$, condition (ii)(a) further implies
\begin{align*}
\DID(u,v)&=E[Y_2(u,v)-Y_2(u,u)\mid O=u,D=v]
\\
&=
E[Y_2(u,v)-Y_2(u,u)]
\\
&=a_v-a_u.
\end{align*}

Next, fix any admissible $q$ in (i). We verify that the reweighted law satisfies Assumption~\ref{ass: invertibility of A}. Since $q$ is admissible, $\mathbb A_q$ is invertible and \eqref{eq: reweighted-stayer-support} holds. Moreover, $E_q|\Delta|\leq\max_{u,v}q_{uv}E|\Delta|/\sum_{x,y}q_{xy}P_{xy}<\infty$ by Assumption~\ref{ass: invertibility of A}(iii). Since $P_{uv}(q)>0$ implies $P_{uv}>0$, we have $\mathcal S(q)\subseteq\mathcal S$ and $\mathcal E(q)\subseteq\mathcal E$. By \eqref{eq: reweighted-cell-means}, the conditional means of $\Delta$ under $q$ equal $m_{uv}$ for every $(u,v)\in\mathcal S(q)$, and for every $(u,v)\in\mathcal E(q)$ the stayer cell $(u,u)$ belongs to $\mathcal S(q)$ by \eqref{eq: reweighted-stayer-support}. Hence the DID contrasts under the reweighted law equal $\DID(u,v)=a_v-a_u$ for every $(u,v)\in\mathcal E(q)$. Lemma~\ref{lem: potential to levels}, applied to the reweighted law, gives
\[
\widetilde\alpha_r(q)=a_r-a_1,\qquad r=1,\ldots,K,
\]
and consequently, for every pair $j,k$,
\[
\widetilde\alpha_k(q)-\widetilde\alpha_j(q)=(a_k-a_1)-(a_j-a_1)=a_k-a_j.
\]

On the other hand, condition (ii)(a), equation~\eqref{eq: all-pairs-mean-potential}, and the normalization of the target weights give
\begin{align*}
\ATE(j,k;w^{\mathrm{ATE}}(q))&=\sum_{(u,v)\in\mathcal S}P_{uv}(q)E[Y_2(j,k)-Y_2(j,j)\mid O=u,D=v]\\
&=\sum_{(u,v)\in\mathcal S}P_{uv}(q)E[Y_2(j,k)-Y_2(j,j)]\\
&=(a_k-a_j)\sum_{(u,v)\in\mathcal S}P_{uv}(q)\\
&=a_k-a_j,
\end{align*}
where $\sum_{(u,v)\in\mathcal S}P_{uv}(q)=1$ because $P_{uv}(q)=0$ for $(u,v)\notin\mathcal S$.
\end{proof}

\begin{corollary}[ATT recovery]\label{cor: appendix-att-recovery}
Under Assumptions~\ref{ass: invertibility of A} and~\ref{ass: common trend}, adjusted AKM contrasts recover the ATT on every supported move if and only if every signed DID cycle has zero sum. %
\end{corollary}
\begin{proof}
Equation~\eqref{eq: did equals att} identifies $\ATT$ with $\DID$ on supported moves. The claim then follows from Theorem~\ref{thm: did characterization} and Lemma~\ref{lem: potential and cycles}. %
\end{proof}

\begin{proposition}[Observable compatibility of common trends and mean constancy]
\label{prop: appendix-observable-compatibility}
Every joint distribution of $(Y_1,Y_2,O,D)$ satisfying Assumption~\ref{ass: invertibility of A}(ii)--(iii) admits potential outcomes $(Y_1(j),Y_2(j,k))_{1\leq j,k\leq K}$ that reproduce the observed outcomes, i.e., $Y_1=Y_1(O)$ and $Y_2=Y_2(O,D)$, and that satisfy Assumption~\ref{ass: common trend} and condition (ii)(a) of Theorem~\ref{prop:local-ate-recovery-all-pairs}. Thus these two restrictions alone impose no further restrictions on the observed distribution.
\end{proposition}
\begin{proof}
Recall the definition of the cell means $m_{uv}=E[\Delta\mid O=u,D=v]$, $(u,v)\in\mathcal S$, at the beginning of this subsection. Choose constants $c_{od}$ with $c_{oo}=0$, $c_{od}=\DID(o,d)$ on supported moves, and arbitrary finite values on unsupported distinct pairs. Construct potential outcomes on the observed probability space as follows. Set $Y_1(j)=Y_1$ for every $j$. For each worker define a baseline $H_j$ by
\[
H_j=\begin{cases}
Y_2,&j=O=D,\\
Y_1+m_{jj},&j=O\ne D,\\
Y_1,&j\ne O.
\end{cases}
\]
Set $Y_2(j,j)=H_j$. For $o\ne d$, set $Y_2(o,d)=H_o+c_{od}$ except at the worker's realized mover pair $(o,d)=(O,D)$, where set $Y_2(o,d)=Y_2$. By construction, $Y_1(O)=Y_1$ and $Y_2(O,D)=Y_2$. At a mover cell $(u,v)$, $Y_2(u,u)-Y_1(u)=m_{uu}$; among stayers at $u$ its mean is also $m_{uu}$. Hence Assumption~\ref{ass: common trend}(ii) holds. For a fixed pair $o\ne d$, the treatment gain $Y_2(o,d)-Y_2(o,o)$ is identically $c_{od}$ in every cell except possibly $(o,d)$. If that cell is supported, its conditional mean is
\[
E[Y_2-Y_1-m_{oo}\mid O=o,D=d]=\DID(o,d)=c_{od}.
\]
Every pair's conditional mean gain is therefore constant across supported cells, and equals its unconditional mean. Every constructed potential wage change $Y_2(j,k)-Y_1(j)$ equals $\Delta$ plus a constant or a constant, so it has a finite first moment under Assumption~\ref{ass: invertibility of A}(iii), and so does every treatment gain. Thus Assumption~\ref{ass: common trend}(i) holds.
\end{proof}

This compatibility result concerns mean constancy and common trends, not the remaining cycle or path conditions. In particular, equality to the sums of \emph{every} path between an unsupported distinct pair also requires those sums to agree, which can itself restrict observable cycles. Without zero-cycle restrictions, append an arbitrary closed detour to a path between that pair to see why the universal path quantifier has observable content.

The next result shows that, once additivity is invoked, the cycle restrictions are the only observable content.

\begin{proposition}[Observable content of the all-pairs conditions]
\label{prop: appendix-all-pairs-compatibility}
A joint distribution of $(Y_1,Y_2,O,D)$ satisfying Assumption~\ref{ass: invertibility of A}(ii)--(iii) admits potential outcomes $(Y_1(j),Y_2(j,k))_{1\leq j,k\leq K}$ that reproduce the observed outcomes, i.e., $Y_1=Y_1(O)$ and $Y_2=Y_2(O,D)$, and that satisfy Assumption~\ref{ass: common trend} and conditions (ii)(a) and (ii)(b) of Theorem~\ref{prop:local-ate-recovery-all-pairs} if and only if $S(C,\DID)=0$ for every signed cycle $C$ in $G^d$.
\end{proposition}
\begin{proof}
Suppose first that such potential outcomes exist, and write $\eta(j,k):=E[Y_2(j,k)-Y_2(j,j)]$. By condition (ii)(a), $\ATT(u,v)=\eta(u,v)$ for every supported move, and Assumption~\ref{ass: common trend} gives $\DID(u,v)=\ATT(u,v)$ by \eqref{eq: did equals att}. Condition (ii)(b) with $k=j$ gives $\eta(j,\ell)=-\eta(\ell,j)$, and with $\ell=1$ it gives $\eta(j,k)=\eta(1,k)-\eta(1,j)$. Hence $\DID(u,v)=\phi_v-\phi_u$ with $\phi_j:=\eta(1,j)$, and every signed cycle sum vanishes because the potential telescopes.

Conversely, suppose that every signed cycle sum vanishes. Applying Lemma~\ref{lem: potential and cycles} to each connected component of $G^u$ gives $\phi\in\mathbb R^K$ with $\DID(u,v)=\phi_v-\phi_u$ for every $(u,v)\in\mathcal E$. Use the construction in the proof of Proposition~\ref{prop: appendix-observable-compatibility} with $c_{od}:=\phi_d-\phi_o$ for all firms $o,d$; these constants satisfy $c_{oo}=0$ and equal $\DID(o,d)$ on supported moves, as that construction requires. It delivers Assumption~\ref{ass: common trend} and condition (ii)(a), with $E[Y_2(o,d)-Y_2(o,o)]=c_{od}$ for every pair of firms $o,d$. Since $c_{od}=c_{o\ell}+c_{\ell d}$ for all firms $o,d,\ell$, condition (ii)(b) also holds.
\end{proof}

\subsection{Sharpness of the bounds in Section~\ref{sec: firm-dispersion-bounds}}
\label{app: dispersion-bounds}

\begin{definition}
\label{def: dispersion-compatible}
Given a distribution of $(Y_1,Y_2,O,D)$ satisfying Assumption~\ref{ass: invertibility of A}, with $\mathcal S$ and $\mathcal E$ its sets of supported cells and supported moves, and finite nonnegative tolerance arrays $\{\varepsilon^a_{jkod}\}_{1\leq j,k\leq K,\,(o,d)\in\mathcal S}$, $\{\varepsilon^b_{jk\ell}\}_{1\leq j,k,\ell\leq K}$, and $\{\varepsilon^c_{od}\}_{(o,d)\in\mathcal E}$, we say that a joint distribution of $((Y_1(j),Y_2(j,k))_{1\leq j,k\leq K},O,D)$ is \emph{compatible} if $E|Y_1(j)|<\infty$ and $E|Y_2(j,k)|<\infty$ for all $j,k$, if $(Y_1(O),Y_2(O,D),O,D)$ has the same distribution as $(Y_1,Y_2,O,D)$, and if the following conditions hold:
\begin{enumerate}[(R1)]
  \item $\bigl|E[Y_2(j,k)-Y_2(j,j)\mid O=o,D=d]-E[Y_2(j,k)-Y_2(j,j)]\bigr|\leq\varepsilon^a_{jkod}$ for all $j,k$ and $(o,d)\in\mathcal S$;
  \item $\bigl|E[Y_2(j,k)-Y_2(j,j)]-E[Y_2(j,\ell)-Y_2(j,j)]-E[Y_2(\ell,k)-Y_2(\ell,\ell)]\bigr|\leq\varepsilon^b_{jk\ell}$ for all $j,k,\ell$;
  \item $\bigl|E[\Delta(o,o)\mid O=o,D=d]-E[\Delta(o,o)\mid O=o,D=o]\bigr|\leq\varepsilon^c_{od}$ for all $(o,d)\in\mathcal E$.
\end{enumerate}
\end{definition}

\begin{definition}
  For fixed, nonnegative firm weights $\omega:=(\omega_1,\ldots,\omega_K)$ summing to one, the \emph{identified set} $\Theta$ is the set of values of $\var_\omega(E[Y_2(1,k)])$ over all compatible distributions.
\end{definition}

\begin{lemma}[Sharp bounds on dispersion in mean potential wages]
\label{lem: dispersion-sharpness}
Fix a distribution of $(Y_1,Y_2,O,D)$ with $E[|Y_1|+|Y_2|]<\infty$ that satisfies Assumption~\ref{ass: invertibility of A}. Fix finite nonnegative arrays $\{\varepsilon^a_{jkod}\}_{1\leq j,k\leq K,\,(o,d)\in\mathcal S}$, $\{\varepsilon^b_{jk\ell}\}_{1\leq j,k,\ell\leq K}$, and $\{\varepsilon^c_{od}\}_{(o,d)\in\mathcal E}$, and nonnegative firm weights $\omega_1,\ldots,\omega_K$ summing to one. Let $\mathcal G$ denote the set of arrays $G$ satisfying \eqref{eq: dispersion-diagonal}--\eqref{eq: dispersion-common-trends}. Then:
\begin{enumerate}[(i)]
  \item An array $G$ belongs to $\mathcal G$ if and only if it is the array \eqref{eq: dispersion-effect-definitions} of some compatible distribution.
  \item If $\mathcal G$ is empty, then $\Theta$ is empty. Otherwise, the minimum $V_{\min}$ and the maximum $V_{\max}$ in \eqref{eq: dispersion-population-objective} are attained and $\Theta=[V_{\min},V_{\max}]$.
\end{enumerate}
\end{lemma}

\begin{proof}
The proof first translates (R1)--(R3) into the constraints of the program, then constructs a compatible distribution for every array in $\mathcal G$, and finally establishes attainment of the variance bounds.

\noindent{\bf Step 1: The restrictions in terms of arrays.} Fix a joint distribution of $((Y_1(j),Y_2(j,k))_{1\leq j,k\leq K},O,D)$ such that $E|Y_1(j)|<\infty$ and $E|Y_2(j,k)|<\infty$ for all $j,k$ and $(Y_1(O),Y_2(O,D),O,D)$ has the distribution of $(Y_1,Y_2,O,D)$. Let $G$ be its array defined in \eqref{eq: dispersion-effect-definitions}; its entries are finite because the potential outcomes are integrable and $P_{od}>0$ for $(o,d)\in\mathcal S$. We show that (R1)--(R3) hold if and only if $G\in\mathcal G$. Since $(Y_1(O),Y_2(O,D),O,D)$ has the distribution of $(Y_1,Y_2,O,D)$, the cell probabilities are $P_{od}$, and $E[Y_2(O,D)-Y_1(O)\mid O=o,D=d]=E[\Delta\mid O=o,D=d]$ for every $(o,d)\in\mathcal S$. By the law of total expectation, $g_{jk}=E[Y_2(j,k)-Y_2(j,j)]$. Hence \eqref{eq: dispersion-diagonal} holds trivially, and (R1) and (R2) are \eqref{eq: dispersion-mean-homogeneity} and \eqref{eq: dispersion-additivity}. Now fix $(o,d)\in\mathcal E$; then $(o,o)\in\mathcal S$ by Assumption~\ref{ass: invertibility of A}(ii). On $\{O=o,D=d\}$, $\Delta(o,o)=Y_2(O,D)-Y_1(O)-(Y_2(o,d)-Y_2(o,o))$, and on $\{O=o,D=o\}$, $\Delta(o,o)=Y_2(O,D)-Y_1(O)$. Taking conditional expectations gives
\begin{align*}
&E[\Delta(o,o)\mid O=o,D=d]-E[\Delta(o,o)\mid O=o,D=o]\\
&\qquad=E[\Delta\mid O=o,D=d]-G_{od| od}-E[\Delta\mid O=o,D=o]
=\DID(o,d)-G_{od| od},
\end{align*}
so (R3) is \eqref{eq: dispersion-common-trends}. Thus (R1)--(R3) hold if and only if $G\in\mathcal G$. In particular, every compatible distribution generates an array in $\mathcal G$. Moreover, $g_{1k}=E[Y_2(1,k)]-E[Y_2(1,1)]$, so $\var_\omega(E[Y_2(1,k)])=V_\omega(G)$.

\noindent{\bf Step 2: Every array in $\mathcal G$ is generated by a compatible distribution.} Fix $G\in\mathcal G$. On the probability space of $(Y_1,Y_2,O,D)$, define, for all $j,k$,
\begin{align*}
Y_1(j)&:=Y_1,\\
H_j&:=\ind\{O=j\}\Bigl(Y_2-\sum_{(o,d)\in\mathcal S}\ind\{O=o,D=d\}G_{od| od}\Bigr),\\
Y_2(j,k)&:=H_j+\sum_{(o,d)\in\mathcal S}\ind\{O=o,D=d\}G_{jk| od}.
\end{align*}
On the event $\{(O,D)\notin\mathcal S\}$, which has probability zero, both sums vanish, so $H_j=\ind\{O=j\}Y_2$ and $Y_2(j,k)=H_j$ there. On $\{O=o,D=d\}$ with $(o,d)\in\mathcal S$, $Y_2(o,d)=Y_2-G_{od| od}+G_{od| od}=Y_2$. Hence $Y_1(O)=Y_1$ and $Y_2(O,D)=Y_2$ everywhere, so $(Y_1(O),Y_2(O,D),O,D)$ has the distribution of $(Y_1,Y_2,O,D)$. Each potential outcome is integrable, since $|Y_1(j)|=|Y_1|$ and $|Y_2(j,k)|\leq|Y_2|+2\max_{j',k',(o,d)}|G_{j'k'| od}|$. By \eqref{eq: dispersion-diagonal}, $Y_2(j,k)-Y_2(j,j)=\sum_{(o,d)\in\mathcal S}\ind\{O=o,D=d\}G_{jk| od}$, so the treatment gain equals the constant $G_{jk| od}$ on $\{O=o,D=d\}$, and the array \eqref{eq: dispersion-effect-definitions} of the constructed distribution is $G$. Since $G\in\mathcal G$, Step~1 shows that (R1)--(R3) hold. The constructed distribution is therefore compatible, and its target variance is $V_\omega(G)$.

Steps~1 and~2 prove (i) and show that $\Theta=\{V_\omega(G):G\in\mathcal G\}$. In particular, $\Theta$ is empty if $\mathcal G$ is empty.

\noindent{\bf Step 3: Attainment and the identified set.} Suppose $\mathcal G$ is nonempty. Each $g_{jk}$ is a linear function of $G$, so every constraint in \eqref{eq: dispersion-diagonal}--\eqref{eq: dispersion-common-trends} is a linear equality or a pair of non-strict linear inequalities. Hence $\mathcal G$ is a closed convex subset of a finite-dimensional Euclidean space.

To show that $\mathcal G$ is bounded, fix $G\in\mathcal G$. For each $(u,v)\in\mathcal E$, \eqref{eq: dispersion-common-trends} and \eqref{eq: dispersion-mean-homogeneity} with $(j,k,o,d)=(u,v,u,v)$ give
\[
|g_{uv}|\leq|\DID(u,v)|+\varepsilon^c_{uv}+\varepsilon^a_{uvuv}.
\]
Moreover, $g_{vv}=0$ by \eqref{eq: dispersion-diagonal}, so \eqref{eq: dispersion-additivity} with $(j,k,\ell)=(v,v,u)$ gives $|g_{vu}+g_{uv}|\leq\varepsilon^b_{vvu}$, and hence $|g_{vu}|\leq|g_{uv}|+\varepsilon^b_{vvu}$. Thus, for every edge $\{u,v\}$ of $G^u$, both $|g_{uv}|$ and $|g_{vu}|$ are bounded by constants that depend only on the $\DID$s and the tolerances. The graph $G^u$ is connected by Assumption~\ref{ass: invertibility of A}(i) and Lemma~\ref{lem: connectedness}(i). For $k\neq1$, choose firms $1=j_0,j_1,\ldots,j_L=k$ such that $\{j_{s-1},j_s\}$ is an edge of $G^u$ for every $s=1,\ldots,L$. By \eqref{eq: dispersion-additivity} with $(j,k,\ell)=(1,j_s,j_{s-1})$, $|g_{1j_s}|\leq|g_{1j_{s-1}}|+|g_{j_{s-1}j_s}|+\varepsilon^b_{1j_s j_{s-1}}$. Since $g_{11}=0$, induction over $s$ gives
\[
|g_{1k}|\leq\sum_{s=1}^{L}|g_{j_{s-1}j_s}|+\sum_{s=1}^{L}\varepsilon^b_{1j_s j_{s-1}}.
\]
For all $j,k$, \eqref{eq: dispersion-additivity} with $(j,k,\ell)=(1,k,j)$ and \eqref{eq: dispersion-mean-homogeneity} give
\[
|g_{jk}|\leq|g_{1k}|+|g_{1j}|+\varepsilon^b_{1kj},\qquad |G_{jk| od}|\leq|g_{jk}|+\varepsilon^a_{jkod}.
\]
Every coordinate of $G$ is therefore bounded by a constant that does not depend on $G$, so $\mathcal G$ is compact.

Since $V_\omega$ is continuous, it attains its minimum $V_{\min}$ at some $G^{\min}\in\mathcal G$ and its maximum $V_{\max}$ at some $G^{\max}\in\mathcal G$. Thus $\Theta=\{V_\omega(G):G\in\mathcal G\}\subseteq[V_{\min},V_{\max}]$. Conversely, by convexity, $G(t):=(1-t)G^{\min}+tG^{\max}\in\mathcal G$ for every $t\in[0,1]$, and $t\mapsto V_\omega(G(t))$ is continuous. By the intermediate value theorem, every value in $[V_{\min},V_{\max}]$ equals $V_\omega(G(t))$ for some $t$ and therefore belongs to $\Theta$. Hence $\Theta=[V_{\min},V_{\max}]$.
\end{proof}

\section{Origin and Destination Firm Effects}
\label{app: two-way}

The regression in \eqref{eq: AKM model differences simplified} controls for time through a single intercept $\beta$ and constrains the coefficient on the destination indicator $\ind\{D=j\}$ to be minus the coefficient on the origin indicator $\ind\{O=j\}$. This appendix drops both restrictions: it allows a separate coefficient on each origin indicator and on each destination indicator, which in the two-period differenced model is exactly a specification with firm-specific time trends. The result is a version of Lemma~\ref{lem: arbitrary no of firms} in which the stayer-trend term $\tau_j$ is absent: the origin indicators absorb every origin-specific component of mean wage growth, the stayer trend $E[\Delta\mid O=o,D=o]$ among them, so that the destination-effect contrasts are linear combinations of difference-in-difference contrasts with no adjustment at all.

\subsection{The two-way regression and its estimand}

Consider the regression
\begin{equation}\label{eq: two-way regression}
	\Delta
	=
	\sum_{k=1}^K\delta_k\ind\{D=k\}
	+
	\sum_{k=2}^K\gamma_k\ind\{O=k\}
	+
	\varepsilon^{\dagger},
	\qquad
	E\bigl[\varepsilon^{\dagger}X^{\dagger}\bigr]=0,
\end{equation}
where
\[
X^{\dagger}
:=
\begin{pmatrix}\bar J\\ I\end{pmatrix}\in\mathbb R^{2K-1},
\qquad
\bar J:=(\ind\{D=1\},\ldots,\ind\{D=K\})' ,
\]
and $I=(\ind\{O=2\},\ldots,\ind\{O=K\})'$ is as in Section~\ref{sec: preliminaries}. The destination block carries a full set of $K$ indicators and therefore absorbs the intercept; the origin block leaves out firm $1$, i.e.\ $\gamma_1:=0$. Because $\sum_{k=1}^K\ind\{D=k\}=1$, the destination block can equivalently be written as $\delta_1+\sum_{k=2}^K(\delta_k-\delta_1)\ind\{D=k\}$, so $\delta_1$ plays the role of the intercept and $\delta_k-\delta_1$ that of the destination effect of firm $k$. Imposing the restriction $\gamma_k=-(\delta_k-\delta_1)$, $k=2,\ldots,K$, returns \eqref{eq: AKM model differences simplified} with $\beta=\delta_1$ and $\alpha_k=\delta_k-\delta_1$; the regression \eqref{eq: two-way regression} is the unrestricted version of the AKM regression. The object of interest is the contrast between two destination effects,
\[
\delta_k-\delta_j
=
(\delta_k-\delta_1)-(\delta_j-\delta_1),
\]
which is the counterpart of the AKM contrast $\alpha_k-\alpha_j$ and does not depend on which firm is used to normalize the destination block.

Let $f_1,\ldots,f_K$ denote the coordinate vectors of $\mathbb R^{2K-1}$ belonging to the destination block and $g_2,\ldots,g_K$ those belonging to the origin block. Conditional on $O=o$, $D=d$, the regressor $X^{\dagger}$ equals
\begin{equation}\label{eq: xi definition}
	\xi_{od}
	:=
	f_d+\ind\{o>1\}\,g_o ,
\end{equation}
so that the design matrix is
\[
\mathbb B
:=
E\bigl[X^{\dagger}X^{\dagger\prime}\bigr]
=
\sum_{o=1}^K\sum_{d=1}^K P_{od}\,\xi_{od}\xi_{od}' .
\]
As for the AKM regression, the estimand is well defined as soon as $\mathbb B$ is invertible.

\begin{assumption}\label{ass: invertibility of B}
	(i) The matrix $\mathbb B$ is invertible. (ii) $P_{oo}>0$ for every firm $o\in\{1,\ldots,K\}$. (iii) $\Delta$ has a finite first moment.
\end{assumption}

Part (ii) strengthens Assumption~\ref{ass: invertibility of A}(ii), which requires stayers only at firms from which some worker moves; here every firm must have stayers, since otherwise its origin effect $\gamma_o$ is not identified. Given part (ii), part (i) is implied by the connectedness condition already used for the AKM regression.

\begin{lemma}[invertibility of the two-way design matrix]\label{lem: invertibility of B}
	Suppose that $P_{oo}>0$ for every firm $o$ and that the undirected graph $G^{u}$ is connected. Then $\mathbb B$ is invertible. In particular, Assumption~\ref{ass: invertibility of A}(i) and Assumption~\ref{ass: invertibility of B}(ii) together imply Assumption~\ref{ass: invertibility of B}(i).
\end{lemma}

\begin{proof}
	Let $\theta\in\mathbb R^{2K-1}$ satisfy $\mathbb B\theta=0$, and write $\phi_k:=f_k'\theta$, $k=1,\ldots,K$, and $\psi_k:=g_k'\theta$, $k=2,\ldots,K$, with the convention $\psi_1:=0$. By \eqref{eq: xi definition}, $\xi_{od}'\theta=\phi_d+\psi_o$ for all $o,d$, and therefore
	\[
	0
	=
	\theta'\mathbb B\theta
	=
	\sum_{o=1}^K\sum_{d=1}^K P_{od}\bigl(\phi_d+\psi_o\bigr)^2 .
	\]
	Every summand is nonnegative, so $\phi_d+\psi_o=0$ whenever $P_{od}>0$. Evaluating this at the stayer cells, which are supported by assumption, gives $\psi_o=-\phi_o$ for every firm $o$. Substituting into the same identity at a supported move $(u,v)\in\mathcal E$ gives $\phi_v=\phi_u$; since $G^{u}$ is connected, $\phi$ is constant across firms. Finally $\phi_1=-\psi_1=0$, so $\phi\equiv 0$ and hence $\psi\equiv 0$, i.e.\ $\theta=0$. Thus $\mathbb B$ has trivial kernel and is invertible. The last claim follows because Assumption~\ref{ass: invertibility of A}(i) implies that $G^{u}$ is connected by Lemma~\ref{lem: connectedness}(i).
\end{proof}

\subsection{The representation}

Define, in analogy with \eqref{eq: omega definition}, the two-way weights and contrast weights
\begin{equation}\label{eq: omega dagger definition}
	\omega^{\dagger}_{j,od}:=f_j'\mathbb B^{-1}\xi_{od}\,P_{od},
	\qquad
	\kappa^{jk,\dagger}_{od}:=\omega^{\dagger}_{k,od}-\omega^{\dagger}_{j,od},
	\qquad
	j,k,o,d\in\{1,\ldots,K\} .
\end{equation}
Like the AKM weights, these depend on the data only through the mobility shares $(P_{od}\colon o,d\in\{1,\ldots,K\})$, and they vanish on unsupported cells.

\begin{proposition}[edge-level representation with origin and destination effects]\label{prop: two-way representation}
	Suppose Assumption~\ref{ass: invertibility of B} holds. Then the following hold for all $j,k\in\{1,\ldots,K\}$.
	\begin{enumerate}[(i)]
		\item The weights sum to $\ind\{o=1\}$ within each origin row and to $\ind\{j=d\}$ within each destination column:
		\begin{equation}\label{eq: two-way row and column sums}
			\sum_{d=1}^K\omega^{\dagger}_{j,od}=\ind\{o=1\},
			\qquad
			\sum_{o=1}^K\omega^{\dagger}_{j,od}=\ind\{j=d\} .
		\end{equation}

		\item The destination coefficients satisfy
		\begin{equation}\label{eq: two-way level representation}
			\delta_j
			=
			E[\Delta\mid O=1,D=1]
			+
			\sum_{(o,d)\in\mathcal E}\omega^{\dagger}_{j,od}\,\DID(o,d) .
		\end{equation}

		\item The contrast between two destination coefficients satisfies
		\begin{equation}\label{eq: two-way contrast representation}
			\delta_k-\delta_j
			=
			\sum_{(o,d)\in\mathcal E}\kappa^{jk,\dagger}_{od}\,\DID(o,d),
			\qquad
			\sum_{1\leq o,d\leq K}\kappa^{jk,\dagger}_{od}=0 .
		\end{equation}
		In particular, no stayer-trend term appears and no adjustment analogous to Definition~\ref{def: adjusted AKM estimand} is needed.

		\item The contrast weights route one unit of mass from firm $j$ to firm $k$: for every firm $r\in\{1,\ldots,K\}$,
		\begin{equation}\label{eq: two-way unit flow}
			\sum_{(u,v)\in\mathcal E}\kappa^{jk,\dagger}_{uv}\bigl(\ind\{v=r\}-\ind\{u=r\}\bigr)
			=
			\ind\{k=r\}-\ind\{j=r\} .
		\end{equation}
	\end{enumerate}
\end{proposition}

\begin{proof}
	Write $m_{od}:=E[\Delta\mid O=o,D=d]$, so that $\DID(o,d)=m_{od}-m_{oo}$ and $\DID(o,o)=0$.

	\textbf{Step 1: the columns of $\mathbb B$.} The coordinates of $X^{\dagger}$ are $f_k'X^{\dagger}=\ind\{D=k\}$ for $k=1,\ldots,K$ and $g_k'X^{\dagger}=\ind\{O=k\}$ for $k=2,\ldots,K$. Hence, for $k=1,\ldots,K$,
	\begin{equation}\label{eq: B column destination}
		\mathbb Bf_k
		=
		E\bigl[X^{\dagger}(X^{\dagger\prime}f_k)\bigr]
		=
		E\bigl[X^{\dagger}\ind\{D=k\}\bigr]
		=
		\sum_{o=1}^K P_{ok}\,\xi_{ok},
	\end{equation}
	and, for $k=2,\ldots,K$,
	\begin{equation}\label{eq: B column origin}
		\mathbb Bg_k
		=
		E\bigl[X^{\dagger}\ind\{O=k\}\bigr]
		=
		\sum_{d=1}^K P_{kd}\,\xi_{kd} .
	\end{equation}

	\textbf{Step 2: row and column sums.} By \eqref{eq: omega dagger definition} and \eqref{eq: B column destination},
	\[
	\sum_{o=1}^K\omega^{\dagger}_{j,od}
	=
	f_j'\mathbb B^{-1}\sum_{o=1}^K P_{od}\,\xi_{od}
	=
	f_j'\mathbb B^{-1}\mathbb Bf_d
	=
	f_j'f_d
	=
	\ind\{j=d\},
	\]
	which is the second identity in \eqref{eq: two-way row and column sums}. Similarly, for $o=2,\ldots,K$, \eqref{eq: B column origin} gives
	\begin{equation}\label{eq: two-way row sum away from base}
		\sum_{d=1}^K\omega^{\dagger}_{j,od}
		=
		f_j'\mathbb B^{-1}\mathbb Bg_o
		=
		f_j'g_o
		=
		0,
	\end{equation}
	because $f_j$ and $g_o$ are distinct coordinate vectors. Summing the column identity over $d$ yields
	\[
	\sum_{o=1}^K\sum_{d=1}^K\omega^{\dagger}_{j,od}
	=
	\sum_{d=1}^K\ind\{j=d\}
	=
	1,
	\]
	and subtracting \eqref{eq: two-way row sum away from base} for $o=2,\ldots,K$ leaves $\sum_{d=1}^K\omega^{\dagger}_{j,1d}=1$, which together with \eqref{eq: two-way row sum away from base} is the first identity in \eqref{eq: two-way row and column sums}.

	\textbf{Step 3: the level representation.} Since $\mathbb B$ is invertible, the solution of \eqref{eq: two-way regression} is $\mathbb B^{-1}E[X^{\dagger}\Delta]$, and by the law of iterated expectations $E[X^{\dagger}\Delta]=E[X^{\dagger}E(\Delta\mid O,D)]=\sum_{o,d}P_{od}m_{od}\xi_{od}$. Therefore
	\[
	\delta_j
	=
	f_j'\mathbb B^{-1}\sum_{o=1}^K\sum_{d=1}^K P_{od}m_{od}\xi_{od}
	=
	\sum_{o=1}^K\sum_{d=1}^K\omega^{\dagger}_{j,od}\,m_{od} .
	\]
	Writing $m_{od}=m_{oo}+\DID(o,d)$ and using the first identity in \eqref{eq: two-way row and column sums},
	\[
	\delta_j
	=
	\sum_{o=1}^K m_{oo}\sum_{d=1}^K\omega^{\dagger}_{j,od}
	+
	\sum_{o=1}^K\sum_{d=1}^K\omega^{\dagger}_{j,od}\,\DID(o,d)
	=
	m_{11}
	+
	\sum_{o=1}^K\sum_{d=1}^K\omega^{\dagger}_{j,od}\,\DID(o,d) .
	\]
	The remaining double sum reduces to a sum over $\mathcal E$ because $\DID(o,o)=0$ on the diagonal and $\omega^{\dagger}_{j,od}=0$ whenever $P_{od}=0$. This is \eqref{eq: two-way level representation}.

	\textbf{Step 4: the contrast.} Subtracting \eqref{eq: two-way level representation} for $j$ from the same identity for $k$ cancels the term $m_{11}$ and gives the first identity in \eqref{eq: two-way contrast representation}. The second follows from Step 2, since the weights of $\omega^{\dagger}_{k,\cdot}$ and of $\omega^{\dagger}_{j,\cdot}$ both sum to one over all cells.

	\textbf{Step 5: the flow identity.} Fix $r$ and sum over all cells. Splitting the summand and using Step 2 twice,
	\[
	\sum_{o=1}^K\sum_{d=1}^K\kappa^{jk,\dagger}_{od}\bigl(\ind\{d=r\}-\ind\{o=r\}\bigr)
	=
	\underbrace{\sum_{o=1}^K\kappa^{jk,\dagger}_{or}}_{=\,\ind\{k=r\}-\ind\{j=r\}}
	-
	\underbrace{\sum_{d=1}^K\kappa^{jk,\dagger}_{rd}}_{=\,\ind\{r=1\}-\ind\{r=1\}\,=\,0} .
	\]
	The diagonal cells contribute nothing to the left-hand side because $\ind\{d=r\}-\ind\{o=r\}$ vanishes when $o=d$, and the unsupported cells contribute nothing because their weights are zero. The left-hand side therefore equals the sum over $\mathcal E$ in \eqref{eq: two-way unit flow}.
\end{proof}

The identity \eqref{eq: two-way unit flow} is exactly the unit-flow property \eqref{eq:one-unit-flow} that drives the path decomposition, with source $j$ and sink $k$. The proof of Lemma~\ref{lem: cycle correction} uses the contrast weights $\kappa^{jk}$ only through that property and through the fact that they are supported on $\mathcal E$. Repeating that proof with $\kappa^{jk,\dagger}$ in place of $\kappa^{jk}$ and then applying the folding argument in the proof of Theorem~\ref{thm: akm path decomposition} gives the following.

\begin{corollary}[path decomposition of the destination-effect contrast]\label{cor: two-way path decomposition}
	Suppose Assumption~\ref{ass: invertibility of B} holds and fix two distinct firms $j,k\in\{1,\ldots,K\}$. Then there exist a finite nonempty collection $\mathcal W^{jk,\dagger}$ of signed paths from $j$ to $k$ and positive constants $\{\lambda^{jk,\dagger}(W)\}_{W\in\mathcal W^{jk,\dagger}}$, depending only on the mobility shares, such that
	\[
	\sum_{W\in\mathcal W^{jk,\dagger}}\lambda^{jk,\dagger}(W)=1
	\qquad\text{and}\qquad
	\delta_k-\delta_j
	=
	\sum_{W\in\mathcal W^{jk,\dagger}}\lambda^{jk,\dagger}(W)\,S(W,\DID) .
	\]
\end{corollary}

\begin{proof}
The contrast identity \eqref{eq: two-way contrast representation} and unit-flow identity \eqref{eq: two-way unit flow} give exactly the two inputs used in the proof of Lemma~\ref{lem: cycle correction}. Orient each nonzero contrast weight by its sign, apply Corollary~\ref{cor:unit-flow-decomposition}, and translate the resulting arcs into signed moves. This gives positive simple-path weights summing to one and positive simple-cycle corrections. The graph is connected: otherwise destination coefficients equal to a nonzero component indicator, together with its negative as the origin coefficients, would produce a nonzero vector in the kernel of $\mathbb B$ (choose a component not containing the normalized origin firm). Thus the cycle-folding argument applies and yields the asserted path representation. All choices use only mobility shares.
\end{proof}

The intermediate representation of Lemma~\ref{lem: cycle correction} transfers in the same way: the peeling step applied to $\kappa^{jk,\dagger}$ returns simple signed paths from $j$ to $k$ together with simple signed cycles, and the cycle corrections are in general not zero, so the folding argument in the proof of Theorem~\ref{thm: akm path decomposition} is still needed. What changes is only which routes carry mass. The two decompositions can put mass on different routes, and in particular on routes that the AKM decomposition traverses in the opposite direction or does not use at all; a move $(u,v)$ enters the auxiliary graph of Section~\ref{app: proofs section descriptive} in its own direction when its weight is positive and in the reverse direction when its weight is negative, and the sign of $\kappa^{jk,\dagger}_{uv}$ need not agree with that of $\kappa^{jk}_{uv}$.

The characterization of Section~\ref{sec: did characterization section} carries over as well, and with the same equivalent conditions.

\begin{corollary}[exact description with origin and destination effects]\label{cor: two-way exact description}
	Suppose Assumption~\ref{ass: invertibility of B} holds. Then the three equivalent conditions of Proposition~\ref{thm: did characterization} are also equivalent to
	\begin{equation}\label{eq: two-way did recovery}
		\delta_d-\delta_o=\DID(o,d)
		\qquad\text{for all }(o,d)\in\mathcal E .
	\end{equation}
\end{corollary}

\begin{proof}
	Suppose first that Proposition~\ref{thm: did characterization}(\ref{item: did-recovery}) holds. As in Step 1 of the proof of that proposition, setting $a_1:=0$ and $a_j:=\widetilde \alpha_j$ for $j=2,\ldots,K$ gives $\DID(u,v)=a_v-a_u$ on $\mathcal E$. Then, by \eqref{eq: two-way contrast representation} and \eqref{eq: two-way unit flow}, for any $(o,d)\in\mathcal E$,
	\[
	\delta_d-\delta_o
	=
	\sum_{(u,v)\in\mathcal E}\kappa^{od,\dagger}_{uv}(a_v-a_u)
	=
	\sum_{r=1}^K a_r\sum_{(u,v)\in\mathcal E}\kappa^{od,\dagger}_{uv}\bigl(\ind\{v=r\}-\ind\{u=r\}\bigr)
	=
	a_d-a_o
	=
	\DID(o,d),
	\]
	which is \eqref{eq: two-way did recovery}. Conversely, if \eqref{eq: two-way did recovery} holds, then $a:=(\delta_1,\ldots,\delta_K)'$ satisfies $\DID(u,v)=a_v-a_u$ on $\mathcal E$, so the signed sum along any path equals the difference between its endpoint values, yielding part (\ref{item: did-path-independence}) of Proposition~\ref{thm: did characterization}.
\end{proof}

\subsection{Discussion}

Three points are worth recording.

First, the stayer-trend term disappears, and \eqref{eq: two-way row sum away from base} says precisely why. The weights of every origin row other than the left-out firm's sum to zero, so any component of the cell means that depends on the cell only through the origin firm is annihilated. The stayer trend $E[\Delta\mid O=o,D=o]$ is such a component: the origin indicators span all origin-specific functions, so the regression absorbs it. Only the row of the left-out firm survives, with weight one, and the resulting term $E[\Delta\mid O=1,D=1]$ is the same for every $j$ and therefore cancels in the contrast $\delta_k-\delta_j$. This is what makes \eqref{eq: two-way contrast representation} an exact statement about difference-in-difference contrasts, whereas Lemma~\ref{lem: arbitrary no of firms} requires either subtracting $\tau_j$ by hand, as in Definition~\ref{def: adjusted AKM estimand}, or assuming a common stayer trend.

Second, the two regressions do not give the same answer. The contrast weights $\kappa^{jk,\dagger}$ solve a different system from the AKM contrast weights $\kappa^{jk}_{od}$ of \eqref{eq: adjusted AKM kappa contrast representation}, so in general $\delta_k-\delta_j\neq\widetilde\alpha_k-\widetilde\alpha_j$: the two are convex averages of signed path sums over the same network, but with different weights on the routes. By Corollary~\ref{cor: two-way exact description} and Proposition~\ref{thm: did characterization}, both reproduce $\DID(o,d)$ for every supported move exactly when the difference-in-difference contrasts are path independent. They can also agree with one another without reproducing every DID; equality of the two specifications alone is therefore not a test of path independence.

Third, the price of the flexible trend is paid in parameters and in support. The regression \eqref{eq: two-way regression} has $2K-1$ coefficients rather than $K$, and Assumption~\ref{ass: invertibility of B}(ii) requires every firm to have stayers, not only those from which workers move.

\section{Data construction and reference tables} \label{app:data}

  \noindent {\bf Source.} The underlying data are the Veneto Worker Histories
  (VWH), a matched employer--employee panel of Italian workers and firms observed from 1985 through 2001. Three source Stata files are joined by \texttt{scripts/build\_matched\_employer\_employee.py}: contracts (spell-level records --- weeks / months / days remunerated, contract type, qualification, remuneration in 2003 euros), worker demographics (sex, year and place of birth, nationality, residence), and the firm registry (industry code \texttt{ateco81}, municipality, province, establishment / cessation dates). The join is by worker id (\texttt{cod\_pgr}) and firm id (\texttt{matr\_az}) after whitespace-stripping both keys and casting them to strings.

  \noindent {\bf Outcome variable.} The outcome $y$ used everywhere
  downstream (AKM firm effects, KSS variance decomposition, every $\kappa$-contrast) is the \emph{log daily wage},
  \begin{align*}
  \log w_{it} \;:=\; \log\!\Bigl(
    \tfrac{\text{annual remuneration (2003 EUR)}}
          {\text{days remunerated in that year}}
  \Bigr),
  \end{align*}
  constructed from the contracts file's \texttt{retrib03} (annual remuneration in 2003 EUR, price-adjusted by the VWH provider) and \texttt{gior\_r} (days remunerated in the year), inherited from KSS's public \texttt{build.do}. The \texttt{log\_dailywages} column in \texttt{py\_final\_1985\_2001\_veneto\_only.parquet} is the raw log daily wage in real (2003 EUR) terms --- mean $\approx 4.71$, per-year means ranging from $4.677$ (1986) to $4.740$ (1991) with no trend across the panel, so it is \emph{not} pre-residualised for a year effect. $\mathrm{Var}(y) \approx 0.18$ in the KSS Table 2 replication (Table~\ref{tab:kss-variance}) is simply the natural variance of log real daily wages on this sample (already deflated to 2003 EUR), not evidence of a pre-adjustment. The year fixed effect $\delta$ of the KSS Table 2 model $y_{it} = \alpha_i + \psi_{J(i,t)}
  + \delta \, \ind\{t{=}2001\} + \varepsilon_{it}$
  is estimated by us, inside \texttt{estimate\_akm\_with\_year\_effect\_cg}, jointly with $\alpha$ and $\psi$ via first-difference CG.

  \noindent {\bf Sample construction.} We build the analysis sample in two
  stages: the KSS-cleaned pre-sample (external to this repository, inherited from Kline, Saggio, and S\o lvsten's public \texttt{build.do}) and our own on-top filters (\texttt{scripts/build\_sample.py}). In order:

  \begin{enumerate}
    \item \textbf{Dominant job per worker--year.} Keep only the primary employment spell per worker--year (highest weeks worked), so a worker appears at most once per year.
    \item \textbf{Working age.} Restrict to workers aged 19--63.
    \item \textbf{Veneto provinces only.} Drop observations whose province of work is outside the Veneto region.
    \item \textbf{Drop pathological wages.} Drop workers who at any point in the panel had a daily wage $\leq 5$ euros, zero days worked in a year, or a year-over-year $|\Delta \log w| \geq 1$.
    \item \textbf{Drop public sector} spells.
    \item \textbf{Drop multi-jobbers.} Drop workers with $\geq 10$ parallel jobs in any year.
    \item \textbf{Drop unknown gender.} Drop workers whose sex is missing or unclassified.
    \item \textbf{Drop residual missings.} Drop observations still missing \texttt{female}, \texttt{age}, or \texttt{year}.
    \item \textbf{Two-year cut (T=2).} Keep only years 1999 and 2001
      --- the pooled T=2 panel that matches KSS Table I. Workers
      observed in only one of the two years are kept (we do not impose a balanced-panel restriction).
    \item \textbf{Derived variables.} Compute \texttt{real\_age\_at\_entry} $=$ $\min(\text{age})$ per worker over the full 1985--2001 history; normalise age as $(\text{age} - 40)/40$ with cubic powers; assign \texttt{type\_worker} as the sortable \texttt{group(year, real\_age\_at\_entry, female)} cell index.
    \item \textbf{No further row filtering} beyond these steps. In particular, we do \emph{not} drop singleton pairs or restrict to the balanced sub-panel at this stage; those cuts happen only when a downstream analysis demands them (e.g. LCS extraction).
  \end{enumerate}

  Steps 1--9 re-apply KSS's sample definition --- steps 1--7 are inherited from KSS's public \texttt{build.do} on the pre-cleaned panel, and steps 8--9 re-apply KSS's \texttt{estim.do} residual missings drop and the KSS Table~I T=2 window (1999+2001) in \texttt{scripts/build\_sample.py}. Steps 10--11 are our downstream conveniences (age controls, \texttt{type\_worker} cell id, and the explicit no-further-filtering contract). The output is \texttt{data/veneto/sample.parquet} carrying the columns \texttt{worker\_id, firm\_id, year, age, age\_norm(\_sq/\_cu), female, real\_age\_at\_entry, type\_worker, log\_dailywages}.

  \noindent {\bf Mover graph and sample cuts.} The mover graph is the
  directed multigraph whose nodes are firms and whose edges record worker moves between firms in the T=2 panel. All AKM-style analyses restrict to the largest connected set (LCS) of this graph so that firm effects are identified relative to a single reference firm.

  \noindent {\bf Which sample is used where.} The pooled T=2 sample of
  step~9 is used in three guises, and it is worth mapping them once so every table below reads unambiguously:

  \begin{itemize}
    \item \textbf{Unbalanced pooled T=2} --- workers observed in \emph{either} 1999 or 2001.
    \item \textbf{Dual-period subsample} --- workers observed in \emph{both} 1999 and 2001 --- is used by the CG (conjugate gradient) solver. The CG solver operates on first-differences, so it requires each worker to contribute a $\Delta y$ observation, which only exists for dual-period workers.
    \item \textbf{KSS leave-one-out (LO) subsample} --- the dual-period subsample further restricted to the leave-one-out connected set of the T=2 mover graph, i.e.\ the largest set of firms that stays connected after removing any single mover. This is the sample the leave-one-out variance correction of Kline--Saggio--S\o lvsten needs to be well-posed, and it is the \emph{single} analysis panel every $\kappa$- / $\omega$- / path- / Oaxaca-style decomposition reads from. In the codebase it is the artefact \texttt{data/veneto/main\_sample.parquet}.
  \end{itemize}

  The LCS sizes under the first two guises (unbalanced pooled T=2 vs.\ CG dual-period) show how many observations the dual-period restriction alone loses (roughly a third of the unbalanced obs.\ count); the further step down to the LO subsample the analysis actually runs on is a smaller additional cut on top of that. KSS themselves use the unbalanced pooled panel for their Table~I descriptive stats and the LO subsample for their Table~2 variance decomposition.

\noindent {\bf Industry.} Every firm carries a single \emph{industry}
  label: the \texttt{ind\_v2} column of the analysis panel, a 6-way sector coding we derive from the first digit of the ATECO81 code recorded for the firm in the Italian firm registry (\texttt{ateco81} in the raw archive). The mapping lives in \texttt{scripts/industry\_codes.py} and is applied once, at sample-build time, so the same attribute reaches every table and figure in this paper. The nine ATECO81 1-digit groups collapse into six sectors:

  \begin{itemize}
    \item \textbf{1 = Metal} --- metal manufacture, machinery, precision instruments (ATECO~3).
    \item \textbf{2 = Light} --- food, textile, wood, paper; the light-manufacturing base of the Veneto industrial districts (ATECO~4).
    \item \textbf{3 = Heavy} --- extraction, non-metallic minerals, chemicals, energy and water (ATECO~1--2).
    \item \textbf{4 = Constr} --- construction (ATECO~5).
    \item \textbf{5 = Trade} --- wholesale, retail, hotels and catering (ATECO~6).
    \item \textbf{6 = Svc} --- transport and communications, financial and insurance, business and other services (ATECO~7--9).
  \end{itemize}

  The coding is firm-level (all workers at the same firm inherit the firm's sector) and does not vary within a firm across the panel. Every off-diagonal cell of the resulting $6 \times 5$ directed grid has $\geq 480$ observed crossers on the Veneto leave-one-out panel, so no design in this paper has to work around an empty cell.

  \begin{table}[htbp]
    \centering
    \caption{Comparison of sample characteristics with KSS}
    \label{tab:kss-counts}
    \figurenote{The table compares our leave-one-out estimation sample for 1999 and 2001 with the pooled leave-one-out sample in Table I of \citet{Kline:2020aa}. Observations are person-years. Movers are workers employed at different firms in the two years. Wage moments are calculated across person-year observations. Dashes indicate statistics not reported in the reference table.}
    \vspace{0.6ex}
    \begin{tabular}{l rr}
      \toprule
       & {Our sample} & {\citet{Kline:2020aa}} \\
      \midrule
      Observations
        & \num{1320034} & \num{1319972} \\
      Workers
        & \num{660017} & {---} \\
      Movers
        & \num{164196} & \num{164203} \\
      Firms
        & \num{42492} & \num{42489} \\
      \midrule
      Mean log daily wage
        & 4.8066 & 4.8066 \\
      Variance of log daily wage
        & 0.1843 & 0.1843 \\
      \bottomrule
    \end{tabular}
  \end{table}

\begin{table}[htbp]
\centering
\caption{The six industries in the estimation sample}
\label{tab:industry-counts}
\small
\begin{tabular}{@{}llrrr@{}}
\toprule
Industry & Activities (ATECO81) & Firms & Workers & Person-years \\
\midrule
Heavy        & extraction, non-metallic minerals, chemicals, energy & 1,868  & 48,285  & 86,064 \\
Service      & transport, financial, business and other services     & 6,226  & 117,469 & 210,561 \\
Trade        & wholesale, retail, hotels and catering                & 10,550 & 131,975 & 237,577 \\
Metal        & metal manufacturing, machinery, precision instruments & 9,100  & 198,141 & 362,933 \\
Light        & food, textiles, wood and paper                        & 10,337 & 194,307 & 356,920 \\
Construction & construction                                          & 4,411  & 37,689  & 65,979 \\
\midrule
All          &                                                       & 42,492 & 660,017 & 1,320,034 \\
\bottomrule
\end{tabular}
\figurenote{Leave-one-out connected set, 1999 and 2001 (Panel A of Table~\ref{tab:path-cross-industry}). The six industries group ATECO81 one-digit codes. Workers are counted in each industry in which they work in 1999 or 2001; the 67,849 workers who change industry appear in two rows, so the column sums to more than the total.}
\end{table}

\begin{table}[htbp]
    \centering
    \caption{Comparison of AKM variance decompositions with KSS}
    \label{tab:kss-variance}
    \figurenote{The table compares our plug-in AKM estimates with the pooled plug-in estimates reported by \citet{Kline:2020aa}, using the respective leave-one-out samples. The model includes worker effects $\alpha_i$, firm effects $\psi_{J(i,t)}$, and a year-2001 indicator with coefficient $\delta$. The wage variance is reported in KSS Table I; the variance components and correlation are from Table II. These are plug-in estimates, without the leave-out bias correction. Dashes indicate quantities not reported in the reference table.}
    \vspace{0.6ex}
    \begin{tabular}{l rr}
      \toprule
       & {Our estimate} & {\citet{Kline:2020aa}} \\
      \midrule
      $\operatorname{Var}(y)$
        & 0.1843 & 0.1843 \\
      $\operatorname{Var}(\alpha)$
        & 0.1321 & 0.1321 \\
      $\operatorname{Var}(\psi)$
        & 0.0358 & 0.0358 \\
      $\operatorname{Cov}(\alpha,\psi)$
        & 0.0039 & 0.0039 \\
      $\operatorname{Corr}(\alpha,\psi)$
        & 0.0560 & 0.0565 \\
      $\hat\delta$
        & 0.0322 & {---} \\
      \bottomrule
    \end{tabular}
  \end{table}

\section{Examples for the diagnostic checks}
\label{app: diagnostic examples}

\subsection{Event study diagnostic}
\label{app: event-study example}

\vspace{-.4cm}
\begin{examplecontinued}{ex: four-firm-illustration}
Suppose the five firms in our example are assigned to two groups (e.g., by the mean wage of stayers) such that
\[
G_L=\{1,2,4\},\qquad G_H=\{3,5\}.
\]
The supported moves from $G_L$ to $G_H$ are $(2,3)$, $(2,5)$, $(4,3)$, and $(4,5)$, while the only supported move from $G_H$ to $G_L$ is $(3,4)$. Assume again, for simplicity, that $E[\Delta\mid O=o,D=o]=0$ for every $o$, and consider the event-study diagnostic \eqref{eq:event-study2} with $G_j=G_H$ and $G_k=G_L$. The reverse move $(4,3)$ is supported, so all terms in the decomposition are well-defined for this choice of groups.

Using the mobility shares in Table~\ref{tab:did-values}, direct calculation gives
\begin{align*}
E\!\left[
S((O,+1,D,+1,O),\DID)
\mid O\in G_H,D\in G_L
\right]
&=0.05 \delta,\\
E[\DID(O,D)\mid O\in G_L,D\in G_H]
-E[\DID(D,O)\mid O\in G_H,D\in G_L]
&\approx 0.08-0.18\delta.
\end{align*}
Thus, \eqref{eq:event-study2} equals approximately $0.08-0.13\delta$. The diagnostic is nonzero at $\delta=0$, when the necessary and sufficient conditions hold, and is zero at $\delta=5/8$, when those conditions fail.

At $\delta=0$, the nonzero diagnostic is entirely due to differences in the composition of firm pairs across the two directional mover populations. The low-to-high conditional mean averages over $(2,3)$, $(2,5)$, $(4,3)$, and $(4,5)$ and equals $57/440\approx0.13$. In contrast, the high-to-low population contains only $(3,4)$, so its mean is $-0.05$. Their sum is approximately $0.08$, even though $\DID(3,4)+\DID(4,3)=0$.

In this example, the compositional difference can be avoided by examining the two-move cycle directly:
\[
\DID(3,4)+\DID(4,3)=0.05\delta.
\]
This signed sum is zero if and only if $\delta=0$. Checking this restriction therefore characterizes exactly when the necessary and sufficient conditions hold in this example. In general, however, zero signed sums on two-move cycles need not imply zero signed sums on longer cycles.
\end{examplecontinued}

\subsection{Binscatter diagnostic}
\label{app: binscatter example}
\vspace{-.4cm}
\begin{examplecontinued}{ex: four-firm-illustration}
Table~\ref{tab:binscatter} groups the supported mover edges into the following intervals according to their adjusted AKM contrasts: $B_1=[-0.05,0.05]$, $B_2=(0.05,0.15]$, and $B_3=(0.15,0.25]$.  Within each bin, the table reports the averages of the adjusted AKM contrasts and the DIDs, weighted by the mobility shares $P_{od}$ in Table~\ref{tab:did-values}.

The first bin contains $(1,2)$, $(3,4)$, and $(4,3)$; the second contains $(2,3)$, $(3,5)$, and $(4,5)$; and the third contains only $(2,5)$. The individual adjusted AKM contrasts in Table~\ref{tab:did-values} are independent of $\delta$. Since the mobility shares are also fixed, neither the assignment of moves to bins nor the within-bin averages of the adjusted AKM contrasts changes with $\delta$. These averages are $0.01$, $7/64$, and $0.20$, respectively.

Within each of the first two bins, the deviations of the DIDs from the adjusted AKM contrasts cancel when weighted by the mobility shares:
\[
0.06(-0.10\delta)+0.04(0.15\delta)=0,
\qquad
0.05(0.24\delta)+0.03(-0.40\delta)=0.
\]
The deviations for the remaining moves $(1,2)$, $(2,3)$, and $(2,5)$ are zero. Consequently, the within-bin averages of the DIDs equal the fixed adjusted AKM contrast averages in all three bins for every value of $\delta$.  The plotted bin averages therefore remain exactly on the 45-degree line, even though the necessary and sufficient conditions in Theorem~\ref{thm: did characterization} fail whenever $\delta\neq0$. In particular, the three fundamental signed cycle sums reported in Table~\ref{tab:fundamental-cycles} are $0.05\delta$, $0.24\delta$, and $-0.50\delta$. Their magnitudes grow without bound as $|\delta|$ increases, while neither the adjusted AKM contrasts nor any plotted bin average changes. This construction varies the three cycle sums together through $\delta$; it does not require assigning their values independently.

\begin{table}[t!]
\centering
\begin{tabular}{lcc}
\toprule
bin &
$E[\widetilde\alpha_D-\widetilde\alpha_O
   \mid \text{bin}]$ &
$E[\DID(O,D)\mid \text{bin}]$ \\
\midrule
$B_1$  & $0.01$ & $0.01$ \\
$B_2$  & $0.11$ & $0.11$ \\
$B_3$  & $0.20$ & $0.20$ \\
\bottomrule
\end{tabular}
\caption{Binscatter of the running example.}
\figurenote{Supported mover edges are grouped by their adjusted AKM contrasts into the bins in the first column. Both remaining columns report conditional averages over these mover edges, with weights proportional to the mobility shares $P_{od}$ in Table~\ref{tab:did-values}. The exact within-bin averages in both columns are $0.01$, $7/64$, and $0.20$, respectively, for every value of $\delta$. Values in the table are rounded to two decimal places. The second (third) column shows the within-bin average of the adjusted AKM contrasts (DIDs).}
\label{tab:binscatter}
\end{table}

Moreover, a weighted least-squares regression of $\DID(o,d)$ on a constant and $\widetilde\alpha_d-\widetilde\alpha_o$, using weights $P_{od}$ over the supported mover edges, has intercept zero and slope one for every value of $\delta$. Thus, neither the plotted bin averages nor the slope coefficient detects the cycle violations in this example.
\end{examplecontinued}

}
\end{spacing}

\newpage
\bibliographystyle{ims}
\bibliography{./ref}

\end{document}